\pdfoutput=1
\documentclass[onecolumn,prd,floats,aps,amsmath,amssymb,nofootinbib, superscriptaddress,preprintnumbers,10pt]{revtex4-2}
\usepackage{comment}
\usepackage[utf8]{inputenc}
\usepackage{mathtools}
\usepackage{bbm}
\usepackage{cancel} %
\usepackage{dcolumn}
\newcolumntype{d}[1]{D{.}{.}{#1}}
\usepackage{bm}%
\usepackage{blindtext}
\usepackage{xcolor}
\usepackage{hyperref}%
\usepackage{graphicx}
\usepackage{amsthm}
\usepackage{bm}
\usepackage{verbatim}
\usepackage{amsmath}
\usepackage{siunitx}
\usepackage{physics}
\usepackage{amssymb}
\usepackage{slashed}
\usepackage{tikz}
\usepackage{cleveref}
\usepackage{booktabs}
\usepackage{dsfont}
\usepackage[normalem]{ulem}

\newcommand{\LinfamufullIoneFLAG}{598.1~(4.8)_{\rm stat} (2.9)_{\rm syst} (2.7)_L [6.2]}
\newcommand{\LinfamufullIoneWP}{593.4~(4.9)_{\rm stat} (3.9)_{\rm syst} (2.7)_L [6.8]}

\newcommand{\amufullIoneFLAG}{571.0~(4.8)_{\rm stat} (2.9)_{\rm syst} [5.6]}
\newcommand{\amufullIoneWP}{566.3~(4.9)_{\rm stat} (3.9)_{\rm syst} [6.2]}
\newcommand{\amuLDIoneFLAG}{342.8~(4.7)_{\rm stat} (2.6)_{\rm syst} [5.4]}
\newcommand{\amuLDIoneWP}{338.4~(4.7)_{\rm stat} (3.6)_{\rm syst} [5.9]}
\newcommand{\amuIWIoneFLAG}{185.43~(27)_{\rm stat} (19)_{\rm syst} [34]}
\newcommand{\amuIWIoneWP}{185.11~(28)_{\rm stat} (26)_{\rm syst} [38]}
\newcommand{\amuSDIoneFLAG}{43.11~(10)_{\rm stat} (12)_{\rm syst} [15]}
\newcommand{\amuSDIoneWP}{43.12~(10)_{\rm stat} (13)_{\rm syst} [16]}
\newcommand{\amufulloctFLAG}{107.3~(1.7)_{\rm stat} (1.3)_{\rm syst} [2.1]}
\newcommand{\amufulloctWP}{106.7~(1.6)_{\rm stat} (1.3)_{\rm syst} [2.1]}
\newcommand{\amuLDoctFLAG}{46.1~(1.7)_{\rm stat} (1.4)_{\rm syst} [2.2]}
\newcommand{\amuLDoctWP}{45.6~(1.6)_{\rm stat} (1.5)_{\rm syst} [2.2]}
\newcommand{\amuIWoctFLAG}{46.980~(72)_{\rm stat} (2)_{\rm syst} [72]}
\newcommand{\amuIWoctWP}{46.902~(69)_{\rm stat} (46)_{\rm syst} [83]}
\newcommand{\amuSDoctFLAG}{13.836~(23)_{\rm stat} (31)_{\rm syst} [38]}
\newcommand{\amuSDoctWP}{13.835~(23)_{\rm stat} (32)_{\rm syst} [39]}
\newcommand{\amufullIzeroFLAG}{121.6~(1.7)_{\rm stat} (1.3)_{\rm syst} [2.1]}
\newcommand{\amufullIzeroWP}{121.1~(1.6)_{\rm stat} (1.3)_{\rm syst} [2.1]}
\newcommand{\amuLDIzeroFLAG}{46.1~(1.7)_{\rm stat} (1.4)_{\rm syst} (0.2)_{c\rm-disc} [2.2]}
\newcommand{\amuLDIzeroWP}{45.6~(1.6)_{\rm stat} (1.5)_{\rm syst} (0.2)_{c\rm-disc} [2.2]}
\newcommand{\amuIWIzeroFLAG}{49.82~(8)_{\rm stat} (3)_{\rm syst} (6)_{c\rm-disc} [11]}
\newcommand{\amuIWIzeroWP}{49.76~(8)_{\rm stat} (6)_{\rm syst} (6)_{c\rm-disc} [12]}
\newcommand{\amuSDIzeroFLAG}{25.28~(9)_{\rm stat} (10)_{\rm syst} (1)_{c\rm-disc} [14]}
\newcommand{\amuSDIzeroWP}{25.32~(10)_{\rm stat} (12)_{\rm syst} (1)_{c\rm-disc} [16]}
\newcommand{\amufulludsdiscFLAG}{-13.0~(1.8)_{\rm stat} (1.3)_{\rm syst} (0.2)_{c\rm-disc} [2.2]}
\newcommand{\amufulludsdiscWP}{-13.0~(1.8)_{\rm stat} (1.4)_{\rm syst} (0.2)_{c\rm-disc} [2.3]}

\newcommand{\amuSDcFLAG}{11.44~(9)_{\rm stat} (10)_{\rm syst} [14]}
\newcommand{\amuSDcWP}  {11.48~(10)_{\rm stat} (12)_{\rm syst} [16]}
\newcommand{\amuWcFLAG}{2.839~(40)_{\rm stat} (34)_{\rm syst} [53]}
\newcommand{\amuWcWP}  {2.857~(44)_{\rm stat} (39)_{\rm syst} [59]}
\newcommand{\amuLDcFLAG}{0.01288~(40)_{\rm stat} (42)_{\rm syst} [58]}
\newcommand{\amuLDcWP}  {0.01302~(44)_{\rm stat} (46)_{\rm syst} [64]}
\newcommand{\amufullcFLAG}{14.30\,(12)_{\rm stat} (13)_{\rm syst} [18]}
\newcommand{\amufullcWP}{14.37~(12)_{\rm stat} (16)_{\rm syst} [20]}
\newcommand{\amuSDsFLAG}{9.040~(20)_{\rm stat} (27)_{\rm syst} [33]}
\newcommand{\amuSDsWP}{9.038~(20)_{\rm stat} (28)_{\rm syst} [34]}

\newcommand{\amufullsFLAG}{53.80~(35)_{\rm stat} (19)_{\rm syst} [40]}
\newcommand{\amufullsWP}{53.71~(35)_{\rm stat} (19)_{\rm syst} [40]}

\newcommand{\LinfamuLDIoneFLAGT}      { 368.4(4.7)(3.6) [5.9]}
\newcommand{\LinfamuLDIoneWPT}      { 364.0(4.7)(4.4) [6.4]}

\newcommand{\LinfamuIWIoneFLAGT}      { 186.58(27)(44) [52]}
\newcommand{\LinfamuIWIoneWPT}      { 186.26(28)(48) [55]}

\newcommand{\LinfamuSDIoneFLAGT}      { 43.17(10)(13) [16]}
\newcommand{\LinfamuSDIoneWPT}      { 43.18(10)(14) [17]}

\newcommand{\LinfamuLDoctFLAGT}       { 46.1(1.7)(1.4) [2.2]}
\newcommand{\LinfamuLDoctWPT}       { 45.6(1.6)(1.5) [2.2]}

\newcommand{\LinfamuIWoctFLAGT}       { 46.980(72)(2) [72]}
\newcommand{\LinfamuIWoctWPT}       { 46.902(69)(46) [83]}

\newcommand{\LinfamuSDoctFLAGT}       { 13.836(23)(31) [38]}
\newcommand{\LinfamuSDoctWPT}       { 13.835(23)(32) [39]}

\newcommand{\LinfamuLDIzeroFLAGT}     { 46.1(1.7)(1.4) [2.2]}
\newcommand{\LinfamuLDIzeroWPT}     { 45.6(1.6)(1.5) [2.2]}

\newcommand{\LinfamuIWIzeroFLAGT} {49.82(8)(7) [11]}
\newcommand{\LinfamuIWIzeroWPT} {49.76(8)(8) [12]}

\newcommand{\LinfamuSDIzeroFLAGT} {25.28(9)(11) [14]}
\newcommand{\LinfamuSDIzeroWPT} {25.32(10)(12) [16]}

\newcommand{\amuLDdiscFLAGT}{ -12.2(1.8)(1.5) [2.3] }
\newcommand{\amuLDdiscWPT}{ -12.1(1.7)(1.6) [2.3] }

\newcommand{\amuIWdiscFLAGT}{ -1.15(18)(17) [25] }
\newcommand{\amuIWdiscWPT}{ -1.16(19)(20) [28] }

\newcommand{\amuSDdiscT}{ 0.000 (12) }
\newcommand{\amuSDcFLAGT}{11.44(9)(10) [14]}
\newcommand{\amuSDcWPT}  {11.48(10)(12) [16]}

\newcommand{\amuWcFLAGT}{2.839(40)(34) [53]}
\newcommand{\amuWcWPT}  {2.857(44)(39) [59]}

\newcommand{\amuLDcFLAGT}{0.01288(40)(42) [58]}
\newcommand{\amuLDcWPT}  {0.01302(44)(46) [64]}

\newcommand{\amuSDsFLAGT}{9.040(20)(27) [33]}
\newcommand{\amuSDsWPT}{9.038(20)(28) [34]}

\newcommand{\amuWsFLAGT}{27.40(16)(14) [21]}
\newcommand{\amuWsWPT}{27.37(17)(17) [24]}

\newcommand{\amuLDsFLAGT}{17.35(27)(11) [29]}
\newcommand{\amuLDsWPT}{17.25(27)(11) [29]}

\skip\footins=24pt plus 4pt minus 8pt

\usetikzlibrary{positioning, shapes, arrows.meta, decorations.pathmorphing}

\newcommand{\Bern}{Institute for Theoretical Physics, Albert Einstein Center for Fundamental Physics,\\University of Bern, Sidlerstrasse 5, CH-3012 Bern, Switzerland}
\newcommand{\hiskp}{HISKP (Theory), Rheinische Friedrich-Wilhelms-Universit\"at Bonn,\\Nussallee 14-16, 53115 Bonn, Germany}
\newcommand{\hpca}{High Performance Computing and Analytics Lab, Rheinische Friedrich-Wilhelms-Universit\"at Bonn,\\ Friedrich-Hirzebruch-Allee 8, 53115 Bonn, Germany}
\newcommand{\CyprusU}{Department of Physics, University of Cyprus, 20537 Nicosia, Cyprus}
\newcommand{\CyprusI}{Computation-based Science and Technology Research Center, The Cyprus Institute,\\20 Konstantinou Kavafi Street, 2121 Nicosia, Cyprus}

\newcommand{\Romadue}{Dipartimento di Fisica and INFN, Universit\`a di Roma ``Tor Vergata",\\Via della Ricerca Scientifica 1, I-00133 Roma, Italy}
\newcommand{\Romatre}{Dipartimento di Matematica e Fisica, Universit\`a Roma Tre and INFN, Sezione di Roma Tre,\\Via della Vasca Navale 84, I-00146 Rome, Italy}
\newcommand{\RomatreINFN}{Istituto Nazionale di Fisica Nucleare, Sezione di Roma Tre,\\Via della Vasca Navale 84, I-00146 Rome, Italy}

\newcommand{\BUW}{Department of Physics, Bergische Universität Wuppertal, Gaußstraße 20, Wuppertal, 42119, Germany}
\newcommand{\HIM}{Helmholtz-Institut Mainz, Johannes Gutenberg-Universit{\"a}t Mainz}
\newcommand{\GSI}{GSI Helmholtz Centre for Heavy Ion Research, 64291 Darmstadt, Germany}

\newcommand{\be}{\begin{equation}}
\newcommand{\ee}{\end{equation}}
\newcommand{\bea}{\begin{eqnarray}}
\newcommand{\eea}{\end{eqnarray}}

\begin{document}

\setlength\abovedisplayskip{14pt}
\setlength\belowdisplayskip{14pt}

\setlength{\parskip}{14pt}
\setlength{\parindent}{0pt}

\title{The hadronic-vacuum-polarization contribution to the muon anomalous magnetic moment in isospin-symmetric lattice QCD with twisted-mass fermions}

\author{C.~Alexandrou}\affiliation{\CyprusU}\affiliation{\CyprusI}
\author{S.~Bacchio}\affiliation{\CyprusI}
\author{A.~De~Santis}\affiliation{\HIM}\affiliation{\GSI}
\author{A.~Evangelista}\affiliation{\CyprusU}
\author{J.~Finkenrath}\affiliation{\BUW}
\author{R.~Frezzotti}\affiliation{\Romadue} 
\author{G.~Gagliardi}\affiliation{\RomatreINFN}
\author{M.~Garofalo}\affiliation{\hiskp}
\author{N.~Kalntis}\affiliation{\Bern}
\author{B.~Kostrzewa}\affiliation{\hpca}
\author{V.~Lubicz}\affiliation{\Romatre}
\author{L.~Maio}\affiliation{\Romadue}
\author{F.~Margari}\affiliation{\Romadue}
\author{F.~Pittler}\affiliation{\CyprusI}
\author{S.~Romiti}\affiliation{\Bern}
\author{F.~Sanfilippo}\affiliation{\RomatreINFN}
\author{C.~Schneider}\affiliation{\CyprusU}
\author{S.~Simula}\affiliation{\RomatreINFN}
\author{N.~Tantalo}\affiliation{\Romadue}
\author{C.~Urbach}\affiliation{\hiskp}
\author{U.~Wenger}\affiliation{\Bern}

\begin{abstract}
\vspace{0.05cm}
\centerline{\includegraphics[height=3.7cm]{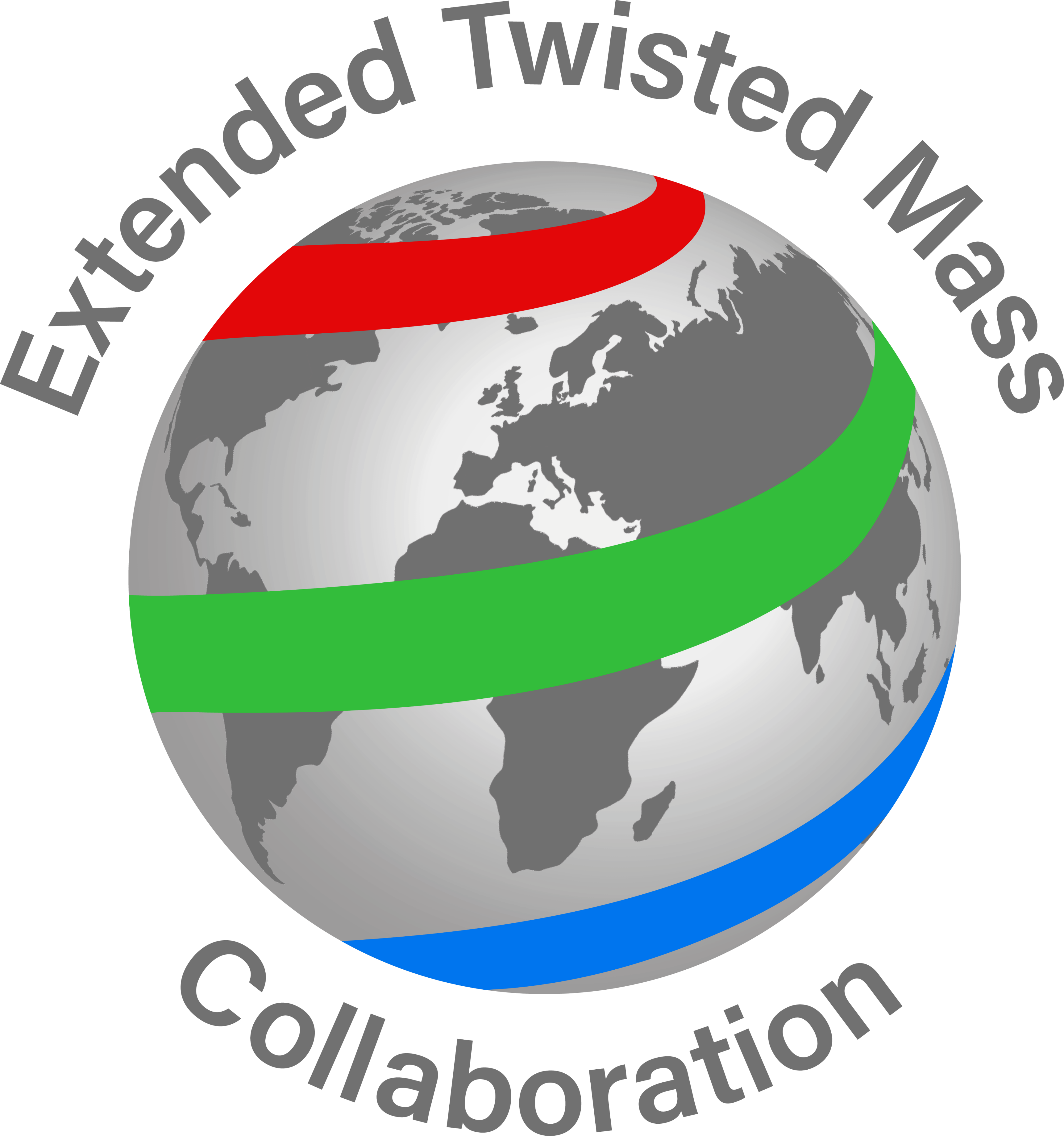}}
\vspace{0.4cm}
We present a lattice-QCD calculation of the leading-order hadronic-vacuum-polarization (HVP) contribution to the anomalous magnetic moment of the muon in isospin-symmetric QCD. We employ the gauge configurations generated by the Extended Twisted Mass Collaboration (ETMC) with $N_f = 2 + 1 + 1$ flavors of Wilson-clover twisted-mass quarks at four lattice spacings and at values of the quark-mass parameters matching the definition of isospin-symmetric QCD given in the FLAG review 2024 and in the 2025 edition of the Muon $g-2$ Theory Initiative White Paper (WP25). After retuning the simulation parameters to precisely match these two definitions of isospin-symmetric QCD, and after carrying out an extrapolation to the continuum limit based on the data at lattice spacings $a \simeq 0.049, 0.057, 0.068$ and  $0.079$~fm, and spatial lattice sizes up to $L \simeq 7.7$~fm, we obtain 
$a_\mu^{\rm HVP, FLAG}(\rm{iso}) = (720.0 \pm 6.9) \times 10^{-10}$ and $ 
a_\mu^{\rm HVP, WP25}(\rm{iso}) = (714.8 \pm 7.4) \times 10^{-10}$.
Our results, which have a relative uncertainty of about $1 \%$, agree %
well with the corresponding determinations by other lattice groups. After adding to our determination of $a_{\mu}^{\rm HVP, WP25}(\rm{iso})$ the leading isospin-breaking corrections provided in the WP25, namely $a_{\mu}^{\rm HVP, WP25}({\rm IB}) = 0.1(3.4)\times 10^{-10}$, we obtain for the full HVP the value $a_{\mu}^{\rm HVP}= (714.9  \pm 8.1)\times 10^{-10}$. This result translates into a Standard Model prediction for the muon anomalous magnetic moment that is fully consistent with the latest result by the Fermilab Muon $g-2$ experiment and strengthens the tension between lattice QCD and data-driven (pre-CMD3) dispersive determinations.

\end{abstract}

\maketitle

\section{Introduction}
\label{sec:introduction}

The anomalous magnetic moment of the muon, $a_\mu = (g_\mu-2)/2$, constitutes a cornerstone observable for testing the Standard Model (SM) owing to the extraordinary precision of its experimental determination. The Fermilab Muon $g-2$ experiment (E989) has completed its full data-taking program and recently reported a new result based on a dataset with 2.5 times more statistics than previously available~\cite{Muong-2:2025xyk}. This final measurement is consistent with the results obtained in earlier runs at Fermilab~\cite{Muong-2:2021ojo,Muong-2:2023cdq} as well as with the Brookhaven National Laboratory (BNL) E821 measurement~\cite{Muong-2:2006rrc}, thereby establishing a stable world average for $a_\mu^\mathrm{exp}$ and achieving a combined uncertainty of 124~ppb.
A completely independent measurement, which will likely lead to a further reduction of the total experimental error, is planned at J-PARC~\cite{Abe:2019thb}.

Matching the experimental precision on the theoretical side is a formidable challenge. While $a_{\mu}$ is dominated by electromagnetic contributions, the leading source of uncertainty in the SM prediction arises from hadronic effects, most notably from the leading-order hadronic-vacuum-polarization (HVP) contribution, $a_\mu^{\rm HVP}$.  In the first White Paper published by the Muon $g_\mu-2$ Theory Initiative~\cite{Aoyama:2020ynm}, the reference theoretical value for $a_\mu^{\rm HVP}$ was obtained from dispersive data-driven analyses based on the experimental determination of the ratio of hadron production to muon pair production in $e^+ e^-$ collisions ($R$-ratio). This approach yielded a data-driven determination of 
 $a_{\mu}^{\rm HVP}$ which, when combined with the remaining SM contributions, led to a prediction for $a_{\mu}$ smaller than the experimental value by about four standard deviations, giving rise to the so-called $g_\mu-2$ Puzzle and to a possible indirect hint of the existence of New Physics (NP).  However, recent high-precision measurements of the $\pi^+\pi^-(\gamma)$-channel contribution to the $R$-ratio by the CMD-3 Collaboration~\cite{CMD-3:2023alj} have revealed strong tensions with earlier determinations (e.g., by CMD-2~\cite{CMD-2:2003gqi,Aulchenko:2006dxz,CMD-2:2005mvb, CMD-2:2006gxt}, KLOE~\cite{KLOE-2:2017fda,KLOE:2008fmq,KLOE:2010qei,KLOE:2012anl}, BaBar~\cite{BaBar:2009wpw,BaBar:2012bdw}\footnote{See also~\cite{Polat:2026ysh} for a recent update by the BaBar Collaboration.}), to a level that prevents building an average 
 of experimental data to obtain a precise evaluation of $a_\mu^{\rm HVP}$
 through the well-known dispersive approach.
Notably, indications of possible issues with the dispersive determination had
already emerged before the CMD-3 measurement. The first lattice QCD+QED
calculation of $a_{\mu}^{\rm HVP}$ with subpercent precision, performed by
the BMW Collaboration~\cite{Borsanyi:2020mff}, found a value larger than the
dispersive determinations which, once combined with the remaining SM
contributions \cite{Aoyama:2012wk,Volkov:2019phy,Volkov:2024yzc,Aoyama:2024aly,Parker:2018vye,Morel:2020dww,Fan:2022eto,Czarnecki:2002nt,Gnendiger:2013pva,Ludtke:2024ase,Hoferichter:2025yih,RBC:2018dos,Giusti:2019xct,Borsanyi:2020mff,Lehner:2020crt,Wang:2022lkq,Aubin:2022hgm,Ce:2022kxy,ExtendedTwistedMass:2022jpw,RBC:2023pvn,Kuberski:2024bcj,Boccaletti:2024guq,Spiegel:2024dec,RBC:2024fic,Djukanovic:2024cmq,ExtendedTwistedMass:2024nyi,MILC:2024ryz,FermilabLatticeHPQCD:2024ppc,Keshavarzi:2019abf,DiLuzio:2024sps,Kurz:2014wya,Colangelo:2015ama,Masjuan:2017tvw,Colangelo:2017fiz,Hoferichter:2018kwz,Eichmann:2019tjk,Bijnens:2019ghy,Leutgeb:2019gbz,Cappiello:2019hwh,Masjuan:2020jsf,Bijnens:2020xnl,Bijnens:2021jqo,Danilkin:2021icn,Stamen:2022uqh,Leutgeb:2022lqw,Hoferichter:2023tgp,Hoferichter:2024fsj,Estrada:2024cfy,Deineka:2024mzt,Eichmann:2024glq,Bijnens:2024jgh,Hoferichter:2024bae,Holz:2024diw,Cappiello:2025fyf,Colangelo:2014qya,Blum:2019ugy,Chao:2021tvp,Chao:2022xzg,Blum:2023vlm,Fodor:2024jyn}, leads to a prediction for $a_{\mu}$ compatible with the experimental
value within about two standard deviations. Moreover, further indications came from the first lattice computation of the energy-smeared $R$-ratio~\cite{ExtendedTwistedMassCollaborationETMC:2022sta}
in isospin-symmetric QCD and from lattice studies of the so-called intermediate Euclidean-time window contribution to $a_{\mu}^{\rm HVP}$~\cite{RBC:2018dos}.
For this latter observable, which is technically simpler to compute than the full $a_{\mu}^{\rm HVP}$, several lattice calculations~\cite{Borsanyi:2020mff,Ce:2022kxy,ExtendedTwistedMass:2022jpw,RBC:2023pvn,FermilabLatticeHPQCD:2023jof,MILC:2024ryz} systematically
obtained accurate results with significantly larger values than those from data-driven evaluations, hinting either at unresolved inconsistencies in the experimental input underlying the dispersive analyses or at possible NP contributions affecting the $e^{+}e^{-}$ data employed in the comparison.

The latest White Paper of the Muon $g-2$ Theory Initiative
\cite{Aliberti:2025beg} (hereafter WP25) explicitly acknowledges these issues
and adopts as the preferred baseline for the SM
prediction of $a_\mu^{\rm HVP}$ the lattice-QCD approach, which in the
meantime---owing to impressive theoretical and numerical advances---has made
available a few independent first-principles determinations of  $a_{\mu}^{\rm HVP}$ and of its numerically dominant
contribution, with percent-level precision or better~\cite{RBC:2024fic,Djukanovic:2024cmq,FermilabLatticeHPQCD:2024ppc}. The increase in the central value induced by adopting the lattice results for $a_{\mu}^{\rm HVP}$ has drastically reduced the long-standing tension between theory and experiment. Nonetheless, the theoretical uncertainty—still significantly larger than the experimental one—remains dominated by the HVP contribution, the determination of which continues to require further refinements and independent lattice calculations. 

We also mention the recent hybrid determination of
$a_{\mu}^{\rm HVP}$ with a precision of $0.48\%$
\cite{Boccaletti:2024guq}, which combines an updated version of the BMW
lattice QCD+QED analysis with a small long-distance contribution obtained
from low-energy $e^+e^-$ data. The resulting central value is consistent with
the lattice average reported in WP25~\cite{Aliberti:2025beg}.

In this work, we present a high-precision lattice-QCD determination of the leading-order HVP contribution to the muon anomalous magnetic moment in isospin-symmetric QCD -- which hereafter as far as lattice results
are concerned will be for short referred to as HVP (with the qualification ``leading order'' understood). The calculation is based on gauge ensembles with $N_f = 2+1+1$ flavors of Wilson–clover twisted-mass quarks generated by the Extended Twisted Mass Collaboration (ETMC). We provide first-principles results for both the isotriplet ($I=1$) and isoscalar ($I=0$) components of $a_\mu^{\rm HVP}$ (see the next section for definitions). To mitigate potential confirmation bias in the analysis, a blinding procedure was implemented by applying a suitable time-dependent shift to the light-quark vector correlation function, which is the most relevant nonperturbative input required for this calculation (see Appendix~\ref{sec:blinding}). The results obtained in this work are primarily based on simulations at the physical pion mass, corresponding to four different lattice spacings down to $a \simeq 0.049~{\rm fm}$ and spatial volumes up to $L \simeq 7.7~{\rm fm}$.

When comparing results obtained in isospin-symmetric QCD, i.e., without including isospin-breaking corrections, it is important to match the definition of the pure QCD world considered. To enable a clean comparison between our results and existing or future lattice determinations, a central aspect of this work is the implementation of two different schemes for isospin-symmetric QCD. The first one, corresponding to the prescription suggested in the latest FLAG review~\cite{FlavourLatticeAveragingGroupFLAG:2024oxs} (hereafter referred to as the FLAG scheme), consists in the following choice of input parameters:
\begin{align}
\label{eq:FLAG_scheme}
f_{\pi}^{\rm FLAG} = 130.5~{\rm MeV},\quad m_{\pi}^{\rm FLAG} = 135.0~{\rm MeV},\quad m_{K}^{\rm FLAG} = 494.6~{\rm MeV},\quad m_{D_{s}}^{\rm FLAG} = 1967~{\rm MeV}.
\end{align}
The second scheme, corresponding to that adopted in the WP25~\cite{Aliberti:2025beg} (hereafter referred to as the WP25 scheme), is defined by
\begin{align}
\label{eq:WP25_scheme}
w_{0}^{\rm WP25} = 0.17236~{\rm fm},\quad m_{\pi}^{\rm WP25} = 135.0~{\rm MeV},\quad m_{K}^{\rm WP25} = 494.6~{\rm MeV},\quad m_{D_{s}}^{\rm WP25} = 1967~{\rm MeV},
\end{align}
where $w_{0}$ denotes the gradient-flow scale introduced in Ref.~\cite{BMW:2012hcm}. The two schemes differ only in the choice of a single input quantity: $f_{\pi}$ in the FLAG scheme versus $w_{0}$ in the WP25 scheme.

For the reader's convenience, we summarize below our main results for the HVP
in isospin-symmetric QCD:
\begin{align}
a_\mu^{\rm HVP,FLAG}({\rm iso})
&= 720.0(5.5)(4.2)[6.9]\times 10^{-10},
\qquad
a_\mu^{\rm HVP,WP25}({\rm iso})
= 714.8(5.5)(4.9)[7.4]\times 10^{-10},
\label{eq:isospin_symmetric_HVP}
\end{align}
corresponding, respectively, to the FLAG and WP25 schemes. The first uncertainty is statistical, the second is systematic, and the value in brackets is the final uncertainty, obtained by adding the two in quadrature. We further report
the individual isotriplet and isoscalar contributions:
\begin{align}
a_\mu^{\rm HVP,FLAG}(I=1)
&= 598.1(4.8)(4.0)[6.2]\times 10^{-10},
\qquad
a_\mu^{\rm HVP,FLAG}(I=0)
= 121.6(1.7)(1.3)[2.1]\times 10^{-10},
\nonumber \\[8pt]
a_\mu^{\rm HVP,WP25}(I=1)
&= 593.4(4.9)(4.8)[6.8]\times 10^{-10},
\qquad
a_\mu^{\rm HVP,WP25}(I=0)
= 121.1(1.6)(1.3)[2.1]\times 10^{-10}.
\end{align}
The total isospin-symmetric HVP is obtained as
$a_\mu^{\rm HVP}({\rm iso})
=a_\mu^{\rm HVP}(I=1)+a_\mu^{\rm HVP}(I=0)+a_\mu^{\rm HVP}(b)$,
where $a_\mu^{\rm HVP}(I=1)$ and $a_\mu^{\rm HVP}(I=0)$ denote the isotriplet and isoscalar contributions
in the $N_f=2+1+1$ theory, while the numerically small bottom-quark contribution, $a_{\mu}^{\rm HVP}(b)=0.32\times 10^{-10}$~\cite{ExtendedTwistedMass:2022jpw},
is included only in the total.\footnote{The small bottom-quark contribution is estimated perturbatively using the RHAD software~\cite{Harlander:2002ur}.} The isotriplet contribution is directly proportional to the light-quark connected
contribution, with a proportionality
factor of $9/10$. The precision of our isospin-symmetric result is $0.96\%$ in the FLAG scheme, and $1.03\%$ in the WP25 scheme.

\begin{figure}[t!]
\centering
\includegraphics[width=0.50\linewidth]{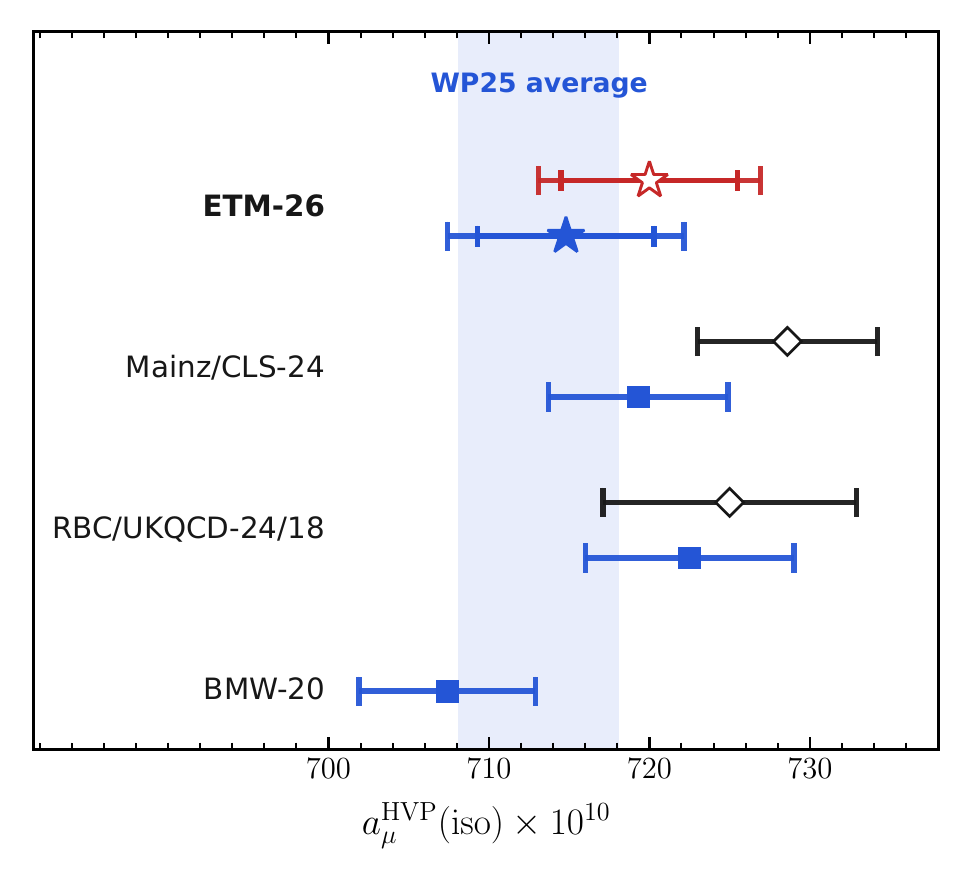}\hspace{-0.3cm}
\includegraphics[width=0.50\linewidth]{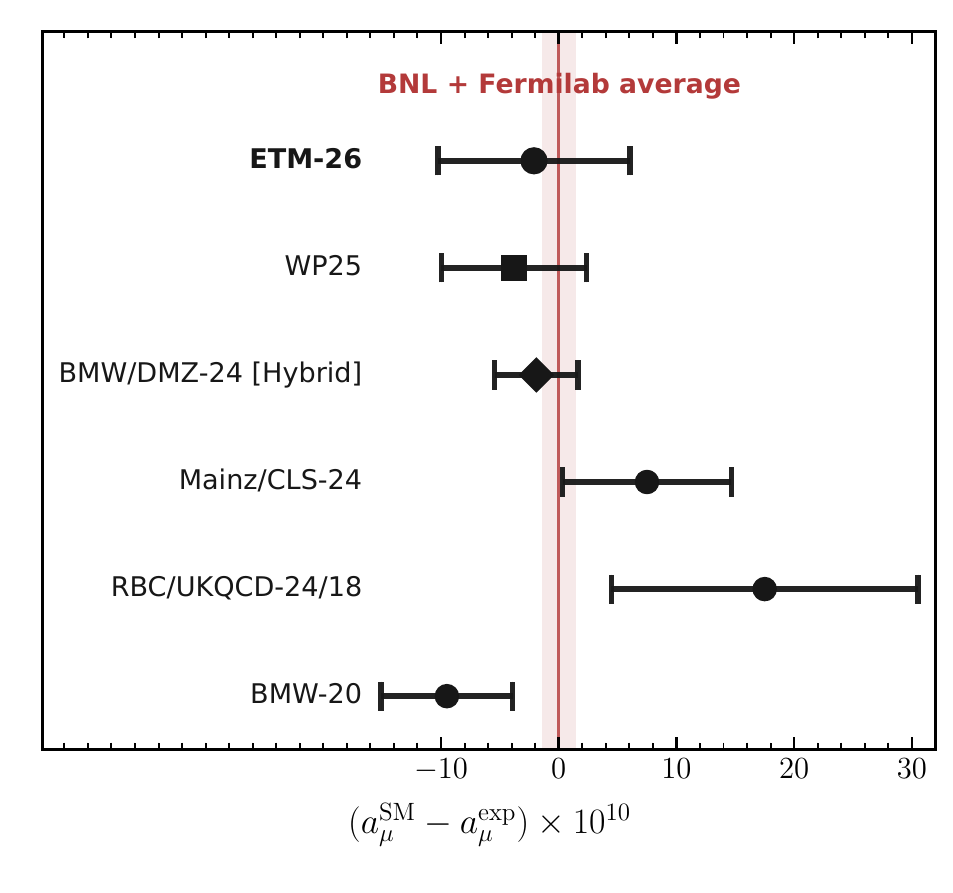}
\caption{\textit{Left}: determinations of $a_\mu^{\rm HVP}({\rm iso})$ in different schemes of isospin-symmetric QCD. Our results are highlighted by an open star in the FLAG scheme and a filled star in the WP25 scheme. Other determinations quoted in the WP25 scheme are shown as filled squares, while results obtained in other schemes are represented by open diamonds; see the corresponding references for their precise definitions. Red, blue, and black provide an additional indication of the FLAG, WP25, and other schemes, respectively. The blue band represents the WP25 average. \textit{Right}: differences between the SM predictions, $a_\mu^{\rm SM}$, and the experimental value, $a_\mu^{\rm exp}$, where $a_\mu^{\rm exp}$ corresponds to the BNL+Fermilab average. Individual lattice-based predictions are shown by filled circles, while the WP25 average is represented by a filled square. The BMW/DMZ-24 result, shown by a filled diamond, is based on a hybrid approach combining lattice QCD with $e^+e^-\to{\rm hadrons}$ data. The SM predictions are obtained by combining the corresponding HVP determinations, including isospin-breaking effects in QCD+QED, with the remaining SM contributions taken from WP25~\cite{Aliberti:2025beg} -- see also Refs.~\cite{Aoyama:2012wk,Volkov:2019phy,Volkov:2024yzc,Aoyama:2024aly,Parker:2018vye,Morel:2020dww,Fan:2022eto,Czarnecki:2002nt,Gnendiger:2013pva,Ludtke:2024ase,Hoferichter:2025yih,RBC:2018dos,Giusti:2019xct,Borsanyi:2020mff,Lehner:2020crt,Wang:2022lkq,Aubin:2022hgm,Ce:2022kxy,ExtendedTwistedMass:2022jpw,RBC:2023pvn,Kuberski:2024bcj,Boccaletti:2024guq,Spiegel:2024dec,RBC:2024fic,Djukanovic:2024cmq,ExtendedTwistedMass:2024nyi,MILC:2024ryz,FermilabLatticeHPQCD:2024ppc,Keshavarzi:2019abf,DiLuzio:2024sps,Kurz:2014wya,Colangelo:2015ama,Masjuan:2017tvw,Colangelo:2017fiz,Hoferichter:2018kwz,Eichmann:2019tjk,Bijnens:2019ghy,Leutgeb:2019gbz,Cappiello:2019hwh,Masjuan:2020jsf,Bijnens:2020xnl,Bijnens:2021jqo,Danilkin:2021icn,Stamen:2022uqh,Leutgeb:2022lqw,Hoferichter:2023tgp,Hoferichter:2024fsj,Estrada:2024cfy,Deineka:2024mzt,Eichmann:2024glq,Bijnens:2024jgh,Hoferichter:2024bae,Holz:2024diw,Cappiello:2025fyf,Colangelo:2014qya,Blum:2019ugy,Chao:2021tvp,Chao:2022xzg,Blum:2023vlm,Fodor:2024jyn}. The red band centered at zero represents the experimental uncertainty.
References to the quoted papers are as follows:
BMW-20~\cite{Borsanyi:2020mff},
BMW/DMZ-24~\cite{Boccaletti:2024guq},
Mainz/CLS-24~\cite{Djukanovic:2024cmq},
RBC/UKQCD-18~\cite{RBC:2018dos},
RBC/UKQCD-24~\cite{RBC:2024fic}, 
WP25~\cite{Aliberti:2025beg}.
}
\label{fig:final_plots}
\end{figure}

At the current level of precision, isospin-breaking (IB) corrections represent an important component of the theoretical prediction for $a_\mu^{\rm HVP}$. A first-principles ETMC determination of these corrections is currently in progress and will be presented in a future publication. For the present analysis, the WP25 report~\cite{Aliberti:2025beg} provides separately the IB correction evaluated within the WP25 scheme. We can therefore combine our result obtained in isospin-symmetric QCD in the same scheme with the value
\begin{align}
a_{\mu}^{\rm HVP, WP25}({\rm IB}) = 0.1(3.4)\times 10^{-10}~,
\end{align}
reported in Ref.~\cite{Aliberti:2025beg}.
We then obtain for the leading-order hadronic-vacuum-polarization contribution to the muon anomalous magnetic moment
\begin{align}
a_\mu^{\rm HVP} = a_{\mu}^{\rm HVP}({\rm iso}) + a_{\mu}^{\rm HVP}({\rm IB}) = 714.9~(5.5)(6.0)[8.1]\times 10^{-10}.
\end{align}
This value agrees well with the WP25 average,
while showing a deviation from the dispersive determination reported in the first edition of the White Paper~\cite{Aoyama:2020ynm}, $a_{\mu}^{\rm HVP-WP20}=693.1(4.0)\times 10^{-10}$, at the level of about $2.4\,\sigma$. When combined with all other SM contributions, our result leads to a theoretical prediction for the muon anomalous magnetic moment that is fully consistent with the BNL+Fermilab experimental average.

Our main results are summarized in Fig.~\ref{fig:final_plots}. The left panel compares our FLAG- and WP25-scheme determinations of $a_{\mu}^{\rm HVP}({\rm iso})$ with recent lattice results and with the WP25 average~\cite{Aliberti:2025beg}. The comparison shows an overall good agreement among current lattice determinations, while also illustrating the residual dependence on the prescription used to define isospin-symmetric QCD. The right panel shows the corresponding predictions for $a_{\mu}^{\rm SM}-a_{\mu}^{\rm exp}$ after including isospin-breaking effects and combining with the remaining SM contributions taken from WP25~\cite{Aliberti:2025beg} -- see also Refs.~\cite{Aoyama:2012wk,Volkov:2019phy,Volkov:2024yzc,Aoyama:2024aly,Parker:2018vye,Morel:2020dww,Fan:2022eto,Czarnecki:2002nt,Gnendiger:2013pva,Ludtke:2024ase,Hoferichter:2025yih,RBC:2018dos,Giusti:2019xct,Borsanyi:2020mff,Lehner:2020crt,Wang:2022lkq,Aubin:2022hgm,Ce:2022kxy,ExtendedTwistedMass:2022jpw,RBC:2023pvn,Kuberski:2024bcj,Boccaletti:2024guq,Spiegel:2024dec,RBC:2024fic,Djukanovic:2024cmq,ExtendedTwistedMass:2024nyi,MILC:2024ryz,FermilabLatticeHPQCD:2024ppc,Keshavarzi:2019abf,DiLuzio:2024sps,Kurz:2014wya,Colangelo:2015ama,Masjuan:2017tvw,Colangelo:2017fiz,Hoferichter:2018kwz,Eichmann:2019tjk,Bijnens:2019ghy,Leutgeb:2019gbz,Cappiello:2019hwh,Masjuan:2020jsf,Bijnens:2020xnl,Bijnens:2021jqo,Danilkin:2021icn,Stamen:2022uqh,Leutgeb:2022lqw,Hoferichter:2023tgp,Hoferichter:2024fsj,Estrada:2024cfy,Deineka:2024mzt,Eichmann:2024glq,Bijnens:2024jgh,Hoferichter:2024bae,Holz:2024diw,Cappiello:2025fyf,Colangelo:2014qya,Blum:2019ugy,Chao:2021tvp,Chao:2022xzg,Blum:2023vlm,Fodor:2024jyn}. 
The BMW/DMZ-24 hybrid determination~\cite{Boccaletti:2024guq}, which combines lattice QCD with $e^+e^-\to{\rm hadrons}$ data, is also shown. The resulting SM predictions are compatible with the BNL+Fermilab experimental average within the quoted uncertainties.

The paper is organized as follows: In Sec.~\ref{sec:definitions}, we briefly summarize the time--momentum representation framework underlying our calculation and describe the isospin decomposition of the HVP. Sec.~\ref{sec:setup} outlines the lattice setup and numerical techniques employed in this work. Sec.~\ref{sec:full_HVP} presents a detailed description of our results for the $I=1$ and $I=0$ components of the HVP, including a dedicated subsection on finite-volume effects, and a comparison with existing lattice determinations. In Sec.~\ref{sec:windows}, we report our results for the Euclidean-time windows of the HVP introduced by the RBC/UKQCD Collaboration~\cite{RBC:2018dos}, and compare them with the results of other lattice groups. Finally, in Sec.~\ref{sec:conclusions} we draw our conclusions and discuss future directions. Technical details concerning the blinding procedure, the renormalization of isospin-symmetric QCD, the evaluation of the gradient-flow scale $w_0$ in the FLAG scheme and of the pion decay constant $f_\pi$ in the WP25 scheme, as well as the computation of the derivatives of the HVP with respect to simulation parameters are collected in the Appendices.

\section{Time-momentum representation and isospin decomposition}
\label{sec:definitions}

Following our previous works\,\cite{Giusti:2017jof,Giusti:2018mdh,Giusti:2019xct, ExtendedTwistedMass:2022jpw,ExtendedTwistedMass:2024nyi}, we adopt the time momentum representation\,\cite{Bernecker:2011gh} and, in continuum notation, evaluate the HVP contribution to the muon anomalous magnetic moment $a_{\mu}^{\rm HVP}$ as
\be
    \label{eq:amu_HVP}
    a_{\mu}^{\rm{HVP}} = 2 \alpha_\mathrm{em}^2 \int_0^\infty ~ dt \, t^2 \, K(m_\mu t) \,V(t) ~ , ~  
\ee
where $t$ is the Euclidean time and the kernel function $K(m_{\mu} t)$ is defined as\footnote{The leptonic kernel $K(z)$ is proportional to $z^2$ at small values of $z$ and it approaches $1$ as $z \to \infty$.}
\be
    \label{eq:kernel}
    K(z) = 2 \int_0^1 dy ( 1- y) \left[ 1 - j_0^2 \left(\frac{z}{2}\frac{y}{\sqrt{1 - y}} \right) \right]~,\qquad j_{0}(y) = \frac{\sin{(y)}}{y} ~ . ~
\ee
The Euclidean vector correlator $V(t)$ is defined as
 \be
     \label{eq:VV}
     V(t) \equiv \frac{1}{3} \sum_{i=1,2,3} \int d^3{x} ~ \langle J_i(\vec{x}, t) J_i^\dagger(0) \rangle 
 \ee
with $J_i(x)$ being the e.m.~current operator
 \be
      \label{eq:Jmu}
     J_i(x) \equiv \sum_{f = u, d, s, c, ...} q_{{\rm em},f} j_i^f(x)\;,
     \qquad
     j_i^f(x)= \overline{\psi}_f(x) \gamma_i \psi_f(x)\;,
     \qquad
     J_i(x) =- J_i^\dagger(x)\;,
     \qquad
     i=1,2,3\;,
 \ee
 and $q_{\mathrm{em},f}$ the electric charge for the quark flavor $f$ (in units of the positron charge). 

The fermionic Wick contractions appearing on the right-hand side of
Eq.~\eqref{eq:VV} give rise to two distinct topologies of Feynman diagrams,
namely the quark-connected and quark-disconnected contributions~\footnote{
We shall often use the shorthands ``light-connected'', ``strange-connected'' and ``charm-connected'' instead of ``light-quark connected'', ``strange-quark connected'' and ``charm-quark connected'' -- and similarly in cases where ``disconnected'' rather than ``connected'' occurs.}. Connected contributions are flavor diagonal, while the disconnected ones contain both
diagonal and off-diagonal flavor components. 
In our previous
works~\cite{ExtendedTwistedMass:2022jpw,ExtendedTwistedMass:2024nyi}, we
adopted the following flavor-based decomposition of $a_\mu^{\rm HVP}$:
\begin{align}
\label{eq:amu_HVP_f}
a_{\mu}^{\rm HVP}
&= a_{\mu}^{\rm HVP}(\ell)
 + a_{\mu}^{\rm HVP}(s)
 + a_{\mu}^{\rm HVP}(c)
 + a_{\mu}^{\rm HVP}({\rm disc})
 + a_{\mu}^{\rm HVP}(b)
 + a_{\mu}^{\rm HVP}({\rm IB})
\nonumber \\[8pt]
&= a_{\mu}^{\rm HVP}({\rm iso})
 + a_{\mu}^{\rm HVP}({\rm IB})~,
\end{align}
where the first three terms in the first line correspond to the
quark-connected contributions from mass-degenerate up and down ($\ell$),
strange ($s$), and charm ($c$) quarks, respectively, while
$a_\mu^{\rm HVP}({\rm disc})$ includes all quark-disconnected contributions,
both flavor diagonal and off-diagonal. The small bottom-quark contribution, $a_{\mu}^{\rm HVP}(b)$ is
estimated separately in perturbation theory and included in
$a_{\mu}^{\rm HVP}({\rm iso})$ as already discussed in the Introduction. The term
$a_{\mu}^{\rm HVP}({\rm IB})$ corresponds to the
$O(\alpha_{\rm em}^{3})$ isospin-breaking corrections, which we do not
explicitly evaluate in the present work and which will be addressed in a
forthcoming publication. While the individual contributions
$a_{\mu}^{\rm HVP}({\rm iso})$ and $a_{\mu}^{\rm HVP}({\rm IB})$ depend on
the specific scheme used to define isospin-symmetric QCD, their
sum---corresponding to the full QCD+QED value of
$a_{\mu}^{\rm HVP}$---does not.

In this work, we reorganize the $N_f=2+1+1$ isospin-symmetric HVP in terms of
definite isospin components ($I=1$ and $I=0$) as
\begin{align}
\label{eq:amu_HVP_I}
a_{\mu}^{\rm HVP}(I=1)
&= \tfrac{9}{10}a_{\mu}^{\rm HVP}(\ell)~,
\qquad\quad
a_{\mu}^{\rm HVP}(I=0)
= \tfrac{1}{10}a_{\mu}^{\rm HVP}(\ell)
 + a_{\mu}^{\rm HVP}({\rm disc})
 + a_{\mu}^{\rm HVP}(s)
 + a_{\mu}^{\rm HVP}(c)~.
\end{align}
Thus,
$a_{\mu}^{\rm HVP}({\rm iso})
=a_{\mu}^{\rm HVP}(I=1)+a_{\mu}^{\rm HVP}(I=0)
+a_{\mu}^{\rm HVP}(b)$.
As pointed out in Refs.~\cite{Ce:2022kxy,Djukanovic:2024cmq}, separating the $I=1$ and
$I=0$ components is advantageous. In the $I=0$ channel, the sizable
finite-size effects and pion-mass dependence affecting the light-connected
contribution are largely cancelled by the disconnected terms. The
cancellation arises because, in isospin-symmetric QCD, the two-pion
intermediate states, responsible for the dominant finite-size and pion-mass
dependence, contribute only to the $I=1$ channel and not to $I=0$. Moreover,
the $I=1$ and $I=0$ components of the Euclidean vector correlator $V(t)$ are
each positive definite, allowing the application of the so-called bounding method~\cite{Borsanyi:2016lpl,RBC:2018dos},
described in Subsec.~\ref{seq:analysis-and-bounding}, independently to the two channels. For
these reasons, we adopt the isospin decomposition above and analyze the
$I=1$ and $I=0$ contributions separately.

Furthermore, for the practical evaluation of $a_\mu^{\rm HVP}(I=0)$, it is convenient to write this contribution in the form~\cite{Djukanovic:2024cmq}
\begin{align}
\label{eq:I0}
a_\mu^{\rm HVP}(I=0)
=
a_\mu^{\rm HVP}({\rm oct})
+
a_\mu^{\rm HVP}(c)
+
\delta a_\mu^{c,{\rm disc}}~,
\end{align}
where  %
$a_{\mu}^{\rm HVP}({\rm oct})$ is obtained from the two-point correlator of the current
$ j_{\mu}^{(8)}(x) = (1/6) [\, j_\mu^u (x)+ j_\mu^d(x) - 2 j_\mu^s(x) \, ] \, $, while $\delta a_\mu^{c,{\rm disc}}$ collects the residual contributions involving disconnected charm loops. Explicitly the octet contribution is given by the following flavor combination
\begin{align}
\label{eq:octet_decomposition}
a_{\mu}^{\rm HVP}({\rm oct}) = \frac{1}{10}a_{\mu}^{\rm HVP}(\ell) + a_{\mu}^{\rm HVP}(s) + a_{\mu}^{\rm HVP}({\rm disc}, uds)~,
\end{align}
where $a_{\mu}^{\rm HVP}({\rm disc}, uds)$ is the disconnected contribution containing only light- and strange-quark loops.
More details concerning the lattice discretization of the $I=1$ and $I=0$ currents and the decomposition adopted for the $I=0$ channel are deferred to Sec.~\ref{sec:LatJ} and Sec.~\ref{seq:analysis-and-bounding}.

Following the suggestion of the RBC/UKQCD collaboration\,\cite{RBC:2018dos}, each of the terms appearing in Eqs.~\eqref{eq:amu_HVP_f}-\eqref{eq:I0} can further be decomposed by multiplying the integration kernel $K(m_\mu t)$ appearing in Eq.~\eqref{eq:amu_HVP} with suitably smoothed Heaviside step-functions, namely
\be
    \label{eq:amu_w}
    a_\mu^{\rm w} = 2 \alpha_\mathrm{em}^2 \int_0^\infty ~ dt \, t^2 \, K(m_\mu t) \, \Theta^{\rm w}(t)\,V(t) ~ \qquad {\rm w} = \{\rm SD, W, LD \} ~ , ~ 
\ee
where the time-modulating functions $\Theta^{\rm w}(t)$ are given by
\bea
      \label{eq:Mt_SD}
      \Theta^{\rm SD}(t) & \equiv & 1 -  \frac{1}{1 + e^{- 2 (t - t_0) / \Delta}} ~ , ~ \\[2mm]
      \label{eq:Mt_W}
      \Theta^{\rm W}(t) & \equiv & \frac{1}{1 + e^{- 2 (t - t_0) / \Delta}} -  \frac{1}{1 + e^{- 2 (t - t_1) / \Delta}} ~ , ~ \\[2mm]      
      \label{eq:Mt_LD}
      \Theta^{\rm LD}(t) & \equiv & \frac{1}{1 + e^{- 2 (t - t_1) / \Delta}} ~  ,
\eea
with the parameters $t_0, t_1, \Delta$ chosen\,\cite{RBC:2018dos} to be equal to
\be
    \label{eq:parms}
    t_0 = 0.4 ~ \mathrm{fm} ~ , ~ \qquad t_1 = 1 ~ {\rm fm} ~ , ~ \qquad \Delta = 0.15~{\rm fm} ~ . ~    
\ee
The resulting time-modulating functions $\Theta^{\rm SD, W, LD}(t)$ are shown, e.g., in Fig.\,1 of Ref.\,\cite{ExtendedTwistedMass:2022jpw}.

In this work, in addition to the total contributions $a_{\mu}^{\rm HVP}(I=1)$ and $a_{\mu}^{\rm HVP}(I=0)$, we compute the three Euclidean-time window observables $a_\mu^{\rm w}(I)$, with ${\rm w} \in {\mathrm{SD},\mathrm{W},\mathrm{LD}}$, for both the isotriplet ($I=1$) and isoscalar ($I=0$) channels. The short-distance (${\rm w}=\mathrm{SD}$) and intermediate (${\rm w}=\mathrm{W}$) windows, for which the precision of the lattice calculation reaches the few-permille level, probe different energy regions of the hadronic $R$-ratio measured in $e^{+}e^{-}$ annihilation. These quantities are experimentally accessible and, since their SM prediction is extremely accurate, constitute by themselves high-precision observables that can serve as sensitive probes of possible NP effects. Our numerical results for the window observables are presented and discussed in Sec.~\ref{sec:windows}.

\section{Setup and numerical techniques}
\label{sec:setup}
\subsection{Scale setting and mistuning corrections}
\label{sec:scale setting}
\begin{table}[t]
\begin{tabular}{lccccccc}
ensemble & $V/a^4$ & $a~[\rm fm]$ & $L~[\rm fm]$ &  $am_{\ell}$ & $am_{s}$ & $am_{c}$ & $am_\mathrm{0}$  \\
\hline 
\multicolumn{8}{c}{\rule[-0.9em]{0pt}{2.2em} FLAG scheme} \\
\hline \\
B64 ~&~$64 \times128$&~$0.07968(13) $~&~$5.10$~&~$0.0006683(28)$~&~$0.018338(58)$~&~$0.23219(85)$~&~$-0.4138874(40)$\\
B96 ~&~$96 \times192$&~$0.07968(13) $~&~$7.65$~&~$0.0006683(28)$~&~$0.018338(58)$~&~$0.23219(85)$~&~$-0.4138874(40)$\\[4pt]
C80 ~&~$80 \times160$&~$0.06801(10) $~&~$5.44$~&~$0.0005834(22)$~&~$0.016006(46)$~&~$0.19735(53)$~&~$-0.3964552(33)$\\[4pt]
D96 ~&~$96 \times192$&~$0.05690(13) $~&~$5.46$~&~$0.0004955(30)$~&~$0.013594(50)$~&~$0.16467(58)$~&~$-0.3760894(26)$\\[4pt]
E112~&~$112\times224$&~$0.048809(75)$~&~$5.47$~&~$0.0004282(18)$~&~$0.011752(30)$~&~$0.14128(37)$~&~$-0.3613163(10)$\\[8pt]
\hline
\multicolumn{8}{c}{\rule[-0.9em]{0pt}{2.2em} $f_{\pi}^{131}$-scheme} \\
\hline \\
B64 ~&~$64 \times128$&~$0.079226(99)$~&~$5.07$~&~$0.0006606(20)$~&~$0.018123(43)$~&~$0.23011(79)$~&~$-0.4138874(40)$\\
B96 ~&~$96 \times192$&~$0.079226(99)$~&~$7.61$~&~$0.0006606(20)$~&~$0.018123(43)$~&~$0.23011(79)$~&~$-0.4138874(40)$\\[4pt]
C80 ~&~$80 \times160$&~$0.06764(10) $~&~$5.41$~&~$0.0005773(22)$~&~$0.015826(47)$~&~$0.19575(54)$~&~$-0.3964552(33)$\\[4pt]
D96 ~&~$96 \times192$&~$0.05659(11) $~&~$5.43$~&~$0.0004901(25)$~&~$0.013442(38)$~&~$0.16330(52)$~&~$-0.3760894(26)$\\[4pt]
E112~&~$112\times224$&~$0.04853(10) $~&~$5.44$~&~$0.0004234(24)$~&~$0.011613(46)$~&~$0.14007(46)$~&~$-0.3613163(10)$\\[8pt]
\hline
\end{tabular}
\caption{ETMC gauge ensembles used to compute $a_{\mu}^{\rm HVP}(I=1)$ and $a_{\mu}^{\rm HVP}(I=0)$. The values of the lattice spacings and of the bare quark masses are fine-tuned in order to match specific target definitions of isospin-symmetric QCD. This is achieved by starting from the simulated gauge ensembles and by taking into account the reweighting factors needed to correct the small mistunings of the simulated bare parameters as discussed in the text. The bare parameters corresponding to the FLAG scheme are reported in the upper part of the table, while in the lower part we give
those corresponding to the $f_{\pi}^{131}$-scheme (which is not identified with the WP25 scheme because the error on $f_\pi$ is not propagated onto the quoted parameters: see text for details).
\label{tab:iso_def}}
\end{table}

For the calculation of $a_{\mu}^{\rm HVP}$ we employ the five ETMC $N_{f}=2+1+1$ gauge ensembles listed in Table~\ref{tab:iso_def}, corresponding to four lattice spacings in the range $a\in [0.048,0.08]~{\rm fm}$ and spatial volumes $L^{3}$ with $L\in [5,7.7]~{\rm fm}$.

In order to match a specific definition of isospin-symmetric QCD, e.g., the FLAG scheme, one has to account for the unavoidable mistuning of the parameters used in the Monte Carlo simulations. Such mistunings must be corrected in order to reproduce the target values of the input quantities defining a given scheme. In the case of the FLAG scheme, the target dimensionless ratios are constructed from the FLAG input values in Eq.~\eqref{eq:FLAG_scheme} as
\begin{align}
R^{\rm FLAG}_\pi =
\left[\frac{m_{\pi}^{2}}{f_{\pi}^{2}}\right]^{\rm FLAG},
\quad
R^{\rm FLAG}_K =
\left[\frac{m_{K}^{2}}{f_{\pi}^{2}}\right]^{\rm FLAG},
\quad
R^{\rm FLAG}_{D_s} =
\left[\frac{m_{D_{s}}}{f_{\pi}}\right]^{\rm FLAG}
\label{eq:R-FLAG} \, .
\end{align}
We denote by $R^{\rm sim}_\mathcal{O}$ ($\mathcal{O}=\pi,\,K,\,D_s$) the corresponding ratios obtained on the simulated ensembles, at the simulated values of the quark masses and gauge coupling $\beta$. Exploiting the fact that each $R^{\rm sim}_\mathcal{O}$ differs from the respective $R^{\rm FLAG}_\mathcal{O}$ only at the few-percent level, we solve, on each gauge ensemble, the linearized system
\begin{align}
\label{eq:syst_A}
\vec{R}^{\rm \, sim} + (m_{0}-m_{0}^{\rm sim})\frac{d\vec{R}}{dm_{0}}\bigg|_{\rm sim}
+ \sum_{f=\ell,s,c}(m_{f}-m_{f}^{\rm sim})\frac{d\vec{R}}{dm_{f}}\bigg|_{\rm sim}
=\vec{R}^{\rm \, FLAG}~,
\end{align}
where
$\vec{R}\equiv\left(R_\pi,R_K,R_{D_s}\right)$,
together with
\begin{align}
\label{eq:syst_B}
m_{\rm PCAC}^{\rm sim}
+ (m_{0}-m_{0}^{\rm sim})
\frac{dm_{\rm PCAC}}{dm_{0}}\bigg|_{\rm sim}
=0~,
\end{align}
which is used to correct for residual deviations from the maximal-twist condition $m_{\rm PCAC}=0$ (see Appendix A of Ref.~\cite{ExtendedTwistedMass:2024nyi} for the definition of the PCAC mass).
In practice, we first
solve the above equation for the unknown $am_0$ and then, using the resulting $am_0$ value,
we solve the system in Eq.~\eqref{eq:syst_A} for the three unknowns $\{am_{\ell},am_{s},am_{c}\}$.\footnote{As explained in Appendix~B of Ref.~\cite{ExtendedTwistedMass:2024nyi}, the tuning to maximal twist can be performed independently from the tuning of the quark masses in a linear approach, as the second-order effects of $m_f$ mistunings on $m_0$ affect observables only by negligible lattice artifacts.}
We remark that $am_0$ represents our determination of the critical mass,
usually denoted by $am_{\rm cr}$, obtained by imposing the condition in
Eq.~\eqref{eq:syst_B} consistently at all lattice spacings.
 
Once the solutions of the system are known, they must be used to correct any observable of interest $\mathcal{O}$, computed on the lattice (together with estimates of its first derivatives with respect to $m_0$, $m_{\ell}$, $m_s$ and $m_c$), via the formula 
\begin{flalign}
    \label{eq:mist_correct_obs}
    \mathcal{O}^{\rm FLAG} = 
    \mathcal{O}^{\rm \, sim} + (m_{0}-m_{0}^{\rm sim})\frac{d\mathcal{O}}{dm_{0}}\bigg|_{\rm sim}
    + \sum_{f=\ell,s,c}(m_{f}-m_{f}^{\rm sim})\frac{d\mathcal{O}}{dm_{f}}\bigg|_{\rm sim}~,
\end{flalign}
which is analogous to Eq.~\eqref{eq:syst_A}, but where now $m_f$ ($f=\ell,s,c$) and $m_0$ are known parameters. For generic observables, the leading dependence on the residual
critical-mass mistuning is linear ($O(am_{\rm PCAC}))$. For some quantities, most notably
$f_\pi$ and $m_\pi$, the linear term is instead found to be numerically
negligible and the leading dependence is quadratic in $m_{0}-m_{0}^{\rm sim}$. In these cases, the
$m_0$ term in Eq.~\eqref{eq:mist_correct_obs} is replaced by
$-\frac{1}{2}(m_0-m_0^{\rm sim})
\,d\mathcal{O}/dm_0|_{\rm sim}$. The 
numerical validation of this prescription is discussed in Appendix~\ref{sec:scale_setting_details}.

Then, the lattice spacing is determined using the FLAG prescription for the pion decay constant, viz.,
\begin{align}
\label{eq:syst_C}
a = \frac{\left(af_{\pi}\right)^{\rm FLAG}}{f_{\pi}^{\rm FLAG}}~,
\end{align}
where $(af_\pi)^{\rm FLAG}$ is the value of the pion decay constant in lattice units properly corrected by means of Eq.~\eqref{eq:mist_correct_obs}.

The quark-mass and $m_0$ derivatives entering the previous equations receive
both valence- and sea-quark contributions, which we separate as
\begin{align}
\frac{d\mathcal O}{dm_f}\bigg|_{\rm sim}
&=
\frac{d\mathcal O}{dm_f}\bigg|_{{\rm val},{\rm sim}}
+
\frac{d\mathcal O}{dm_f}\bigg|_{{\rm sea},{\rm sim}}\,,
\qquad f=\ell,s,c, 0\,.
\label{eq:val_sea_derivatives}
\end{align}
The main strategy to compute both valence and sea derivatives follows that introduced in Appendix~B of
Ref.~\cite{ExtendedTwistedMass:2024nyi}. Further details on the present
determination, including the main improvements with respect to 
the analysis of Ref. ~\cite{ExtendedTwistedMass:2024nyi},
are provided in Appendix~\ref{sec:scale_setting_details}.

The valence contribution to the derivative with respect to a given parameter
is obtained at fixed gauge background by recomputing the correlation
functions defining $\mathcal O$ with propagators evaluated at values of that
parameter slightly shifted from its simulated value. The derivative is then
determined from the correlated finite difference with respect to the result
obtained at the simulated value.\footnote{The only exception is represented by the valence $m_{0}$ derivatives, which are evaluated through pseudoscalar-density insertions on the valence quark lines defining the observable $\mathcal O$. See e.g. Eq.~\eqref{eq:m0_val_derivatives}.} The valence derivatives are evaluated with
sufficiently high statistics that their uncertainties are subleading compared
with those of the observables at the simulation point.

The sea derivatives are instead obtained by differentiating the fermion
determinant and are given by the gauge-connected correlations between the
observable and the corresponding fermionic loop (see Appendix~B of
Ref.~\cite{ExtendedTwistedMass:2024nyi}):
\begingroup
\newcommand{\scalarloopf}{%
\raisebox{-0.40\height}{%
\includegraphics[height=2.9em]{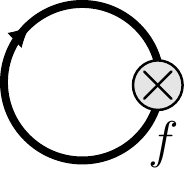}}}
\newcommand{\pseudoscalarloopf}{%
\raisebox{-0.40\height}{%
\includegraphics[height=2.9em]{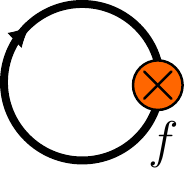}}}
\begin{align}
\frac{d\mathcal O}{dm_f}\bigg|_{{\rm sea},{\rm sim}}
&=
\left\langle
\mathcal O\,\times \, \scalarloopf
\right\rangle_{\rm sim}
-
\left\langle\mathcal O\right\rangle_{\rm sim} \times
\left\langle\scalarloopf\right\rangle_{\rm sim}~,\qquad f=\ell,s,c~,
\nonumber\\[11pt]
\frac{d\mathcal O}{dm_0}\bigg|_{{\rm sea},{\rm sim}}
&=
\sum_{f=\ell, s,c}
-ir_{f} \left[
\left\langle 
\mathcal O\, \times \, \pseudoscalarloopf
\right\rangle_{\rm sim}
-
\left\langle\mathcal O  \right\rangle_{\rm sim} \times 
\left\langle\pseudoscalarloopf\right\rangle_{\rm sim}
\right].
\label{eq:sea_derivatives}
\end{align}
\endgroup
The gray and red blobs denote scalar and pseudoscalar insertions, respectively. The loop carries a fixed flavor index $f$, whereas the derivative with respect to $m_0$ involves a sum over all sea-quark flavors. For $f=\ell$, the insertions represent the sum of the $u$- and $d$-quark contributions. In the preceding equation, $r_f$ denotes the twisted-Wilson $r$-parameter associated with the quark field of flavor $f$. We recall that $r_u=-r_d=1$. For the heavier flavors $f=s,c$, we average over the two choices $r_s=\pm1$ and $r_c=\pm1$, respectively, in order to avoid introducing $O(a)$ discretization effects.

The sea-$m_{\ell}$ and sea-$m_0$ derivatives are computed on all the physical-point
ensembles using the full statistics available for the HVP analysis (see Subsec.~\ref{sec:stat_noise_red}). The
corresponding determination of the sea-strange and sea-charm derivatives is too
noisy to be useful on these large volumes. Since their statistical
uncertainties scale approximately as $\sqrt{L^3}$, we determine these two
derivatives using dedicated smaller-volume ensembles comprising several
thousand gauge configurations. Further details are given in
Appendix~\ref{sec:scale_setting_details}.

\begin{figure}
    \centering
    \includegraphics[width=\linewidth]{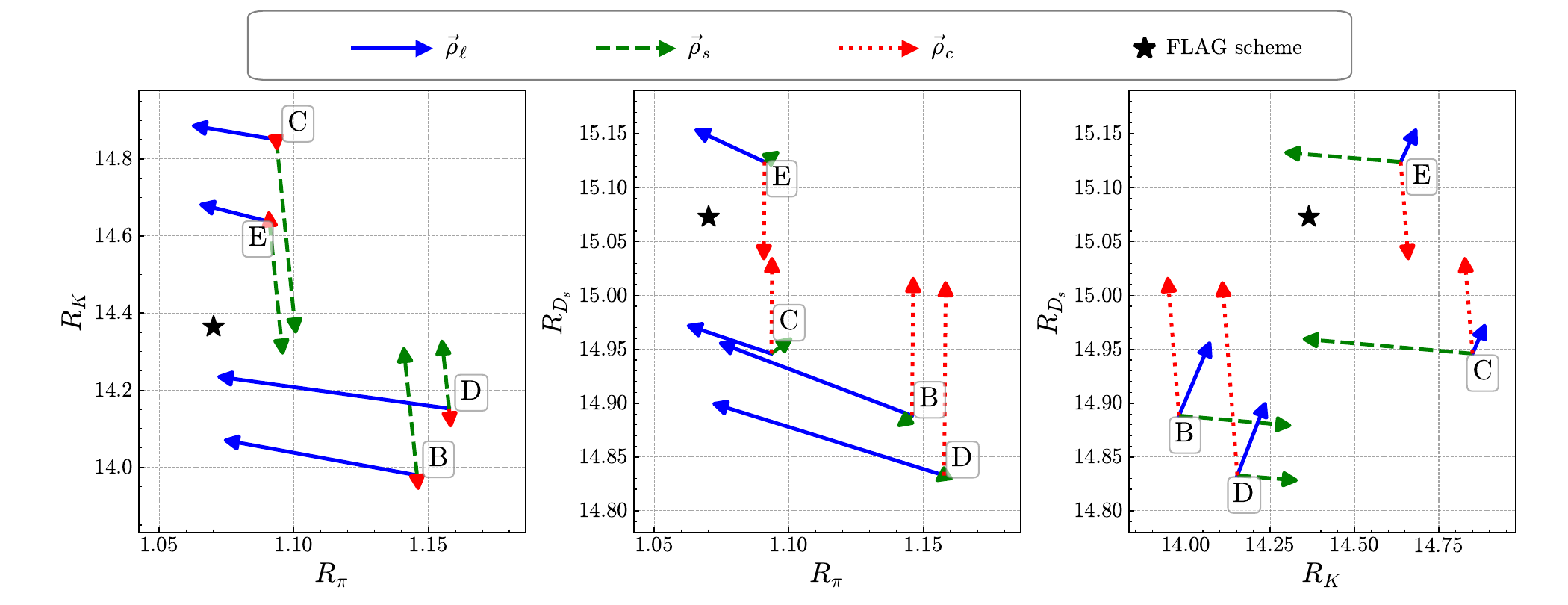}
    \caption{Graphical representation of the simulated ensembles and of the mistuning corrections in the $(R_\pi,R_K,R_{D_s})$ space, projected onto the $(R_\pi,R_K)$, $(R_\pi,R_{D_s})$, and $(R_K,R_{D_s})$ planes. The vectors $\vec \rho_f$, defined in Eq.~\eqref{eq:mistuning_vectors}, represent the projections onto the corresponding plane of the ensemble shifts induced by the mistuning of the quark mass $m_f$. Their starting points coincide with the coordinates of the maximally twisted simulation points of each ensemble. The star is placed on the coordinates of $\displaystyle\vec R^{\rm \, FLAG}$ in the corresponding plane. For each ensemble, the vector sum $\sum_f \vec \rho_f$ connects the simulation point to $\displaystyle \vec R^{\rm \, FLAG}$, thereby providing a graphical representation of the solutions of the mistuning system defined in Eq.~\eqref{eq:syst_A}.}
    \label{fig:simu_point_aggregate}
\end{figure}

Using this strategy, we are able to tune the bare parameters of the theory with high precision to match the FLAG-scheme definition of isospin-symmetric QCD. The resulting values of the lattice spacings and quark masses, obtained by solving the system of Eqs.~(\ref{eq:syst_A}--\ref{eq:syst_C}), are reported in the upper part of Table~\ref{tab:iso_def}. 
A graphical summary of the scale-setting procedure is given in Fig.~\ref{fig:simu_point_aggregate}.
The figure shows the simulated ensembles in $\vec R$-space through their projections onto the
$(R_\pi,R_K)$, $(R_\pi,R_{D_s})$, and $(R_K,R_{D_s})$ planes.
The colored vectors correspond, for each gauge ensemble, to the shift in
$\vec{R}$-space induced by correcting for quark-mass mistuning effects, and are given by
\begin{align}\label{eq:mistuning_vectors}
\vec \rho_f =
\left(
\frac{d}{dm_f}\left(\frac{m_\pi}{f_\pi}\right)^2,
\frac{d}{dm_f}\left(\frac{m_K}{f_\pi}\right)^2, 
\frac{d}{dm_f}\left(\frac{m_{D_s}}{f_\pi}\right)
\right)_{\rm sim}(m_f-m_f^{\rm sim}),
\qquad
f=\ell,s,c~.
\end{align}
These corrections are applied to the values of $\vec{R}$ from the simulated gauge ensembles after correcting for the small deviations from the maximal twist condition Eq.~\eqref{eq:syst_B}. By construction, the vector sum $\sum_{f} \vec \rho_{f}$, connects $\vec{R}$ at the simulation points (corrected by critical-mass mistunings) to the $\vec R^{\rm \, FLAG}$ point, shown in Fig.~\ref{fig:simu_point_aggregate} as a black star.

Following Eq.~\eqref{eq:mist_correct_obs} the mistuning corrections are applied to $a_\mu^{\rm HVP}$ and to the Euclidean-time windows. This requires the computation of the derivatives of $V(t)$ with respect to the simulation parameters\footnote{
Strictly speaking, the integration kernel $K(m_\mu\,t)$ depends on the simulation parameters through the scale, $a\equiv a(m_\ell,m_s,m_c,m_0)$, which enters implicitly in its argument.
However, throughout this work we evaluate the kernel using the (mistuning corrected) lattice spacing determined via Eq.~\eqref{eq:syst_C} and taking into account its statistical fluctuations. Consequently, $K$ is treated as independent of $\hat m$ when taking the derivative in Eq.~\eqref{eq:HVP_derivatives}, thereby avoiding a double counting of the mistuning corrections.}
\begin{flalign}
    \label{eq:HVP_derivatives}
    \frac{\partial}{\partial \hat{m}}
    \,
    a_{\mu}^{\rm{HVP}} = 2 \alpha_\mathrm{em}^2 \int_0^\infty ~ dt \, t^2 \, K(m_\mu t) \,\frac{\partial}{\partial \hat{m}}\,V(t) ~ , ~  
\end{flalign}
where $\hat{m}$ is, in turn, $m_\ell$, $m_s$, $m_c$ and $m_0$. Each derivative is then multiplied by the corresponding parameter mistuning to obtain the associated correction to 
$a_{\mu}^{\rm{HVP}}$. These corrections are evaluated following the
general strategy described above for a generic observable $\mathcal O$.
Further details specific to the HVP are given in
Appendix~\ref{sec:app_der_HVP}. With the few exceptions discussed below,
mistuning corrections are included for all contributions to
$a_\mu^{\rm HVP}$ and for the corresponding window observables.

For the isotriplet contribution, the dominant light-quark-mass and $m_0$
corrections are applied directly to the time-dependent correlator before the
bounding procedure described in
Subsec.~\ref{seq:analysis-and-bounding} is performed. The strange- and charm-sea
corrections are instead estimated on the dedicated smaller-volume ensembles
and included after bounding; both are very small and compatible with zero.
For $a_\mu^{\rm HVP}({\rm oct})$, which is determined with a precision of
about $2\%$, we apply the relevant mistuning corrections to the connected
contributions $\frac{1}{10}a_\mu^{\rm HVP}(\ell)$ and
$a_\mu^{\rm HVP}(s)$ in the decomposition of
Eq.~\eqref{eq:octet_decomposition}, except for the light-quark-mass correction
discussed below. We neglect the $m_0$ and strange- and charm-sea-quark mass corrections to
$a_\mu^{\rm HVP}({\rm disc},uds)$, whose uncertainty is about $15-20\%$ and for
which these effects are negligible at the present precision. The light-quark-mass mistuning correction to
$a_\mu^{\rm HVP}({\rm oct})$ requires separate consideration. The
light-quark-mass dependence of
$\frac{1}{10}a_\mu^{\rm HVP}(\ell)$ is mainly generated by the strong
$m_{\ell}$ dependence of the two-pion contribution. Since this contribution
is absent in the octet channel, the shift in
$\frac{1}{10}a_\mu^{\rm HVP}(\ell)$ induced by the mistuning of $m_{\ell}$
is expected to be largely cancelled by the corresponding shift in the
disconnected contribution. Since the light-quark-mass correction to
$\frac{1}{10}a_{\mu}^{\rm HVP}(\ell)$ alone is already approximately three
to four times smaller than the statistical uncertainty of the per-ensemble
octet determination, we apply no light-quark-mass correction to either
component. This cancellation is explicitly verified in
Subsec.~\ref{sec:full_HVP_I0}.

Overall, mistuning corrections to the octet
contribution turn out to be effectively negligible at the present level of
precision, while for the isotriplet contribution the largest correction applied on a single ensemble amounts to a shift of about $1.3\%$.\footnote{ For the charm-connected contribution
$a_\mu^{\rm HVP}(c)$, all mistuning corrections are included as in the case of the isotriplet contribution. The residual contribution
$\delta a_\mu^{c,{\rm disc}}$, containing disconnected diagrams with a
charm-quark loop, is itself negligible (see Subsec.~\ref{sec:full_HVP_I0}). We therefore do not evaluate
mistuning corrections for this contribution.}

We now discuss the implementation of the WP25 scheme of isospin-symmetric QCD, see Eq.~\eqref{eq:WP25_scheme}, in our lattice setup. The WP25 prescription differs from the FLAG scheme only
for the use of the gradient-flow scale $w_{0}$ \cite{Luscher:2010iy,BMW:2012hcm} rather than the pion decay constant as an input quantity.
As detailed in Appendix~\ref{sec:fpi_WP25}, we determine the value of the pion decay constant corresponding to the WP25 isospin-symmetric world, i.e., the value of $f_{\pi}$ obtained when the input parameters of the theory are fixed according to the WP25 prescription~\eqref{eq:WP25_scheme}. The analysis presented in Appendix~\ref{sec:fpi_WP25} yields
\begin{align}
\label{eq:fpi_WP25}
f_{\pi}^{\rm WP25}=131.00(27)~{\rm MeV}~\,.
\end{align}

Eq.~\eqref{eq:fpi_WP25} establishes a one-to-one correspondence between the original $w_{0}$-based formulation and a fully equivalent $f_{\pi}$-based characterization of the WP25 isospin-symmetric world. Exploiting this correspondence is advantageous for us from a numerical viewpoint. In our twisted-Wilson lattice regularization we observe that, for $a_{\mu}^{\rm HVP}$, using $w_{0}$ as an input quantity leads to significantly larger cutoff effects and final errors 
as compared to using $f_{\pi}$.\footnote{A similar behavior has also been observed in Ref.~\cite{Djukanovic:2024cmq}, which employs the Wilson-clover lattice quark regularization.} For this reason, rather than carrying out the full analysis of $a_{\mu}^{\rm HVP}$ directly in the original $w_{0}$-based formulation of the WP25 scheme, we perform the calculation within the corresponding $f_{\pi}$-based description.

However, as the value of $f_{\pi}^{\rm WP25}$ in Eq.~\eqref{eq:fpi_WP25} carries an uncertainty, propagating this error into $a_{\mu}^{\rm HVP}$ requires determining the dependence of $a_{\mu}^{\rm HVP}$ on the input value of $f_{\pi}$. Therefore, in addition to the scheme associated with the central value in Eq.~\eqref{eq:fpi_WP25}, which we refer to as the $f_{\pi}^{131}$-scheme, we consider a few neighboring $f_{\pi}$-based schemes with $f_\pi$ set
to $130.5$~MeV (FLAG-scheme value), $130.8$~MeV and $131.1$~MeV.
In each of these schemes, using its defining $f_\pi$ value: a) we evaluate the analogs of the input ratios in Eq.~\eqref{eq:R-FLAG} and solve the renormalization system analogous to that in Eq.~\eqref{eq:syst_A}, 
together with Eq.~\eqref{eq:syst_B}; 
b) we compute the mistuning corrections to the correlators relevant for the $I=1$ and $I=0$ terms of $a_{\mu}^{\rm HVP}$ and determine the lattice spacing through the analogs of  Eq.~\eqref{eq:mist_correct_obs} and~\eqref{eq:syst_C}. 
As a representative example, the resulting values of all the quark mass parameters in lattice units as well as the lattice spacing (in~${\rm fm}$), are quoted in  Table~\ref{tab:iso_def} for the case of the $f_{\pi}^{131}$-scheme.

The results for the $I=1$ and $I=0$ contributions to $a_{\mu}^{\rm HVP}$ in the target WP25 scheme are then obtained by matching via linear interpolation to Eq.~\eqref{eq:fpi_WP25}, while the uncertainty associated with the error on $f_{\pi}^{\rm WP25}$ is propagated using the dependence of $a_{\mu}^{\rm HVP}$ on $f_{\pi}$ determined from the set of neighboring schemes just mentioned above. More details are given in the following sections where our results are presented.

\subsection{Lattice discretization of the vector correlator}
\label{sec:LatJ}

Here we describe how the formal definition of the Euclidean correlator
$V(t)$ relevant for $a_\mu^{\rm HVP}$, given in Eq.~\eqref{eq:VV}, is
implemented on the lattice in terms of the contributions introduced in
Eqs.~\eqref{eq:amu_HVP_f} and~\eqref{eq:amu_HVP_I}. The key step is to
decompose $V(t)$ into a sum of correlators whose integrals against the kernel
$K(m_\mu t)$ reproduce the individual terms in
Eq.~\eqref{eq:amu_HVP_I}, and to evaluate these correlators separately using
a convenient mixed-action lattice setup. This setup closely follows that
described in
Refs.~\cite{ExtendedTwistedMass:2022jpw,ExtendedTwistedMass:2024nyi},
to which we refer for full details. Here we discuss only the aspects that are
specific to the present calculation. Throughout this subsection, all
correlators and expectation values are understood to be evaluated in the
isospin-symmetric QCD theory.

In the formal continuum isospin-symmetric QCD, starting from the decomposition of the electromagnetic current $J_k(x)$ in terms of the individual quark-flavor currents and charges, given in Eq.~\eqref{eq:Jmu}, the two-point correlator entering $V(t)$ can be rewritten, for $k=1,2,3$, as
\begin{equation}
\label{eq:JJfull}
\begin{split}
\langle J_k(x)J_k^\dagger(0)\rangle
&=
-\langle J_k(x)J_k(0)\rangle
\\
&=
-\langle j_k^{(3)}(x)j_k^{(3)}(0)\rangle
-
\left\langle
\left[j_k^{(8)}+\frac{2}{3}j_k^c\right](x)
\left[j_k^{(8)}+\frac{2}{3}j_k^c\right](0)
\right\rangle \, ,
\end{split}
\end{equation}
where in the last step we made use of isospin symmetry.
The isotriplet and octet currents are given by
\begin{equation}
\label{eq:I0_oct}
j_k^{(3)}
=
\frac{1}{2}\left(j_k^u-j_k^d\right) ,
\qquad
j_k^{(8)}
=
\frac{1}{6}\left(j_k^u+j_k^d-2j_k^s\right) . 
\end{equation}
Introducing the flavor-changing currents
\begin{equation}
j_k^{(+)}
=
\frac{1}{\sqrt{2}}\bar u\gamma_k d,
\qquad
j_k^{(-)}
=
\frac{1}{\sqrt{2}}\bar d\gamma_k u
=
-\left[j_k^{(+)}\right]^\dagger ,
\end{equation}
isospin symmetry implies
\begin{equation}
\label{eq:I1ident}
\langle j_k^{(3)}(x)j_k^{(3)}(0)\rangle
=
\langle j_k^{(+)}(x)j_k^{(-)}(0)\rangle .
\end{equation}
Eq.~\eqref{eq:JJfull} can therefore be written as
\begin{equation}
\label{eq:JJiso}
\begin{split}
\langle J_k(x)J_k^\dagger(0)\rangle
&=
-\langle j_k^{(+)}(x)j_k^{(-)}(0)\rangle
\\[8pt]
&\quad
-
\left\langle
\left[j_k^{(8)}+\frac{2}{3}j_k^c\right](x)
\left[j_k^{(8)}+\frac{2}{3}j_k^c\right](0)
\right\rangle .
\end{split}
\end{equation}
All currents appearing on the right-hand side of
Eq.~\eqref{eq:JJiso} are understood to be properly normalized and chiral
covariant up to discretization effects. The different contributions can be
evaluated separately using suitable lattice regularizations for the
valence-quark fields. The term
$-\langle j_k^{(+)}(x)j_k^{(-)}(0)\rangle$ contains only quark-connected
diagrams. The terms
$-(2/3)\langle j_k^{(8)}(x)j_k^c(0)\rangle$ and
$-(2/3)\langle j_k^c(x)j_k^{(8)}(0)\rangle$ contain only
quark-disconnected diagrams, while both quark-connected and
quark-disconnected diagrams contribute to
$-\langle j_k^{(8)}(x)j_k^{(8)}(0)\rangle$ and
$-(4/9)\langle j_k^c(x)j_k^c(0)\rangle$.

The mixed-action setup used to evaluate on the lattice the various contributions on the
right-hand side of Eq.~\eqref{eq:JJiso} is described in detail in
Ref.~\cite{ExtendedTwistedMass:2024nyi}, in particular in
Eqs.~(A1)--(A5), with the difference that the present calculation requires an
extended set of valence-quark fields and their corresponding ghost fields.
With reference to Eq.~(A4) of
Ref.~\cite{ExtendedTwistedMass:2024nyi}, the relevant valence-quark fields
are
\begin{align}
&q_u,\ q_d,\ q_u':
\qquad\quad\,\,\,\,\,
r_u=-r_d=r_u'=1,
\nonumber\\
&q_s,\ q_s',\ q_s'',\ q_s''':
\qquad
r_s=-r_s'=r_s''=r_s'''=1,
\nonumber\\
&q_c,\ q_c',\ q_c'',\ q_c''':
\qquad
r_c=-r_c'=r_c''=-r_c'''=1.
\label{eq:valq}
\end{align}
These fields are used below to construct the lattice correlators of
interest. We adopt the notation of Appendix~A of
Ref.~\cite{ExtendedTwistedMass:2024nyi} and express the lattice correlators
in terms of the light-quark doublet $\Psi_\ell=(u,d)^T$ and the OS
valence-quark fields listed in Eq.~\eqref{eq:valq}.

For the purely quark-connected $I=1$ contribution, we employ two
regularizations sharing the same continuum limit, hereafter denoted as TM and
OS:
\begin{align}
\langle j_k^{(+)}(x)j_k^{(-)}(0)\rangle
&\longrightarrow
Z_A^2\,\frac{1}{2}
\left\langle
\bar\Psi_\ell\gamma_k\tau^{(+)}\Psi_\ell(x)\,
\bar\Psi_\ell\gamma_k\tau^{(-)}\Psi_\ell(0)
\right\rangle
&&
\text{(TM)}~,
\nonumber\\[8pt]
\langle j_k^{(+)}(x)j_k^{(-)}(0)\rangle
&\longrightarrow
Z_V^2\,\frac{1}{2}
\left\langle
\bar q_u\gamma_k q_u'(x)\,
\bar q_u'\gamma_k q_u(0)
\right\rangle
&&
\text{(OS)}~,
\label{eq:V33}
\end{align}
where
$\tau^{(\pm)}=\frac{1}{2}(\tau^1\pm i\tau^2)$, with
$\tau^{1,2,3}$ denoting the Pauli matrices. For the correlator of two $j_k^{(8)}$ currents (octet correlator), we use the
regularization
\begin{equation}
\langle j_k^{(8)}(x)j_k^{(8)}(0)\rangle
\longrightarrow
Z_V^2
\left\langle
j_k^{(8)\,{\rm OS}}(x)j_k^{(8)\,{\rm OS}}(0)
\right\rangle
\qquad
\text{(hereafter denoted as OS)}.
\label{eq:V88}
\end{equation}
The current $j_k^{(8)\,{\rm OS}}$ is chosen to reproduce the required
combination of quark-connected and quark-disconnected contributions.
Appropriate averages over the Wilson $r$ parameters are also performed to
suppress residual $O(a^2)$ lattice artifacts:
\begin{equation}
j_k^{(8)\,{\rm OS}}
=
\frac{1}{6}
\left(
\bar\Psi_\ell\gamma_k\Psi_\ell
-
\bar q_s\gamma_k q_s
-
\bar q_s'\gamma_k q_s'
-
\bar q_s''\gamma_k q_s''
+
\bar q_s'''\gamma_k q_s'''
\right).
\label{eq:J8_OS_current}
\end{equation}
For the quark-connected correlator of two charm currents, we employ two
valence-quark regularizations sharing the same continuum limit, hereafter
denoted as TM and OS:
\begin{align}
\langle j_k^c(x)j_k^c(0)\rangle^{\rm conn}
&\longrightarrow
Z_A^2
\left\langle
\bar q_c\gamma_k q_c'(x)\,
\bar q_c'\gamma_k q_c(0)
\right\rangle
&&
\text{(TM)}~,
\nonumber\\[8pt]
\langle j_k^c(x)j_k^c(0)\rangle^{\rm conn}
&\longrightarrow
Z_V^2
\left\langle
\bar q_c\gamma_k q_c''(x)\,
\bar q_c''\gamma_k q_c(0)
\right\rangle
&&
\text{(OS)}~.
\label{eq:Vcc-con}
\end{align}
The correlator contributions containing two quark-disconnected charm loops
are implemented through
\begin{equation}
\begin{split}
\langle j_k^c(x)j_k^c(0)\rangle^{\rm disc}
\longrightarrow
Z_V^2\,\frac{1}{4}
\Big\langle&
\left[
\bar q_c\gamma_k q_c
+
\bar q_c'\gamma_k q_c'
\right](x)
\\
&\times
\left[
\bar q_c''\gamma_k q_c''
+
\bar q_c'''\gamma_k q_c'''
\right](0)
\Big\rangle
\qquad
\text{(hereafter denoted as OS)}.
\end{split}
\label{eq:Vcc-dis}
\end{equation}
Similarly, contributions containing one quark-disconnected charm loop are
implemented as
\begin{equation}
\begin{split}
\langle j_k^{(8){\rm OS}}(x)j_k^c(0)\rangle^{\rm disc}
\longrightarrow
Z_V^2\,\frac{1}{2}
\Big\langle&
j_k^{(8){\rm OS}}(x)
\\
&\times
\left[
\bar q_c''\gamma_k q_c''
+
\bar q_c'''\gamma_k q_c'''
\right](0)
\Big\rangle
\qquad
\text{(hereafter denoted as OS)}.
\end{split}
\label{eq:V8c-dis}
\end{equation}
The residual contribution $\delta a_\mu^{c,{\rm disc}}$, obtained from the
sum of the terms containing either two or one quark-disconnected charm loops,
i.e., those in Eqs.~\eqref{eq:Vcc-dis}--\eqref{eq:V8c-dis}, is evaluated
only on a representative gauge ensemble, since it turns out to be numerically
negligible, as also observed in
Refs.~\cite{Borsanyi:2020mff,Djukanovic:2024cmq}. Technical details
concerning the numerical evaluation of the quark-disconnected contributions
entering the decomposition in Eq.~\eqref{eq:JJfull} are deferred to
Subsec.~\ref{sec:stat_noise_red}.

The renormalization constants $Z_V$ and $Z_A$ are the UV-finite normalization
factors required to make the light-quark TM currents
$\bar\Psi_\ell\gamma_\mu\gamma_5\tau^{1,2,3}\Psi_\ell$ and
$\bar\Psi_\ell\gamma_\mu\tau^{1,2,3}\Psi_\ell$, respectively, chiral
covariant. See, for example, the discussion of $Z_A$ below Eq.~(A15) of
Ref.~\cite{ExtendedTwistedMass:2024nyi}. Precise nonperturbative
determinations of $Z_V$ and $Z_A$ are obtained using the same strategy based
on exact Ward identities and universality as in
Refs.~\cite{ExtendedTwistedMass:2022jpw,ExtendedTwistedMass:2024nyi}. Since
the procedure is unchanged in the present work, we refer to those references
for further details and report in Table~\ref{tab:RCs} the values used in this
calculation.

Combining all contributions, we obtain the following decomposition of the
zero-spatial-momentum correlator $V(t)$, in which each contribution is
separately chiral covariant and $O(a)$ improved:
\begin{equation}
\begin{split}
V(t)
&=
V^{I=1}(t)
+
V^{8}(t)
+
\frac{4}{9}
\left[
V^{c,c;{\rm conn}}(t)
+
V^{c,c;{\rm disc}}(t)
\right] +
\frac{2}{3}
\left[
V^{8,c;{\rm disc}}(t)
+
V^{c,8;{\rm disc}}(t)
\right]\,,
\end{split}
\label{eq:Vtot}
\end{equation}
where $V^{I=1}$, $V^{8}$, $V^{c,c;{\rm conn}}$,
$V^{c,c;{\rm disc}}$, and $V^{8,c;{\rm disc}}$ are obtained from the
corresponding correlators in Eqs.~\eqref{eq:V33},
\eqref{eq:V88}, \eqref{eq:Vcc-con}, \eqref{eq:Vcc-dis}, and
\eqref{eq:V8c-dis}, respectively, by integrating over the spatial
coordinates, averaging over the three spatial components, $k=1,2,3$, and including
the overall minus sign in Eq.~\eqref{eq:JJiso}. The contribution
$V^{c,8;{\rm disc}}$ is obtained analogously to
$V^{8,c;{\rm disc}}$, with the two currents interchanged. As anticipated
above, for the infinite-volume and continuum extrapolations we find it
convenient to analyze separately the $I=1$ contribution and the $I=0$
contribution, the latter being given by the sum of all terms on the
right-hand side of Eq.~\eqref{eq:Vtot} except the first.

\begin{table}[t!]
\centering
\begin{tabular}{ lcc }
ensemble & $Z_{V}$ & $Z_{A}$ \\[4pt]
\colrule
\\
\textrm{B64} & ~ $0.706354(54)$  ~  & ~ $0.74296(19)$  ~  \\
\textrm{B96} &  ~  $0.706406(52)$  ~  &  ~  $0.74261(19)$   ~  \\
\textrm{C80} &  ~  $0.725440(33)$  ~  &  ~  $0.75814(13)$  ~ \\
\textrm{D96} & ~  $0.744132(31)$  ~  &  ~  $0.77367(10)$  ~   \\
\textrm{E112} &   ~  $0.758238(18)$  ~  &  ~  $0.78548\,\,\,(9)$  ~  \\[8pt]
\hline
\end{tabular}
 \caption{The values of $Z_{V}$ and $Z_{A}$ used in this work for each of the ETMC ensembles of Table~\ref{tab:iso_def}, determined by employing the hadronic method described in Appendix B of Ref.~\cite{ExtendedTwistedMass:2022jpw}.}
    \label{tab:RCs} 

\end{table}

\subsection{Statistics and noise reduction techniques}
\label{sec:stat_noise_red}
In this subsection, we discuss the application of low-mode-averaging (LMA) techniques to obtain an high-precision determination of the isotriplet correlator $V^{I=1}(t)$ on all the ensembles listed in Table~\ref{tab:iso_def}, as well as a generalisation of the one-end trick (OET) used to improve the precision of quark-disconnected vector correlators. Further details on the implementation of LMA techniques and a comprehensive discussion of their performance on twisted-mass ensembles can be found in Ref.~\cite{Alexandrou:2026rca}.

Before discussing the noise-reduction techniques, we summarise the statistics used in this work. For the quark-connected light correlators on the ensembles with $L\simeq 5.1-5.5~\mathrm{fm}$, a new set of correlators employing exact LMA has been generated. The number of gauge configurations, stochastic sources, and deflated modes is reported in the leftmost panel of Table~\ref{tab:statistics}. For the B96 ensemble, which has a larger physical volume than the other ensembles (see Table~\ref{tab:iso_def}), a multigrid implementation of LMA has been employed to alleviate the $V^2$ scaling of exact LMA. The corresponding statistics are reported in the middle panel of the table. The light-strange quark-disconnected correlators have been computed using the generalised OET (described below), with the corresponding statistics listed in the rightmost panel. For the strange- and charm-quark connected correlators, we use the same statistics as in Ref.~\cite{ExtendedTwistedMass:2024nyi}.

Finally, we comment on the charm-quark disconnected contribution, which we refer to as $\delta a_\mu^{\rm c, disc}$. To estimate its magnitude, we computed charm-quark loops on the B ensemble using the same gauge configurations employed for the light-strange quark loops. These loops were evaluated using hierarchical probing~\cite{Stathopoulos:2013aci} with a probing distance of 4. For each gauge configuration, approximately 4600 vectors were used, corresponding to the product of the number of spin-color diluted stochastic vectors and the number of Hadamard vectors (see Refs.~\cite{Abdel-Rehim:2016pjw,Alexandrou:2022osy}).
\begin{table}
\centering
\begin{tabular}{lrcr}
Ensemble & $N_U$ & $N^{\mathrm{IR}}_{\rm ev}$ & $N_\eta$  \\[4pt]
\colrule
\\
B64  & 1300 & 400 & 1024  \\
C80  & 870  & 530 & 1600  \\
D96  & 670  & 530 &  960  \\ 
E112 & 680  & 530 & 1344  \\[8pt] 
\hline
\end{tabular}
\hspace{15pt}
\begin{tabular}{lrcr}
Ensemble  & $N_U$ & $N^\mathrm{MG}_\eta$ & $N_\eta$\\[4pt]
\colrule
\\
B96 & 490 & 11200 & 768\\[8pt]
\hline
\end{tabular}
\hspace{15pt}
\begin{tabular}{lrr}
Ensemble & $N_U$ & $N_\eta$  \\[4pt]
\colrule
\\
B64  & 773 &  100\\
B96  & 298 &  200\\[2pt]
C80  & 790 &  80\\
D96  & 800 &  150\\
E112 & 595 &  80\\[8pt]
\hline
\end{tabular}
\caption{Statistics for the two-point correlation functions using the ETMC ensembles of Table~\ref{tab:iso_def}, where LMA and generalisation of the OET noise-reduction techniques are applied. In the leftmost table we report the number of configurations $N_U$, the number of exactly deflated low modes $N^{\mathrm{IR}}_{\rm ev}$ which were used in the LMA setup to reconstruct the IR part, and the number of time-wall stochastic sources $N_\eta$ used to compute the full correlation function, respectively. The central table shows the number of gauge configurations, the number of time-wall stochastic sources $N^\mathrm{MG}_\eta$ for the high-statistics estimation of the multigrid IR correlator and the number of stochastic sources $N_\eta$ employed to compute the full correlation function. The rightmost table shows the number of configurations and the number of volume stochastic sources $N_{\eta}$ used in the 
OET based computation of quark-disconnected terms.}
\label{tab:statistics}
\end{table}

\subsubsection{Low-mode averaging}
We employ two different implementations of LMA: exact LMA (exLMA), which is used on the B64,
C80, D96, and E112 ensembles, and multigrid LMA (mgLMA) which is used on the
larger-volume B96 ensemble. The two implementations can be both  formulated through the following
control-variate estimator\footnote{To ease the notation, we denote $V^{I=1}(t)$
simply by $V(t)$ throughout the subsection.}
\begin{equation}
V^\mathrm{LMA}_\eta(t)
=
V_\eta(t)
+
\alpha(t)
\left[
V^\mathrm{cv}_{\mathrm{impr.}}(t)
-
V^\mathrm{cv}_\eta(t)
\right],
\qquad
\alpha(t)
=
\frac{
{\rm Cov}\!\left[
V_\eta(t),V^\mathrm{cv}_\eta(t)
\right]
}{
{\rm Var}\!\left[
V^\mathrm{cv}_\eta(t)
\right]
}~.
\label{eq:corr_defl_var}
\end{equation}
Here $V_\eta(t)$ is the standard stochastic estimate of the vector correlator $V(t)$. Instead $V^\mathrm{cv}(t)$ is
a control  correlator and
$V^\mathrm{cv}_\eta(t)$ its stochastic estimate, evaluated using the same
stochastic sources as $V_\eta(t)$. The quantity
$V^\mathrm{cv}_{\mathrm{impr.}}(t)$ is instead an improved estimate of
the same control correlator with reduced (or even vanishing) stochastic uncertanties. The two implementations, exLMA and mgLMA,
differ both in the choice of $V^\mathrm{cv}(t)$ and in the way
$V^\mathrm{cv}_{\mathrm{impr.}}(t)$ is obtained, as detailed below.
In both cases, $V^\mathrm{cv}_{\mathrm{impr.}}(t)$ and
$V^\mathrm{cv}_\eta(t)$ have the same stochastic expectation value at
fixed gauge configuration. Therefore, $V_\eta^{\rm LMA}(t)$ is an
unbiased estimator of $V(t)$ for any  value of $\alpha(t)$.

The variance reduction is achieved by choosing a control correlator whose
stochastic estimate $V^\mathrm{cv}_\eta(t)$ is correlated with
$V_\eta(t)$. The coefficient $\alpha(t)$ in
Eq.~\eqref{eq:corr_defl_var} is the optimal choice that minimizes the
variance of $V_\eta^{\rm LMA}(t)$. Further details are given in
Ref.~\cite{Alexandrou:2026rca}.

We begin with the exLMA implementation, used on the ensembles with spatial
extent $L\simeq 5.1$--$5.5~\mathrm{fm}$. Starting from
\begin{equation}
Q\equiv D\gamma_5,
\qquad
Qv_j=\lambda_jv_j~,
\end{equation}
where $D$ is the massless Wilson-clover Dirac operator, while $\lambda_j$ and
$v_j$ are, respectively, the eigenvalues and corresponding eigenvectors
of the Hermitian operator $Q$, ordered by increasing
$|\lambda_j|$, we define $P^{\rm IR}$ as the projector onto the
subspace spanned by the lowest $N_{\rm ev}^{\rm IR}$ eigenvectors:
\begin{equation}
P^{\rm IR}
=
\sum_{j=1}^{N_{\rm ev}^{\rm IR}}v_jv_j^\dagger~.
\label{eq:IR_proj}
\end{equation}
We further denote by $D(m,r)$ the Wilson-clover twisted-mass Dirac operator for a
quark with mass $m$ and Wilson parameter $r = \pm 1$, 
and by $S(m,r)$ its
inverse. A useful property of the twisted-mass Dirac-operator
spectrum~\cite{Alexandrou:2026rca} is that the projection of the
propagator $S(m,r)$ onto the $P^{\rm IR}$ subspace is simply given by
\begin{equation}
S^\mathrm{IR}(m,r)
=
D^{-1}(m,r)P^{\rm IR}
=
\sum_{j=1}^{N_{\rm ev}^{\rm IR}}
\frac{1}{\lambda_j+irm}\,
\gamma_5v_jv_j^\dagger~.
\label{eq:S_IR_exLMA}
\end{equation}
In the exLMA implementation, the control correlator $V^{\rm cv}(t)$ is
obtained by replacing, in the construction of $V(t)$, each full quark
propagator $S(m,r)$ with the corresponding infrared-projected propagator
$S^{\rm IR}(m,r)$.\footnote{Note that the TM and OS isotriplet
correlators defined in terms of the two currents in
Eq.~\eqref{eq:V33} differ in the choice of the Wilson $r$-parameter
carried by the quark and antiquark. In both cases, all the required
infrared-projected propagators can be constructed from the same set of
eigenvalues $\lambda_j$ and eigenvectors $v_j$ through
Eq.~\eqref{eq:S_IR_exLMA}, without additional eigenvector computations.}
The improved estimate $V^{\rm cv}_{\rm impr.}(t)$ is evaluated in an all-to-all fashion, by explicitly constructing $S^{\rm IR}(m,r)$ from the lowest $N_{\rm ev}^{\rm IR}$ eigenvalues and eigenvectors of $Q$ according to Eq.~(\ref{eq:IR_proj}). It is therefore free of stochastic noise on each gauge configuration, and we denote it by $V^{\rm IR}(t)$ in the following. The stochastic estimate of the control correlator, $V^{\rm cv}_{\eta}(t)$, is instead obtained by contracting propagators built from the IR projection of the same stochastic sources used in $V_\eta(t)$. We denote this estimate by $V_\eta^{\rm IR}(t)$.

In the standard implementation of exLMA, $\alpha(t)$ is set to unity,
so that the stochastic infrared contribution is simply replaced by its exact
value:
\begin{equation}
V^\mathrm{exLMA}_\eta(t)
=
V_\eta(t)
-
V^\mathrm{IR}_\eta(t)
+
V^\mathrm{IR}(t)~.
\label{eq:V_exLMA}
\end{equation}
As detailed in Ref.~\cite{Alexandrou:2026rca}, we find that the
optimized value of $\alpha(t)$ in Eq.~\eqref{eq:corr_defl_var} deviates from unity by about
$10\%$ or less at large Euclidean times, while the resulting additional
reduction of the statistical uncertainty is only about $2\%$. We
therefore use simply $\alpha(t)=1$ for the production of the correlators improved through exLMA.

\begin{figure}
\centering
\includegraphics[width=0.49\textwidth]{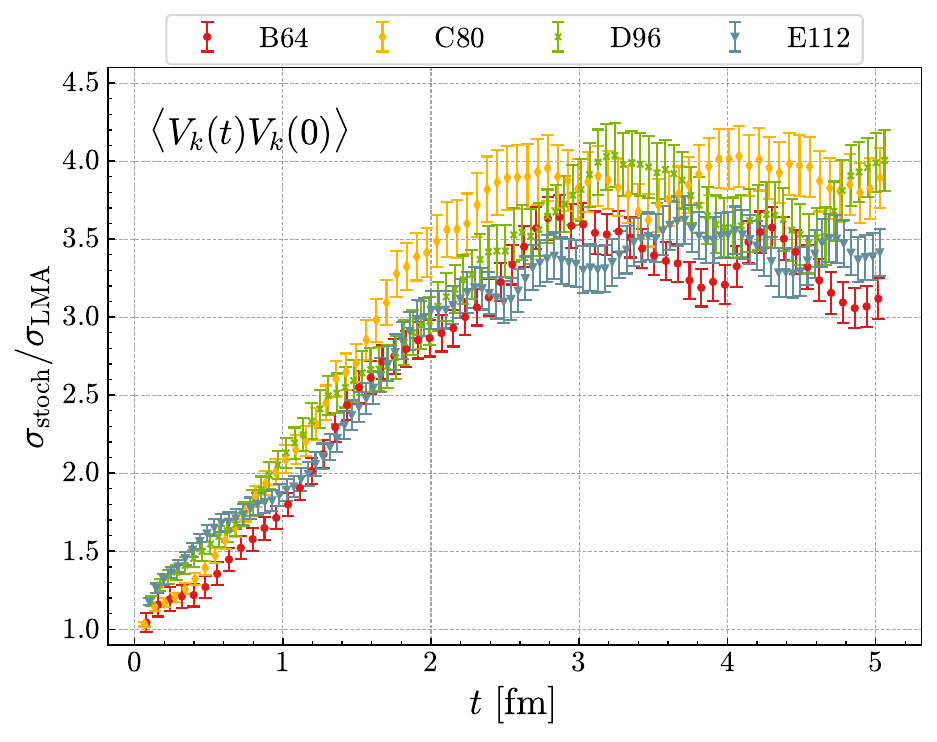}
\includegraphics[width=0.49\textwidth]{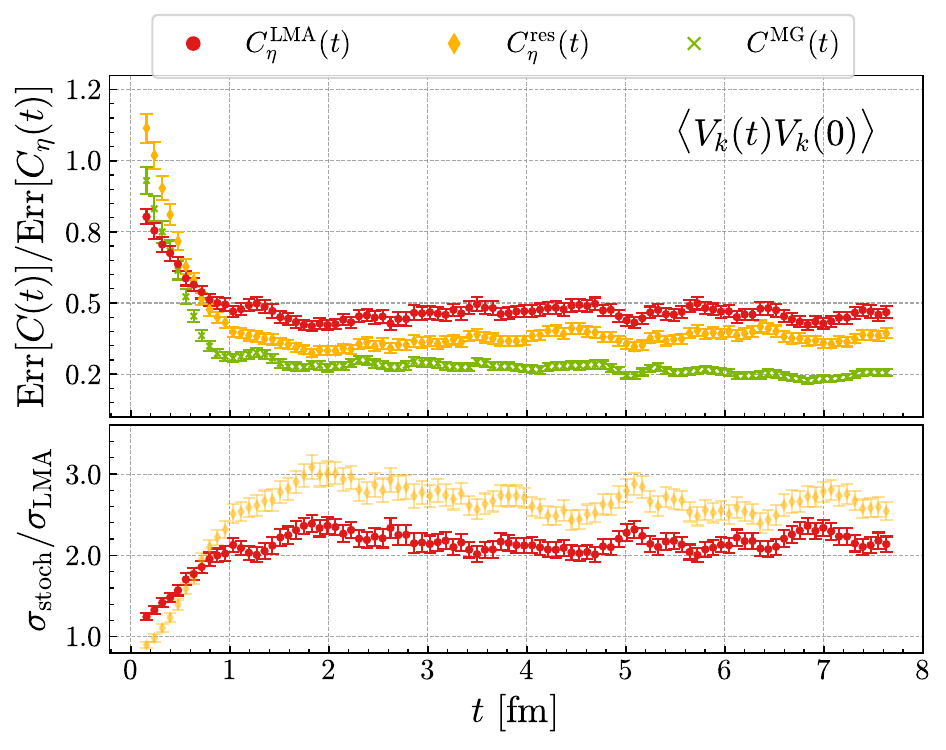}
\caption{\textit{Left}: Gain in the signal-to-noise ratio obtained with
exact LMA for the B64, C80, D96, and E112 ensembles, which have
$L\simeq 5.1$--$5.5~\mathrm{fm}$. Each color corresponds to a
different ensemble. \textit{Right}: In the upper panel, error reduction
for the complete mgLMA correlator (red), the residual contribution
(yellow), and the high-statistics multigrid contribution (green), as
defined in Eq.~\eqref{eq:V_MG_components}. In the lower panel, total
gain obtained with the available statistics (red) and gain expected in
the infinite-statistics limit of the high-statistics multigrid estimator
(yellow).}
\label{fig:gain_LMA}
\end{figure}

The number of eigenmodes $N_{\rm ev}^{\rm IR}$ is chosen such that the
variance reduction saturates at large Euclidean times. We use $400$
eigenmodes for the B64 ensemble, with $L\simeq 5.1~\mathrm{fm}$, and
$530$ eigenmodes for the C80, D96, and E112 ensembles, with
$L\simeq 5.5~\mathrm{fm}$, as summarized in
Table~\ref{tab:statistics}. This difference reflects the approximate
scaling with the physical volume of the number of eigenmodes required
to achieve a comparable noise reduction~\cite{Alexandrou:2026rca}.
The left panel of Fig.~\ref{fig:gain_LMA} shows the resulting gain.
Comparable performance is indeed observed for all four ensembles, with the
stochastic uncertainty at large Euclidean times reduced by a factor of
approximately $3$--$4$.

For the B96 ensemble, with $L\simeq 7.7~\mathrm{fm}$, computing the
number of exact eigenmodes required by exLMA becomes prohibitively
expensive. We therefore employ the mgLMA implementation described in
Ref.~\cite{Alexandrou:2026rca}, which we briefly summarize here. The
starting point is the multigrid coarse-grid operator
\begin{equation}
D_c(m,r)
=
P^\dagger D(m,r)P~,
\end{equation}
where $P$ denotes the multigrid prolongation operator (we refer to~\cite{Alexandrou:2026rca} for details on the properties of $P$). An approximate
infrared-projected propagator is obtained by projecting $D_{c}(m,r)$ onto the subspace spanned by 
a fixed number of low modes on the coarse grid and prolonging back
the resulting coarse-grid projected propagator, $S_{c}^{\rm IR}(m,r)$, via
\begin{equation}
S^\mathrm{MG,IR}(m,r)
=
P S_c^\mathrm{IR}(m,r)P^\dagger~.
\label{eq:S_MG_IR}
\end{equation}
The mgLMA control correlator, which we denote by
$V^\mathrm{MG,IR}(t)$, is obtained by replacing the complete propagators
entering $V(t)$ with $S^\mathrm{MG,IR}(m,r)$. Its stochastic estimate
$V^\mathrm{MG,IR}_\eta(t)$ is constructed using the same stochastic
sources and contractions as $V_\eta(t)$. Since an exact all-to-all
evaluation of $V^\mathrm{MG,IR}(t)$ remains computationally too expensive,
the improved estimate is obtained in this case through high-statistics
sampling:
\begin{equation}
V^\mathrm{MG,IR}_{\mathrm{impr.}}(t)
=
\frac{1}{N^\mathrm{MG}_\eta}
\sum_{s=1}^{N^\mathrm{MG}_\eta}
V^\mathrm{MG,IR}_{\eta_s}(t)~.
\label{eq:V_MG_impr}
\end{equation}
Thus, unlike $V^\mathrm{IR}(t)$ in exLMA,
$V^\mathrm{MG,IR}_{\mathrm{impr.}}(t)$ is not the exact all-to-all
correlator.

The quality of the multigrid approximation determines the correlation
between $V^\mathrm{MG,IR}_\eta(t)$ and $V_\eta(t)$, and hence the
achievable variance reduction. Moreover, since the coarse-grid
projection modifies the relative normalization of the two correlators,
the optimal coefficient $\alpha(t)$ is significantly different from unity.

On the B96 ensemble,
$V^\mathrm{MG,IR}_{\mathrm{impr.}}(t)$ is computed using
$N^\mathrm{MG}_\eta=11200$ stochastic sources, compared with
$N_\eta=768$ sources for the full correlator, as reported in
Table~\ref{tab:statistics}. The larger statistics is chosen such that
the residual stochastic uncertainty on $V^{\rm MG, IR}_{\rm impr.}(t)$ is sensibly smaller than that on $V^{\rm MG,IR}_{\eta}(t)$, while maintaining a balanced computational
cost~\cite{Alexandrou:2026rca}. To discuss the performance of mgLMA,
we write
\begin{equation}
V^\mathrm{mgLMA}_\eta(t)
=
V^\mathrm{res,MG}_\eta(t)
+
V^\mathrm{MG}(t)~,
\end{equation}
where
\begin{equation}
V^\mathrm{res,MG}_\eta(t)
\equiv
V_\eta(t)
-
\alpha(t)V^\mathrm{MG,IR}_\eta(t),
\qquad
V^\mathrm{MG}(t)
\equiv
\alpha(t)V^\mathrm{MG,IR}_{\mathrm{impr.}}(t)~.
\label{eq:V_MG_components}
\end{equation}
The right panel of Fig.~\ref{fig:gain_LMA} shows the resulting error
reduction. The upper panel compares the uncertainties of the complete
mgLMA correlator and of the two contributions in
Eq.~\eqref{eq:V_MG_components}. The lower panel shows the total gain
obtained with the available statistics, together with the gain expected
in the infinite-statistics limit of the improved estimator
($N_\eta^{\rm MG}\to\infty$). With the statistics used in this work,
mgLMA reduces the uncertainty by approximately a factor of two at large
Euclidean times.

\subsubsection{Generalized one-end trick for quark-disconnected contributions}
\label{sec:OET}
To compute the disconnected contribution to $a_\mu^{\rm HVP}({\rm oct})$, the fundamental quantity to be determined is the difference between light- and strange-quark vector loops, as follows from the flavor structure of the octet current in Eq.~\eqref{eq:I0_oct}. We evaluate this difference using a generalisation of the OET, which can be regarded as the twisted-mass realization of the split-even technique developed in Ref.~\cite{Giusti:2019kff} for Wilson fermions.

Consistently with the notation introduced in the previous subsection, we denote by $S(m,r;x,y)$ the quark propagator with mass $m$ and Wilson parameter $r$, with $S(m,r)=D^{-1}(m,r)$, and define
\begin{equation}
B_i(m,r;z)\equiv\Tr\!\left[\gamma_i S(m,r;z,z)\right],
\label{eq:B_loop_def}
\end{equation}
where the trace runs over spin and color.

To construct the generalized-OET estimator for the light-strange loop difference, we exploit an integrated Ward identity (WI) associated with a nonsinglet axial transformation of a valence $(\ell,s)$ twisted-mass doublet. For a fixed choice of $r$, we consider the valence doublet $\psi_r=(\ell,s)^T$, with $\widehat m={\rm diag}(m_\ell,m_s)$ and $r_\ell=-r_s=r$, and introduce the nonsinglet operators
\begin{equation}
V^{\mu,a}\equiv\Bar\psi_r\gamma^\mu\tau^a\psi_r,\qquad
P^a\equiv\frac{i}{2}\Bar\psi_r\gamma^5\tau^a\psi_r,\qquad
A^{\mu,a}\equiv\Bar\psi_r\gamma^\mu\gamma^5\tau^a\psi_r,
\label{eq:OET_operators}
\end{equation}
where $\tau^a$ are the Pauli matrices acting in the $(\ell,s)$ flavor space.

The WI follows from an infinitesimal transformation of the valence fermionic integration variables at fixed gauge background. We consider the global nonsinglet axial transformation $\delta\psi_r=i\omega\gamma_5\tau^1\psi_r$ and $\delta\Bar\psi_r=i\omega\Bar\psi_r\gamma_5\tau^1$, with $\omega$ an infinitesimal constant. In the massless limit $m_{\ell}=m_{s}=0$, the valence-quark action $\mathcal{S}_{\rm val}$ of the field $\psi_r$ is invariant under this transformation, while the presence of the mass term results in $\delta\mathcal S_{\rm val}=2\omega(m_\ell+m_s)\sum_x P^1(x)$. Under the same transformation, the nonsinglet axial current $A^{i,2}(z)$ varies as $\delta A^{i,2}(z)=2\omega V^{i,3}(z)$, which results in the following WI
\begin{equation}
\expval{V^{i,3}(z)}_F=(m_\ell+m_s)\sum_x\expval{P^1(x)A^{i,2}(z)}_F,
\label{eq:disco_WI}
\end{equation}
where $\expval{\cdot}_F$ denotes the path integral over the valence fermionic fields at fixed gauge background.

At the propagator level, Eq.~\eqref{eq:disco_WI} becomes~\cite{DiPalma:2025iud}
\begin{align}
B_i(m_\ell,r;z)-B_i(m_s,-r;z)
={}&
\frac{m_\ell+m_s}{2}\sum_x
\Tr\Big[
\gamma^5 S(m_\ell,r;x,z)\gamma^i\gamma^5 S(m_s,-r;z,x)
\nonumber\\
&\hspace{19mm}
-\gamma^5 S(m_s,-r;x,z)\gamma^i\gamma^5 S(m_\ell,r;z,x)
\Big].
\label{eq:disco_WI_prop}
\end{align}
For our stochastic estimator, we use the representation of the light-strange loop difference given by the right-hand side of Eq.~\eqref{eq:disco_WI_prop}. Stochastic volume sources are placed at the pseudoscalar-density insertion and used to estimate the sum over $x$ for all points $z=(\mathbf z,t)$. The spatial sum over $\mathbf z$ at fixed time, which is the quantity needed to determine the quark-disconnected contribution to the octet correlator, can therefore be evaluated exactly without  stochastic estimates. This is the origin of the reduction in stochastic variance. In a direct evaluation of the left-hand side of Eq.~\eqref{eq:disco_WI_prop}, instead, the volume sources are placed at the vector-current insertion and the required spatial sum over $\mathbf z$ is therefore evaluated stochastically. Correspondingly, placing the stochastic sources at the axial-current insertion on the right-hand side of Eq.~\eqref{eq:disco_WI_prop} reproduces, source by source, the direct stochastic estimator of the left-hand side, therefore leading to no variance reduction.
Eq.~\eqref{eq:disco_WI_prop} thus provides the twisted-mass realization of the split-even estimator of Ref.~\cite{Giusti:2019kff}. The standard twisted-mass OET relation~\cite{ETM:2008zte} can be recovered in our framework by replacing the strange quark with a replica $\ell'$ of the light flavor satisfying $r_{\ell'}=-r_\ell$, and by considering the nonsinglet scalar density instead of the nonsinglet axial current.

The quark-disconnected contribution generated by the OS octet current in Eq.~\eqref{eq:J8_OS_current} requires an average over opposite values of the twisted-Wilson parameter for both the light and strange quarks. This average does not require any additional calculation. Indeed, the $\gamma_5$-hermiticity relation $S^\dagger(m,r;x,y)=\gamma_5S(m,-r;y,x)\gamma_5$ implies $B_i(m,-r;z)=-[B_i(m,r;z)]^\star$. Therefore, once the light-strange difference is evaluated for one choice, e.g., $r_\ell=+1$ and $r_s=-1$, the result for the opposite choice follows exactly by complex conjugation. Equivalently, the $r$-averaged loop entering the OS current is $i\,{\rm Im}[B_i(m,+1;z)]=i\,{\rm Im}[B_i(m,-1;z)]$.

We further combine the generalized OET with the frequency-splitting technique of Ref.~\cite{Giusti:2019kff}. Introducing $N_\ell$ intermediate masses $m_\ell=m_0<m_1<\cdots<m_{N_\ell}<m_{N_\ell+1}=m_s$, the light-strange vector-loop difference can be decomposed exactly as
\begin{equation}
\overline B_i(m_\ell;z)-\overline B_i(m_s;z)
=
\sum_{j=0}^{N_\ell}
\left[
\overline B_i(m_j;z)-\overline B_i(m_{j+1};z)
\right],
\label{eq:frequency_splitting}
\end{equation}
where $\overline B_i(m;z)\equiv[B_i(m,+1;z)+B_i(m,-1;z)]/2$. Each term on the right-hand side is evaluated with the generalized OET described above. In the general frequency-splitting strategy of Ref.~\cite{Giusti:2019kff}, different numbers of stochastic sources can be assigned to the different mass intervals to optimize the performance. In our implementation, we use the same number of stochastic volume sources for all terms in Eq.~\eqref{eq:frequency_splitting}.

The quark-disconnected contribution to the octet correlator $V^{8}(t)$ used in this work has been produced with $N_\ell=3$  intermediate masses
\begin{equation}
m_i/m_\ell=2,\,3,\,5,\qquad i=1,2,3.
\label{eq:OET_intermediate_masses}
\end{equation}

\begin{figure}
\centering
\includegraphics[width=0.49\textwidth]{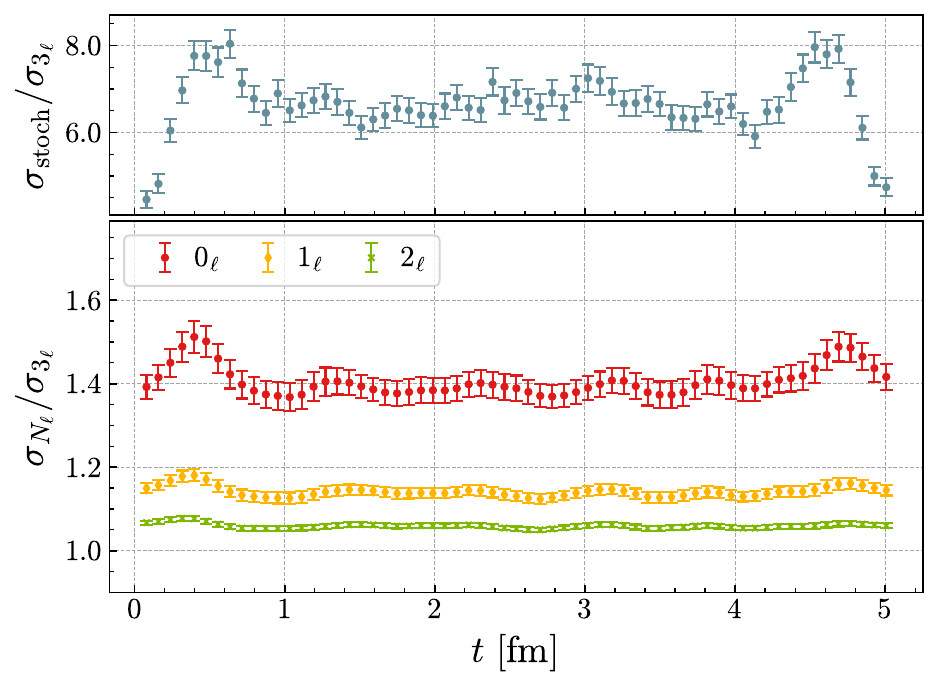}
\includegraphics[width=0.49\textwidth]{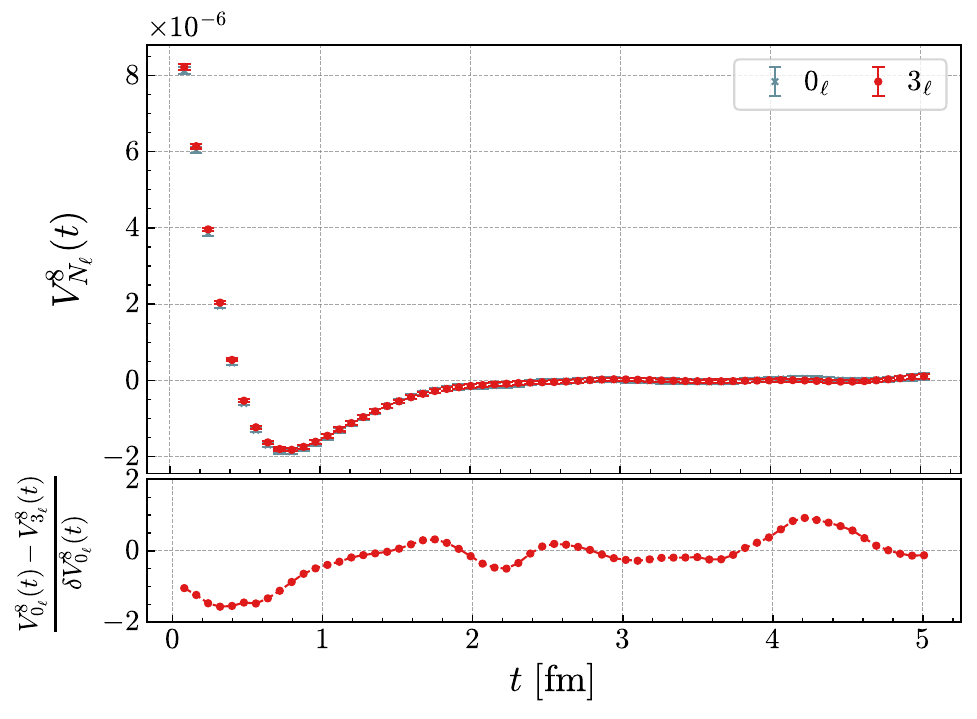}
\caption{
\textit{Left}: Top panel: gain in statistical precision for the octet quark-disconnected correlator obtained with the generalized OET (and $N_{\eta}=100$ volume sources) relative to the standard stochastic estimate obtained in Ref.~\cite{ExtendedTwistedMass:2022jpw} using about $3\times 10^{3}$ volume sources. Bottom panel: comparison of the statistical precision obtained using frequency splitting with $N_\ell=0,1,2$ intermediate masses, at fixed number of stochastic sources for each term.
\textit{Right}: Top panel: comparison of the octet quark-disconnected correlator obtained with $N_\ell=0$ and $N_\ell=3$. Bottom panel: corresponding difference normalized to the statistical uncertainty.
The different values of $N_\ell$ correspond to different computational costs, and the comparison is therefore not intended as a fixed-cost performance benchmark.
}
\label{fig:disco_corr}
\end{figure}

Representative results are shown in Fig.~\ref{fig:disco_corr} for the B64 ensemble, using $N_U\simeq750$ gauge configurations and $N_\eta=100$ stochastic volume sources, with $am_\ell=7.2\times10^{-4}$ and $m_s/m_\ell=25.4$. We denote by $0_\ell$ the result obtained without intermediate masses, while $1_\ell$, $2_\ell$, and $3_\ell$ correspond to frequency-splitting estimates with one, two, and three intermediate masses, respectively. The top-left panel shows the reduction of the statistical uncertainty obtained with the generalized OET relative to the result obtained in Ref.~\cite{ExtendedTwistedMass:2022jpw} using a naive stochastic estimate of the vector loops (the number of effective volume sources used in that work was about $3\times 10^{3}$). The bottom-left panel compares the statistical uncertainties obtained with the different frequency-splitting decompositions, while the right panels shows that the correlators obtained with $N_\ell=0$ and $N_\ell=3$ are statistically compatible. Since the different decompositions have different computational costs and the number of stochastic sources has not been optimized separately for the various terms, we do not use these comparisons to assess their relative computational efficiency.

With the setup adopted for production, namely $N_\ell=3$ and $N_\eta=\mathcal O(100)$ volume sources for each term, the stochastic noise of the vector loops is subleading and the variance is dominated by the gauge noise. Relative to the precision of our previous determination of the vector loops from Ref.~\cite{ExtendedTwistedMass:2022jpw}, obtained with naive stochastic estimators, the setup used in the present analysis reduces the statistical uncertainty by a factor of about seven.

\subsection{Analysis strategy and bounding method}
\label{seq:analysis-and-bounding}
Having described the numerical strategy used to evaluate the vector correlator
on the ETMC ensembles, we now explain how to obtain $a_{\mu}^{\rm HVP}$ from
the knowledge of $V(t)$. Starting from the decomposition introduced in
Eqs.~(\ref{eq:amu_HVP_f})-(\ref{eq:I0}), we first consider the isotriplet contribution,
$a_{\mu}^{\rm HVP}(I=1)$, and the octet contribution,
$a_{\mu}^{\rm HVP}({\rm oct})$. Following the notation of Subsec.~\ref{sec:LatJ}, we denote by $V^{I=1}(t)$ and $V^{8}(t)$ the
corresponding Euclidean correlation functions, obtained by replacing the
electromagnetic current in Eq.~\eqref{eq:VV} with the isotriplet current $j_{\mu}^{(3)}$ and the octet $j_{\mu}^{(8)}$ current in Eq.~\eqref{eq:I0_oct}.

In principle, their contribution to $a_{\mu}^{\rm HVP}$ could be obtained by
performing the numerical integration with the kernel function in Eq.~\eqref{eq:amu_HVP},
summing over all time slices available on a given lattice. However, this
strategy turns out to be suboptimal, since it leads to sizeable statistical
uncertainties originating from the large-time behavior of the correlation
functions, where the signal-to-noise ratio (S/N) decreases exponentially.

To tame this S/N problem, we adopt the so-called \textit{bounding method}.
This strategy, first applied in Ref.~\cite{Borsanyi:2016lpl} and subsequently refined in
Ref.~\cite{RBC:2018dos}, exploits the positivity of the spectral representation of the two
correlation functions $V^{I=1}(t)$ and $V^{8}(t)$ to derive rigorous bounds on their long-distance tail.
For $t>t_{\rm cut}$, one has
\begin{align}
V^{X}(t_{\rm cut})\, e^{-m_{\rm eff}^{X}(t^*)(t-t_{\rm cut})}
\leq
V^{X}(t)
\leq
V^{X}(t_{\rm cut})\, e^{-E_{0}^{X}(t-t_{\rm cut})},
\qquad
X=\{I=1,8\}~,
\end{align}
where $t^*<t_{\rm cut}$ is an arbitrary timeslice and
$m_{\rm eff}^{X}(t^*)$ denotes the effective mass extracted from the
correlator $V^{X}(t)$ at $t=t^*$. The quantity $E_{0}^{X}$ is any energy satisfying $E_{0}^{X} \leq E_{\rm{ground}}^{X}$, where $E_{\rm{ground}}^{X}$  denotes the energy of the lowest state contributing to $V^{X}(t)$.

The above inequalities can be used to construct upper- and lower-bounding
correlators,
\begin{align}
\label{eq:bounding-V-upper}
V_{\rm upper}^{X}(t, t_{\rm cut}) &=
\begin{cases}
V^{X}(t), & t\leq t_{\rm cut}, \\
V^{X}(t_{\rm cut})e^{-E_{0}^{X}(t-t_{\rm cut})}, & t>t_{\rm cut} ,
\end{cases}
\\[0.2cm]
\label{eq:bounding-V-lower}
V_{\rm lower}^{X}(t, t_{\rm cut}) &=
\begin{cases}
V^{X}(t), & t\leq t_{\rm cut}, \\
V^{X}(t_{\rm cut})e^{-m_{\rm eff}^{X}(t^*)(t-t_{\rm cut})}, & t>t_{\rm cut} ,
\end{cases}
\end{align}
which satisfy
\begin{align}
V_{\rm lower}^{X}(t,t_{\rm cut})
\leq
V^{X}(t)
\leq
V_{\rm upper}^{X}(t,t_{\rm cut}).
\end{align}
By inserting $V_{\rm upper}^{X}(t,t_{\rm cut})$ and $V_{\rm lower}^{X}(t,t_{\rm cut})$ into
Eq.~\eqref{eq:amu_HVP}, exploiting the fact that the kernel function $K(z)$ in Eq.~\eqref{eq:kernel} is strictly positive, one obtains corresponding upper and lower estimates for the HVP
contribution. As the value of $t_{\rm cut}$ is increased, the two estimates converge
towards each other and eventually become statistically compatible. The final
result is then extracted from a constant fit in this region, where the upper and lower bounds have converged within uncertainties.\footnote{We consider the upper and lower bounds to have converged when the central value of each bound lies within the errorbars of the other.} 

\begin{figure}
\centering
\includegraphics[width=0.48\linewidth]{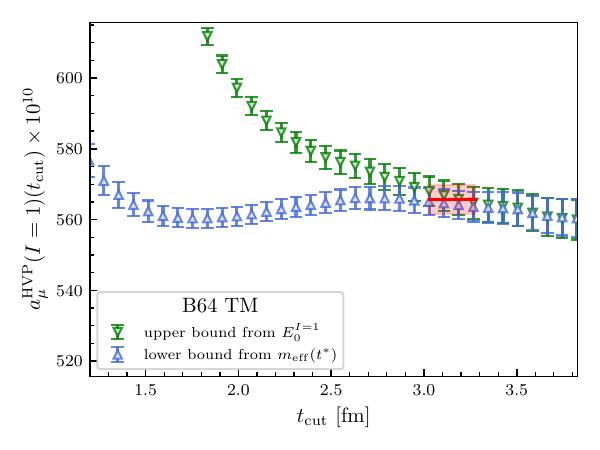}
\includegraphics[width=0.48\linewidth]{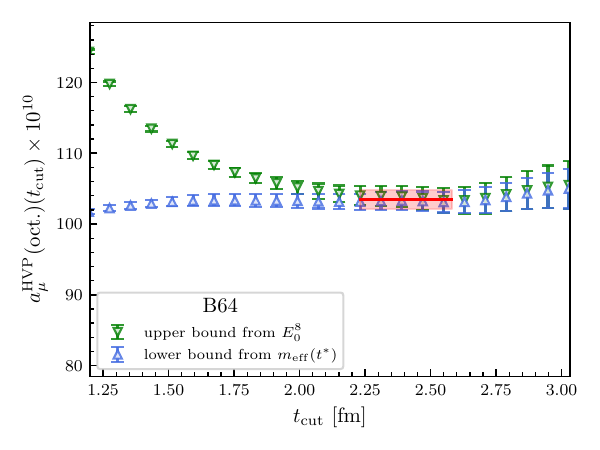}
\caption{\textit{Left}: bounding procedure used to determine $a_\mu^\mathrm{HVP}(I=1)$ in the TM regularization on the B64 ensemble. The green and blue triangles  correspond to the upper and lower bounds, respectively. The horizontal red band indicates our final determination obtained from a constant fit over a range of $0.25~{\rm fm}$. \textit{Right}: same as in the left panel, but for $a_\mu^\mathrm{HVP}({\rm oct})$ on the B64 ensemble. The horizontal red band indicates our final determination obtained from a constant fit over a range of about $0.32~{\rm fm}$. The illustrative results shown in this figure correspond to the determination in the FLAG scheme. All mistuning corrections have been included in the data shown.}
\label{fig:bounding}
\end{figure}

In Fig.~\ref{fig:bounding}, we illustrate the application of the bounding method to
$a_{\mu}^{\rm HVP}(I=1)$ (left panel) and
$a_{\mu}^{\rm HVP}({\rm oct})$ (right panel). For the isotriplet
contribution, we set $t^{*}=1.8~{\rm fm}$ and choose $E_{0}^{I=1}=0.98\,E_{0,\rm free}^{I=1}$, where
\begin{align}
E_{0,\rm free}^{I=1}
=
2\sqrt{m_{\pi}^{2}+\left(\frac{2\pi}{L}\right)^{2}}
\label{eq:E0f1_OS}
\end{align}
is the energy of the lowest noninteracting $P$-wave two-pion state in a finite box of spatial extent $L$. This expression is used for the OS regularization (see Eq.~\eqref{eq:V33}). In the TM regularization, however, the corresponding vector current in Eq.~\eqref{eq:V33} is effectively a charged current, so that the lowest two-pion state it interpolates is of the $\pi^{+}\pi^{0}$ type. At finite lattice spacing, twisted-mass QCD exhibits an isospin breaking between charged and neutral pions: while the charged pion (with mass $m_{\pi}$) is an exact Goldstone boson in the chiral limit even at finite lattice spacings ($m_{\pi}^{2} \propto m_{\ell}$), the neutral pion has a different mass $m_{\pi^{0}}$, differing from $m_{\pi}$ by cutoff effects,
\begin{align}
\label{eq:neutral_pion}
(m_{\pi^0})^{2}
=
m_{\pi}^{2}
+
Ca^{2}
+
O(a^{4})~,
\end{align}
with $C\simeq -0.044 \pm 0.008~{\rm GeV^{4} }$.\footnote{This value of $C$ results from a linear fit in $a^2$ of ETMC data on ensembles B64, C80, D96 for $m_{\pi^0}^2 - m_\pi^2$, which within errors (as large as 10\% -- 15\% on $m_{\pi^0}$) are well consistent with Eq.~(\ref{eq:neutral_pion}).}
To account for this effect, in the case of the TM regularization we have used instead
\begin{align}
E_{0,\rm free}^{I=1}
=
\sqrt{m_{\pi}^{2}+\left(\frac{2\pi}{L}\right)^{2}} + 
\sqrt{m_{\pi^{0}}^{2}+\left(\frac{2\pi}{L}\right)^{2}}~.
\label{eq:E0f1_TM}
\end{align}
In both the TM and OS case, the $2\%$ reduction in $E_{0}^{I=1}$ with respect to
the noninteracting value is motivated by the attractive nature of the
$\pi\pi$ interaction, which lowers the energy of the finite-volume states.
For the lowest-lying level, this effect is known to be very modest; in
particular, within the Meyer-Lellouch-L\"uscher-Gounaris--Sakurai (MLLGS) model~\cite{Meyer:2011um,Luscher:1990ux,Lellouch:2000pv,Gounaris:1968mw} (see Subsec.~\ref{sec:FSE} for an extended discussion of the model) it corresponds to
a downward shift of only about $1.4\%$ on boxes with $L\simeq 5.4~{\rm fm}$. The choice
$E_{0}^{I=1}=0.98\,E_{0,\rm free}^{I=1}$ therefore provides a conservative
lower bound on the energy of the lowest state contributing to the
correlator. The results displayed in the left panel of Fig.~\ref{fig:bounding}
correspond to the TM regularization on the B64 ensemble. The red band represents our final determination of $a_{\mu}^{\rm HVP}(I=1)$, while its horizontal extent indicates the time interval over which a constant fit has been performed, which is always chosen to be of about $0.25~{\rm fm}$. In the isotriplet channel, the upper and lower bounds become compatible
within uncertainties at $t_{\rm cut}\simeq 3~{\rm fm}$ on the ensembles
with spatial extents in the range $L\simeq 5$--$5.5~{\rm fm}$. On the
larger B96 ensemble, with $L\simeq 7.6~{\rm fm}$, the smaller energy of the
lowest-lying two-pion state leads to a slower convergence, and compatibility is reached only at
$t_{\rm cut}\simeq 3.7~{\rm fm}$.

In the case of the octet contribution, $a_{\mu}^{\rm HVP}({\rm oct})$, which was analyzed independently by a different subgroup, a very similar strategy was adopted. A minor difference concerns the choice of $t^{*}$, which was determined as the largest time for which the statistical uncertainty on the effective mass remained below $5\%$. Similarly to the case of the isotriplet contribution, the costant fit in this case is performed over an interval of about $0.3-0.35~{\rm fm}$. 

A second, more important, difference concerns the choice of the energy $E_{0}^{8}$. In the isospin-symmetric limit, the current $j_{\mu}^{8}$ does not couple to two-pion states (it is the $\omega$ channel, which decays into three pions in the isospin limit).\footnote{In our lattice setup, although isospin and parity are broken
at finite lattice spacing, charge conjugation remains exact and the
coupling of the spatial octet current to a $\pi^+\pi^-$ state in a
$P$-wave is forbidden by an exact discrete parity-flavor symmetry of
the twisted-mass regularization. In the $N_f=2$ theory, with only the
$u$ and $d$ quarks in the sea and
$j_\mu^{(8)}\propto\bar{q}_{u}\gamma_{\mu} q_{u}+\bar q_{d}\gamma_{\mu} q_{d}$, this symmetry is
$\mathcal P_F^{1,2}=\mathcal P\times(u\leftrightarrow d)$, where
$\mathcal P$ denotes parity and $u\leftrightarrow d$ denotes the interchange of the up- and down-quark fields.. In our $N_f=2+1+1$ setup, the $\mathcal P_{F}^{1,2}$ transformation must be augmented by a
$\tau^3$ transformation of the nondegenerate strange--charm doublet in the sea sector,
and in the valence sector by the exchange
$q_s\leftrightarrow q_s'$ of the valence quarks contributing to the full octet current
$j_\mu^{(8)\,{\rm OS}}$ in Eq.~\eqref{eq:J8_OS_current}. The $q_s''$ and $q_s'''$ valence fields entering
$j_\mu^{(8)\,{\rm OS}}$ have identical masses and Wilson-$r$ values and their contribution cancel in matrix elements to purely pionic states. The same selection
rule therefore applies to the full $N_f=2+1+1$ setup.} Consequently, the lightest multiparticle states contributing to the
correlator are three-pion states. In the continuum limit and neglecting
final-state interactions, the energy of the lowest-lying $\pi\pi\pi$
state is expected to be~\cite{Ce:2022eix}
\begin{align}
E_{3\pi} =
2\sqrt{ m_{\pi}^{2} + \left(\frac{2\pi}{L}\right)^{2}}
+
\sqrt{m_{\pi}^{2} + 2\left(\frac{2\pi}{L}\right)^{2}}~.
\end{align}
This expression follows from the requirement that the three-pion state carry the quantum numbers $J^{PC}=1^{--}$.

However, isospin and parity symmetries are broken at finite lattice spacing in twisted-mass QCD and, moreover, for the ensembles with
$L\in[5,5.5]~{\rm fm}$ one has $E_{3\pi}>m_\omega$. To remain
conservative, we therefore employ for $E_0^8$ a value lower than both
$E_{3\pi}$ and $m_\omega$ on all ensembles, given by
\begin{align}
E_{0}^{8} =2\sqrt{ m_{\pi}^{2} + \left(\frac{2\pi}{L}\right)^{2}}
+
m_{\pi^{0}}~,
\end{align}
where we remind that $m_{\pi^{0}}$ is the twisted-mass neutral pion mass (see Eq.~\eqref{eq:neutral_pion}). In the case of the octet contribution, the lower and upper bounds become compatible at values of $t_{\rm cut}$ in the range $2.2-2.5~{\rm fm}$ (see Fig.~\ref{fig:bounding} right), depending on the ensemble considered. 

For the charm-connected contribution, $a_{\mu}^{\rm HVP}(c)$, whose spectral decomposition also satisfies the positivity properties required by the bounding method, the same strategy could be applied in principle. However, the signal-to-noise ratio of the corresponding Euclidean correlator decreases only slightly with Euclidean time. As a result, a direct numerical evaluation of the integral in Eq.~\eqref{eq:amu_HVP} achieves essentially the same statistical precision as the bounding method. We therefore adopt the former approach, which is numerically simpler.

For the remaining contribution to $a_{\mu}^{\rm HVP}(\rm iso)$,
$\delta a_{\mu}^{c,\mathrm{disc}}$, which is expected to be dominated by
short-distance physics, we also rely on direct numerical integration. To
assess its numerical relevance, we evaluate this contribution on a
representative gauge ensemble and study the convergence of the corresponding
integral as a function of the upper integration limit in Euclidean time. We
find that the integral rapidly reaches a plateau and yields a contribution
that is negligible at the current level of precision (see discussion in Subsec.~\ref{sec:full_HVP_I0}). 

Finally, we note that for the short-distance and intermediate-window contributions, $a_{\mu}^{\rm SD}$ and $a_{\mu}^{\rm W}$, respectively, the long-distance region is exponentially suppressed by the time-modulating functions $\Theta^{\rm SD}(t)$ and $\Theta^{\rm W}(t)$. Therefore, the bounding method is unnecessary even for the isotriplet and octet contributions. For these quantities, we rely, for simplicity, on standard numerical integration.

\section{Results for the $I=1$ and $I=0$ contributions to $a_{\mu}^{\rm HVP}({\rm iso})$}
\label{sec:full_HVP}
In this section, we present our analyses of the isotriplet and isosinglet
contributions to $a_{\mu}^{\rm HVP}({\rm iso})$, starting from the
ensemble results obtained with the strategy described in
Sec.~\ref{seq:analysis-and-bounding}. In
Subsec.~\ref{sec:full_HVP_I1} we first determine the $I=1$ contribution
in the continuum limit at a fixed reference volume, separately in the
FLAG and WP25 schemes. The subsequent extrapolation from the reference
volume to infinite volume, which is particularly important in this
channel, is discussed in Subsec.~\ref{sec:FSE}. In
Subsec.~\ref{sec:full_HVP_I0} we determine the $I=0$ contribution
through its decomposition into the octet term, the charm-connected
contribution, and the residual disconnected contribution containing at
least one charm loop, denoted by $\delta a_{\mu}^{c,{\rm disc}}$.
Finally, in Subsec.~\ref{sec:full_HVP_summary} we summarize the results
and compare our determinations of the $I=1$, octet, and charm-connected
contributions with those obtained by other lattice collaborations.

\subsection{The $I=1$ contribution}
\label{sec:full_HVP_I1}

Our first goal is to disentangle the continuum-limit extrapolation from the
sizable finite-volume effects affecting the isotriplet channel. These
effects arise because the infinite-volume continuous $I=1$ two-pion spectrum is replaced in finite volume by a discrete
tower of $\pi\pi$ energy levels. We therefore proceed in two steps. We
first extrapolate the lattice results to the continuum limit at a fixed spatial
volume, chosen to be
\begin{align}
V_{\rm ref}=L_{\rm ref}^{3}\,,
\qquad
L_{\rm ref}=5.46~{\rm fm}\,.
\label{eq:Lref}
\end{align}
In the FLAG scheme, this value coincides with the spatial extents of the C80,
D96, and E112 ensembles to better than $0.4\%$.\footnote{\label{fn:Lref_scheme}Changing the input
value of $f_{\pi}$ used for scale setting induces small shifts in the lattice
spacings and, consequently, in the physical volumes of the ensembles; the corresponding values in the $f_{\pi}^{131}$-scheme are reported in Table~\ref{tab:iso_def}. As a result, the reference volumes at which the continuum extrapolations are carried out in the
different fixed-$f_{\pi}$ schemes explored below differ slightly, by at most $0.04~{\rm fm}$ in linear extent. After taking the
continuum limit in each scheme, we correct the result to the common value
$L_{\rm ref}=5.46~{\rm fm}$. The required finite-volume shift, estimated using
the MLLGS representation discussed in Subsec.~\ref{sec:FSE}, is at most $0.5\times10^{-10}$, corresponding
to less than $0.1\sigma$, and is therefore numerically negligible.}
Once the continuum result at $L_{\rm ref}$ is determined, the correction needed to extrapolate from $L_{\rm ref}$ to the infinite-volume limit is evaluated in
Subsec.~\ref{sec:FSE}.

The value of $a_{\mu}^{\rm HVP}(I=1)$ on each ensemble is extracted with
the bounding method described in
Subsec.~\ref{seq:analysis-and-bounding}. As discussed in
Subsec.~\ref{sec:LatJ}, the isotriplet correlator is evaluated using
both the TM and OS discretizations of the vector current. These two regularizations differ only by
cutoff effects. We
therefore jointly extrapolate both regularizations to the continuum limit and
constrain them to a common value at $a=0$.

Before describing the continuum-limit extrapolation at
$L=L_{\rm ref}$, we subtract the tree-level lattice artifacts generated
by the short-distance part of the time integral in
Eq.~\eqref{eq:amu_HVP}. The integration of the region $t\sim O(a)$
generates a potentially harmful term proportional to $a^{2}\log(a)$,
whose dominant part originates from the tree-level correlator. We remove
this dominant tree-level artifact in a controlled way by computing, for
each regularization and set of quark masses, the difference between the
lattice and continuum tree-level correlators and subtracting its
contribution to $a_{\mu}^{\rm HVP}$. If left unsubtracted, this
tree-level $a^{2}\log(a)$ artifact can mimic a regular $a^{2}$
dependence over the limited range of lattice spacings available in the
calculation and thereby bias the continuum extrapolation
~\cite{DellaMorte:2008xb,Ce:2021xgd,
ExtendedTwistedMass:2022jpw}. For the full isotriplet contribution, however, this bias is numerically
small compared with its uncertainty: in
Ref.~\cite{ExtendedTwistedMass:2022jpw}, it was shown to amount to about
$1\times10^{-10}$.\footnote{For the short-distance window itself, where
the statistical uncertainty is substantially smaller, these cutoff
effects become comparatively much more important. In the dedicated
analysis of Sec.~\ref{sec:windows}, we therefore employ two
complementary strategies. In the first, the continuum extrapolation is
supplemented by additional ansatzes containing terms proportional to
$a^{2}\alpha_s^n(1/a)\log(a)$, with $n=2,3$. In the second, an
ultraviolet regulator is introduced in the time integral and the
continuum limit is taken at fixed regulator, thereby avoiding the
logarithmically enhanced cutoff effects generated by the
$t\sim O(a)$ region; the missing short-distance contribution is
subsequently restored using continuum perturbation theory. The agreement
between the two determinations provides a stringent validation of the
treatment of short-distance cutoff effects.}

For regularization
${\rm r}={\rm TM},{\rm OS}$, we define the tree-level integrand difference
\begin{align}
W^{I=1,{\rm r}}_{{\rm TL}}(t;a,m_\ell )
={}&
t^{2}K(m_{\mu}t)\,\Theta^{\rm SD}(t)
\left[
V^{ I=1,{\rm r}}_{{\rm TL, lat}}(t;a,m_\ell)
- V^{ I=1}_{{\rm TL, cont}}(t;m_\ell )
\right]~,
\label{eq:TL_integrand}
\end{align}
where ``TL'' denotes the free-field contribution at order
$O(\alpha_s^0)$,
$V^{ I=1,{\rm r}}_{{\rm TL, lat}}$ and
$V^{ I=1}_{{\rm TL, cont}}$ are the lattice and continuum tree-level
correlators, respectively, and $m_\ell$ denotes the (bare) light-quark
mass. The associated
correction is evaluated directly as the discrete sum
\begin{align}
\Delta_{\rm TL}a_{\mu}^{\rm HVP,r}(I=1, a)
={}&
2\alpha_{\rm em}^{2}\,a
\sum_{n=1}^{T/(2a)}
W^{I=1,{\rm r}}_{{\rm TL}}(na;a,m_\ell)\,.
\label{eq:TL_improvement}
\end{align}
The
ensemble value entering the continuum-limit extrapolation is then obtained by performing the following subtraction
\begin{align}
a_{\mu}^{\rm HVP,r}(I=1,a)
\rightarrow a_{\mu}^{\rm HVP,r}(I=1,a) - \Delta_{\rm TL}a_{\mu}^{\rm HVP,r}(I=1, a)\,.
\label{eq:TL_improved_observable}
\end{align}
In the remainder of this section all the results presented include the tree-level improvement above. This procedure follows exactly the strategy employed in our
calculation of the strange and charm contributions to the HVP
~\cite{ExtendedTwistedMass:2024nyi}.

\begin{figure}[t]
\centering
\includegraphics[width=\linewidth]{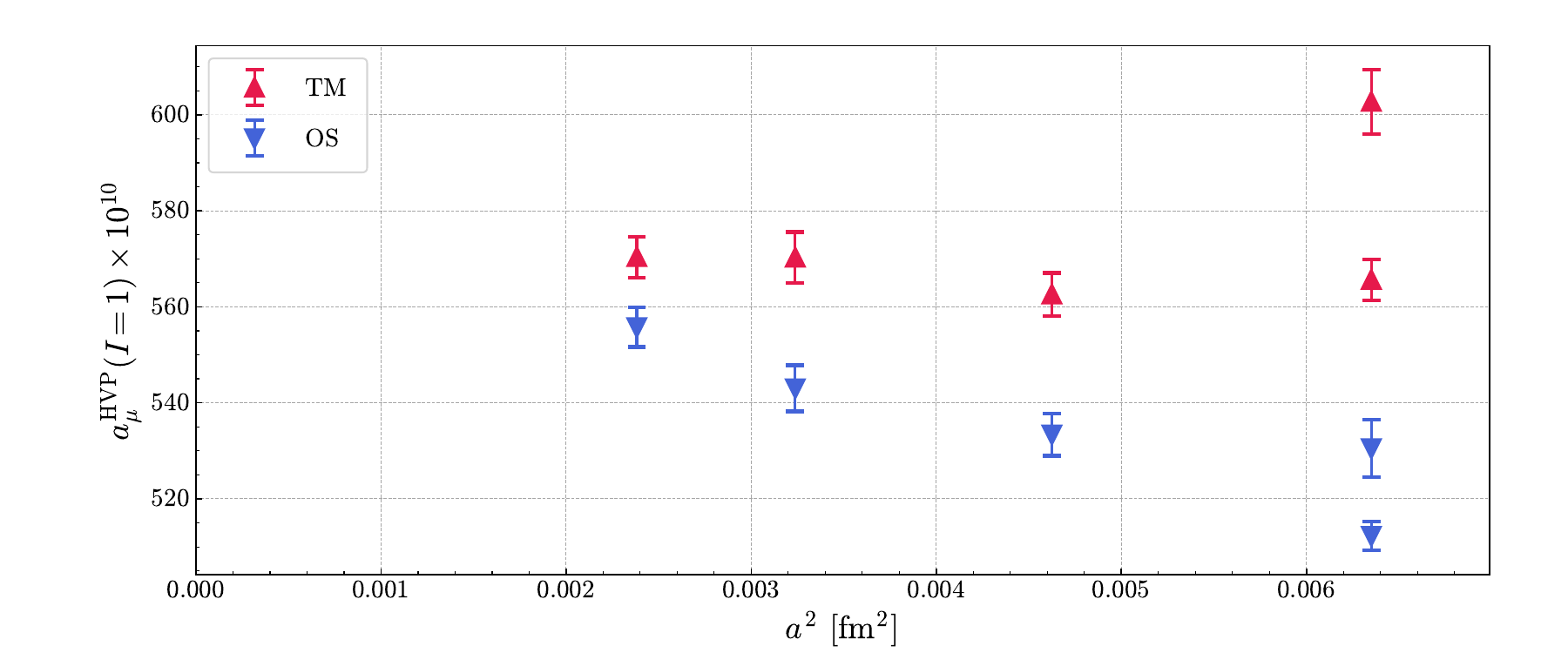}
\caption{Results for $a_{\mu}^{\rm HVP}(I=1)$ in the FLAG scheme as a
function of $a^{2}$. At the coarsest lattice spacing, two points are shown for each discretization, corresponding to the B64 (lower points) and B96 (higher points) ensembles.
The ensemble values are obtained using the bounding method
described in Sec.~\ref{seq:analysis-and-bounding}.}
\label{fig:I1_FLAG_ensembles}
\end{figure}

The ensemble results in the FLAG scheme are shown in
Fig.~\ref{fig:I1_FLAG_ensembles} as a function of the lattice spacing
squared. The TM and OS determinations are displayed separately.
The TM data are remarkably flat over the range of lattice spacings
considered, with no statistically significant dependence on the cutoff, while
the OS data exhibit a somewhat more visible, although still moderate,
dependence on $a^{2}$. The B64 result has the smallest statistical
uncertainty, reflecting the larger statistics employed on this ensemble,
as reported in Table~\ref{tab:statistics}.

The C80, D96, and E112 results can be used directly in the fixed-volume
continuum analysis, since their spatial extents are already very close to
$L_{\rm ref}$. At the coarsest lattice spacing, instead, the two
available ensembles bracket the reference volume: B64 has
$L\simeq5.1~{\rm fm}$, while B96 has $L\simeq7.7~{\rm fm}$. Since these
two ensembles have the same lattice spacing and bare parameters, they
allow us to resolve the volume dependence at fixed cutoff and to
interpolate the B-ensemble result to $L_{\rm ref}$. This interpolation is
performed independently for the TM and OS regularizations.

\begin{figure}[t]
\centering
\includegraphics[width=0.48\linewidth]{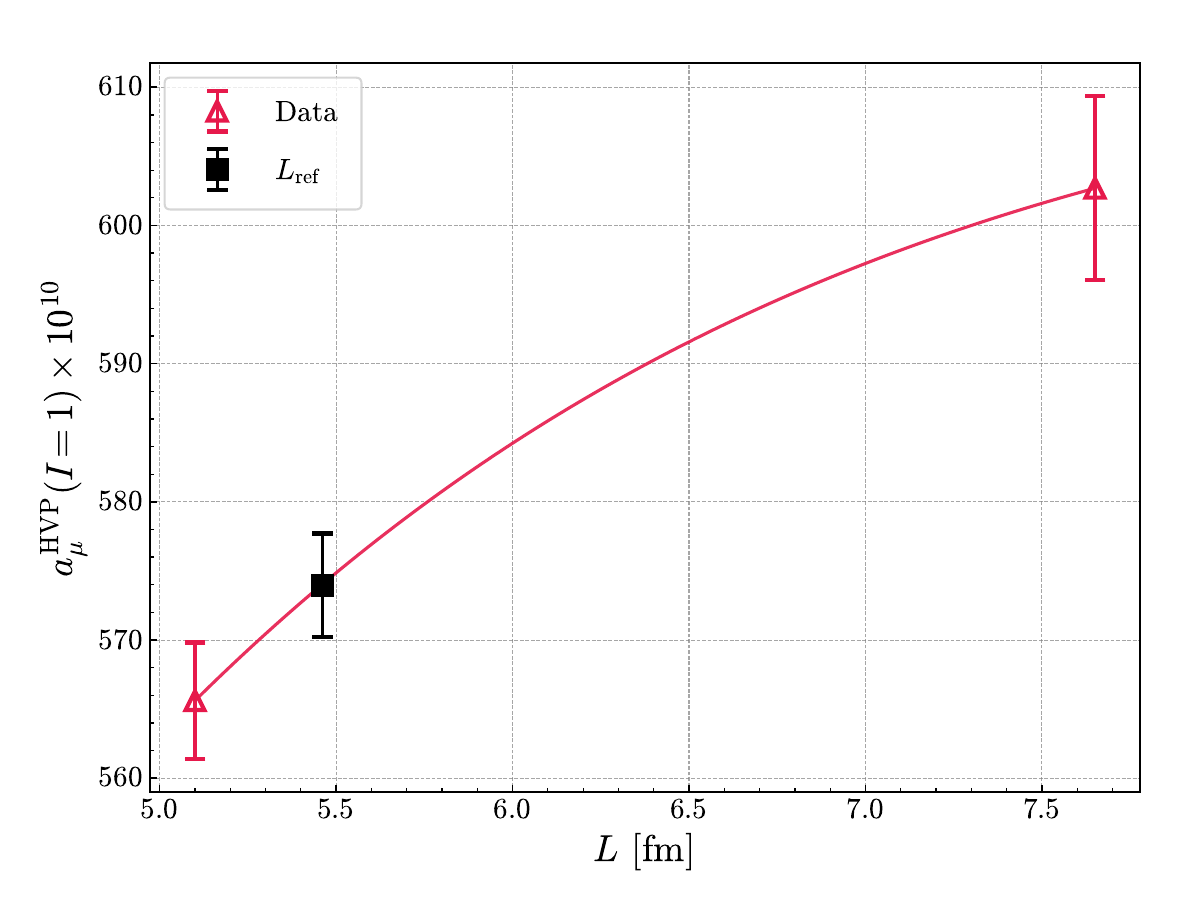}
\includegraphics[width=0.48\linewidth]{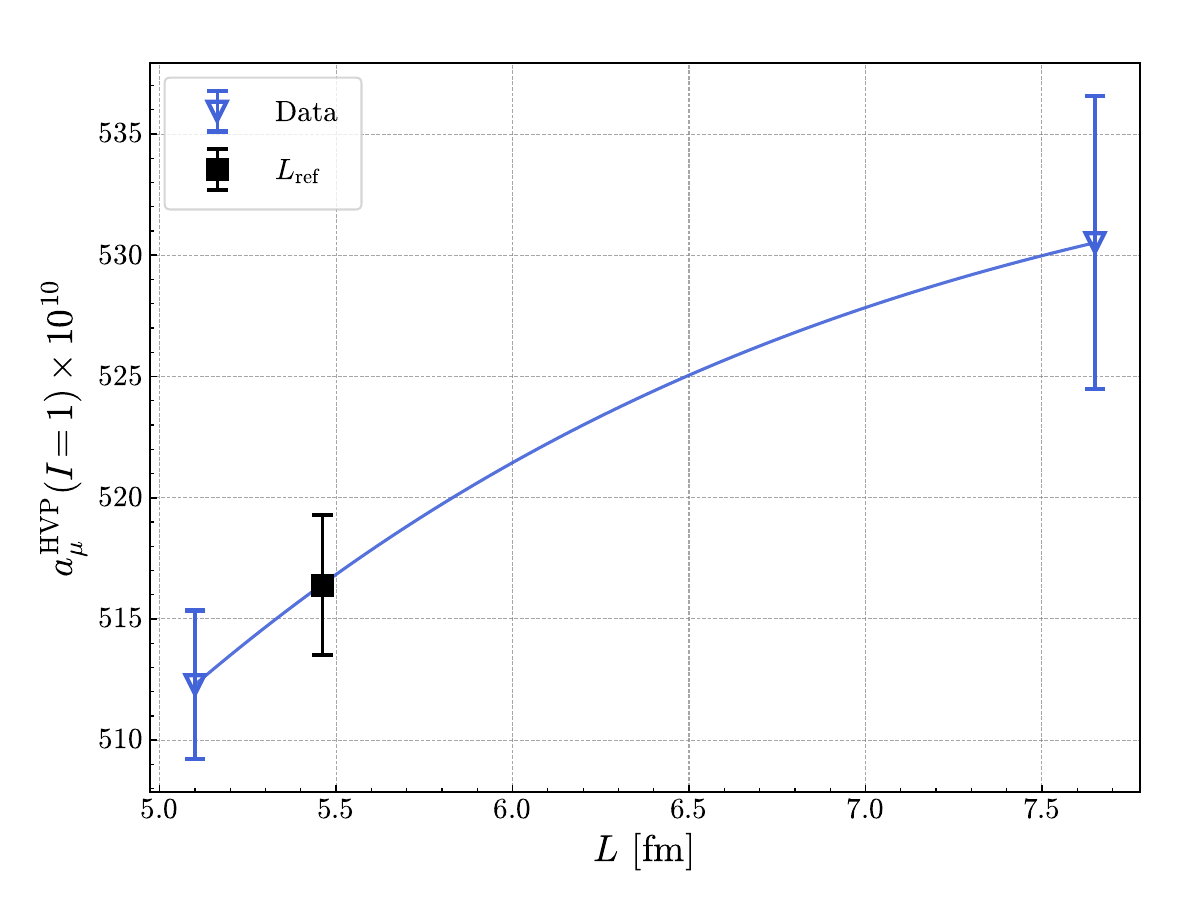}
\caption{Interpolation of the B64 and B96 determinations of
$a_{\mu}^{\rm HVP}(I=1;a_B,L)$ to the reference volume
$L_{\rm ref}=5.46~{\rm fm}$ for the TM ({\it left}) and OS ({\it right})
regularizations. The interpolation is performed using the functional form in Eq.~\eqref{eq:I1_L_interpolation}. The black squares indicate the values obtained at
$L=L_{\rm ref}$.}
\label{fig:interpolation_volume}
\end{figure}

To interpolate the results obtained on the B64 and B96 ensembles to the
reference volume, we employ a simple parameterization inspired by the
finite-volume behavior predicted by the MLLGS representation, which
will be discussed in detail in Subsec.~\ref{sec:FSE}. In the range of
volumes spanned by the two B ensembles,
$L\in[5.1,7.7]~{\rm fm}$, the MLLGS finite-volume dependence can be well described by the functional form\footnote{The same functional form provides an equally good description when the MLLGS model is evaluated using unequal charged- and neutral-pion masses, to account for the $O(a^2)$ charged--neutral pion mass splitting arising at finite lattice spacing in twisted-mass QCD, as discussed extensively in Subsec.~\ref{sec:FSE}. The corresponding coefficients $c_0$ and $c_1$ are, of course, different.}
\begin{align}
a_{\mu}^{\rm HVP}(I=1;L)
=
c_{0}+c_{1}L\,e^{-m_{\pi}L}\,.
\label{eq:I1_L_interpolation}
\end{align}
For each regularization separately, the coefficients $c_{0}$
and $c_{1}$ are determined by fitting the B64 and B96 lattice results, and the
fitted function is subsequently evaluated at $L=L_{\rm ref}$. The interpolated value is tightly
constrained by the precise B64 determination and is only mildly sensitive
to the assumed volume dependence between B64 and B96. In
Fig.~\ref{fig:interpolation_volume}, we show the volume interpolation for
the TM and OS regularizations in the left and right panels, respectively. The interpolated points at $L=L_{\rm ref}$ are indicated by
the black square markers.

The value interpolated to $L_{\rm ref}$ replaces the B64 and B96
determinations in the continuum analysis. The resulting fixed-volume data
set therefore consists of one B, one C, one D, and one E point for each
regularization, for a total of eight data points. Correlations are retained
between the TM and OS determinations at the same lattice spacing, whereas
data points corresponding to different lattice spacings are
uncorrelated. The TM--OS correlation coefficient is approximately $0.2$ at the coarsest lattice spacing and increases, as expected, towards the continuum,
reaching approximately $0.7$ on the E112 ensemble.

The continuum extrapolation at $L=L_{\rm ref}$ has been carried out
independently by three analysis groups, which adopted slightly different
analysis choices and implementations. These alternative strategies and
their mutual consistency are discussed in Appendix~\ref{sec:blinding}. The strategy described below is the common final analysis on
which the three groups converged after comparing their independent
implementations and discussing the methodological differences, and it is the
one used to obtain the results quoted in the main text.

\begin{figure}[t]
\centering
\includegraphics[width=\linewidth]{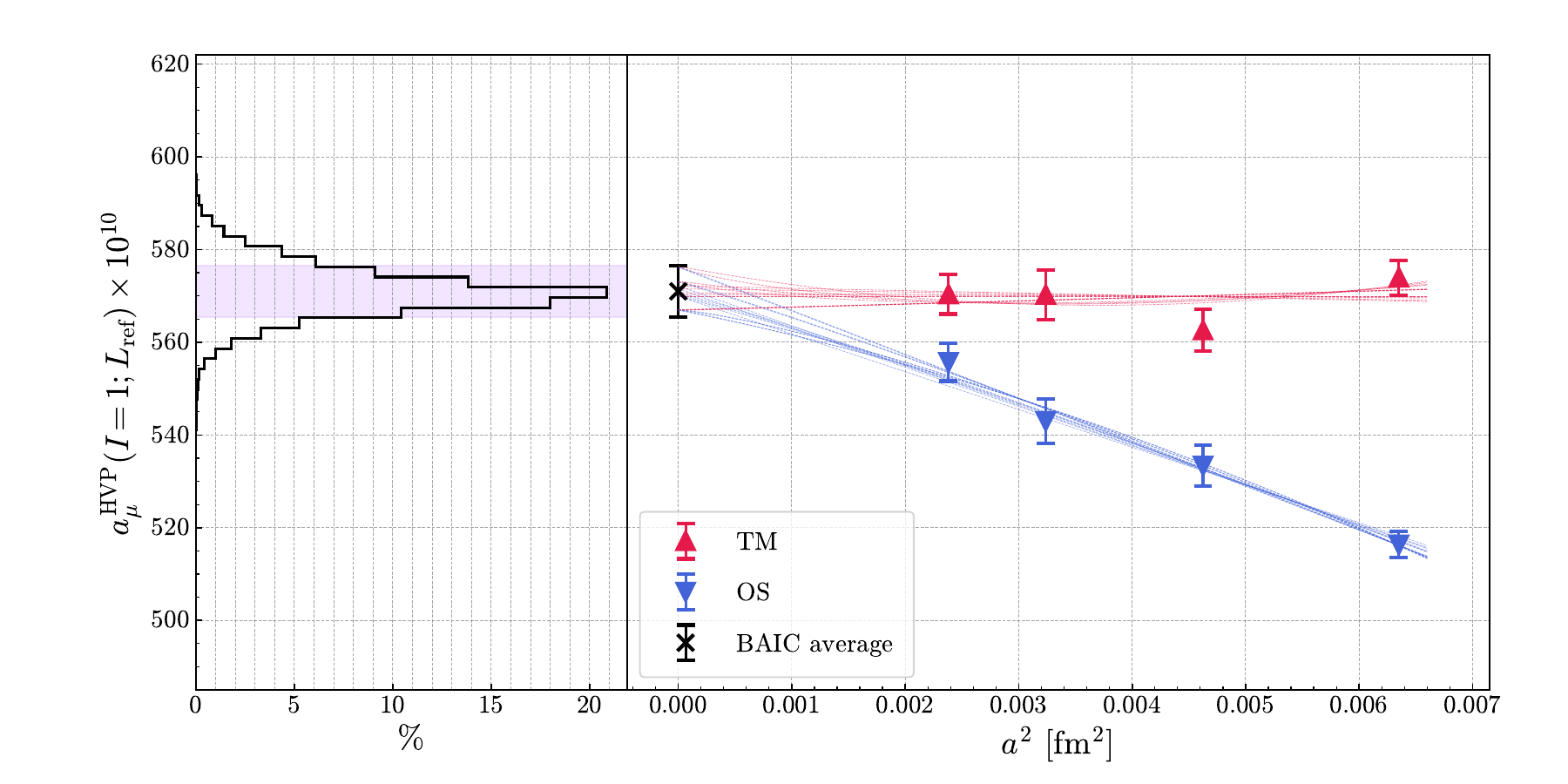}
\caption{Continuum extrapolation of
$a_{\mu}^{\rm HVP}(I=1;L_{\rm ref})$ in the FLAG scheme. The TM and OS
data are fitted simultaneously with a common continuum limit. A
representative subset of the fit functions entering the model average is displayed
with dashed lines. The continuum BAIC average is shown at $a^{2}=0$,
while the distribution on the left represents the corresponding
model-averaged probability distribution.}
\label{fig:I1_FLAG_cont}
\end{figure}

As discussed in Subsec.~\ref{sec:LatJ}, the TM and OS discretizations provide two lattice regularizations of the same continuum observable. Universality therefore requires them to converge to a common continuum limit. We exploit this theoretical constraint by fitting simultaneously the TM and OS data, imposing a common value at $a=0$ while allowing for regularization-dependent cutoff effects. Our fit functions are taken to be of the form
\begin{align}
a_{\mu}^{{\rm HVP,r}}(I=1;a,L_{\rm ref})
={}&
a_{\mu}^{\rm HVP}(I=1;L_{\rm ref})
+c_{1}^{\rm r}a^{2}
+c_{1L}^{\rm r}
\frac{a^{2}}
{\left|\log(a^{2}\Lambda_{0}^{2})\right|^{\gamma}}
+c_{2}^{\rm r}a^{4}\,,
\qquad
{\rm r}={\rm TM},{\rm OS}\; ,
\label{eq:I1_continuum_ansatz}
\end{align}
where the first term is the common continuum value, while $c_{1}^{\rm r}$,
$c_{1L}^{\rm r}$, and $c_{2}^{\rm r}$ parameterize, respectively, a regular
$a^{2}$ term, a logarithmically modified $a^{2}$ contribution,
and an $a^{4}$ term for the regularization
${\rm r}={\rm TM},{\rm OS}$. We set $\Lambda_{0}=300~{\rm MeV}$ and fix $\gamma=0.42$ as an
effective choice for the logarithmically modified $O(a^2)$ term. The
analyses of Refs.~\cite{Husung:2022kvi,Husung:2024cgc} in the Symanzik
effective theory indicate, for the leading nonvanishing logarithmic
corrections, an exponent of about $0.21$ associated with the
dimension-six operators entering the Symanzik effective action, and an
exponent of about $0.42$ associated with the dimension-five operators
entering the Symanzik expansion of the current, after a naive
extrapolation to $N_f=4$. Since the case $\gamma=0$, corresponding to a
regular $a^2$ dependence, is already explicitly included through the
$c_1^{\rm r} a^2$ term, we use $\gamma=0.42$ as a representative nonzero
logarithmic exponent. The smaller value $\gamma\simeq0.21$ produces
an intermediate behavior between $\gamma=0$ and $\gamma=0.42$.

Individual fit functions are obtained from Eq.~\eqref{eq:I1_continuum_ansatz} by setting selected coefficients to zero. We never allow all coefficients to vary simultaneously: all fits contain at most four free parameters, including the common continuum-limit value. This restriction avoids overfitting and appears appropriate in view of the mild cutoff effects observed in our lattice data, see Fig.~\ref{fig:I1_FLAG_cont}. Particularly notable is the behavior of the TM data, which are nearly flat in $a^2$ within the quoted statistical uncertainties.

Given the moderate amount of lattice artifacts we observe and in order not to lose statistical information, all the fits entering our final analysis are performed on the complete eight-point data set.\footnote{Fits based on reduced data sets, obtained by removing the
coarsest or the second-coarsest lattice spacing, were also considered by
analysis group B in Appendix~\ref{sec:A1}. The corresponding
model-averaged result is compatible with the final determination. Given
the mild cutoff dependence observed here, and in particular the nearly
flat behaviour of the TM data, we have decided to retain all four
lattice spacings in the final analysis, since removing one of them would
leave only three lattice spacings to determine such a small cutoff
dependence, making the extrapolation more sensitive to statistical
fluctuations of the remaining data points.} The fit pool consists of the following ansatzes, ordered by the number of fit parameters:
\begin{itemize}
\item the two fits with two free parameters in which the TM data are fitted to a constant, while the OS data contain either a regular $a^2$ term or a logarithmically modified $a^2$ term;
\item the four fits with three free parameters in which each regularization contains one cutoff term, chosen independently to be either proportional to $a^2$ or to $a^2/|\log(a^2\Lambda_0^2)|^\gamma$;
\item the eight fits with four free parameters in which one regularization contains a single cutoff term, proportional either to $a^2$ or to $a^2/|\log(a^2\Lambda_0^2)|^\gamma$, while the other contains a regular $a^2$ term supplemented by either a logarithmically modified $a^2$ term or an $a^4$ term, with the roles of TM and OS interchanged.
\end{itemize}
The resulting pool includes the simplest fit forms that describe the smooth $a^2$ dependence of the data, together with fit functions allowing for a logarithmic modification or curvature in one of the two regularizations, while keeping the number of fit parameters reasonably small.

The results of the different continuum extrapolations are combined using
the Bayesian Akaike Information Criterion (BAIC)~\cite{Neil:2022joj}. For a fit $k$, we define
\begin{align}
\widetilde w_{k}^{\rm BAIC}
=
\exp\left[
-\frac{1}{2}
\left(
\chi_{k}^{2}
+2N_{{\rm par},k}
-2N_{{\rm data},k}
\right)
\right],
\qquad \qquad w_k^{\rm BAIC}=
\frac{\widetilde w_k^{\rm BAIC}}
{\sum_j\widetilde w_j^{\rm BAIC}}\,.
\label{eq:BAIC_weight}
\end{align}
Denoting by $x_k^{(b)}$ the continuum result from a fit $k$ on
bootstrap sample $b$, and by $x_{k}$ its mean value, we define the model-averaged bootstrap samples and
the corresponding central value as
\begin{align}
\overline{x}^{(b)}
=
\sum_k w_k^{\rm BAIC}x_k^{(b)}\,,
\qquad
\overline{x}
=
\left\langle\overline{x}^{(b)}\right\rangle_b\,.
\label{eq:BAIC_bootstrap_average}
\end{align}
The statistical uncertainty is the bootstrap width of the distribution
of $\overline{x}^{(b)}$. This procedure correctly propagates the correlations among
the continuum results obtained from the different fits. The systematic
uncertainty associated with the continuum extrapolation is instead
estimated from the weighted spread of the central fit results,
\begin{align}
\sigma_{\rm syst}^{2}
=
\sum_k w_k^{\rm BAIC}
\left(x_k-\overline{x}\right)^{2}\,.
\label{eq:BAIC_syst}
\end{align}
The statistical and systematic errors are combined in
quadrature.

\begin{figure}[t]
\centering
\includegraphics[width=\linewidth]{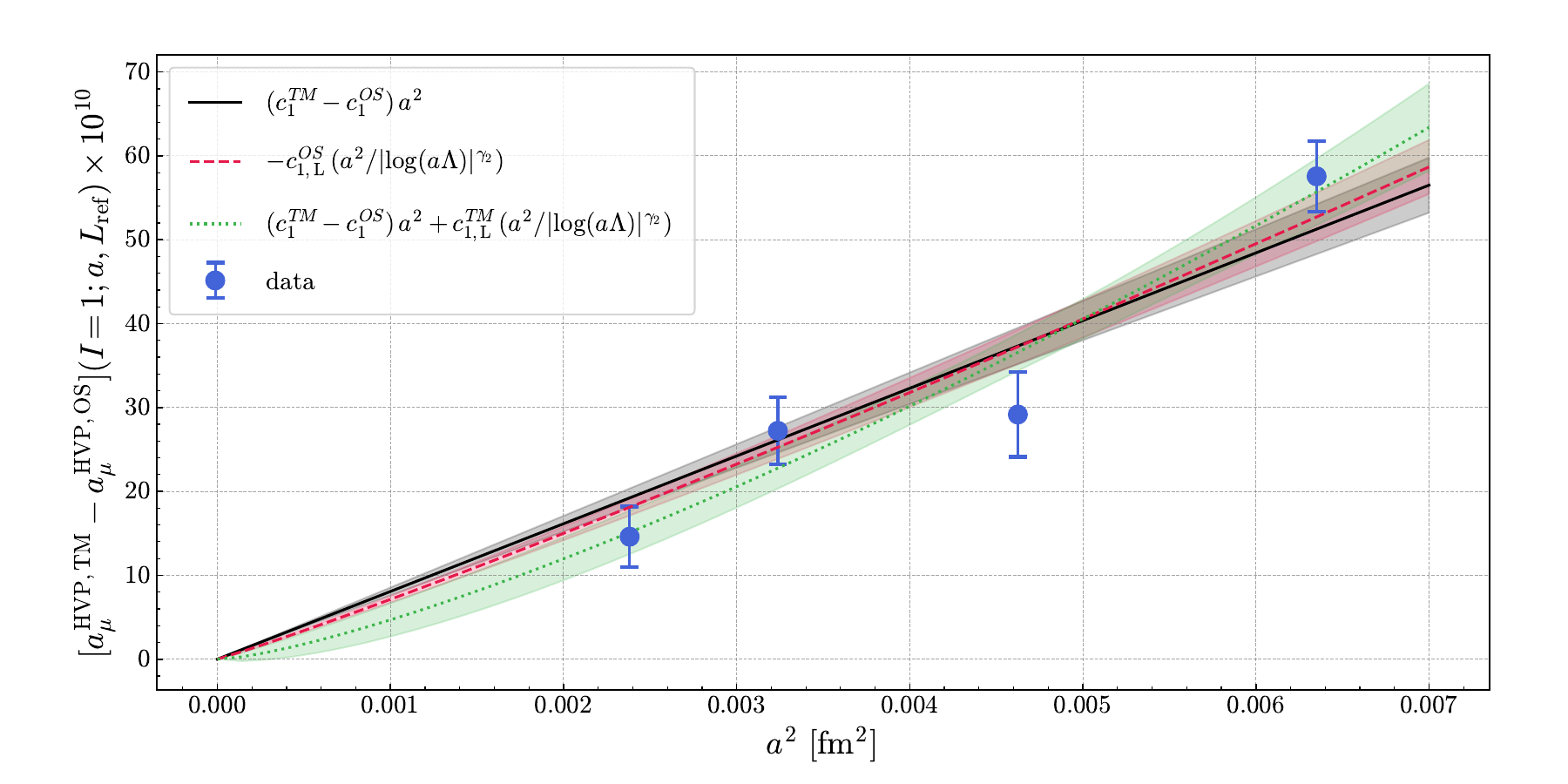}
\caption{Difference
$
[a_{\mu}^{\rm HVP,TM}-a_{\mu}^{\rm HVP,OS}](I=1;a,L_{\rm ref})$
as a function of $a^2$, together with the corresponding $a^2$ dependence obtained from a representative subset of the best-fit functions entering the model average. Since the two regularizations are constrained to a common continuum limit, all curves vanish at $a=0$ by construction.}
\label{fig:I1_diff_fit}
\end{figure}

Fig.~\ref{fig:I1_FLAG_cont} summarizes the continuum-limit extrapolation. The TM and OS data at the four lattice spacings, after the interpolation to $L_{\rm ref}$, are shown together with a representative subset of the fit functions entering the model average, displayed as dashed lines. We recall that all fits share a common continuum limit for the TM and OS regularizations. The resulting BAIC-averaged continuum
value is shown at $a^{2}=0$. On the left side of the figure, we display
the BAIC-weighted distribution of the continuum-limit determinations
obtained from the different fits we performed. More explicitly,
each bootstrap determination $x_k^{(b)}$ obtained with fit $k$ contributes
to the histogram with weight $w_k^{\rm BAIC}$. The resulting distribution
therefore reflects both the statistical error of each fit and
the spread associated with the different continuum extrapolations.

As a complementary illustration of the fit quality, we show in
Fig.~\ref{fig:I1_diff_fit}  the difference between the TM and OS determinations as a function of $a^{2}$, together with the corresponding $a^{2}$ dependence obtained from representative best-fit functions entering the model average. The corresponding curves necessarily vanish in the continuum limit and provide a direct check of how well the fitted regularization-dependent cutoff effects describe the observed TM--OS splitting at finite lattice spacing.

Our final result in the FLAG scheme is
\begin{align}
a_{\mu}^{\rm HVP,FLAG}(I=1;L_{\rm ref})
=
\amufullIoneFLAG \times10^{-10}\,.
\label{eq:I1_FLAG_Lref}
\end{align}
The first uncertainty is statistical, while the second is the systematic
uncertainty associated with the continuum-limit extrapolation; the
uncertainty in square brackets is their combination in quadrature.
The relative uncertainty is slightly below $1\%$.
Eq.~\eqref{eq:I1_FLAG_Lref} is the continuum-extrapolated result at the finite
volume $V=L^{3}_{\rm ref}$; the finite-volume correction required to obtain
the infinite-volume result has not yet been applied and will be discussed
in Subsec.~\ref{sec:FSE}.

We now turn to the determination of the $I=1$ contribution in the WP25 scheme, $a_{\mu}^{\rm HVP, WP25}(I=1;L_{\rm ref})$. As discussed in
Sec.~\ref{sec:scale setting}, the WP25 scheme uses the
gradient-flow scale $w_{0}$, rather than $f_{\pi}$, as
input parameter, while the remaining hadronic inputs are the
same as in the FLAG scheme. In Sec.~\ref{sec:scale setting}, we established
the equivalent $f_{\pi}$-based characterization of the WP25 scheme that corresponds to $f_{\pi}^{\rm WP25}=131.00(27)~{\rm MeV}$ (see Eq.~\eqref{eq:fpi_WP25}). This correspondence allows us to avoid the comparatively large cutoff
effects observed when $w_{0}$ is used directly in the analysis of $a_{\mu}^{\rm HVP}$. In practice, we repeat
the complete continuum analysis in a set of nearby schemes in which the
input value of $f_{\pi}$ is fixed to different values, fit the dependence of the continuum
result on $f_{\pi}$, and finally evaluate the fitted function at
the value in Eq.~\eqref{eq:fpi_WP25}, propagating its uncertainty.

\begin{figure}[t]
\centering
\includegraphics[width=0.49\linewidth]{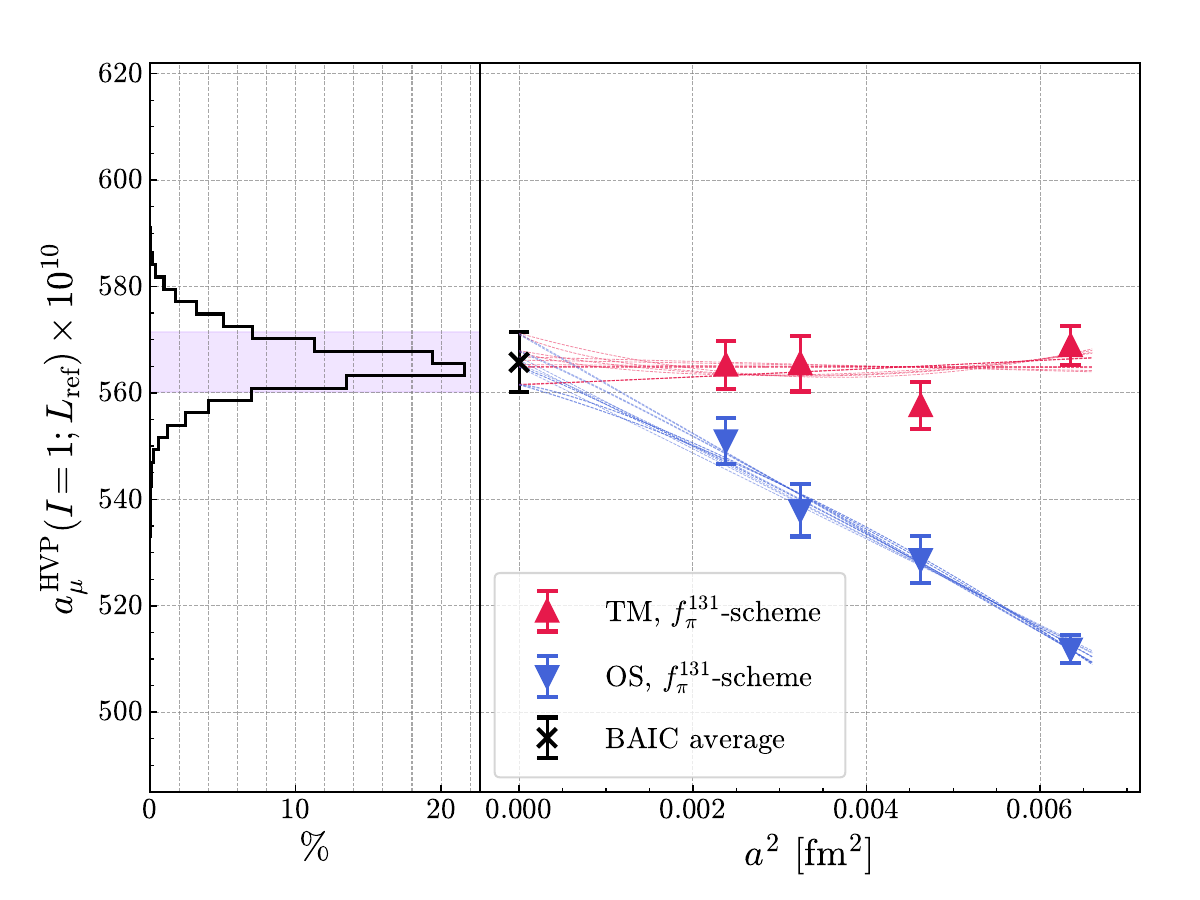}
\includegraphics[width=0.49\linewidth]{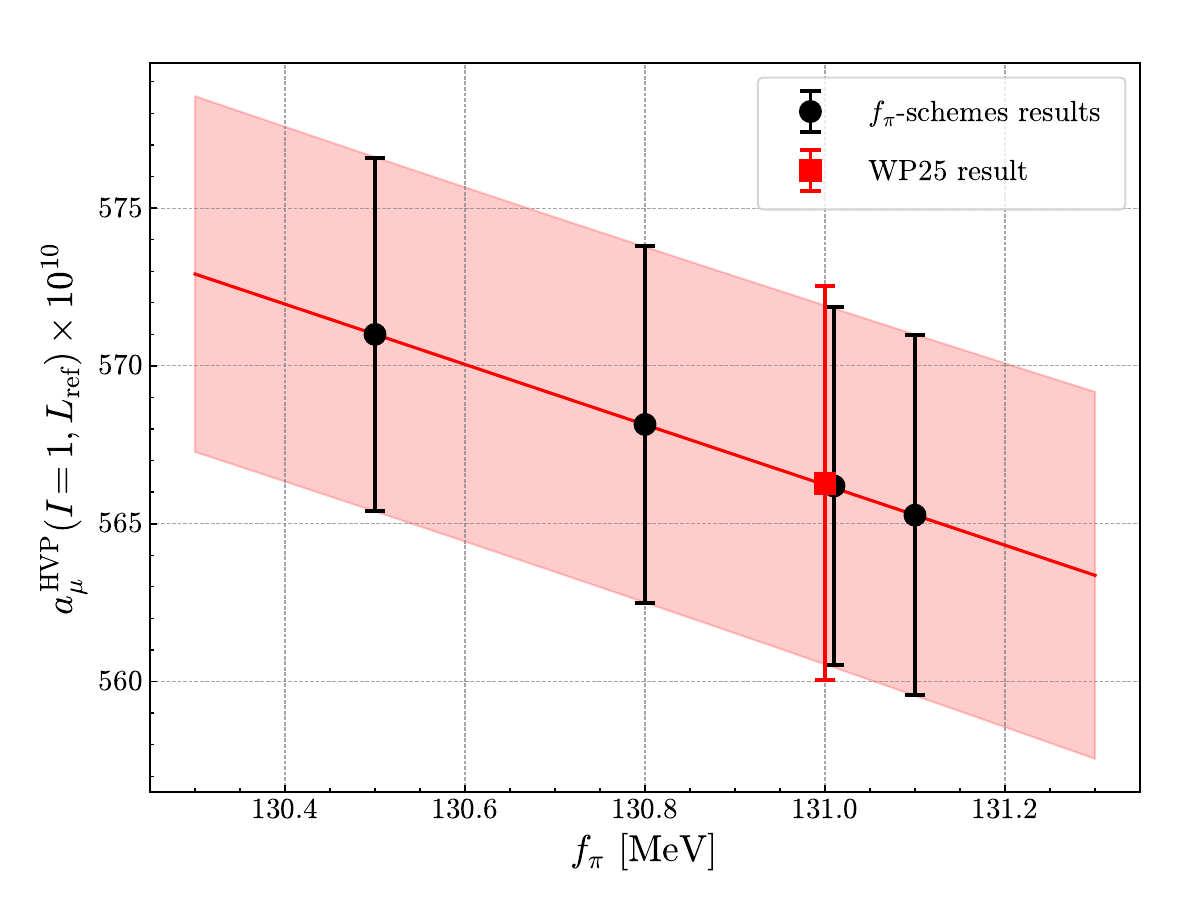}
\caption{
\textit{Left}:
Same as Fig.~\ref{fig:I1_FLAG_cont}, but in the
$f_{\pi}^{131}$ scheme.
\textit{Right}:
Dependence of the continuum result for
$a_{\mu}^{\rm HVP}(I=1;L_{\rm ref})$ on the input value of $f_{\pi}$
used to set the scale. The colored band represents the linear fit in
Eq.~\eqref{eq:I1_fpi_interp}. The data point indicated by the red square corresponds to our final result in the WP25 scheme. The value corresponding to the $f_{\pi}^{131}$-scheme has been slightly shifted horizontally for better visualization.}
\label{fig:I1_fpi131_cont_and_WP_int}
\end{figure}

As a representative example, the left panel of
Fig.~\ref{fig:I1_fpi131_cont_and_WP_int} shows the continuum extrapolation in the
fixed-input $f_{\pi}^{131}$ scheme. The analysis follows exactly the
same procedure as in the FLAG scheme: the B64 and B96 results are first
interpolated to a reference volume, the TM and OS data are then
extrapolated simultaneously to a common continuum limit using the same fit
functions, and the resulting continuum
determinations are combined using the same model-averaging
prescription. As discussed in footnote~\ref{fn:Lref_scheme}, for the fixed-$f_\pi$ schemes with $f_\pi\neq f_\pi^{\rm FLAG}$ the reference volume differs slightly from $L_{\rm ref}=5.46~{\rm fm}$. This small difference is corrected after the continuum extrapolation, so that all results correspond to the common value $L_{\rm ref}=5.46~{\rm fm}$. For presentation, the data and fit curves shown in the left panel of Fig.~\ref{fig:I1_fpi131_cont_and_WP_int} have already been shifted by this small finite-volume correction, so that the plot is directly referred to the common value $L_{\rm ref}=5.46~{\rm fm}$.

Since the input values of $f_{\pi}$ considered in the analysis span an
interval smaller than $1~{\rm MeV}$, we describe the $f_{\pi}$ dependence of
$a_{\mu}^{\rm HVP}(I=1;L_{\rm ref})$ by the linear function
\begin{align}
a_{\mu}^{\rm HVP}(I=1;L_{\rm ref};f_{\pi})
={}&
c_{I=1}
+s_{I=1}
\left(f_{\pi}-f_{\pi}^{\rm FLAG}\right)~,
\label{eq:I1_fpi_interp}
\end{align}
and we recall that $f_{\pi}^{\rm FLAG}=130.5~{\rm MeV}$.
The intercept $c_{I=1}$ of the linear function is the value in the FLAG scheme, while the coefficient $s_{I=1}$ gives the
derivative of the isotriplet contribution with respect to the input value
of $f_{\pi}$. The right panel of Fig.~\ref{fig:I1_fpi131_cont_and_WP_int} shows the continuum
results obtained in the FLAG scheme and in the neighboring
$f_{\pi}$ schemes, together with the fitted linear dependence. The red
square is obtained by evaluating Eq.~\eqref{eq:I1_fpi_interp} at
$f_{\pi}=f_{\pi}^{\rm WP25}$ and propagating the uncertainty of
$f_{\pi}^{\rm WP25}$. It therefore represents our final result in the WP25
scheme.

We obtain
\begin{align}
a_{\mu}^{\rm HVP,WP25}(I=1;L_{\rm ref})
=
\amufullIoneWP \times10^{-10}
\,,
\label{eq:I1_WP25_Lref}
\end{align}
where the systematic error includes the uncertainty on $f_\pi^{\rm WP25}$. This result corresponds to a downward shift with respect to Eq.~\eqref{eq:I1_FLAG_Lref} by about $5\times 10^{-10}$.
The slope determined from the same fit is\footnote{The finite-volume correction from the common reference volume
$L_{\rm ref}$ to infinite volume, determined in
Subsec.~\ref{sec:FSE}, is the same for all the $f_\pi$ schemes
considered here and therefore does not affect the slope with respect to
$f_\pi$. The derivative in Eq.~\eqref{eq:I1_fpi_derivative} can thus be
regarded as the derivative of the infinite-volume result.}
\begin{align}
\frac{\partial a_{\mu}^{\rm HVP}(I=1)}
{\partial f_{\pi}}
=
s_{I=1}
=
-9.5 \, (2.0)
\times10^{-10}~{\rm MeV}^{-1}\,.
\label{eq:I1_fpi_derivative}
\end{align}
This derivative provides a convenient quantity for translating the result
between nearby prescriptions for isospin-symmetric QCD and may therefore
facilitate comparisons among lattice calculations using different inputs
to define the isospin-symmetric theory.

\subsection{Finite-volume correction to the $I=1$ contribution}
\label{sec:FSE}

We now discuss the extrapolation of the
continuum results obtained at $L=L_{\rm ref}$ in
Eqs.~\eqref{eq:I1_FLAG_Lref} and \eqref{eq:I1_WP25_Lref} to infinite
volume.  To this end, we make use of the MLLGS representation
\cite{Luscher:1990ux,Lellouch:2000pv,Meyer:2011um,Gounaris:1968mw}, which provides a model for the two-pion
contribution to the isotriplet vector correlator in finite volume.
Since the leading finite-volume effects are generated by the
low-lying two-pion states, the MLLGS model gives a sensible description of the finite-volume effects that must be removed in order to reach the
infinite-volume limit.  We already adopted this approach in our previous
calculation of the window observables of the HVP
\cite{ExtendedTwistedMass:2022jpw}. The MLLGS model has also been used
in several other lattice studies of the HVP, including recent
calculations by the Mainz/CLS~\cite{Djukanovic:2024cmq},
Fermilab/HPQCD/MILC~\cite{FermilabLatticeHPQCD:2024ppc}, and
BMW~\cite{Borsanyi:2020mff,Boccaletti:2024guq} collaborations.

We now describe the main features of the MLLGS representation of the (finite-volume)
isotriplet vector correlator $V^{I=1}(t,L)$.   For two
degenerate pions of mass $m_\pi$, the
two-pion contribution to the spatial isotriplet vector correlator at vanishing momentum is written as
\begin{align}
 V_{\pi\pi}(t,L)
 =
 \sum_{n=1}^{\infty} \nu_n |A_n|^2 e^{-\omega_n t},
 \qquad
 \omega_n=2\sqrt{m_\pi^2+k_n^2},
\label{eq:MLLGS_standard_corr}
\end{align}
where $\nu_n$ is the number of lattice vectors $\vec{n}= (n_{1},n_{2},n_{3})\in \mathbb{Z}^{3}$ with $|\vec{n}|^{2}=n$. The allowed
momenta $k_{n}$ are determined by the L\"uscher quantization condition
\begin{align}
 \delta_{11}(k_n)+\phi(q_n)=n\pi,
 \qquad
 q_n=\frac{k_nL}{2\pi}.
\label{eq:MLLGS_standard_levels}
\end{align}
In the previous equation $\delta_{11}(k)$ is the elastic $I=1$, $P$-wave phase shift and
$\phi(q)$ is a known kinematical function.  The
relation between finite- and infinite-volume matrix elements is fixed
by the Lellouch--L\"uscher relation
\cite{Lellouch:2000pv}, specialized to the present case of the $\pi\pi$ contribution to the isotriplet correlator
in Ref.~\cite{Meyer:2011um},
\begin{align}
 \nu_n |A_n|^2
 =
 \frac{2k_n^5}{3\pi\omega_n^2}
 \frac{|F_\pi(\omega_n^2)|^2}
 {k_n\delta_{11}'(k_n)+q_n\phi'(q_n)}.
\label{eq:MLLGS_standard_amplitude}
\end{align}
In the previous equation $F_{\pi}(s)$ is the time-like pion form factor, and a primed function denotes the derivative with respect to the argument of the function.

The time-like pion form factor is modelled using the Gounaris-Sakurai (GS) parameterization~\cite{Gounaris:1968mw}, which assumes the dominance of the $\rho$-resonance and expresses $F_{\pi}(s)$ as
\begin{align}
 F_\pi^{\rm GS}(s)
 =
 \frac{D_\pi(0)}{D_\pi(s)},
 \qquad
 D_\pi(s)=M_\rho^2-s-A_{\pi\pi}(s).
\label{eq:MLLGS_standard_GS}
\end{align}
The model enters through the assumed imaginary part of
$A_{\pi\pi}(s)$, or equivalently through the energy-dependent
$\rho\to\pi\pi$ width,
\begin{align}
 \operatorname{Im}A_{\pi\pi}(s)
 &=
 \sqrt{s}\,\Gamma_{\rho\pi\pi}(s),
\nonumber\\[8pt]
 \Gamma_{\rho\pi\pi}(s)
 &=
 \frac{g_{\rho\pi\pi}^2}{6\pi}
 \frac{k^3(s)}{s}\,
 \theta(s-4m_\pi^2),
 \qquad
 k(s)=\sqrt{\frac{s}{4}-m_\pi^2}.
\label{eq:MLLGS_standard_width}
\end{align}
The $k^3$ behavior is the two-body phase-space factor appropriate to a
$P$-wave, while the assumption that its coefficient is described by
one energy-independent coupling $g_{\rho\pi\pi}$ is the central ansatz of the GS parameterization.

Once the imaginary part of $A_{\pi\pi}(s)$ has been specified,
analyticity determines its real part up to two subtraction constants.
Using two subtractions at $s=0$, one has
\begin{align}
 A_{\pi\pi}(s)
 =
 c_0+c_1s+
 \frac{s^2}{\pi}
 \int_{0}^{\infty}ds'\,
 \frac{\operatorname{Im}A_{\pi\pi}(s')}
 {s'^2(s'-s-i0)} .
\label{eq:MLLGS_standard_dispersion}
\end{align}
The twice-subtracted dispersive integral fixes the nonanalytic part of
$A_{\pi\pi}(s)$, while $c_0$ and $c_1$ are two (real) unknowns, which are determined by imposing
the on-shell conditions
\begin{align}
 \operatorname{Re}A_{\pi\pi}(M_\rho^2)=0,
 \qquad
 \left.
 \frac{d}{ds}\operatorname{Re}A_{\pi\pi}(s)
 \right|_{s=M_\rho^2}=0 .
\label{eq:MLLGS_standard_subtractions}
\end{align}
The phase shift $\delta_{11}(k)$ is then equal, due to Watson's theorem, to the phase of $F_{\pi}^{\rm GS}(s)$, and it is given by
\begin{align}
 \cot\delta_{11}(k(s))
 =
 \frac{M_\rho^2-s-\operatorname{Re}A_{\pi\pi}(s)}
 {\operatorname{Im}A_{\pi\pi}(s)}.
\label{eq:MLLGS_standard_phase}
\end{align}
Explicit analytic expressions for $A_{\pi\pi}(s)$ are given in
Appendix~F of Ref.~\cite{ExtendedTwistedMass:2022jpw}. Besides the pion mass, the MLLGS model contains two free parameters: the $\rho$-mass $M_{\rho}$ and the effective coupling $g_{\rho\pi\pi}$.

The corresponding infinite-volume isotriplet correlator is
\begin{align}
 V_{\pi\pi}^{\infty}(t)
 =
 \frac{1}{48\pi^2}
 \int_{2m_\pi}^{\infty}d\omega\,
 \omega^2
 \left(1-\frac{4m_\pi^2}{\omega^2}\right)^{3/2}
 |F_\pi(\omega^2)|^2e^{-\omega t}.
\label{eq:MLLGS_standard_infinite}
\end{align}
The finite-volume correction is obtained from the difference between
the infinite- and finite-volume correlators,
\begin{align}
 \Delta_{\rm FV}a_\mu^{\rm HVP}(I=1;L)
 \equiv{}&
 a_\mu^{\rm HVP}(I=1;\infty)
 -
 a_\mu^{\rm HVP}(I=1;L)
\nonumber\\[6pt]
 ={}&
 2\alpha_{\rm em}^2
 \int_0^\infty dt\,t^2K(m_\mu t)
 \left[
 V_{\pi\pi}^{\infty}(t)-V_{\pi\pi}(t,L)
 \right].
\label{eq:I1_MLLGS_correction}
\end{align}
Before using the MLLGS model to perform the infinite-volume extrapolation, we validate the description of the finite-volume dependence
against the lattice data.  This test requires some care in
twisted-mass QCD.  At nonzero lattice spacing, as already discussed
(see Eq.~\eqref{eq:neutral_pion}), the charged and neutral pions,
$\pi^\pm$ and $\pi^0$, are not degenerate, and only the charged pion
is protected as a Goldstone boson. Moreover, in the mixed-action setup used for
the OS regularization, an additional Osterwalder--Seiler (OS) pion,
$\pi^{\rm OS}$, is also present.  At the lattice spacings considered in
this work, the OS-pion mass is larger than those of the unitary pion states, and moreover one has $m_{\pi^{0}} < m_{\pi}$.
All three pion species become degenerate only in the continuum limit.

The distorted pion spectrum implies that finite-volume effects at
nonzero lattice spacing acquire an indirect cutoff dependence. The spectral content of the OS isotriplet correlator (see Eq.~\eqref{eq:V33}) is particularly
involved.  In addition to the continuum-like $\pi^+\pi^-$ states,
mixed-action states containing OS pions, such as
$\pi^{\rm OS}\pi^{\rm OS}$ and possibly
$\pi^{\rm OS}\pi^0$, may contribute at nonzero lattice spacing.
Identifying the relative weights of these states is difficult and therefore we do not use
the OS correlator to validate the MLLGS model.

The TM regularization provides a considerably cleaner test.  It is
unitary and the isotriplet correlator corresponds to a charged
current correlator. The two-pion states contributing to the TM isotriplet correlator are therefore of a single type, namely  $\pi^+\pi^0$ states. However, since the two pions are nondegenerate at finite lattice
spacing, the MLLGS representation must be evaluated using the exact
kinematics of the $\pi^+\pi^0$ channel, thereby accounting for the
$O(a^2)$ charged-neutral pion mass splitting of the twisted-mass
regularization.  We stress that this does not introduce an additional
finite-volume model: it is the same MLLGS construction, with the same
dynamical assumptions and the same parameters, applied to a
two-particle channel with nondegenerate masses.  The usual
equal-mass representation is recovered as the special case
$m_{\pi^+}=m_{\pi^0}$.\footnote{Here and in the following discussion of
the nondegenerate $\pi^+\pi^0$ channel, we denote the charged-pion
mass by $m_{\pi^+}$; elsewhere in the paper the same quantity is simply
denoted by $m_\pi$.}

For two
pseudoscalars of unequal masses, the vacuum-to-two-pions charged-vector-current matrix element
contains in general both a transverse form factor ($F_{+0}= |F_{+0}|e^{i\delta_{+0}}$) analogous to the equal-mass case, and a longitudinal ($F_{L}$) form factor. However, since the correlator we model is built from the spatial
components of the current at vanishing total momentum, the longitudinal
contribution is projected out and does not contribute. 

For the nondegenerate $\pi^+\pi^0$ channel, the finite-volume
spectrum is obtained by using the corresponding nondegenerate two-particle
kinematics,
\begin{align}
 E_{\pi^+}(k)=\sqrt{m_{\pi^+}^2+k^2},\quad
 E_{\pi^0}(k)=\sqrt{m_{\pi^0}^2+k^2},
\quad
 \omega(k)=E_{\pi^+}(k)+E_{\pi^0}(k).
\label{eq:MLLGS_deformed_kinematics}
\end{align}
Neglecting the coupling to higher partial waves as in
Eq.~\eqref{eq:MLLGS_standard_levels}, at vanishing total momentum the
quantization condition for nondegenerate masses retains the same form as
in the degenerate case~\cite{Leskovec:2012gb},
\begin{align}
 \delta_{+0}(k_n)+\phi(q_n)=n\pi,
 \qquad
 q_n=\frac{k_nL}{2\pi},
\label{eq:MLLGS_deformed_quantization}
\end{align}
while the energies are
$\omega_n=E_{\pi^+}(k_n)+E_{\pi^0}(k_n)$. The finite-volume correlator is again written as
\begin{align}
V_{\pi^{+}\pi^{0}}(t,L) = \sum_{n=1}^{\infty} \nu_{n} |A_{n}^{+0}|^{2} e^{-\omega_{n}t}~.
\end{align}
The finite-volume matrix elements are related to the infinite-volume
form factor through
\begin{align}
 \nu_n|A_n^{+0}|^2
 =
 \frac{k_n^5}
 {6\pi E_{\pi^+}(k_n)E_{\pi^0}(k_n)}
 \frac{|F_{+0}(\omega_n^2)|^2}
 {k_n\delta_{+0}'(k_n)+q_n\phi'(q_n)}~.
 \label{eq:MLLGS_deformed_amplitude}
\end{align}
This is the single-channel, rest-frame specialization of the general
finite-volume relation for vacuum-to-two-particle matrix elements
derived in Ref.~\cite{Briceno:2015tza}. It reduces to
Eq.~\eqref{eq:MLLGS_standard_amplitude} when
$m_{\pi^+}=m_{\pi^0}$.

The GS model itself is not changed when applied to the
$\pi^+\pi^0$ channel.  Its dynamical assumptions remain the
dominance of a single $\rho$ resonance and an energy-independent
effective coupling $g_{\rho\pi\pi}$.  Evaluating the same model for two
nondegenerate particles only changes the two-particle threshold and
the $P$-wave phase-space factor entering the energy-dependent width,
\begin{align}
 \operatorname{Im}A_{+0}(s)
 &=
 \sqrt{s}\,\Gamma_{\rho,+0}(s),
\nonumber\\[8pt]
 \Gamma_{\rho,+0}(s)
 &=
 \frac{g_{\rho\pi\pi}^2}{6\pi}
 \frac{k^3(s)}{s}\,
 \theta\!\left[s-(m_{\pi^+}+m_{\pi^0})^2\right],\quad k(s)=
 \frac{\lambda^{1/2}
 \!\left(s,m_{\pi^+}^2,m_{\pi^0}^2\right)}
 {2\sqrt{s}},
\label{eq:MLLGS_deformed_width}
\end{align}
where $\lambda(x,y,z)=x^{2}+y^{2}+z^{2}-2xy-2xz-2yz$ is the K\"all\'en function.
The real part is then obtained from the same twice-subtracted
dispersion relation as in Eq.~\eqref{eq:MLLGS_standard_dispersion}, and the same two on-shell conditions as before. Finally,\footnote{The unit-normalization condition $F_{+0}^{\rm GS}(0)=1$, while not exact in the case of unequal masses, is extremely accurate. The Ademollo-Gatto theorem forbids linear corrections in the symmetry-breaking parameter, so that deviations from unity start only at second order in the charged--neutral pion mass-squared splitting, $F_{+0}(0)=1+O\left((m_{\pi^+}^2-m_{\pi^0}^2)^2\right)=1 + O(a^{4})$. For $m_{\pi^+}\simeq135~{\rm MeV}$ and $m_{\pi^0}\simeq110~{\rm MeV}$ (the values we will consider below), a one-loop chiral estimate~\cite{Cirigliano:2002ng} gives a correction of order $2\times10^{-4}$ in $F_{+0}(0)$, which is therefore entirely negligible in the present analysis. \label{footnote_AG}} 
\begin{align}
 F_{+0}^{\rm GS}(s)
 =
 \frac{D_{+0}(0)}{D_{+0}(s)},
 \qquad
 D_{+0}(s)=M_\rho^2-s-A_{+0}(s)~.
\label{eq:MLLGS_deformed_formfactor}
\end{align}
Thus, MLLGS evaluated for the nondegenerate $\pi^+\pi^0$ channel
contains the same two parameters, $M_\rho$ and
$g_{\rho\pi\pi}$, and the same dynamical input as in the degenerate
case. No additional parameter is introduced.
The two pion masses enter through the exact two-particle kinematics,
the position of the physical threshold and the associated
phase-space factor.  The standard equal-mass expressions are
recovered identically when $m_{\pi^+}=m_{\pi^0}$.

The numerically dominant difference between the MLLGS model evaluated with equal and unequal masses is the distortion of the finite-volume two-pion
spectrum,
\begin{align}
 2\sqrt{m_{\pi^{+}}^2+k_n^2}
 \longrightarrow
 E_{\pi^+}(k_n)+E_{\pi^0}(k_n),
\end{align}
which enters directly in the exponentials
$e^{-\omega_nt}$. The net effect, since $m_{\pi^{0}} < m_{\pi}^{+}$ on our ensembles, is to increase the size of the finite-volume effects with respect to the degenerate case with two charged pions of mass $m_{\pi^{+}}$. The accompanying modifications of the
finite-volume matrix elements, through the form factor and the unequal-mass kinematical prefactor in
Eq.~\eqref{eq:MLLGS_deformed_amplitude}, are instead numerically subleading.

We validate the MLLGS representation for nondegenerate pion kinematics using the TM results obtained on the B64 and B96 ensembles. These ensembles have the same bare parameters and lattice spacing but different spatial extents, $L_{\rm B64}\simeq 5.1~{\rm fm}$ and $L_{\rm B96}\simeq 7.65~{\rm fm}$. We compare both the absolute B96--B64 correlator difference and the corresponding relative difference with the MLLGS predictions.

As discussed in Subsec.~\ref{sec:LatJ} (see Eq.~\eqref{eq:V33}), the charged current entering the TM correlator is renormalized by the axial-current renormalization constant $Z_A$. The values of $Z_A$ used throughout this work, as well as in our previous analyses
\cite{ExtendedTwistedMass:2022jpw,ExtendedTwistedMass:2024nyi}, were determined by matching pseudoscalar decay constants computed in the TM and OS regularizations. The matching condition was imposed in the $\eta_{ss'}$ channel, because the strange-pseudoscalar correlators allow for a substantially more precise determination than the corresponding light-pseudoscalar correlators. At nonzero lattice spacing, however, the value of $Z_A$ depends on the specific matching condition through $O(a^2)$ cutoff effects. In particular, imposing the same matching condition in the light-pseudoscalar sector yields a slightly different, but equally legitimate, determination of $Z_A$.

This ambiguity is relevant when comparing the absolute correlator difference with the MLLGS prediction. The GS form factor satisfies the continuum point-like normalization condition at $s=0$ (in the case of unequal masses up to very tiny corrections, see footnote~\ref{footnote_AG}), whereas the normalization of the local lattice current is fixed only up to cutoff effects. Since the absolute B96--B64 difference retains the overall factor $Z_A^2$, its normalization depends on the particular determination of $Z_A$. By contrast, this factor cancels in the relative B96--B64 difference, which is therefore insensitive to the normalization ambiguity.

For the comparison of the absolute difference with MLLGS, we therefore also consider the value of $Z_A$ obtained from the light-pseudoscalar matching condition. On the B ensembles, the difference between the correlator normalizations obtained using the light- and strange-sector determinations amounts to $7\%$. For the comparison, we renormalize the absolute B96--B64 correlator difference using a value of $Z_A$ obtained by averaging the two determinations and assign their half-difference as an additional systematic uncertainty. This prescription is intended to account for possible $O(a^2)$ artifacts in the point-like normalization condition that could affect the comparison with the MLLGS model. No corresponding systematic uncertainty is assigned to the relative B96--B64 difference, since the common factor $Z_A^2$ cancels.

The pion masses entering the MLLGS representation are taken from the B-ensemble determinations, and their uncertainties are propagated to the model prediction. In particular, we use $m_{\pi^0}=110(9)\,{\rm MeV}$. The two free parameters of the MLLGS representation are set to
\begin{align}
M_\rho=0.77(1)\ {\rm GeV},
\qquad
g_{\rho\pi\pi}=5.75(25)~.
\label{eq:pars_MLLGS}
\end{align}
The quoted uncertainties are chosen conservatively, guided by GS
descriptions of the experimental timelike pion form factor and by
lattice determinations of the $\rho$-resonance parameters based on the
finite-volume $I=1$, $P$-wave $\pi\pi$ spectrum
(see, e.g.,~\cite{CMD-2:2006gxt,ExtendedTwistedMass:2019omo,Erben:2019nmx}
and references therein).

\begin{figure}[t]
\centering
\includegraphics[width=0.47\linewidth]{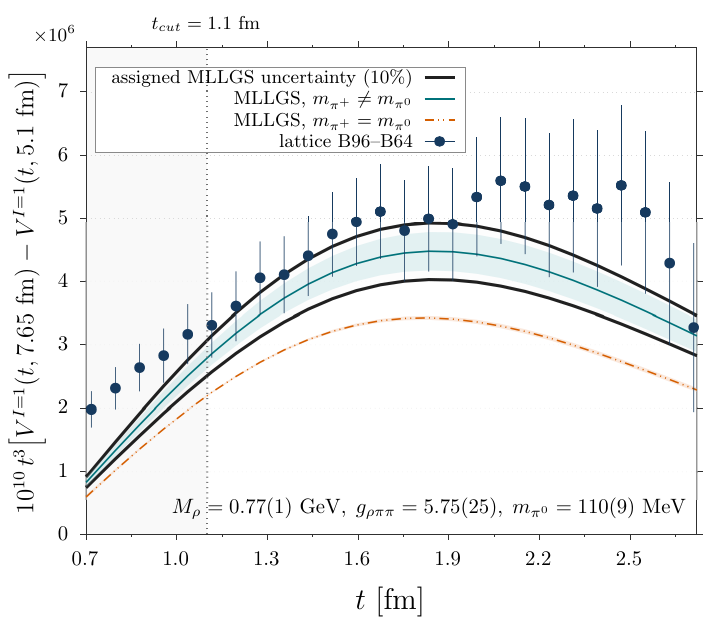}
\includegraphics[width=0.47\linewidth]{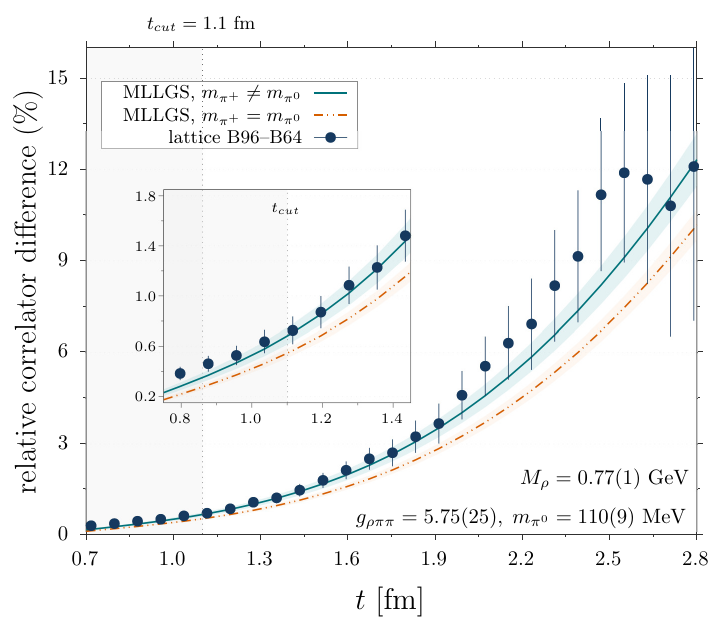}
\caption{Validation of the MLLGS model using the TM
correlators on the B96 and B64 ensembles.  The left panel shows the correlator difference as a function of time, while the right panel shows the relative difference. The MLLGS model evaluated at equal masses employs the value of $m_{\pi^+}$ obtained on the B ensemble in the TM regularization. The MLLGS model evaluated at unequal masses uses both $m_{\pi^{0}}$ and $m_{\pi^+}$ obtained for the same lattice ensemble and regularization.}
\label{fig:I1_MLLGS_validation}
\end{figure}

The comparison is shown in Fig.~\ref{fig:I1_MLLGS_validation}. Both the absolute and relative differences are compatible with the corresponding MLLGS predictions at Euclidean time separations larger than approximately $1.1~{\rm fm}$. This is consistent with previous findings, including those of the Mainz/CLS analysis, which showed that finite-volume effects in the isovector correlator at times larger than $1~{\rm fm}$ are well described by the MLLGS model~\cite{Djukanovic:2024cmq}.

In both panels of Fig.~\ref{fig:I1_MLLGS_validation}, we also display the corresponding equal-mass MLLGS prediction. As the figure demonstrates, accounting for the unequal $\pi^+$ and $\pi^0$ masses in the MLLGS representation is essential for reproducing the finite-volume dependence observed in the lattice data. By contrast, the prediction obtained with degenerate charged-pion masses clearly underestimates the size of the finite-volume effects. We also show a $10\%$ uncertainty band around the prediction obtained with nondegenerate pion masses. Combining this assigned MLLGS uncertainty with the lattice uncertainty, the deviation between the MLLGS prediction and the lattice data for the correlator difference does not exceed $1\sigma$ over the interval $1.1~{\rm fm}\leq t\leq2.0~{\rm fm}$ and amounts to $0.67\sigma$ on average. The observed agreement provides a nontrivial validation of the model and supports its use for estimating finite-volume effects at times larger than $1.1~{\rm fm}$, with an assigned uncertainty of $10\%$.

To extrapolate to the infinite-volume limit our continuum results at $L=L_{\rm ref}$ we therefore proceed as follows. We evaluate the contribution to the finite-volume correction arising from the region $t\geq 1.1~{\rm fm}$ by using Eq.~\eqref{eq:I1_MLLGS_correction} with the time integration restricted to this interval and with $M_\rho$ and $g_{\rho \pi \pi}$ as in Eq.~(\ref{eq:pars_MLLGS}). At the physical-point value $m_\pi=135~{\rm MeV}$ and for $L=L_{\rm ref}$, this contribution amounts to $25.1\times 10^{-10}$. We assign to it a relative uncertainty of $10\%$, corresponding to an absolute uncertainty of $2.5\times 10^{-10}$.

The contribution from the complementary region, $t<1.1~{\rm fm}$, is estimated directly from the lattice data by integrating the measured B96--B64 correlator difference with the HVP kernel up to $1.1~{\rm fm}$. This gives $2.01(27)\times 10^{-10}$. As discussed above, this data-driven estimate is expected to overestimate the corresponding continuum finite-volume effect. Indeed, finite-volume effects are small in this time region, so the B96 ensemble may be regarded as effectively corresponding to infinite volume, whereas the spatial extent of the B64 ensemble is slightly smaller than $L_{\rm ref}$. Moreover, the lighter neutral pion present at finite lattice spacing enhances the finite-volume effects relative to the mass-degenerate continuum case. For comparison, the continuum MLLGS prediction over the region $t<1.1~{\rm fm}$ amounts to approximately $1.0\times 10^{-10}$. This prediction is expected to underestimate the short-distance finite-volume effect, since the comparison with the B96--B64 data shows that MLLGS predicts smaller values for the correlator difference, even when the effects due to the charged/neutral-pion-mass splitting at finite lattice spacing are taken into account. We therefore use the data-driven estimate as the central value of the finite-volume correction in the region $t< 1.1~{\rm fm}$ and assign the difference between the data-driven estimate and the continuum MLLGS prediction over the same time region as a conservative systematic uncertainty on this contribution. This prescription brackets the $t<1.1~{\rm fm}$ contribution to the finite-volume correction since the continuum MLLGS prediction is expected to underestimate it, and the data-driven estimate is expected to overestimate it.

The central value of the total finite-volume correction is obtained by adding the continuum MLLGS prediction for $t\geq 1.1~{\rm fm}$ and the data-driven estimate for $t<1.1~{\rm fm}$. Our final estimate of the finite-volume correction needed to extrapolate from $L_{\rm 
 ref}$ to the infinite-volume limit is therefore 
\begin{align} 
\Delta_{\rm FV}a_\mu^{\rm HVP}(I=1;L_{\rm ref}) = 27.1(2.5)(1.0)[2.7]\times 10^{-10},
\label{eq:FSE} 
\end{align} 
where the first uncertainty is the $10\%$ uncertainty assigned to the
continuum MLLGS prediction in the region $t\geq 1.1~{\rm fm}$, while
the second is the conservative uncertainty assigned to the contribution
from the region $t<1.1~{\rm fm}$, determined from the difference between
the data-driven estimate and the continuum MLLGS prediction. The total uncertainty, shown in square brackets, is obtained by combining these two uncertainties in quadrature. Our final relative uncertainty on the finite-volume correction is very similar to the $9\%$ uncertainty estimated by the Mainz/CLS Collaboration in Ref.~\cite{Djukanovic:2024cmq}.
\subsection{The $I=0$ contribution}
\label{sec:full_HVP_I0}

We now turn to the isoscalar channel. We recall the decomposition
introduced in Eq.~\eqref{eq:I0}, which we adopt as a convenient
organization of the different contributions entering the analysis:
\begin{align}
a_{\mu}^{\rm HVP}(I=0)
=
a_{\mu}^{\rm HVP}({\rm oct})
+a_{\mu}^{\rm HVP}(c)
+\delta a_{\mu}^{c,{\rm disc}}\,.
\label{eq:I0_decomp_secIV}
\end{align}
The first term is obtained from the light-strange octet current in Eq.~\eqref{eq:I0_oct} and
contains both connected and disconnected Wick contractions. As explained
in Sec.~\ref{sec:LatJ}, it is evaluated using a mixed-action setup which provides the positive spectral representation
needed for the application of the bounding method (see Eq.~\eqref{eq:J8_OS_current}). The second term, $a_{\mu}^{\rm HVP}(c)$, denotes the charm-connected contribution. It is evaluated using both the TM and OS discretizations in Eq.~\eqref{eq:Vcc-con}, which differ only by cutoff effects and share the same continuum limit. The final
term, $\delta a_{\mu}^{c,{\rm disc}}$, collects all disconnected
contractions containing at least one charm loop (see the discussion below Eq.~\eqref{eq:V8c-dis}). We first verify directly
that this residual contribution is negligible at the precision of the
present calculation, and then determine the octet and charm-connected
terms in both the FLAG and WP25 schemes.

\begin{figure}[t]
\centering
\includegraphics[width=0.7\linewidth]{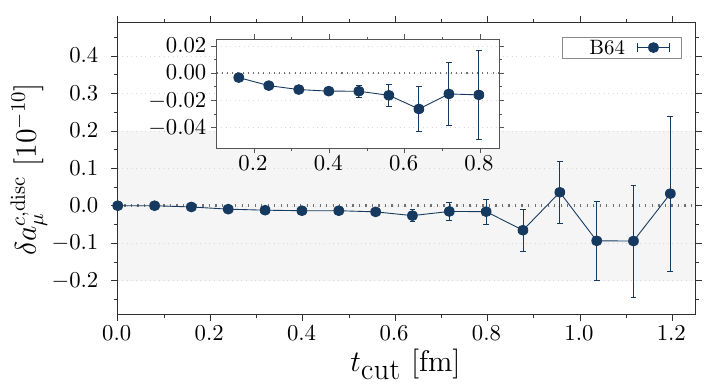}
\caption{Partial time integral of the HVP contribution containing at least one charm loop,
$\delta a_{\mu}^{c,{\rm disc}}$, on the B64 ensemble as a function of the
upper integration time $t_{\rm cut}$. A plateau is reached at very short times $t_{\rm cut}\simeq 0.5~{\rm fm}$ as shown in the figure inset. To stay on the conservative side we attribute an uncertainty to this contribution equal to the statistical error at a later time $t_{\rm cut}=1.2~{\rm fm}$, which is about $0.2\times 10^{-10}$, and which corresponds to the semi-transparent band. }
\label{fig:I0_charm_disc}
\end{figure}

Previous lattice studies have found $\delta a_{\mu}^{c,{\rm disc}}$ to be very small (see, e.g., Refs.~\cite{Borsanyi:2020mff,Djukanovic:2024cmq}). We nevertheless test this explicitly in our setup using the B64 ensemble. Fig.~\ref{fig:I0_charm_disc} shows the partial contribution as a function of the upper limit $t_{\rm cut}$ on the HVP integral in Eq.~\eqref{eq:amu_HVP}. A plateau is observed starting at approximately $t_{\rm cut}\simeq0.5~{\rm fm}$, consistent with the short-distance nature of this contribution. To remain conservative, when constructing our final estimate of the $I=0$ contribution, we assign a conservative uncertainty associated with charm-disconnected loop diagrams equal to the statistical uncertainty observed at $t_{\rm cut}=1.2~{\rm fm}$, namely approximately $0.2\times10^{-10}$. This uncertainty is entirely negligible compared with those of the remaining contributions to the $I=0$ channel, which will turn out to be more than one order of magnitude larger. Computing the charm-disconnected  loop contribution to the window observables further confirm the short-distance nature of this contribution. Using the short-distance window kernel $\Theta^{\rm SD}(t)$, we obtain $1.18(14)\times10^{-12}$, which is the only one for which a statistically significant signal is observed. For the intermediate window, we instead find $-2(6)\times10^{-12}$. In Sec.~\ref{sec:windows}, we will follow the same conservative strategy adopted here for the full contribution and assign additional uncertainties of $1.2\times10^{-12}$ and $6\times10^{-12}$ to the $I=0$ short-distance and intermediate-window contributions, respectively.

Having established that the  charm-disconnected contributions are negligibly small at the present precision, we now consider the octet and
charm-connected terms. Before performing their continuum extrapolations,
we apply the same tree-level subtraction introduced for the isotriplet
contribution in
Eqs.~\eqref{eq:TL_integrand}--\eqref{eq:TL_improved_observable}. For each
contribution, the correction is obtained by replacing the isotriplet
tree-level correlators in those equations with the appropriate lattice
and continuum free-field correlators and by using the relevant bare quark
masses. We therefore do not repeat the explicit formulae here. For the octet contribution, only the light-connected and
strange-connected terms have a nonvanishing tree-level
contribution and are affected by this subtraction. The disconnected
diagrams vanish in the free theory, their perturbative expansion
starting only at $O(\alpha_s^3)$.  For the charm-connected term,
the tree-level correction is evaluated separately for the TM and OS discretizations
using the tuned bare charm mass on each ensemble. In this case, retaining
the charm-mass dependence in the free-field correlator is 
important and substantially improves the effectiveness of the subtraction
in reducing cutoff effects. As in the isotriplet analysis, all the octet
and charm-connected results presented below include this tree-level
improvement.

\begin{figure}[t]
\centering
\includegraphics[width=\linewidth]{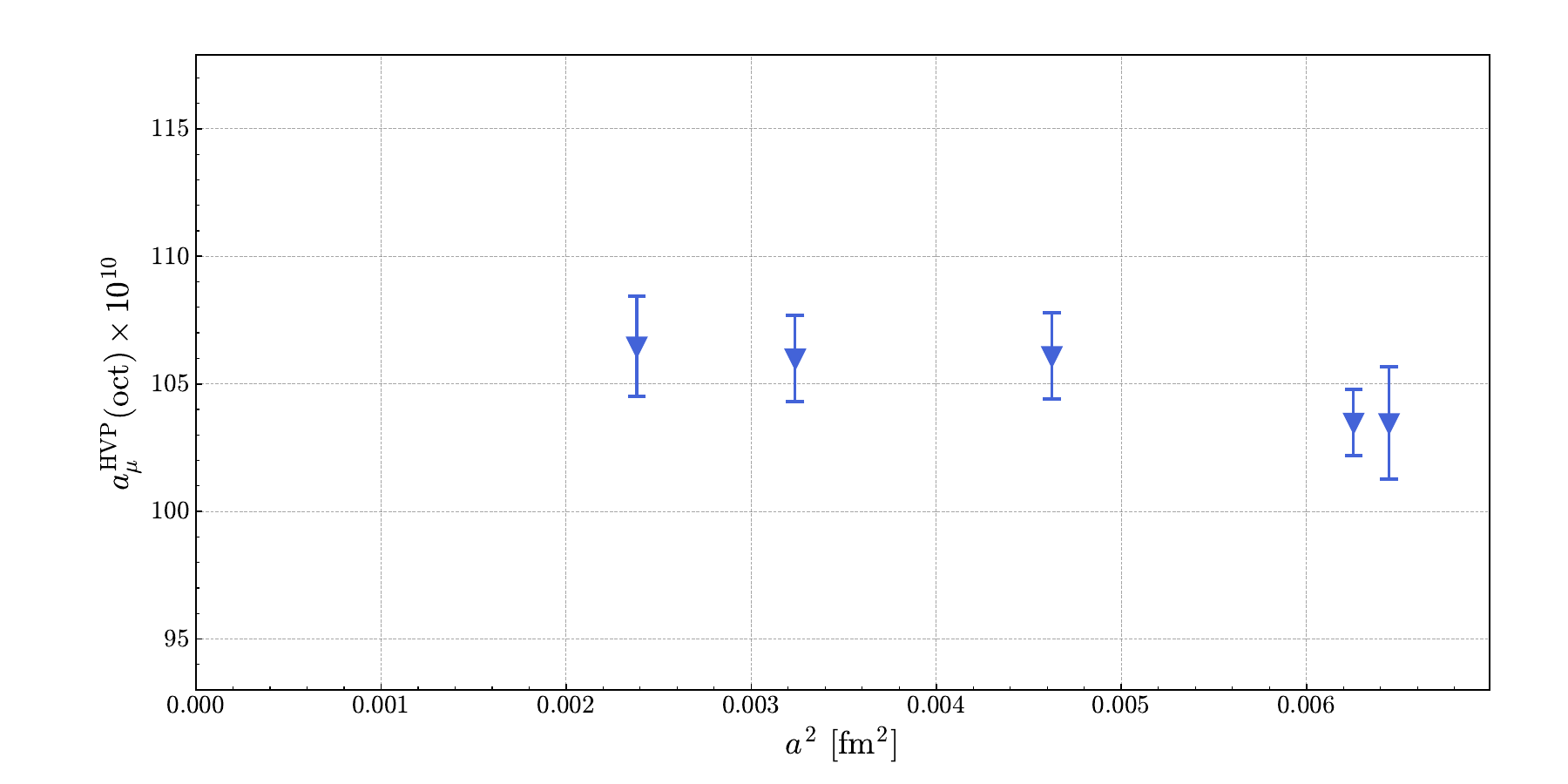}
\caption{Results for $a_{\mu}^{\rm HVP}({\rm oct})$ in the FLAG
scheme as a function of $a^{2}$. The B64 and B96 points share the same
lattice spacing but correspond to different spatial extents and are
slightly displaced horizontally for visualization. Their agreement
within uncertainties indicates that finite-volume effects are negligible
at the current precision.}
\label{fig:oct_FLAG_ensembles}
\end{figure}

We begin by discussing the octet contribution, which, for each ensemble is determined using the bounding method described in
Subsec.~\ref{seq:analysis-and-bounding}. Fig.~\ref{fig:oct_FLAG_ensembles}
shows our results in the FLAG scheme for the five ensembles we analyzed. In contrast to
the isotriplet case, the B64 and B96 results perfectly agree within uncertainties,
despite the substantial difference between their spatial extents. This
behavior is expected because the isovector $P$-wave two-pion states
responsible for the dominant finite-volume effects in the $I=1$ channel
do not contribute to the octet correlator. No statistically significant
volume dependence is observed, and the octet results can therefore be
treated as effectively at infinite volume at the precision of the present
calculation. As in the isotriplet analysis, the B64 point is the most
precise one because of the larger statistics employed on this ensemble;
see Table~\ref{tab:statistics}.

As already discussed in Sec.~\ref{sec:setup}, we did not correct the octet
data for the mistuning effects on the simulated pion masses, which are at
most approximately $4\%$. The reason for this choice was that the mistuning
effect originating from the connected light-quark component of
$a_{\mu}^{\rm HVP}({\rm oct})$, which is given by
$\frac{1}{10}a_{\mu}^{\rm HVP}(\ell)$ in the notation of
Eq.~\eqref{eq:octet_decomposition}, is at most of order
$3$--$4\times 10^{-11}$ and is therefore already much smaller than our
statistical uncertainties on the octet contribution, which are of order
$1.3$--$1.7\times 10^{-10}$. Moreover, we argued that the dominant contribution to the quark-mass
dependence of $a_{\mu}^{\rm HVP}(\ell)$ arises from the two-pion contribution.
Since two-pion states are absent in this channel, we expected the
already marginal correction to $\frac{1}{10}a_{\mu}^{\rm HVP}(\ell)$ to be
compensated by an analogous shift in the opposite direction in the 
disconnected contribution $a_{\mu}^{\rm HVP}({\rm disc},uds)$.

\begin{figure}[t]
  \centering
  \includegraphics[width=0.8\linewidth]{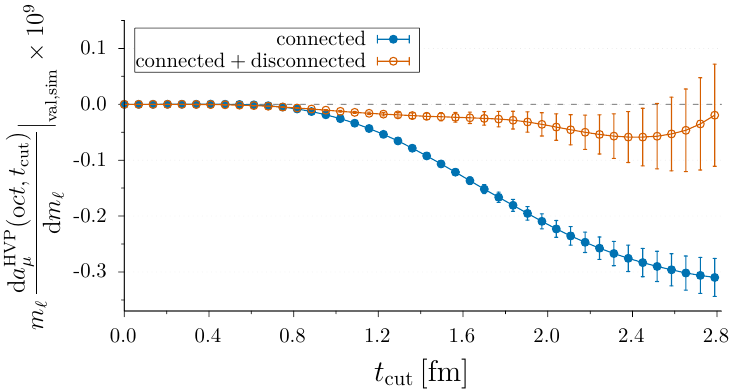}
  \caption{%
  Valence light-quark-mass derivative of the truncated
  $a_{\mu}^{\rm HVP}(\rm oct)$ integral on the C80 ensemble, shown as a function
  of the upper integration limit $t_{\rm cut}$. The connected contribution,
  corresponding to $\frac{1}{10}a_{\mu}^{\rm HVP}(\ell)$, is compared with
  the full result obtained by adding the contribution of the disconnected term
  $a_{\mu}^{\rm HVP}({\rm disc},uds)$. 
  }
  \label{fig:comparison_der}
\end{figure}

To provide a numerical verification of this picture, we evaluated the
dominant valence light-quark-mass derivative of
$a_{\mu}^{\rm HVP}({\rm oct})$ and compared it with that of
$\frac{1}{10}a_{\mu}^{\rm HVP}(\ell)$. The results obtained on the C80
ensemble are shown in Fig.~\ref{fig:comparison_der}. In the figure, we show
both the connected contribution and the sum of the connected and disconnected
contributions to the valence light-quark-mass derivative of the truncated HVP
integral, with $t_{\rm cut}$ denoting the upper limit of the time integral in
Eq.~\eqref{eq:amu_HVP}.
The results indeed confirm this picture. The connected contribution,
originating from $a_{\mu}^{\rm HVP}(\ell)$, is almost entirely cancelled by
the disconnected contribution, such that the full derivative is strongly
suppressed with respect to the connected-only contribution. Moreover, as is
clearly observed in the figure, the connected contribution is dominated by
large Euclidean distances, with full saturation not yet occurring at
$t_{\rm cut}\simeq 2.8~{\rm fm}$, whereas the full derivative reaches a
plateau much earlier. This behavior is consistent with our interpretation
that, after the cancellation of the two-pion contribution, the remaining
quark-mass dependence originates from three-pion and possibly heavier states,
whose signals are more strongly exponentially suppressed at large Euclidean
times.

As shown in the figure, a $10\%$ mistuning of the light-quark mass, which is
larger than the mistunings present in our data, would produce a shift in the
full octet contribution of order
$\simeq 5\times 10^{-12}$. This is almost $30$ times smaller than the
statistical uncertainty of the most precise determination, obtained on the
B64 ensemble, and is therefore entirely negligible.

The octet data shown in Fig.~\ref{fig:oct_FLAG_ensembles} exhibit an extremely mild dependence on $a^{2}$. All determinations are mutually compatible within uncertainties. The largest differences are those between the B64 result and the results obtained on each of the C80, D96, and E112 ensembles, but they remain below $1.2\sigma$ in all cases. In particular, the determinations from the three finest ensembles are in excellent agreement and lie well within one another’s uncertainties.

\begin{figure}[t]
\centering
\includegraphics[width=\linewidth]{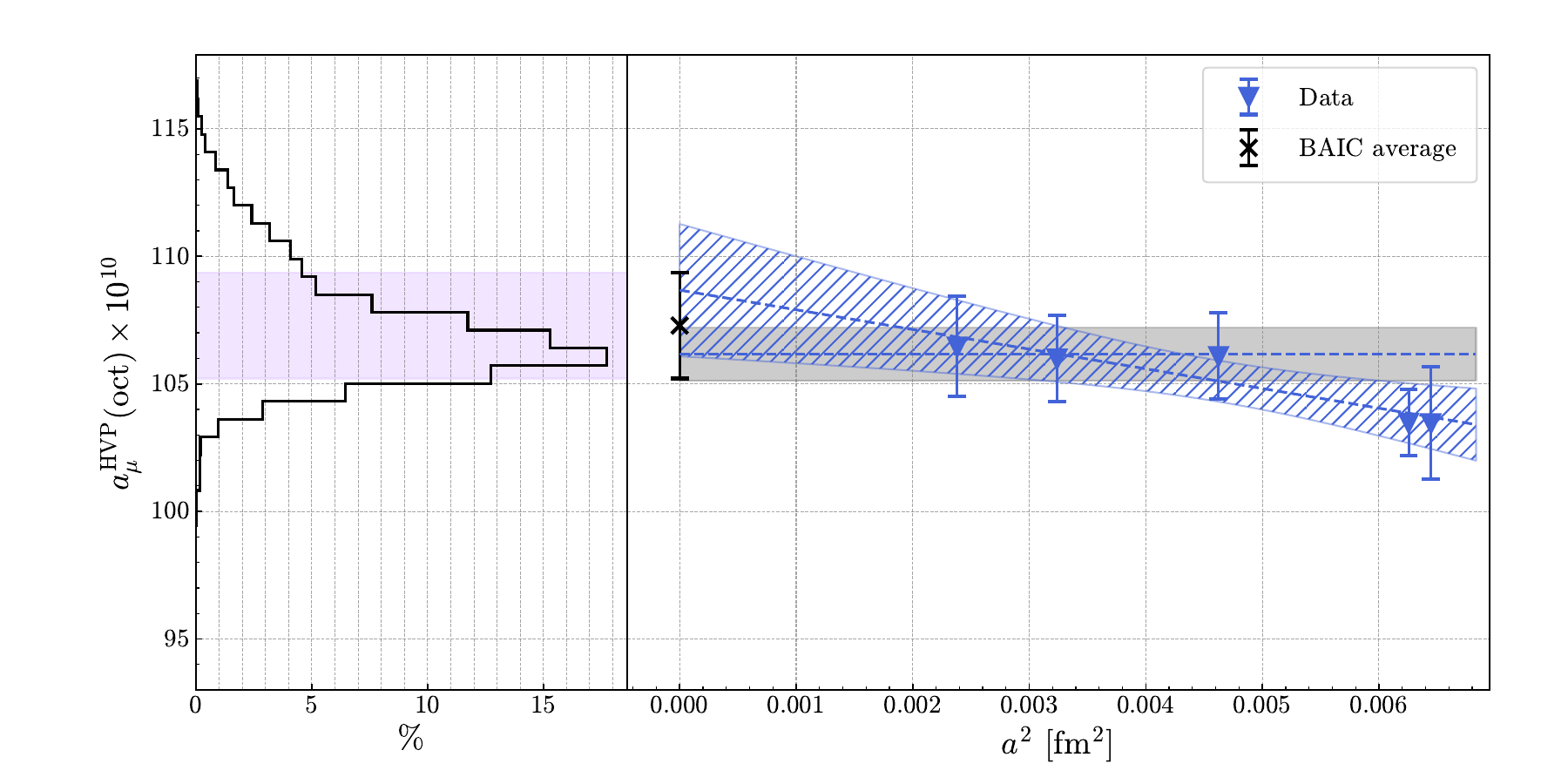}
\caption{Continuum extrapolation of
$a_{\mu}^{\rm HVP}({\rm oct})$ in the FLAG scheme. The two bands
correspond to a linear fit in $a^{2}$ to all ensembles and a constant fit to C80, D96, and E112. The final
continuum value is obtained combining the two fits using the BAIC procedure (see Eqs.~\eqref{eq:BAIC_weight}-\eqref{eq:BAIC_syst}). The
distribution on the left is constructed from the continuum-limit results
of the two fits and their bootstrap samples, weighted according to the
corresponding BAIC weights.}
\label{fig:oct_FLAG_cont}
\end{figure}

Given the observed flat behavior over the explored broad range of lattice
spacings, the data do not support the inclusion of terms beyond linear order
in $a^{2}$ in the continuum extrapolation. We therefore consider only two
simple continuum extrapolations: a fit linear in $a^{2}$ to all ensemble
determinations and a constant fit restricted to the C80, D96, and E112
ensembles, corresponding to the three finest lattice spacings. The two
continuum results are combined using the BAIC procedure of
Eqs.~\eqref{eq:BAIC_weight}--\eqref{eq:BAIC_syst}. Fig.~\ref{fig:oct_FLAG_cont} shows the ensemble data together with the
two fit bands and the corresponding continuum results. On the left side
of the figure, we display the BAIC-weighted distribution of the
continuum-limit result obtained from the two fits and their
bootstrap samples. Each bootstrap result associated with either fit
contributes with the corresponding BAIC weight, so that the distribution
reflects both the statistical fluctuations and the spread between the two
continuum-extrapolated results.

As a further check, we have also considered replacing the linear $a^2$
dependence by a logarithmically modified term proportional to
$a^2/|\log(a^2\Lambda_0^2)|^\gamma$, with $\gamma=0.42$, and repeated the full analysis. The
resulting central value and uncertainty are essentially unchanged. Given
the very small cutoff effects observed in this channel, the linear and logarithmically modified $a^2$ terms lead to practically indistinguishable continuum extrapolations.

The resulting FLAG-scheme value from our central analysis is
\begin{align}
a_{\mu}^{\rm HVP,FLAG}({\rm oct})
=
\amufulloctFLAG \times10^{-10}
\,.
\label{eq:oct_FLAG}
\end{align}
Its relative uncertainty is approximately $2\%$. In absolute terms, this
uncertainty is about a factor of $2.7$ smaller than that of the isotriplet
contribution, indicating that the octet term is a subleading source of
uncertainty in the complete isospin-symmetric HVP result.

\begin{figure}[t]
\centering
\includegraphics[width=0.49\linewidth]{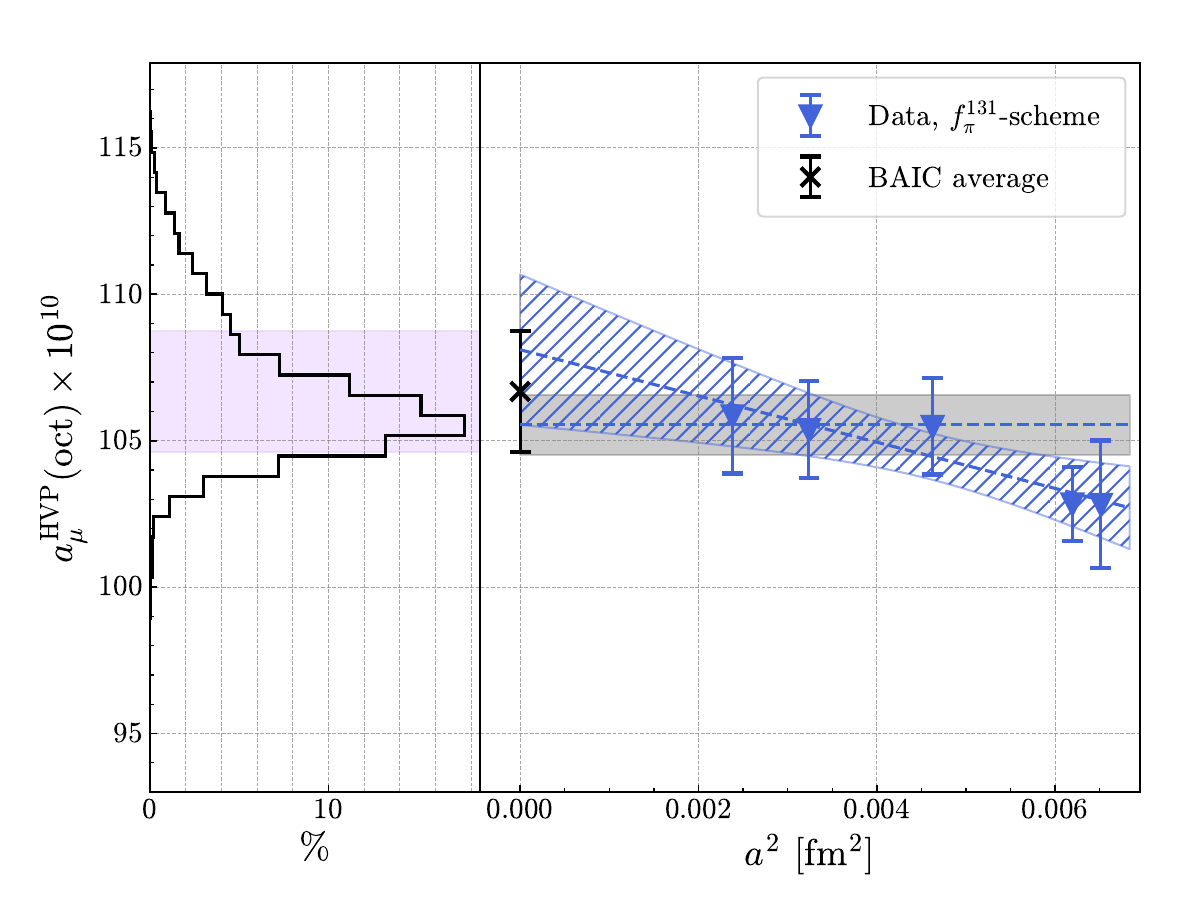}
\includegraphics[width=0.49\linewidth]{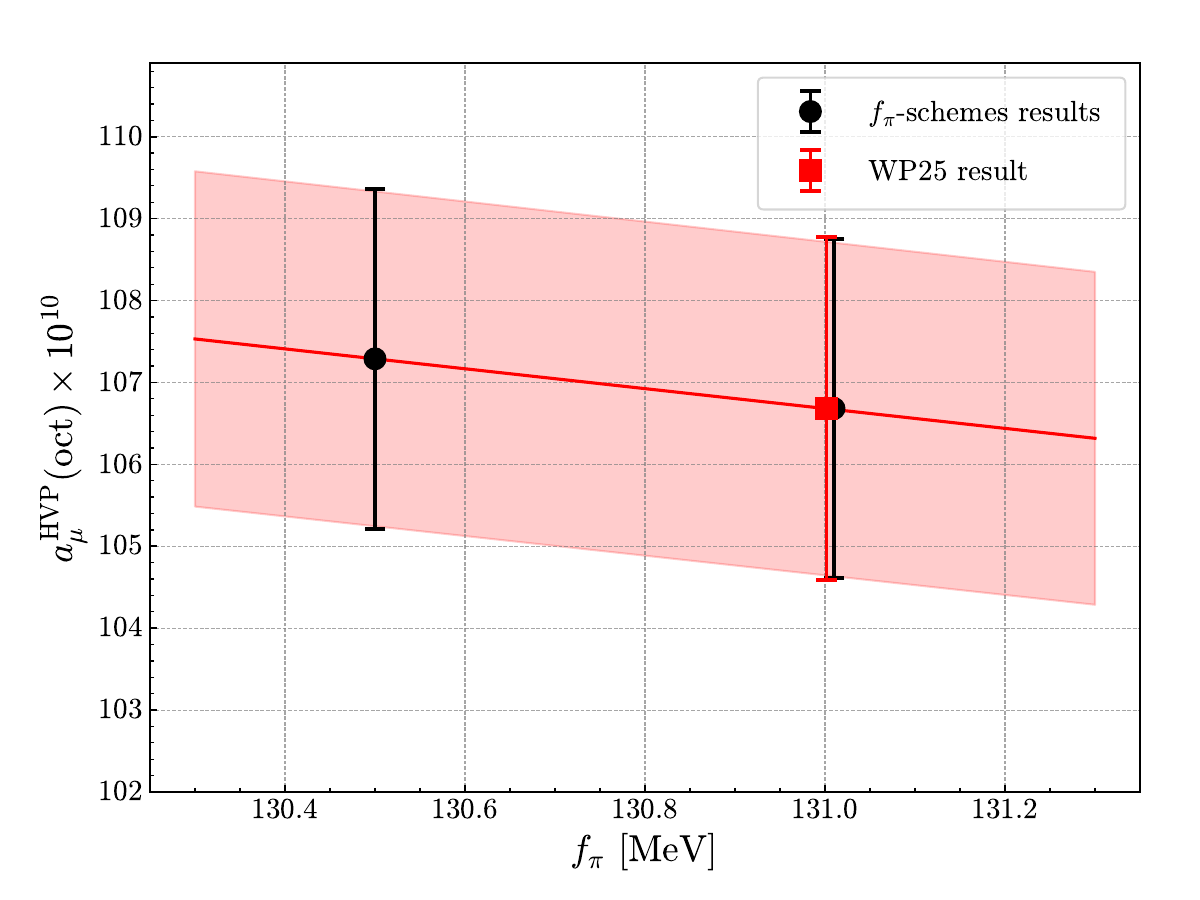}
\caption{
\textit{Left}:
Same as Fig.~\ref{fig:oct_FLAG_cont}, but in the
$f_{\pi}^{131}$ scheme.
\textit{Right}:
Dependence of the continuum octet contribution on the input
value of $f_{\pi}$. The red band corresponds to the linear interpolation performed using the ansatz in
Eq.~\eqref{eq:oct_fpi_interp}. The data point indicated by the red square corresponds to the final result in the WP25 scheme. The value corresponding to the $f_{\pi}^{131}$-scheme has been slightly shifted horizontally for better visualization.
}
\label{fig:oct_fpi131_cont_and_WP_int}
\end{figure}

We next determine the octet contribution in the WP25 scheme. As in the
isotriplet analysis, the original $w_0$-based prescription is implemented
through the equivalent $f_{\pi}$-based characterization given in
Eq.~\eqref{eq:fpi_WP25}. The dependence of the octet contribution on the input value of
$f_{\pi}$ is weaker than for the isotriplet contribution, and it is
therefore sufficient to repeat the continuum analysis at two values: the
FLAG value $f_{\pi}^{\rm FLAG}=130.5~{\rm MeV}$ and
$f_{\pi}=131~{\rm MeV}$. In the $f_{\pi}^{131}$ scheme, the
complete analysis is repeated using the same tree-level subtraction,
the same ansatzes for the continuum-limit
extrapolation, and the same BAIC used for the FLAG-scheme determination. The corresponding continuum extrapolation is shown
in the left panel of Fig.~\ref{fig:oct_fpi131_cont_and_WP_int}. As for the isotriplet contribution,
$a_{\mu}^{\rm HVP}({\rm oct})$ decreases as the input value of
$f_{\pi}$ is increased. In this case, however, the difference between
the FLAG- and WP25-scheme results is more modest, corresponding to a
downward shift of approximately $1\times10^{-10}$.

We linearly interpolate between the results obtained in the FLAG- and $f_{\pi}^{131}$-scheme using 
\begin{align}
a_{\mu}^{\rm HVP}({\rm oct};f_{\pi})
=
a_{\mu}^{\rm HVP,FLAG}({\rm oct})
+s_{\rm oct}
\left(f_{\pi}-f_{\pi}^{\rm FLAG}\right)\,.
\label{eq:oct_fpi_interp}
\end{align}
The value at $f_{\pi}=f_{\pi}^{\rm FLAG}$ is the FLAG result in
Eq.~\eqref{eq:oct_FLAG}, while $s_{\rm oct}$ gives the derivative of the
octet contribution with respect to the input value of $f_{\pi}$.
The right panel of Fig.~\ref{fig:oct_fpi131_cont_and_WP_int} shows the two continuum
determinations and the result of the linear interpolation. The red square data point is
obtained by evaluating Eq.~\eqref{eq:oct_fpi_interp} at
$f_{\pi}=f_{\pi}^{\rm WP25}$ and propagating the uncertainty of
$f_{\pi}^{\rm WP25}$.
This gives
\begin{align}
a_{\mu}^{\rm HVP,WP25}({\rm oct})
=
\amufulloctWP \times10^{-10}
\,.
\label{eq:oct_WP25}
\end{align}
The derivative obtained from the same interpolation is
\begin{align}
\frac{\partial a_{\mu}^{\rm HVP}({\rm oct})}{\partial f_{\pi}}
=
s_{\rm oct}
=
-1.21\,(26)
\times10^{-10}~{\rm MeV}^{-1}\,.
\label{eq:oct_fpi_derivative}
\end{align}
We finally consider the charm-connected contribution. As described in
Subsec.~\ref{sec:LatJ}, this contribution can be evaluated with both the TM and OS regularizations. The two lattice determinations have different cutoff
effects and approach a common continuum limit. 
The finite-volume effects on this contribution have been shown to be negligibly small on lattice volumes with $L\gtrsim 5~{\rm fm}$~\cite{ExtendedTwistedMass:2024nyi}. The cutoff effects in the charm-connected contribution are instead substantially
larger than in the octet channel. The dependence on $a^{2}$ is steeper,
and the data show visible curvature, indicating that higher-order cutoff
effects cannot be neglected. The availability of two regularizations is
therefore especially useful, since their different approaches to a common
continuum limit provide an important constraint on the extrapolation. The continuum analysis closely follows the strategy adopted in our
previous determination of the connected charm-quark
contribution~\cite{ExtendedTwistedMass:2024nyi}. We fit the TM and OS
data simultaneously, imposing a common continuum value and allowing the
cutoff dependence to be regularization dependent. Starting from the general form in Eq.~\eqref{eq:I1_continuum_ansatz}, we construct the fit pool by allowing, independently for each regularization, all combinations of the three cutoff terms proportional to $a^{2}$, $a^{4}$, and $a^{2}/|\log(a^{2}\Lambda_{0}^{2})|^{\gamma}$, with at most two of them retained simultaneously for a given regularization. Thus, all pairwise combinations are included, together with the corresponding lower-parameter variants. The only restriction we impose, to avoid overfitting, is to exclude ansatzes in which all three cutoff terms are simultaneously present for the same regularization. In contrast to
Ref.~\cite{ExtendedTwistedMass:2024nyi}, where the values $\gamma=1,2,3$ were considered, here we use the values $\gamma=0.42,1,2$, for consistency with the isotriplet analysis. All these fits are repeated using the complete data set and data sets in
which the coarsest or next-to-coarsest lattice spacing is omitted from
one or both regularizations.
The resulting continuum determinations are
combined using the BAIC procedure
described in Subsec.~\ref{sec:full_HVP_I1}.

\begin{figure}[t]
\centering
\includegraphics[width=\linewidth]{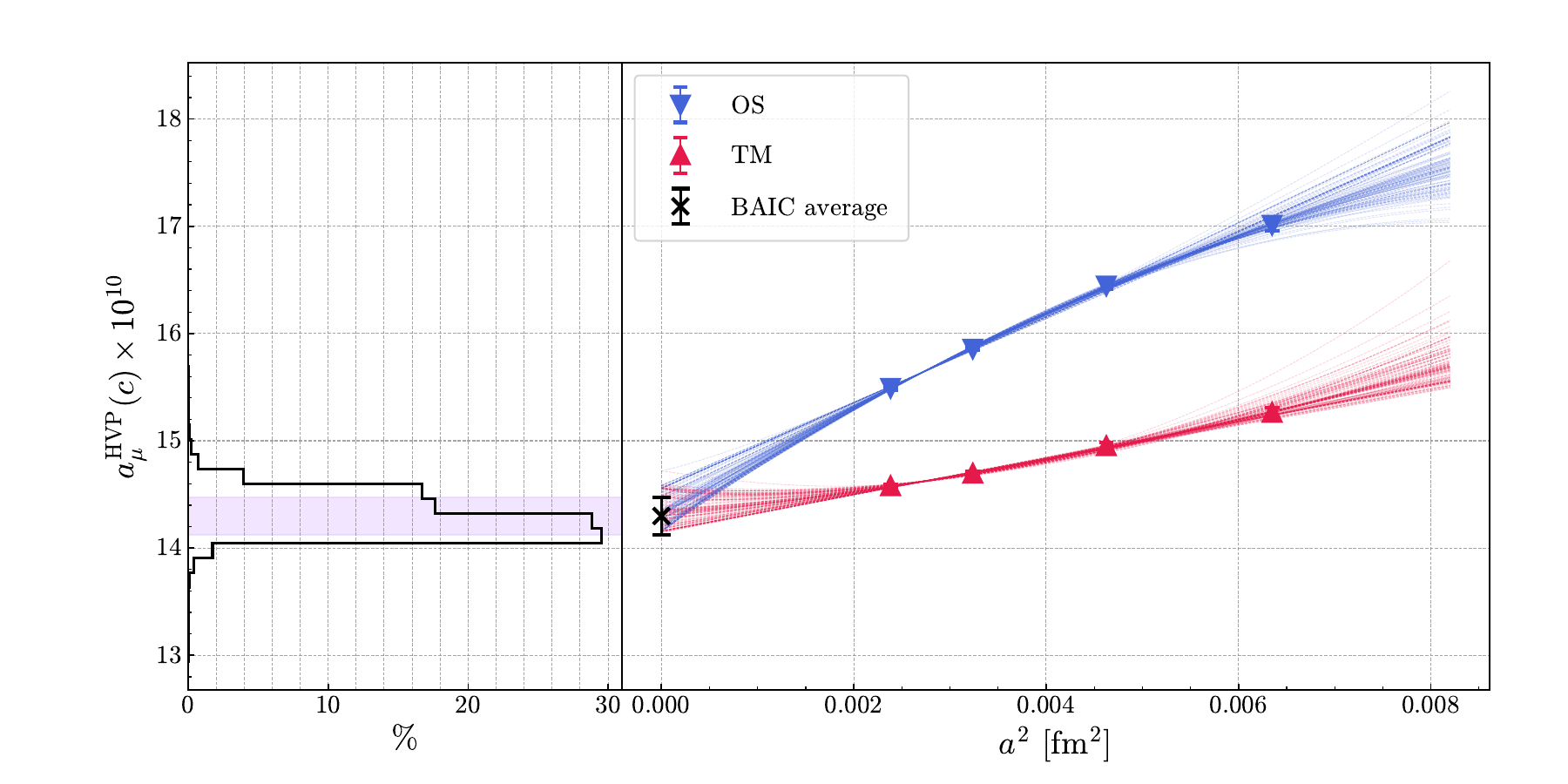}
\caption{Continuum extrapolation of the charm-connected contribution in
the FLAG scheme. The TM and OS data are fitted simultaneously imposing a
common continuum limit, with regularization-dependent cutoff terms. A
representative subset of the fit functions entering the model average is
shown with dashed lines. The distribution on the left is constructed from
the bootstrap samples of each continuum fit, weighted according to the
corresponding BAIC weights.}
\label{fig:charm_FLAG_cont}
\end{figure}

Fig.~\ref{fig:charm_FLAG_cont} shows the TM and OS data together with
a representative subset of the fit functions entering the model average.
All fits impose a common continuum value for the two regularizations. The
resulting BAIC-averaged continuum determination is shown at $a^{2}=0$.
The distribution on the left is the BAIC-weighted distribution of the
continuum-limit results and it is constructed in the same way as in the isotriplet and octet analyses. We
obtain
\begin{align}
a_{\mu}^{\rm HVP, FLAG}(c)
=\amufullcFLAG\times 10^{-10}
\,.
\label{eq:charm_FLAG}
\end{align}
Compared with our previous determination in
Ref.~\cite{ExtendedTwistedMass:2024nyi},
$a_{\mu,{\rm ETM-24}}^{\rm HVP,FLAG}(c)=14.56(13)\times10^{-10}$,
the present central value is lower by $0.26\times10^{-10}$, while the
total uncertainty is somewhat larger. The present analysis differs from
that of Ref.~\cite{ExtendedTwistedMass:2024nyi} mainly through the
improved determination of the mistuning corrections, both in the tuning
of the scale-setting parameters and in the corrections applied directly to
$a_{\mu}^{\rm HVP}(c)$. In particular, the improved determination of the
sea-quark mistuning effects on the observables entering the scale-setting analysis leads to a shift of about $1.4\sigma$ in the
charm-quark mass on the C80 ensemble with respect to the value used in
Ref.~\cite{ExtendedTwistedMass:2024nyi}. In addition, the sea-charm
mistuning correction to $a_{\mu}^{\rm HVP}(c)$ is now determined directly
from first principles using the dedicated small-volume ensembles described
in Appendix~\ref{sec:app_der_HVP}, rather than estimated using the perturbative
scaling relation adopted in Ref.~\cite{ExtendedTwistedMass:2024nyi}.

\begin{figure}[t]
\centering
\includegraphics[width=0.49\linewidth]{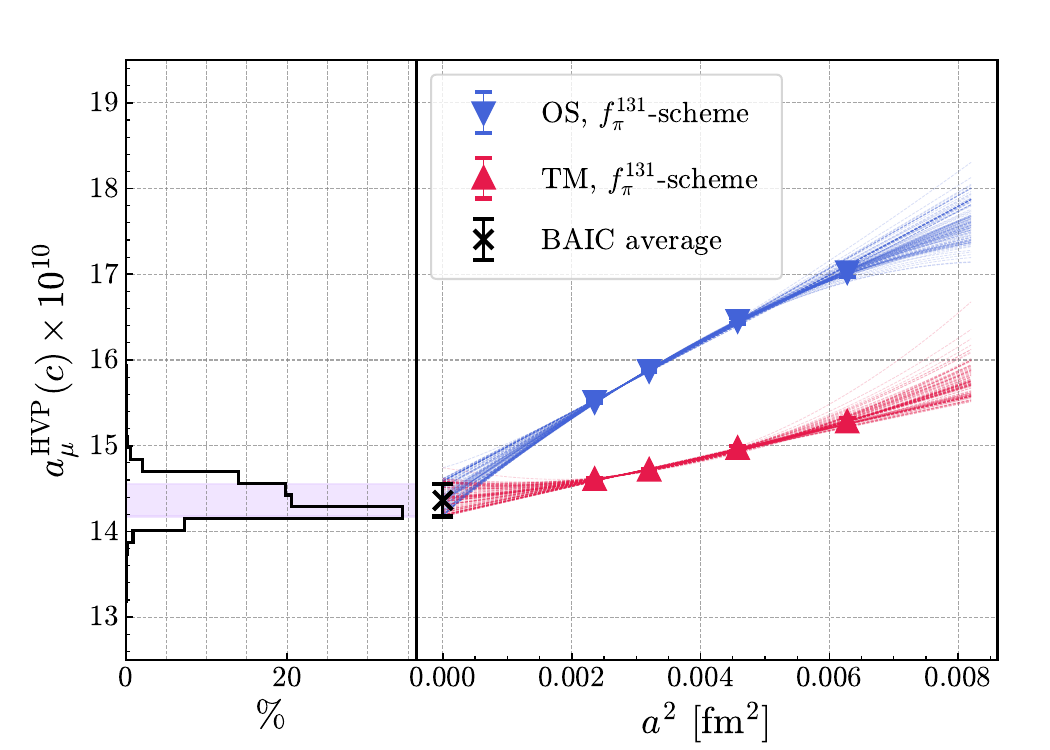}
\includegraphics[width=0.49\linewidth]{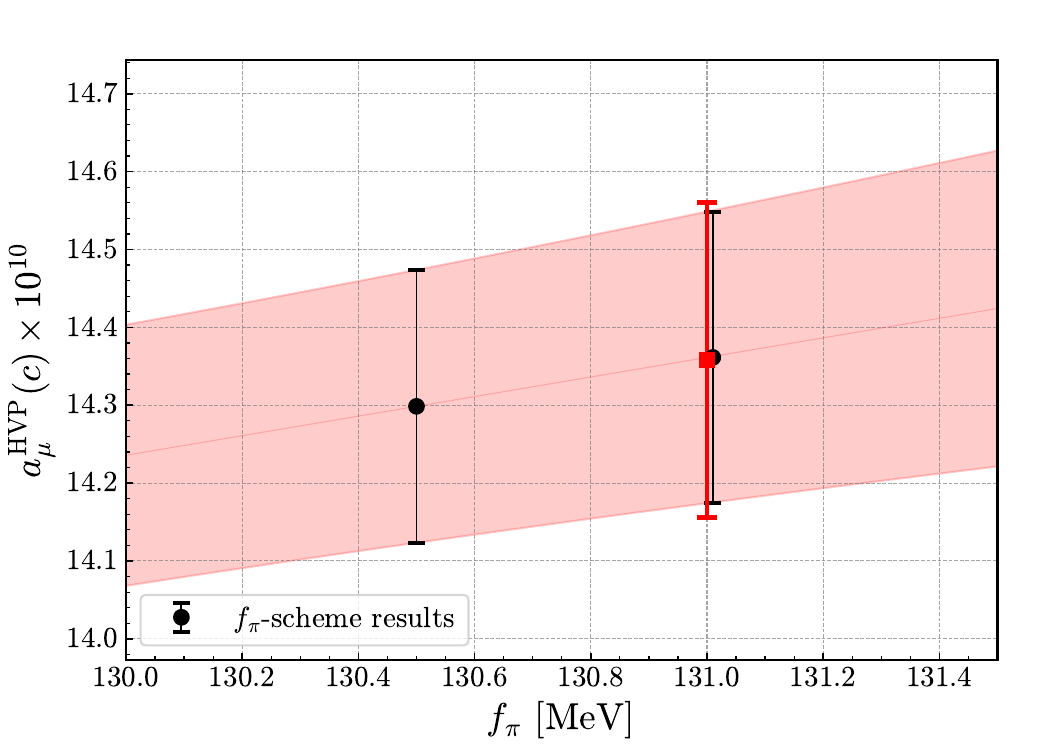}
\caption{\textit{Left}: Same as Fig.~\ref{fig:charm_FLAG_cont} but in the $f_{\pi}^{131}$-scheme. \textit{Right}:  Dependence of the charm-connected contribution on the input
value of $f_{\pi}$. The red band corresponds to the linear interpolation performed using the ansatz in
Eq.~\eqref{eq:charm_fpi_interp}. The data point indicated by the red square corresponds to the final result in the WP25 scheme.
The value corresponding to the $f_{\pi}^{131}$-scheme has been slightly shifted horizontally for better visualization.}
\label{fig:charm_WP25_matching}
\end{figure}

To determine the charm-connected contribution in the WP25 scheme, we
again use the equivalent $f_{\pi}$-based implementation of the original
$w_0$ prescription. The dependence on the input value of $f_{\pi}$ is
mild, and two continuum determinations are sufficient: the FLAG result
and the result obtained in the fixed-input $f_{\pi}^{131}$ scheme. At
$f_{\pi}=131~{\rm MeV}$, the analysis is performed exactly as in the FLAG scheme.
The corresponding continuum extrapolation is shown in the left panel of 
Fig.~\ref{fig:charm_WP25_matching}. We linearly interpolate between the two continuum determinations using
\begin{align}
a_{\mu}^{\rm HVP}(c;f_{\pi})
=
a_{\mu}^{\rm HVP,FLAG}(c)
+s_{c}
\left(f_{\pi}-f_{\pi}^{\rm FLAG}\right)\,.
\label{eq:charm_fpi_interp}
\end{align}
The value at $f_{\pi}=f_{\pi}^{\rm FLAG}$ is the FLAG-scheme result in
Eq.~\eqref{eq:charm_FLAG}, while $s_c$ gives the derivative with respect
to the input value of $f_{\pi}$. The right panel of Fig.~\ref{fig:charm_WP25_matching}
shows the two continuum results as well as the result of the linear interpolation.
The red square data point is obtained by evaluating
Eq.~\eqref{eq:charm_fpi_interp} at
$f_{\pi}=f_{\pi}^{\rm WP25}$ and propagating its uncertainty, and therefore corresponds to our final WP25-scheme determination.
We obtain
\begin{align}
a_{\mu}^{\rm HVP,WP25}(c)=\amufullcWP\times 10^{-10}~.
\label{eq:charm_WP25}
\end{align}
The corresponding derivative is 
\begin{align}
\frac{\partial a_{\mu}^{\rm HVP}(c)}{\partial f_{\pi}}
=
s_c
=
0.139\,(59)
\times10^{-10}~{\rm MeV}^{-1}\,.
\label{eq:charm_fpi_derivative}
\end{align}
As for the isotriplet and octet contributions, this derivative provides a
convenient quantity for translating the result between nearby
prescriptions for isospin-symmetric QCD.

Combining the octet and charm-connected contributions, and adding the conservative $0.0(2)\times 10^{-10}$ estimate for $\delta a_{\mu}^{c,{\rm disc}}$, we obtain
\begin{align}
a_{\mu}^{\rm HVP,FLAG}(I=0)
&=
\amufullIzeroFLAG\times10^{-10},
\label{eq:I0_FLAG_final}\\[8pt]
a_{\mu}^{\rm HVP,WP25}(I=0)
&=
\amufullIzeroWP\times10^{-10}~.
\label{eq:I0_WP25_final}
\end{align}
The relative precision of the complete isoscalar contribution is $1.7\%$.
Its absolute uncertainty is subleading compared with that of the
isotriplet contribution and therefore has only a limited impact on the
uncertainty of the complete isospin-symmetric HVP result.

\subsection{Summary and comparison with existing lattice results}
\label{sec:full_HVP_summary}
Adding the finite-volume correction given in Eq.~\eqref{eq:FSE} to our continuum results at $L=L_{\rm ref}$ for $a_{\mu}^{\rm HVP}(I=1)$, given in Eqs.~\eqref{eq:I1_FLAG_Lref}--\eqref{eq:I1_WP25_Lref}, we obtain the following values for the isotriplet contribution in the FLAG and WP25 schemes:
\begin{align}
a_{\mu}^{\rm HVP,FLAG}(I=1)
&=
\LinfamufullIoneFLAG\times10^{-10}~, \\[8pt]
a_{\mu}^{\rm HVP,WP25}(I=1)
&=
\LinfamufullIoneWP\times10^{-10}~.
\label{eq:final_I1}
\end{align}
The first uncertainty is statistical, the second is the systematic uncertainty associated with the continuum-limit extrapolation (which, in the case of the WP25-scheme determination, also includes the uncertainty in $f_{\pi}^{\rm WP25}$), while the third arises from finite-volume effects. The uncertainty in square brackets is the total uncertainty. We thus determine this contribution with a precision of $1.05\%$ in the FLAG scheme and $1.15\%$ in the WP25 scheme. For the octet, charm-connected, and full $I=0$ contributions, we have shown in Subsec.~\ref{sec:full_HVP_I0}, in agreement with previous studies~\cite{Boccaletti:2024guq,Djukanovic:2024cmq}, that finite-volume effects are negligibly small. Our final estimates are given in Eqs.~\eqref{eq:oct_FLAG}, \eqref{eq:charm_FLAG}, and \eqref{eq:I0_FLAG_final} in the FLAG scheme, and in Eqs.~\eqref{eq:oct_WP25}, \eqref{eq:charm_WP25}, and \eqref{eq:I0_WP25_final} in the WP25 scheme.

\begin{figure*}
\centering
\includegraphics[width=0.8\linewidth]{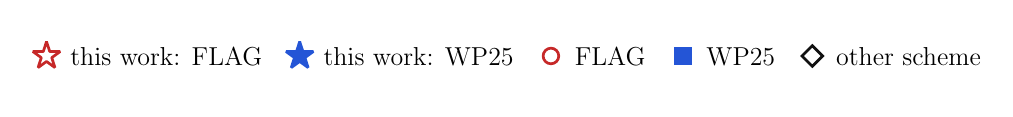}\\[-15pt]
  \resizebox{\textwidth}{!}{%
\includegraphics[]{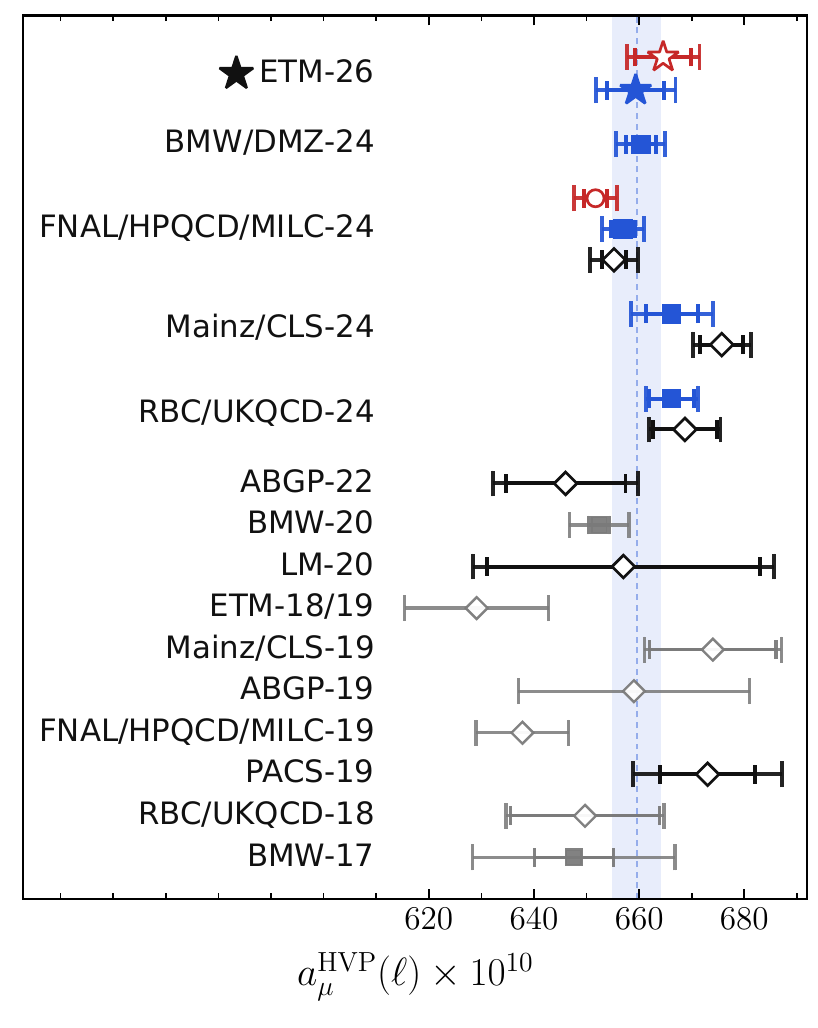}\hspace{-0.25cm}
\includegraphics[]{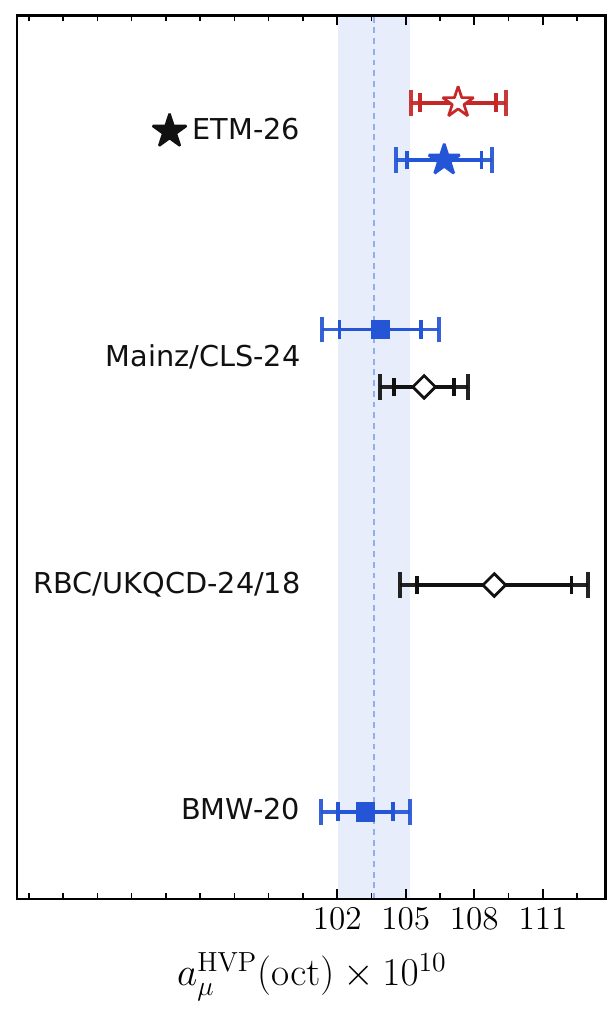}\hspace{-0.25cm}
\includegraphics[]{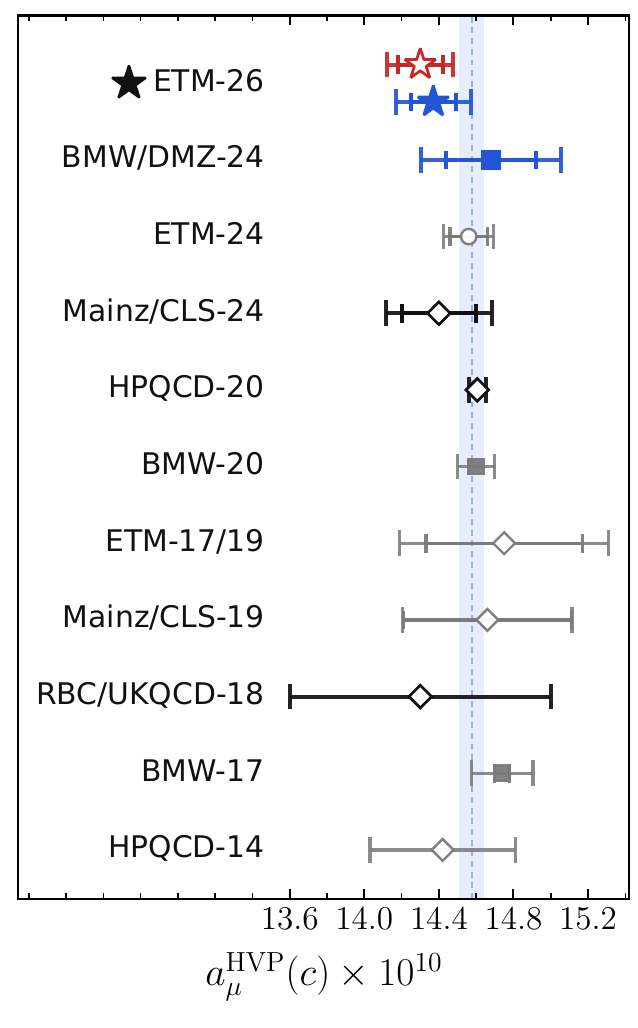}}
\caption{Comparison of lattice QCD determinations of the light-connected
contribution $a_{\mu}^{\rm HVP}(\ell)$ (\textit{left}), the octet contribution
$a_{\mu}^{\rm HVP}({\rm oct})$ (\textit{middle}), and the charm-connected
contribution $a_{\mu}^{\rm HVP}(c)$ (\textit{right}). Our results are highlighted
by star markers, open for the FLAG scheme and filled for the WP25 scheme.
Results from other collaborations are shown as open circles in the FLAG
scheme, filled squares in the WP25 scheme, and open diamonds in other
isospin-symmetric schemes. Colors provide an additional indication of the
scheme, with red used for FLAG, blue for WP25, and black for the other
schemes. Gray symbols denote superseded determinations. The vertical bands correspond to the WP25 averages~\cite{Aliberti:2025beg}.
References to the quoted papers are as follows:
ABGP-19~\cite{Aubin:2019usy},
ABGP-22~\cite{Aubin:2022hgm},
BMW-17~\cite{Budapest-Marseille-Wuppertal:2017okr},
BMW-20~\cite{Borsanyi:2020mff},
BMW/DMZ-24~\cite{Boccaletti:2024guq},
ETM-17~\cite{Giusti:2017jof},
ETM-18~\cite{Giusti:2018mdh},
ETM-19~\cite{Giusti:2019hkz},
ETM-24~\cite{ExtendedTwistedMass:2024nyi},
FNAL/HPQCD/MILC-19~\cite{FermilabLattice:2019dbx},
FNAL/HPQCD/MILC-24~\cite{FermilabLatticeHPQCD:2024ppc},
HPQCD-14~\cite{Chakraborty:2014mwa},
HPQCD-20~\cite{Hatton:2020qhk},
LM-20~\cite{Lehner:2020crt},
Mainz/CLS-19~\cite{Gerardin:2019rua},
Mainz/CLS-24~\cite{Djukanovic:2024cmq},
PACS-19~\cite{Shintani:2019wai},
RBC/UKQCD-18~\cite{RBC:2018dos},
RBC/UKQCD-24~\cite{RBC:2024fic}.
}
\label{fig:comparison_panel1}
\end{figure*}

In Fig.~\ref{fig:comparison_panel1}, we compare our results for the light-connected contribution, $a_{\mu}^{\rm HVP}(\ell)=\frac{10}{9}a_{\mu}^{\rm HVP}(I=1)$, the octet contribution, $a_{\mu}^{\rm HVP}({\rm oct})$, and the charm-connected contribution, $a_{\mu}^{\rm HVP}(c)$, with existing determinations by other lattice groups~\cite{Budapest-Marseille-Wuppertal:2017okr,Borsanyi:2020mff, Boccaletti:2024guq, Gerardin:2019rua, Djukanovic:2024cmq, RBC:2018dos, RBC:2024fic, FermilabLattice:2019dbx, FermilabLatticeHPQCD:2024ppc, Giusti:2018mdh, Giusti:2019hkz, ExtendedTwistedMass:2024nyi, Aubin:2019usy, Aubin:2022hgm, Shintani:2019wai, Lehner:2020crt, Hatton:2020qhk,Chakraborty:2014mwa}. We display our results in both the FLAG and WP25 schemes. Whenever possible, using information provided in the original publications or in the WP25 report~\cite{Aliberti:2025beg}, we quote the results of other groups both in their preferred isospin-symmetric scheme and in the WP25 and/or FLAG schemes.\footnote{For this comparison, we assume, as done in the WP25 report~\cite{Aliberti:2025beg}, the ``\textit{BMW-20 scheme $=$ WP25 scheme}'' correspondence.}

We adopt the following graphical convention. Our results are highlighted by
star markers, open for the FLAG scheme and filled for the WP25 scheme.
Results from other groups are represented by open circles when quoted in the
FLAG scheme, by filled squares when quoted in the WP25 scheme, and by open
diamonds when quoted in a scheme different from both FLAG and WP25. Colors
provide an additional indication of the scheme, with red used for FLAG, blue
for WP25, and black for the other schemes. Gray points indicate superseded
results.

As shown in the left panel of Fig.~\ref{fig:comparison_panel1}, our result for the light-connected contribution in the WP25 scheme is compatible within one standard deviation with all recent lattice determinations performed in the same scheme. These include the FNAL/HPQCD/MILC-24 result~\cite{FermilabLatticeHPQCD:2024ppc}, the Mainz/CLS-24 result~\cite{Djukanovic:2024cmq}, the BMW-20 result~\cite{Borsanyi:2020mff}, the lattice-only BMW/DMZ-24 result~\cite{Boccaletti:2024guq},\footnote{The BMW/DMZ-24 result for $a_{\mu}^{\rm HVP}(\ell)$ shown in the figure is obtained by adding to the fixed-volume pure-lattice determination given in Eq.~(S69) of Ref.~\cite{Boccaletti:2024guq} the finite-volume correction reported in the penultimate column of Table~S21 of the same reference. Since the finite-volume correction given in that table refers to the $I=1$ contribution, we include the usual conversion factor $10/9$.} and the RBC/UKQCD-24 determination~\cite{RBC:2024fic}. The largest differences are observed with respect to the Mainz/CLS-24 and RBC/UKQCD-24 results, whose central values exceed ours by $6.9\times 10^{-10}$. Nevertheless, both remain compatible with our result at the $1\sigma$ level within the quoted uncertainties. The average quoted in WP25, represented by the blue band in the figure, is in excellent agreement with our determination.

The comparison in the FLAG scheme is less straightforward. Besides our result, the only determination that can be identified unambiguously as a FLAG-scheme determination is the $f_{\pi}$-based FNAL/HPQCD/MILC-24 result~\cite{FermilabLatticeHPQCD:2024ppc}, since the central values of all its hadronic inputs coincide with those adopted in the FLAG scheme. Our FLAG-scheme result exceeds the FNAL/HPQCD/MILC-24 determination by approximately $13\times10^{-10}$, corresponding to a significance of about $1.6\sigma$. Moreover, the two calculations find opposite signs for the shift from the FLAG to the WP25 scheme: the correction is positive for FNAL/HPQCD/MILC-24 and negative in our analysis. It is also worth noting that the preferred $f_{\pi}$-based scheme used by Mainz/CLS-24 is numerically very close to the FLAG scheme. Their input value for $f_{\pi}$ is compatible, well within the quoted uncertainties, with $f_{\pi}^{\rm FLAG}=130.5~{\rm MeV}$. Their pion mass, $m_{\pi}=134.9768(5)~{\rm MeV}$, differs from the FLAG value by less than $0.02\%$. Moreover, the input values used for $m_K$ and $m_{D_s}$, which have only a subleading impact on $a_{\mu}^{\rm HVP}(\ell)$, differ from the corresponding FLAG values by only $0.081\%$ and $0.075\%$, respectively. Such small differences are unlikely to be relevant at the present precision of $0.6-1\%$. It is therefore reasonable to regard the $f_{\pi}$-based Mainz/CLS-24 determination as effectively corresponding to the FLAG scheme. Under this well-motivated assumption, the comparison between the $f_{\pi}$-based Mainz/CLS-24 determination and our result follows the same pattern observed in the WP25 scheme: the Mainz/CLS-24 central value is slightly larger than ours. Owing to the smaller uncertainties of the $f_{\pi}$-based determinations, the difference is somewhat more significant in the FLAG scheme, amounting to $1.3\sigma$, which however can still be attributed to a statistical fluctuation. By contrast, the difference between the same Mainz/CLS-24 determination and the FNAL/HPQCD/MILC-24 determination in the FLAG scheme reaches a significance of approximately $3.5\sigma$. 

Therefore, while an overall agreement is observed among the recent
lattice calculations of $a_{\mu}^{\rm HVP}(\ell)$, the spread among some
of them and the different shifts predicted when moving from the WP25 to the
FLAG scheme deserve further investigation. In particular, the agreement
among lattice determinations is better in the WP25 scheme, whereas the
different shifts to the FLAG scheme increase the spread among the results. 

For completeness, we now comment on the earlier ETM-18/19
determination~\cite{Giusti:2018mdh,Giusti:2019hkz}, which is shown as a
superseded result in Fig.~\ref{fig:comparison_panel1}. Our present result for $a_\mu^{HVP}(\ell)$ in the FLAG scheme is
larger than the ETM-18/19 one by approximately $35\times10^{-10}$, corresponding to a difference
of about $2.3\sigma$. That calculation was performed with $N_f=2+1+1$ 
twisted-mass fermions (without clover improvement) using ensembles with pion masses above the physical
value, and therefore required a significant extrapolation to the physical point. Moreover, the MLLGS model was used not only to estimate finite-volume
effects, but also in the determination of
$a_{\mu}^{\rm HVP}(\ell)$: its parameters were fitted to the lattice
correlator, and the resulting MLLGS representation was then used to evaluate
the long-distance contribution to the HVP. In that treatment, the
two-pion contribution was described assuming degenerate pion masses, whereas
for the TM regularization used in that analysis the 
two-pion states at finite lattice spacing are of the $\pi^+\pi^0$ type. Finally, in Refs.~\cite{Giusti:2018mdh,Giusti:2019hkz} a duality model was adopted to describe the vector correlator at short and intermediate time distances. 
While we are not in a position to
quantitatively assess the impact of these effects on the final result, the sizable pion-mass extrapolation and the model-dependent treatment of the
vector correlator in the analysis of Refs.~\cite{Giusti:2018mdh,Giusti:2019hkz} may have contributed to the observed difference.

Turning to the octet contribution, shown in the middle panel of Fig.~\ref{fig:comparison_panel1}, to our knowledge three independent lattice determinations are available for comparison: the direct Mainz/CLS-24 calculation~\cite{Djukanovic:2024cmq}, the BMW-20 result~\cite{Borsanyi:2020mff}, and a result reconstructed from RBC/UKQCD data by combining the determination of $a_{\mu}^{\rm HVP}(\ell)$ from RBC/UKQCD-24~\cite{RBC:2024fic} with the strange-connected and charmless disconnected contributions computed in RBC/UKQCD-18~\cite{RBC:2018dos}. For RBC/UKQCD, we obtain the octet contribution by adding one tenth of the light-connected contribution given in Ref.~\cite{RBC:2024fic} to the sum of the strange-connected and disconnected $uds$ contributions from Ref.~\cite{RBC:2018dos}, combining all quoted uncertainties in quadrature. The BMW-20 result for the octet contribution, shown in the figure, is obtained from the infinite-volume-extrapolated values reported in Table~11 of Ref.~\cite{Aliberti:2025beg}.\footnote{Because finite-volume effects are strongly suppressed in the octet contribution, essentially the same value is obtained by combining the single-flavor and disconnected contributions shown in Fig.~1 of Ref.~\cite{Borsanyi:2020mff}, which correspond to a reference lattice extent  $L=6.272~{\rm fm}$.} We note that the disconnected contribution computed by BMW-20 also includes charm-disconnected diagrams. However, their presence is irrelevant for the present comparison since BMW-20 estimates this effect to be $0.0(1)\times10^{-10}$. As discussed in Subsec.~\ref{sec:full_HVP_I0}, we similarly estimate a very small effect of $0.0(2)\times10^{-10}$. The determinations in the WP25 scheme are in reasonable agreement. Our result is slightly larger than both the Mainz/CLS-24 and BMW-20 determinations, but the differences are not statistically significant, amounting to $0.8\sigma$ and $1.2\sigma$, respectively.\footnote{Shifting the BMW-20 determination of $a_{\mu}^{\rm HVP}({\rm oct})$ by the difference in $\frac{1}{10}a_{\mu}^{\rm HVP}(\ell)$ between the BMW/DMZ-24 and BMW-20 determinations, while keeping the total uncertainty unchanged, improves the compatibility to approximately $0.9\sigma$.} Under the assumption discussed above that the $f_{\pi}$-based Mainz/CLS-24 determination can effectively be regarded as a FLAG-scheme result, we also find good agreement between Mainz/CLS-24 and our determination in the FLAG scheme. Our central value is again slightly larger, but the two results are compatible at  the $0.6\sigma$ level.

The comparison for the charm-connected contribution, shown in the right panel of Fig.~\ref{fig:comparison_panel1}, indicates that our FLAG-scheme result has a smaller central value than the ETM-24 determination, which is now a superseded result. The shift amounts to $0.26\times10^{-10}$, corresponding to about $1.5$ times the total uncertainty of the present determination. The origin of the slightly lower central value has already been discussed in Subsec.~\ref{sec:full_HVP_I0}. We do not observe any significant scheme dependence when moving from the FLAG to the WP25 scheme.

To enable a comparison for the strange-connected and purely disconnected contributions, we have repeated the analysis of the strange-connected contribution $a_{\mu}^{\rm HVP}(s)$ using the same procedure described in Ref.~\cite{ExtendedTwistedMass:2024nyi}. The only difference with respect to that analysis is the improved treatment of quark-mass mistuning effects, made possible by substantially increased statistics and by the use of a dedicated small-volume ensemble to determine the noisy sea strange- and charm-quark mistuning corrections; see the discussion in Sec.~\ref{sec:scale setting} and in Appendix~\ref{sec:app_der_HVP}.

Repeating the analysis of Ref.~\cite{ExtendedTwistedMass:2024nyi}, we obtain the following updated value in the FLAG scheme:
\begin{align}
\label{eq:FLAG_strange_update}
a_{\mu}^{\rm HVP,FLAG}(s)
&=\amufullsFLAG\times10^{-10}~,
\end{align}
and the following result in the WP25 scheme:
\begin{align}
\label{eq:WP25_strange_update}
a_{\mu}^{\rm HVP,WP25}(s)
&=\amufullsWP\times10^{-10}~.
\end{align}
The updated FLAG-scheme result is fully consistent with the ETM-24 value, $a_{\mu,{\rm ETM-24}}^{\rm HVP,FLAG}(s)=53.57(63)\times10^{-10}$, while reducing the uncertainty by approximately $40\%$. This improvement is driven by the refined treatment of the mistuning corrections discussed above. As in the case of $a_{\mu}^{\rm HVP}(c)$, we do not observe any significant difference between the results in the two schemes.

 Finally, we turn to the disconnected contribution, which we estimate as follows. Exploiting Eq.~\eqref{eq:octet_decomposition}, we subtract from the octet contribution one tenth of the light-connected contribution, $a_{\mu}^{\rm HVP}(\ell)$, and the strange-connected contribution, $a_{\mu}^{\rm HVP}(s)$, thereby obtaining the sum of the charmless disconnected $uds$ contributions, denoted as $a_{\mu}^{\rm HVP}({\rm disc}, uds)$ in Eq.~\eqref{eq:octet_decomposition}. To remain conservative, we combine separately in quadrature the statistical and systematic uncertainties of the three quantities. This prescription neglects correlations among the corresponding continuum-extrapolated results. We expect correlations between the disconnected contribution and the connected combination $\frac{1}{10}a_{\mu}^{\rm HVP}(\ell)+a_{\mu}^{\rm HVP}(s)$ to be negligible, consistently with the very different noise properties of the underlying connected and disconnected correlators as functions of Euclidean time. The octet contribution is therefore positively correlated with the connected contribution that is subtracted from it. The corresponding covariance enters the variance of the disconnected contribution with a negative sign and thus reduces its uncertainty. Neglecting this correlation, as we do, therefore provides a conservative estimate of the final uncertainty.

To the resulting $a_{\mu}^{\rm HVP}({\rm disc}, uds)$, we add the estimate $\delta a_{\mu}^{c,{\rm disc}}=0.0(2)\times10^{-10}$ given in Subsec.~\ref{sec:full_HVP_I0}, which has a negligible impact on the final uncertainty. We obtain in this way
\begin{align}
a_{\mu}^{\rm HVP,FLAG}({\rm disc})
&=
\amufulludsdiscFLAG\times10^{-10}~, \\[8pt]
a_{\mu}^{\rm HVP,WP25}({\rm disc})
&=
\amufulludsdiscWP\times10^{-10}~.
\end{align}

\begin{figure}
\centering
\includegraphics[width=0.8\linewidth]
{comparison_plots/legend_new.pdf}\\[-15pt]
\includegraphics[width=0.40\linewidth]{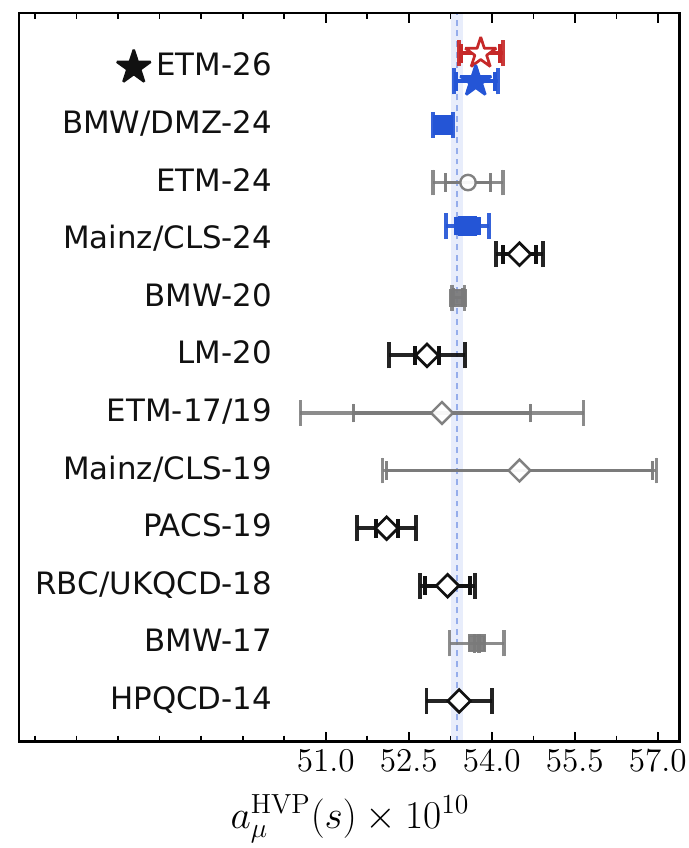}
\hspace{1.2cm}\includegraphics[width=0.40\linewidth]{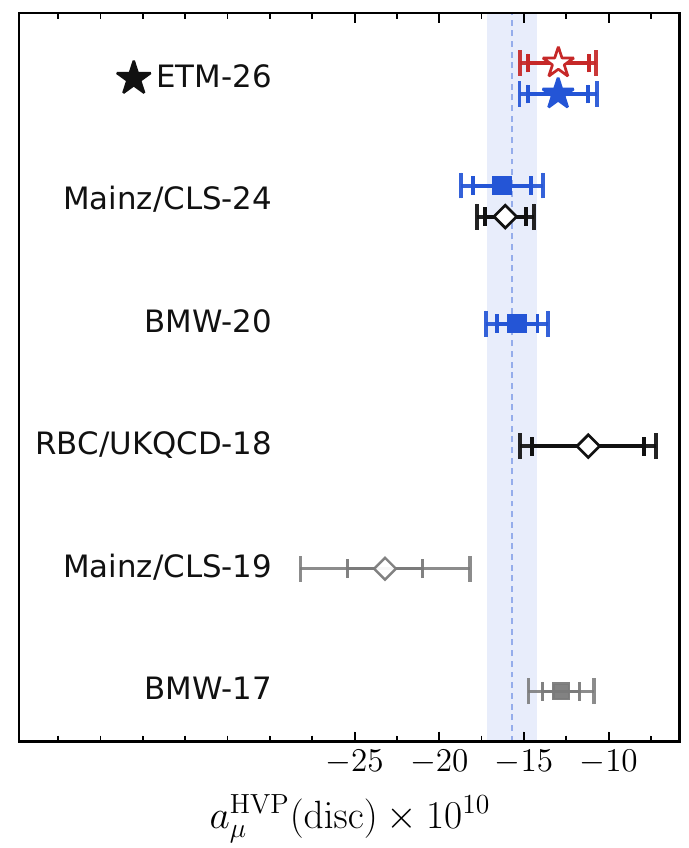}
\caption{Comparison of lattice-QCD determinations of the strange-connected
contribution $a_{\mu}^{\rm HVP}(s)$ (\textit{left}) and the purely disconnected
contribution $a_{\mu}^{\rm HVP}({\rm disc})$ (\textit{right}). The graphical
conventions are the same as in Fig.~\ref{fig:comparison_panel1}. The vertical bands correspond to the WP25 averages~\cite{Aliberti:2025beg}.
References to the quoted papers are as follows:
BMW-17~\cite{Budapest-Marseille-Wuppertal:2017okr},
BMW-20~\cite{Borsanyi:2020mff},
BMW/DMZ-24~\cite{Boccaletti:2024guq},
ETM-17~\cite{Giusti:2017jof},
ETM-19~\cite{Giusti:2019hkz},
ETM-24~\cite{ExtendedTwistedMass:2024nyi},
HPQCD-14~\cite{Chakraborty:2014mwa},
LM-20~\cite{Lehner:2020crt},
Mainz/CLS-19~\cite{Gerardin:2019rua},
Mainz/CLS-24~\cite{Djukanovic:2024cmq},
PACS-19~\cite{Shintani:2019wai},
RBC/UKQCD-18~\cite{RBC:2018dos}.
}
\label{fig:comparison_panel2}
\end{figure}

Having determined both the strange-connected and disconnected contributions,
we compare them with existing lattice determinations in
Fig.~\ref{fig:comparison_panel2}, using the same graphical conventions as in
Fig.~\ref{fig:comparison_panel1}.
The comparison of $a_{\mu}^{\rm HVP}({\rm disc})$ shows a pattern similar to that observed for $a_{\mu}^{\rm HVP}({\rm oct})$, with the BMW-20 and Mainz results having slightly smaller central values, while remaining compatible within approximately $1\sigma$. For $a_{\mu}^{\rm HVP}(s)$, good agreement is found with the most recent lattice determinations. It is worth noting the comparatively large shift between the Mainz/CLS-24
strange-connected result in their $f_{\pi}$-based scheme and the value
attributed to the same calculation in the WP25 scheme in
Ref.~\cite{Aliberti:2025beg}. This shift is larger than the very small difference between our FLAG- and WP25-scheme determinations. Although the kaon mass used by Mainz/CLS is approximately $0.08\%$ larger than the FLAG value, it seems unlikely that this difference alone could account for a shift of about $1.7\%$ in $a_{\mu}^{\rm HVP}(s)$. 

Combining the isotriplet ($I=1$) and isosinglet ($I=0$) contributions in
the FLAG and WP25 schemes, we obtain the final values for the
isospin-symmetric leading-order HVP given in
Eq.~\eqref{eq:isospin_symmetric_HVP}. In combining the statistical
uncertainties, we include a correlation coefficient of $0.25$ between the
$I=1$ and $I=0$ contributions. This corresponds to the ensemble-averaged
correlation between $a_{\mu}^{\rm HVP}(I=0)$ and the determination of
$a_{\mu}^{\rm HVP}(I=1)$ in the OS regularization. The latter is the determination that enters explicitly in the
construction of the $I=0$ contribution through the term
$\frac{1}{9}a_{\mu}^{\rm HVP}(I=1)$. This is a conservative choice, since
the corresponding correlation with the TM determination is substantially
smaller, while the final isotriplet result combines both TM and OS data,
thereby reducing the correlation between the continuum-extrapolated $I=1$
and $I=0$ contributions.

\section{Window observables of the HVP}
\label{sec:windows}
Having computed the different contributions to
$a_{\mu}^{\rm HVP}({\rm iso})$ in the previous section, we now present the
results for the three RBC/UKQCD windows introduced in
Eqs.~\eqref{eq:amu_w}--\eqref{eq:Mt_LD}. The analysis closely follows that of
the full $a_{\mu}^{\rm HVP}$, and we therefore mainly highlight the differences with respect to that analysis,
where present. We determine the short-distance (SD), intermediate (W), and
long-distance (LD) windows independently. Although the latter can also be
obtained by subtracting
$a_{\mu}^{\rm SD}+a_{\mu}^{\rm W}$ from $a_{\mu}^{\rm HVP}$, we determine
it directly through an independent analysis. Verifying a posteriori that
$a_{\mu}^{\rm SD}+a_{\mu}^{\rm W}+a_{\mu}^{\rm LD}$ is well compatible
with the result obtained in the previous section for $a_{\mu}^{\rm HVP}$ provides a cross-check of
the internal consistency of the different analyses. Similarly to the previous section, we use the notation $a_{\mu}^{\rm w, FLAG}$ and $a_{\mu}^{\rm w, WP25}$, with ${\rm w}=\{\rm SD,W,LD\}$, to indicate our results in the FLAG and WP25 scheme, respectively. 

The structure of this section closely follows that of
Sec.~\ref{sec:full_HVP}: we first consider the $I=1$ contribution and then the
$I=0$ contribution. Finally, we summarize our results and compare them with existing lattice determinations.
\subsection{The $I=1$ contribution  }
\label{sec:window_I1}
The isotriplet windows are evaluated from the same TM and OS light-connected
correlators used for the full contribution. As in
Subsec.~\ref{sec:full_HVP_I1}, we first perform the continuum-limit extrapolation at the
reference spatial extent $L_{\rm ref}=5.46~{\rm fm}$. The infinite-volume
extrapolation of the SD, W, and LD contributions are discussed in  Subsec.~\ref{sec:window_FSE}.

We begin with the LD contribution, $a_{\mu}^{\rm LD}(I=1)$, which dominates the isotriplet result and
is analyzed exactly as the full $I=1$ HVP. The B64 and B96 determinations are
interpolated to $L_{\rm ref}$ independently for the TM and OS regularizations
using the same ansatz as in Eq.~\eqref{eq:I1_L_interpolation}.
The resulting data set consisting of eight points (four for each of the two regularizations) is then extrapolated to the continuum by
fitting the two regularizations simultaneously,
using the same family of fit functions  adopted for $a_{\mu}^{\rm HVP}(I=1)$.

\begin{figure}[t]
\centering
\includegraphics[width=\linewidth]{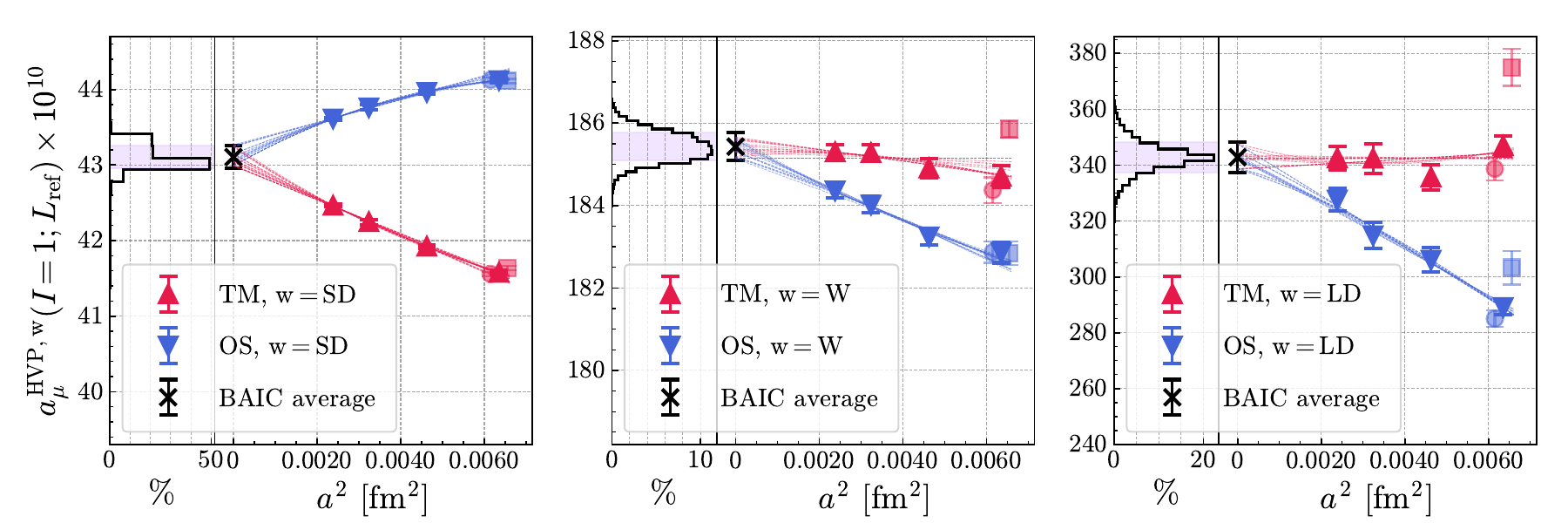}
\caption{Continuum extrapolations of the isotriplet SD (\textit{left}), W (\textit{center}),
and LD (\textit{right}) contributions in the FLAG scheme. The TM and OS data are fitted
simultaneously imposing a common continuum limit and regularization-dependent
cutoff effects. 
At the coarsest lattice spacing, the B64 (circles) and the B96 (squares) results are shown, displaced horizontally by a small offset for readability, in addition to the values interpolated to $L_{\rm ref}$. A
representative subset of the fit functions entering the BAIC average is
displayed together with the corresponding model-averaged probability
distribution. The W and LD data exhibit only mild cutoff effects, while the SD window, as expected, shows a more pronounced lattice-spacing dependence.}
\label{fig:I1_windows_FLAG_cont}
\end{figure}

The intermediate window, $a_{\mu}^{\rm W}(I=1)$, follows the same route. In particular, the B64--B96 interpolation
uses again the $L e^{-m_\pi L}$ functional form of
Eq.~\eqref{eq:I1_L_interpolation}, and the continuum limit is obtained from
the same fit pool used for the LD and full contributions. As will be visible
in the central panel of Fig.~\ref{fig:I1_windows_FLAG_cont}, cutoff effects
are very mild also in this window. The statistical precision of our data points is of about $0.1-0.15\%$. Relative to our previous determination in
Ref.~\cite{ExtendedTwistedMass:2022jpw}, the present analysis includes the
finer E112 ensemble, substantially increased statistics, and corrections for
the mistuning of all action parameters. In
Ref.~\cite{ExtendedTwistedMass:2022jpw}, only the mistuning of the light-quark
mass was taken into account. No tree-level subtraction is applied to either the W or LD contribution. The
improvement of Eqs.~\eqref{eq:TL_integrand}--\eqref{eq:TL_improved_observable}
is designed to remove cutoff effects generated by the integration of the
region $t\sim O(a)$, which is exponentially suppressed by both
$\Theta^{\rm W}(t)$ and $\Theta^{\rm LD}(t)$.

For the SD window, $a_{\mu}^{\rm SD}(I=1)$, the improvements relative to
Ref.~\cite{ExtendedTwistedMass:2022jpw} are the same as for the intermediate window. In contrast to the case of the W and LD contribution, in this case cutoff effects are instead much more pronounced. Their absolute
size is modest, approximately $1\times10^{-10}$ over the range of lattice
spacings considered, yet it is much larger than the statistical uncertainty
of the individual data points. A precise continuum determination therefore
requires enough flexibility to describe terms beyond a simple linear
dependence on $a^2$. In this case, to describe the cutoff dependence, we start from the general simultaneous
TM--OS ansatz of Eq.~\eqref{eq:I1_continuum_ansatz}. Unlike the full, W, and LD
analyses, we retain all combinations containing up to five free parameters,
including the common continuum limit. Fits with six or seven free parameters
are excluded because, with only eight data points, they would leave two or one
degrees of freedom and would be particularly susceptible to overfitting. In
addition to the baseline choice $\gamma=0.42$ for the logarithmically modified
$a^2$ term, we also include fits with $\gamma=1$ and $2$. These variations
allow for residual logarithmically modified cutoff effects dynamically
generated by the short-distance integration beyond the tree-level term that
has already been subtracted, for instance contributions of the form
$a^2\alpha_s^{n}(1/a)\log(a)$ with $n=2,3$. The continuum results are combined
with the BAIC procedure defined in
Eqs.~\eqref{eq:BAIC_weight}--\eqref{eq:BAIC_syst}.

Fig.~\ref{fig:I1_windows_FLAG_cont} collects the FLAG-scheme continuum
extrapolations for all three windows. The contrast among the panels makes the
main features of the lattice-spacing dependence transparent: the W and LD data show only
a mild dependence on $a^{2}$, whereas the much smaller statistical
errors in the SD window make cut-off effects behind a simple linear $a^{2}$-term visible.  In addition to the points interpolated
to $L_{\rm ref}$, the B64 and B96 determinations are also displayed to make the size of the finite-volume dependence
visible. The black markers in each figure correspond to the final continuum-extrapolated values after application of the BAIC procedure, while the dashed lines are selected examples of continuum fits.

Because the continuum extrapolation is most delicate in the SD window, we
subject the previous analysis to an additional validation. Following the
complementary strategy introduced in
Ref.~\cite{ExtendedTwistedMass:2022jpw}, we remove from the Euclidean-time
integral the region below a fixed physical time $t_{\rm min}$ and define
\begin{align}
 a_{\mu}^{\rm SD}(I=1;t_{\rm min})
 ={}&2\alpha_{\rm em}^{2}
 \int_{t_{\rm min}}^{\infty}dt\,t^2 K(m_\mu t)
 \Theta^{\rm SD}(t)V^{I=1}(t)\,.
\label{eq:I1_SD_tmin}
\end{align}
The continuum limit is then taken at fixed $t_{\rm min}$, chosen larger
than the coarsest lattice spacing. Since the region $t\sim O(a)$ is
excluded from the integration, the logarithmically enhanced
short-distance cutoff effects generated by the integration are eliminated
by construction. For each $t_{\rm min}$, we therefore use the same fit
pool as in the standard analysis, except that the additional
logarithmically modified terms with $\gamma=1,2$ are no longer included.
The simultaneous TM--OS fit and the model-average criterion are otherwise
unchanged.

\begin{figure}[t]
\centering
\includegraphics[width=\linewidth]{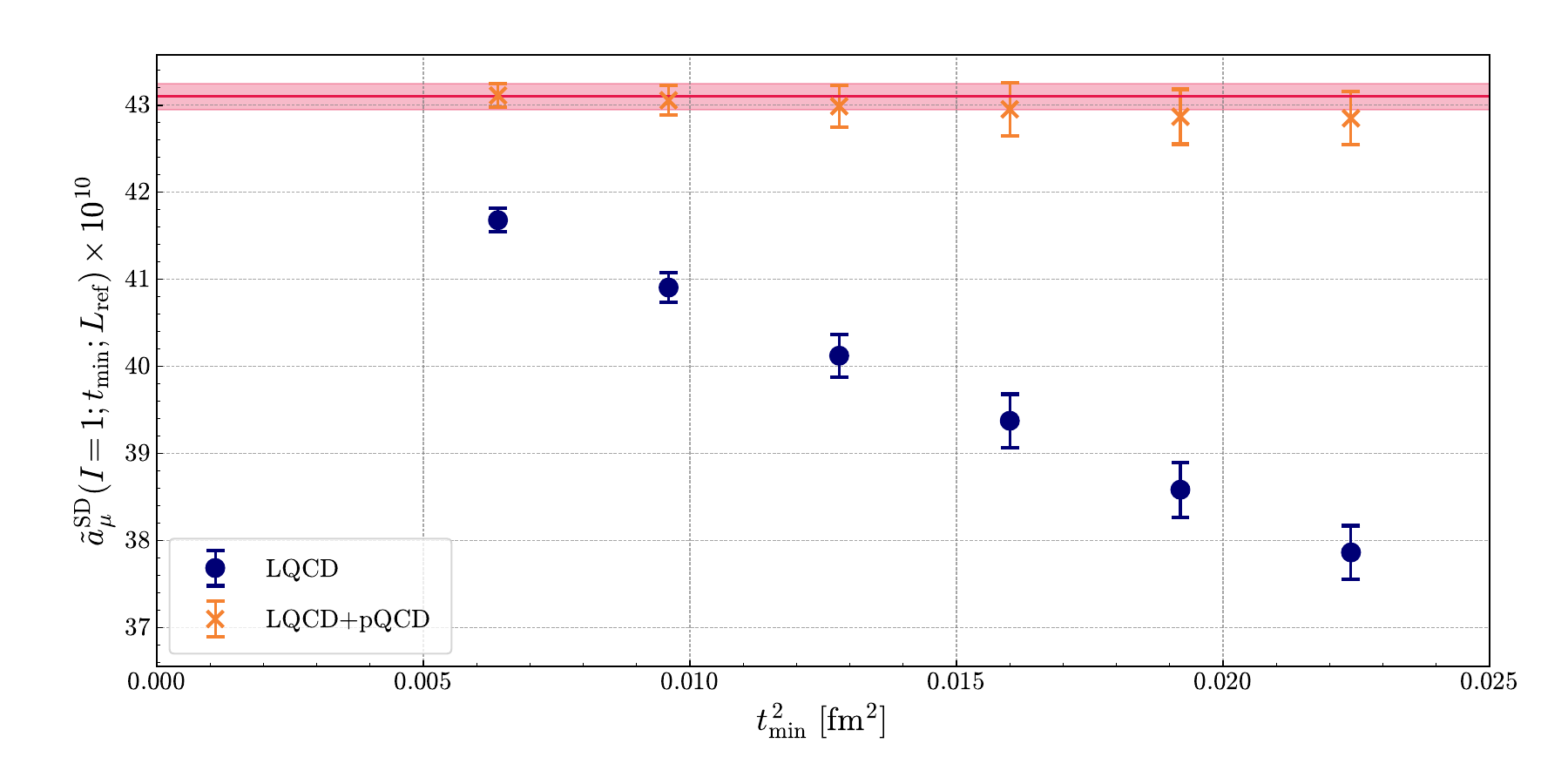}
\caption{Continuum-limit results of the $t_{\rm min}$ analysis for the isotriplet
short-distance contribution at $L=L_{\rm ref}$ in the FLAG scheme. The blue circles show the
truncated contribution
$a_{\mu}^{\rm SD}(I=1;t_{\rm min};L_{\rm ref})$, obtained by taking the
continuum limit at fixed $L_{\rm ref}$. The orange crosses show
$\widetilde a_{\mu}^{\rm SD}(I=1;t_{\rm min};L_{\rm ref})$, which here
denotes the result obtained by adding to
$a_{\mu}^{\rm SD}(I=1;t_{\rm min};L_{\rm ref})$ the N$^{4}$LO massless
continuum perturbative contribution over the interval
$t\in[0,t_{\rm min}]$, evaluated with RHAD. This perturbative contribution
is evaluated in infinite volume rather than at $L=L_{\rm ref}$; the
resulting difference is entirely negligible at the present precision.
The horizontal band shows the result of the standard $t_{\rm min}=0$
analysis at the same reference volume.}
\label{fig:I1_SD_tmin}
\end{figure}

The omitted contribution is then restored using continuum perturbation
theory,
\begin{align}
\widetilde a_{\mu}^{\rm SD}(I=1;t_{\rm min})
={}&a_{\mu}^{\rm SD}(I=1;t_{\rm min})
+\Delta a_{\mu}^{\rm SD,pert}(I=1;t_{\rm min})\,,
\label{eq:I1_SD_tmin_restored}
\\[8pt]
\Delta a_{\mu}^{\rm SD,pert}(I=1;t_{\rm min})
={}&2\alpha_{\rm em}^{2}
\int_{0}^{t_{\rm min}}dt\,t^2 K(m_\mu t)
\Theta^{\rm SD}(t)
V_{\rm N^{4}LO}^{I=1,\rm cont}(t)\,.
\label{eq:I1_SD_tmin_pert}
\end{align}
The continuum correlator in
Eq.~\eqref{eq:I1_SD_tmin_pert} is evaluated at N$^{4}$LO in the massless
limit with the RHAD package~\cite{Harlander:2002ur}, as in
Ref.~\cite{ExtendedTwistedMass:2022jpw}. As shown in
Fig.~\ref{fig:I1_SD_tmin}, the modified determinations approach a stable
plateau as $t_{\rm min}$ is reduced and agree with the standard
$t_{\rm min}=0$ result. This agreement provides a nontrivial check that
the fit pool used in the standard analysis is sufficient to correctly
describe the cutoff dependence of our lattice results.

We can now quote our final FLAG-scheme results corresponding to the crossed data points in black in Fig.~\ref{fig:I1_windows_FLAG_cont},
\begin{align}
 a_{\mu}^{{\rm SD},{\rm FLAG}}(I=1, L_{\rm ref})
 &=
 \amuSDIoneFLAG
 \times10^{-10}\,,
\label{eq:I1_SD_FLAG_Lref}\\[8pt]
 a_{\mu}^{{\rm W},{\rm FLAG}}(I=1;L_{\rm ref})
 &=
 \amuIWIoneFLAG
 \times10^{-10}\,,
\label{eq:I1_W_FLAG_Lref}\\[8pt]
 a_{\mu}^{{\rm LD},{\rm FLAG}}(I=1;L_{\rm ref})
 &=
 \amuLDIoneFLAG
 \times10^{-10}\,.
\label{eq:I1_LD_FLAG_Lref}
\end{align}
The LD window provides the largest contribution to the isotriplet HVP,
whereas the SD contribution is considerably smaller. The latter is,
however, the most sensitive to discretization effects and therefore
requires the most elaborate continuum extrapolation. This is also
reflected in its error budget, for which the systematic uncertainty due to the continuum-limit extrapolation is
slightly larger than the statistical one. The W and SD contributions are
determined with relative precisions of about $0.2\%$ and $0.4\%$,
respectively, while the LD contribution is determined at the level of
about $1.7\%$. The absolute uncertainty of the latter is comparable to
that of the full isotriplet contribution and therefore dominates its
uncertainty. All the results in the previous equation correspond to
$L=L_{\rm ref}$. The finite-volume treatment of all three windows is
deferred to Subsec.~\ref{sec:window_FSE}.

The corresponding results in the WP25 scheme are obtained by repeating the
complete continuum analysis in the different $f_\pi$-schemes and linearly
interpolating the results to $f_{\pi}^{\rm WP25}=131.00(27)~{\rm MeV}$, following the strategy adopted for the full contribution in the previous section. We obtain in this way
\begin{align}
 a_{\mu}^{{\rm SD},{\rm WP25}}(I=1; L_{\rm ref})
 &=
 \amuSDIoneWP
 \times10^{-10}\,,
\label{eq:I1_SD_WP25}\\[8pt]
 a_{\mu}^{{\rm W},{\rm WP25}}(I=1;L_{\rm ref})
 &=
 \amuIWIoneWP
 \times10^{-10}\,,
\label{eq:I1_W_WP25_Lref}\\[8pt]
 a_{\mu}^{{\rm LD},{\rm WP25}}(I=1;L_{\rm ref})
 &=
 \amuLDIoneWP
 \times10^{-10}\,.
\label{eq:I1_LD_WP25_Lref}
\end{align}
For completeness, the slopes describing the dependence of the three windows on the input value of $f_\pi$, as determined from the linear fits, are
\begin{align}
 s_{I=1}^{\rm SD}
 &\equiv
 \frac{d a_{\mu}^{\rm SD}(I=1)}{df_\pi}
 ={0.03}\,(19)\times10^{-10}~{\rm MeV}^{-1}\,, \\[8pt]
 s_{I=1}^{\rm W}
 &\equiv
 \frac{d a_{\mu}^{\rm W}(I=1)}{df_\pi}
={-0.65}\,(26)\times10^{-10}~{\rm MeV}^{-1}\,, \\[8pt]
 s_{I=1}^{\rm LD}
 &\equiv \frac{d a_{\mu}^{\rm LD}(I=1)}{df_\pi}
={-8.9}\,(1.8)\times10^{-10}~{\rm MeV}^{-1}\,.
\label{eq:I1_win_slope}
\end{align}

\subsection{Finite-volume corrections to the $I=1$ Euclidean-window contributions}
\label{sec:window_FSE}
In this subsection, we determine the finite-volume corrections to the
isotriplet SD, W, and LD contributions. These corrections are required to
extrapolate the results obtained at $L=L_{\rm ref}$ in
Eqs.~\eqref{eq:I1_SD_FLAG_Lref}--\eqref{eq:I1_LD_WP25_Lref}
to the infinite-volume limit.

We begin with the SD window, for which finite-volume effects are very small.
At the coarsest lattice spacing, $a_B\simeq0.0796~{\rm fm}$, we use the TM
result obtained on the B96 ensemble, with $L\simeq7.6~{\rm fm}$, as an
estimate of the infinite-volume value. Indeed, finite-volume effects at
short distances are sufficiently small that the $L\simeq7.6~{\rm fm}$
result can be regarded as effectively corresponding to $L\to\infty$.
Its difference from the corresponding determination at
$L=L_{\rm ref}$ is
$0.6(1)\times10^{-11}$. As discussed in
Subsec.~\ref{sec:FSE}, at nonzero lattice spacing the TM regularization is
expected to slightly enhance finite-volume effects relative to the continuum,
owing to the $O(a^2)$ mass splitting between the charged and neutral pions.
Given the smallness of the observed shift, we take its central value as our
estimate of the finite-volume correction from $L_{\rm ref}$ to infinite
volume and conservatively assign to it a $100\%$ relative uncertainty:
\begin{align}
 \Delta_{\rm FV}a_{\mu}^{\rm SD}(I=1;L_{\rm ref})
 \equiv a_{\mu}^{\rm SD}(I=1;\infty)
       -a_{\mu}^{\rm SD}(I=1;L_{\rm ref})
 =0.06(6)\times10^{-10}\,.
\label{eq:I1_SD_FV_correction}
\end{align}
The correction is numerically negligible at our current level of precision:
its central value amounts to only about one third of the uncertainty of the
result at $L=L_{\rm ref}$.

For the intermediate-window contribution, $a_{\mu}^{\rm W}(I=1)$, we
estimate finite-volume effects following the same strategy adopted in
Subsec.~\ref{sec:FSE} for the part of the finite-volume correction to the
full $I=1$ contribution originating from the region
$t\leq1.1~{\rm fm}$. A data-driven estimate is obtained from the difference
between the B96 TM determination and the corresponding result at
$L=L_{\rm ref}$, evaluated in the same regularization and at the same lattice
spacing. This gives a shift of
$1.15(21)\times10^{-10}$. As discussed above, at nonzero
lattice spacing the TM regularization contains $\pi^+\pi^0$ states with
$m_{\pi^0}<m_{\pi^+}$. The resulting enhancement of
finite-volume effects implies that this data-driven estimate is expected to
be somewhat larger than the corresponding continuum correction. Conversely,
the continuum MLLGS model accounts only for two-pion intermediate states and
therefore tends to underestimate the finite-volume effect in the
intermediate window, where contributions from heavier states might still be
important. The MLLGS prediction amounts to approximately
$0.75\times10^{-10}$. We therefore adopt the data-driven determination as our
central value and assign an uncertainty equal to the difference between the
data-driven and MLLGS estimates. This is the same conservative prescription
adopted in Subsec.~\ref{sec:FSE} for the region $t\leq1.1~{\rm fm}$ and
gives
\begin{align}
 \Delta_{\rm FV}a_{\mu}^{\rm W}(I=1;L_{\rm ref})
 \equiv a_{\mu}^{\rm W}(I=1;\infty)
       -a_{\mu}^{\rm W}(I=1;L_{\rm ref})
 =1.15(40)\times10^{-10}\,.
\label{eq:I1_W_FV_correction}
\end{align}
Finally, for the LD contribution we estimate finite-volume effects following
the same strategy adopted for the full $I=1$ contribution
$a_{\mu}^{\rm HVP}(I=1)$. In the region $t\geq1.1~{\rm fm}$, we rely on the
MLLGS-model prediction and assign to it a relative uncertainty of $10\%$.
For $t\leq1.1~{\rm fm}$, we instead use the data-driven estimate and assign
an uncertainty equal to its difference from the corresponding MLLGS
prediction, as discussed in the paragraph preceding
Eq.~\eqref{eq:FSE}. We obtain
\begin{align}
 \Delta_{\rm FV}a_{\mu}^{\rm LD}(I=1;L_{\rm ref})
 &\equiv a_{\mu}^{\rm LD}(I=1;\infty)
       -a_{\mu}^{\rm LD}(I=1;L_{\rm ref})
 \nonumber\\[8pt]
 &=25.6(2.5)(0.3)[2.5]\times10^{-10}\,,
\label{eq:I1_LD_FV_correction}
\end{align}
where the first uncertainty is associated with the finite-volume correction
in the region $t\geq1.1~{\rm fm}$, while the second corresponds to the
complementary region $t\leq1.1~{\rm fm}$. The uncertainty in square brackets
is their sum in quadrature.

The sum of the finite-volume corrections in
Eqs.~\eqref{eq:I1_SD_FV_correction}--\eqref{eq:I1_LD_FV_correction}
is fully consistent with the correction applied to the full
isotriplet contribution and given in Eq.~\eqref{eq:FSE}. The central values differ by
only about $2.9\times10^{-11}$, corresponding to about $0.1\sigma$.
This tiny difference reflects the slightly different procedures used to
estimate finite-volume effects for the individual windows and for their sum.

\subsection{The $I=0$ contribution}
\label{sec:window_I0}

\begin{figure}[t]
\centering
\includegraphics[width=\linewidth]{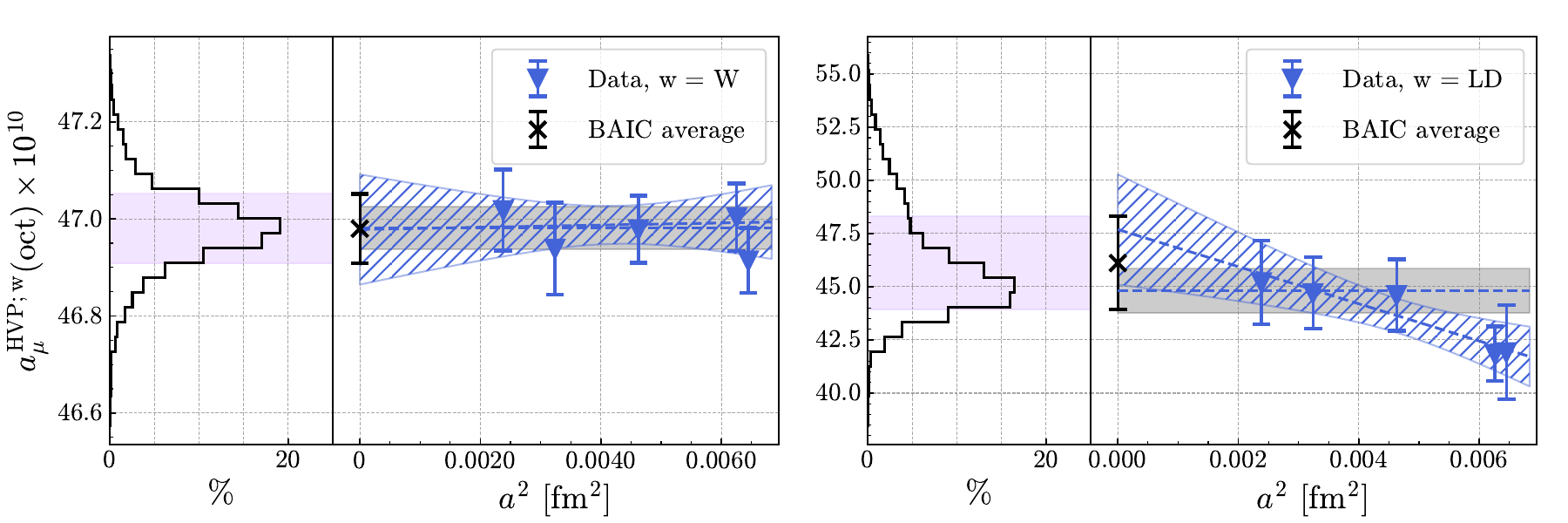}
\caption{Continuum extrapolations of the octet W (\textit{left}) and LD (\textit{right})
contributions in the FLAG scheme. In each panel, the two bands correspond to
a fit linear in $a^2$ to all ensembles and a constant fit to the C80, D96, and
E112 results. The final continuum determination is obtained from their
Bayesian--Akaike model average. The B64 and B96 points are shown separately and
are in very good agreement. The W data are
particularly flat, with no visible cutoff effects over the full range of
lattice spacings explored.}
\label{fig:oct_windows_W_LD_FLAG}
\end{figure}

We now turn to the isoscalar channel. For each window we use the same
decomposition adopted for the full contribution in
Eq.~\eqref{eq:I0_decomp_secIV}, namely
\begin{align}
 a_{\mu}^{\rm w}(I=0)
 =a_{\mu}^{\rm w}({\rm oct})
 +a_{\mu}^{\rm w}(c)
 +\delta a_{\mu}^{c,{\rm disc},{\rm w}}\,,
 \qquad {\rm w}=\{\mathrm{SD},\mathrm{W},\mathrm{LD}\}\,.
\label{eq:I0_window_decomp}
\end{align}
We begin with the octet contribution. The analysis of the LD contribution, $a_{\mu}^{\rm LD}({\rm oct})$, is identical to that of $a_{\mu}^{\rm HVP}({\rm oct})$. As shown in the right panel of
Fig.~\ref{fig:oct_windows_W_LD_FLAG}, the B64 and B96 determinations are in very good agreement within uncertainties as in the case of the full contribution, confirming that finite-volume effects are
negligible at the present precision. The dependence on $a^2$ is also very
mild. We therefore consider the same two continuum fits as in
Subsec.~\ref{sec:full_HVP_I0}: a fit linear in $a^2$ to all ensembles and a
constant fit restricted to C80, D96, and E112. Their continuum values are
combined using the BAIC procedure described in
Eqs.~\eqref{eq:BAIC_weight}--\eqref{eq:BAIC_syst}.

The intermediate window shows an even flatter dependence on the lattice
spacing. The B64 and B96 results again agree within uncertainties, and the
determinations at all four lattice spacings are mutually compatible. The
absence of visible cutoff effects, illustrated in the left panel of
Fig.~\ref{fig:oct_windows_W_LD_FLAG}, is remarkable given the precision of
the data and the broad range of lattice spacings explored. To extrapolate
the result to the continuum limit, we perform the same two continuum fits
employed for the LD and full octet contributions and combine them using the
BAIC procedure. As in the isotriplet analysis, no subtraction of the tree-level cutoff
effects is applied to the W and LD windows.

Our final results in the FLAG scheme are
\begin{align}
 a_{\mu}^{{\rm LD},{\rm FLAG}}({\rm oct})
 &=
\amuLDoctFLAG 
 \times10^{-10}\,,
\label{eq:oct_LD_FLAG}\\[8pt]
 a_{\mu}^{{\rm W},{\rm FLAG}}({\rm oct})
 &=
 \amuIWoctFLAG\times10^{-10}\,.
\label{eq:oct_W_FLAG}
\end{align}
while in the WP25 scheme we obtain
\begin{align}
 a_{\mu}^{{\rm LD},{\rm WP25}}({\rm oct})
 &=
 \amuLDoctWP\times10^{-10}\,,
\label{eq:oct_LD_WP25}\\[8pt]
 a_{\mu}^{{\rm W},{\rm WP25}}({\rm oct})
 &=
 \amuIWoctWP\times10^{-10}\,,
\label{eq:oct_W_WP25}
\end{align}
The W contribution is determined with very high precision, at the level of
a few per mille, whereas the LD contribution has a relative uncertainty of
about $5\%$. For completeness, the slopes describing the dependence of the W and LD
octet contributions on the input value of $f_\pi$, as determined from the
linear fits, are
\begin{align}
 s_{\rm oct}^{\rm W}
 \equiv \frac{d a_{\mu}^{\rm W}({\rm oct})}{d f_\pi}
 &= -0.155(71)\times10^{-10}~{\rm MeV}^{-1}\,,
 \\[8pt]
 s_{\rm oct}^{\rm LD}
 \equiv \frac{d a_{\mu}^{\rm LD}({\rm oct})}{d f_\pi}
 &= -1.09(22)\times10^{-10}~{\rm MeV}^{-1}\,.
\end{align}
The SD octet contribution, $a_{\mu}^{\rm SD}({\rm oct})$, is determined in a different way. At short distances, the $\rm{SU}(3)$ suppression of the $uds$
disconnected correlator is particularly effective because at small Euclidean times the correlator is
only weakly sensitive to quark-mass differences. This is consistent with the
strong suppression of the disconnected $uds$ SD contribution observed in
Ref.~\cite{ExtendedTwistedMass:2022jpw}; see in particular Table~I of that
reference. In the present calculation, we have found that
$a_{\mu}^{\rm SD}({\rm disc},uds)$ ranges from approximately
$-10(4)\times10^{-14}$ on the coarsest ensemble to
$2(5)\times10^{-14}$ on the finest one. Given the smallness of this contribution, we prefer to simply quote a conservative value for this contribution, that we take to be $a_{\mu}^{\rm SD}({\rm disc}, uds)=0(3)\times 10^{-13}$.

Since the SD octet contribution is given by
\begin{align}
 a_{\mu}^{\rm SD}({\rm oct})
 =\frac{1}{9}a_{\mu}^{\rm SD}(I=1)
 +a_{\mu}^{\rm SD}(s)
 +a_{\mu}^{\rm SD}({\rm disc},uds)\,,
\label{eq:oct_SD_decomposition}
\end{align}
given our conservative estimate of $a_{\mu}^{\rm SD}({\rm disc}, uds)$, and having determined
the isotriplet contribution in the previous subsections, the only missing ingredient to determine $a_{\mu}^{\rm SD}({\rm oct})$ is the strange-connected contribution $a_{\mu}^{\rm SD}(s)$. We update the analysis of Ref.~\cite{ExtendedTwistedMass:2024nyi}.
The treatment of sea-quark-mass mistuning effects is improved through
increased statistics and the use of additional small-volume ensembles
to determine the strange- and charm-sea contributions. The continuum
fits are based on the same fit pool adopted for the
isotriplet SD window.\footnote{As in the case of the isotriplet
contribution, we use $\gamma=0.42,1,2$ for the logarithmically enhanced
$a^{2}$ term, whereas in Ref.~\cite{ExtendedTwistedMass:2024nyi} we used
$\gamma=1,2,3$.}

\begin{figure}[t]
\centering
\includegraphics[width=0.49\linewidth]{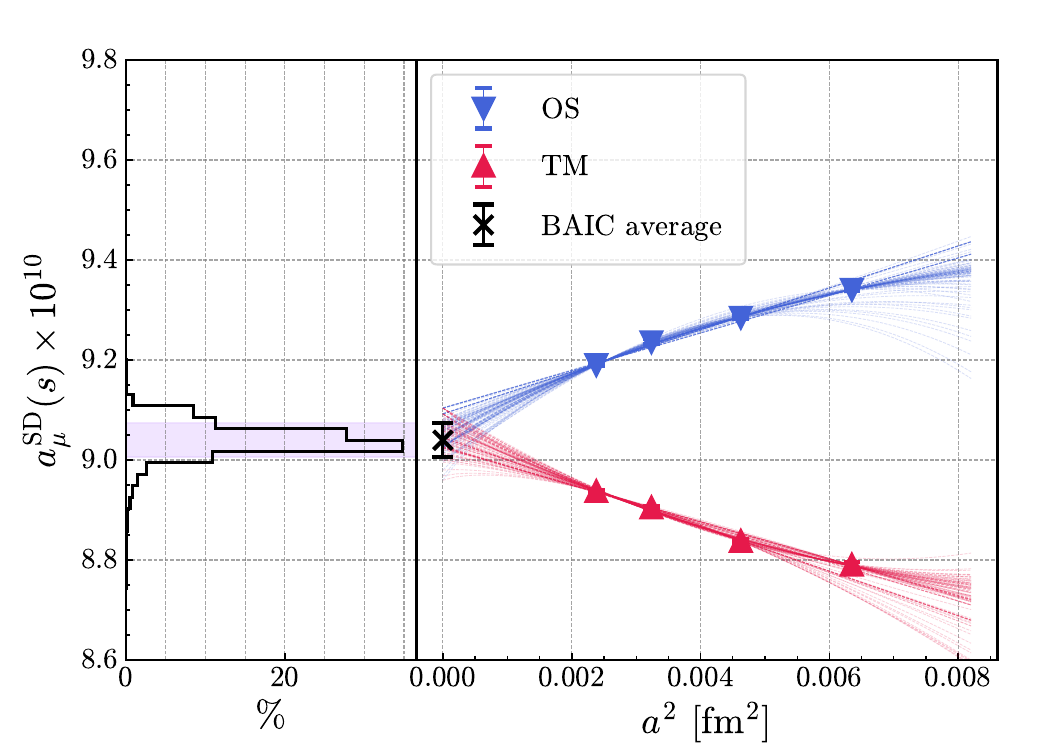}
\includegraphics[width=0.49\linewidth]{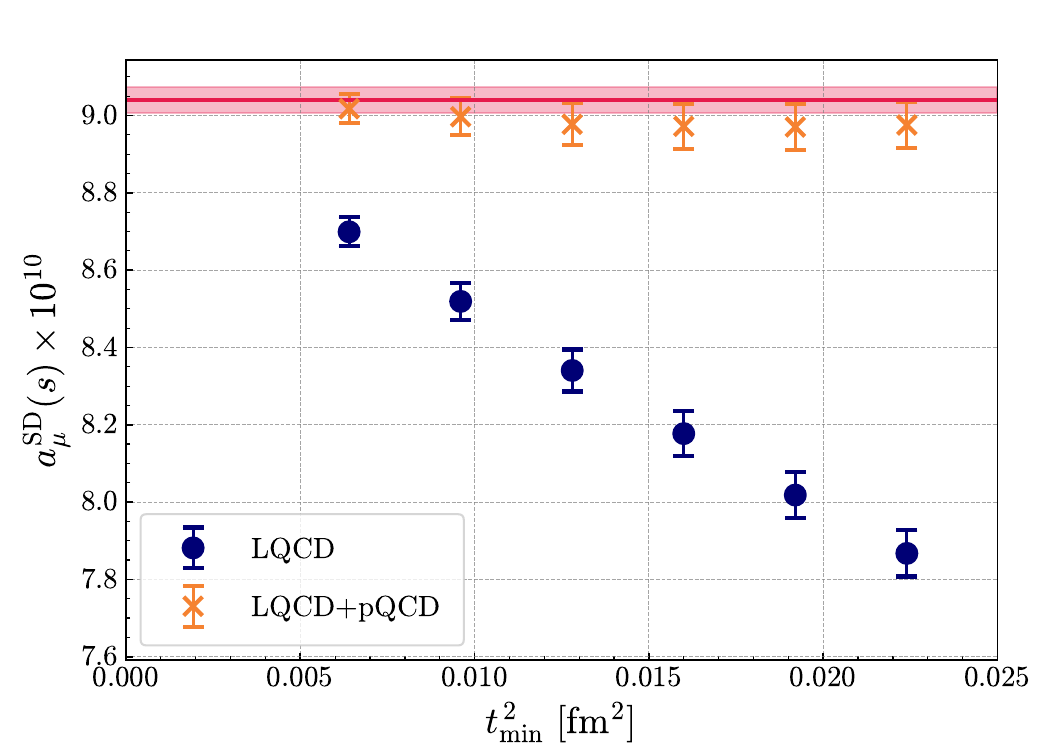}
\caption{Strange-connected contribution to the short-distance window in the FLAG scheme.
{\it{Left}}: Continuum-limit extrapolation of $a_{\mu}^{\rm SD}(s)$ from combined fits of the TM
and OS data using the same fit pool as for the isotriplet SD contribution.
{\it{Right}}: $t_{\rm min}$ analysis of the short-distance contribution in the FLAG scheme. The blue circles show the
truncated contribution $a_{\mu}^{\rm SD}(s;t_{\rm min})$, while the orange crosses show
$\widetilde a_{\mu}^{\rm SD}(s;t_{\rm min})$, obtained by restoring the omitted interval
$t\in[0,t_{\rm min}]$ using the N$^{4}$LO massless continuum perturbative prediction
from RHAD. The horizontal band shows the result of the standard analysis.
The agreement with the standard result provides a nontrivial crosscheck
of the continuum extrapolation.}
\label{fig:strange_SD_FLAG_cont}
\end{figure}

The left panel of Fig.~\ref{fig:strange_SD_FLAG_cont} shows our continuum-limit extrapolation in the FLAG scheme. The figure is analogous to that presented for the intermediate and long-distance window contributions to the octet term. Finite-volume effects are negligibly small in this case for lattice volumes
with $L\gtrsim5~{\rm fm}$, as already shown in our previous studies
\cite{ExtendedTwistedMass:2022jpw,ExtendedTwistedMass:2024nyi}, and the
ensemble results shown in the figure can therefore be regarded as
effectively corresponding to the infinite-volume limit. The right panel of
Fig.~\ref{fig:strange_SD_FLAG_cont} shows the complementary
$t_{\rm min}$ analysis, performed in exactly the same way as for
$a_{\mu}^{\rm SD}(I=1)$. The determinations obtained by first removing the
interval $[0,t_{\rm min}]$, taking the continuum limit, and then restoring
it with continuum perturbation theory agree very well with the standard
$t_{\rm min}=0$ result throughout the full range of $t_{\rm min}$ values
explored.

\begin{figure}[t]
\centering
\includegraphics[width=0.49\textwidth]{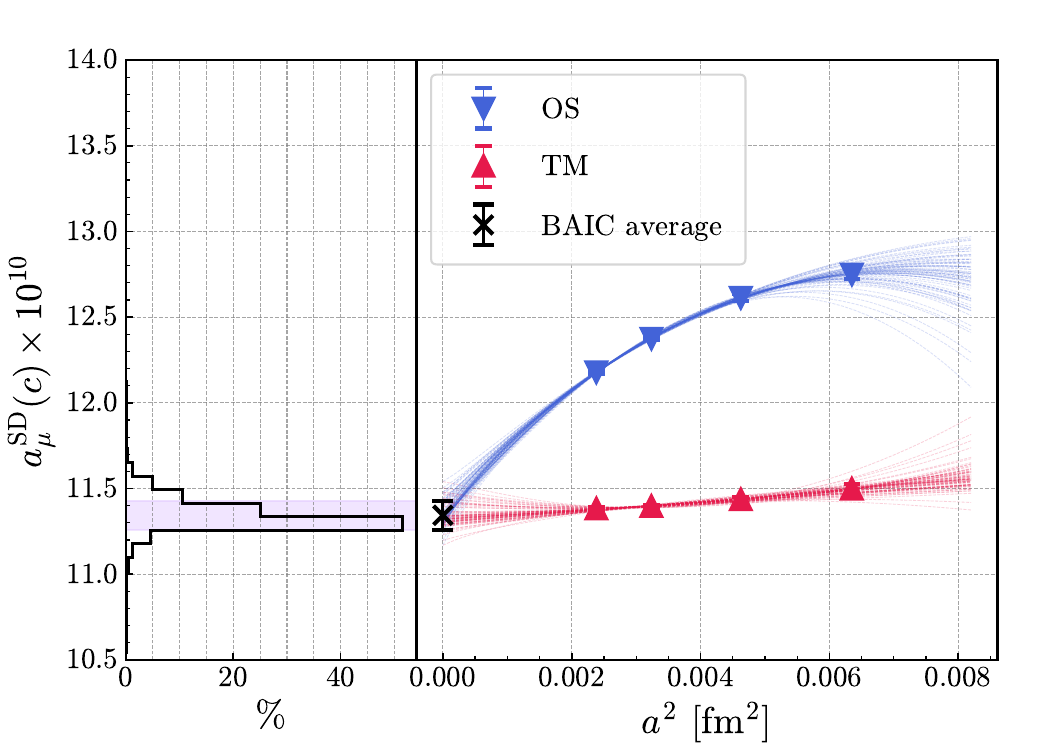}
\includegraphics[width=0.49\textwidth]{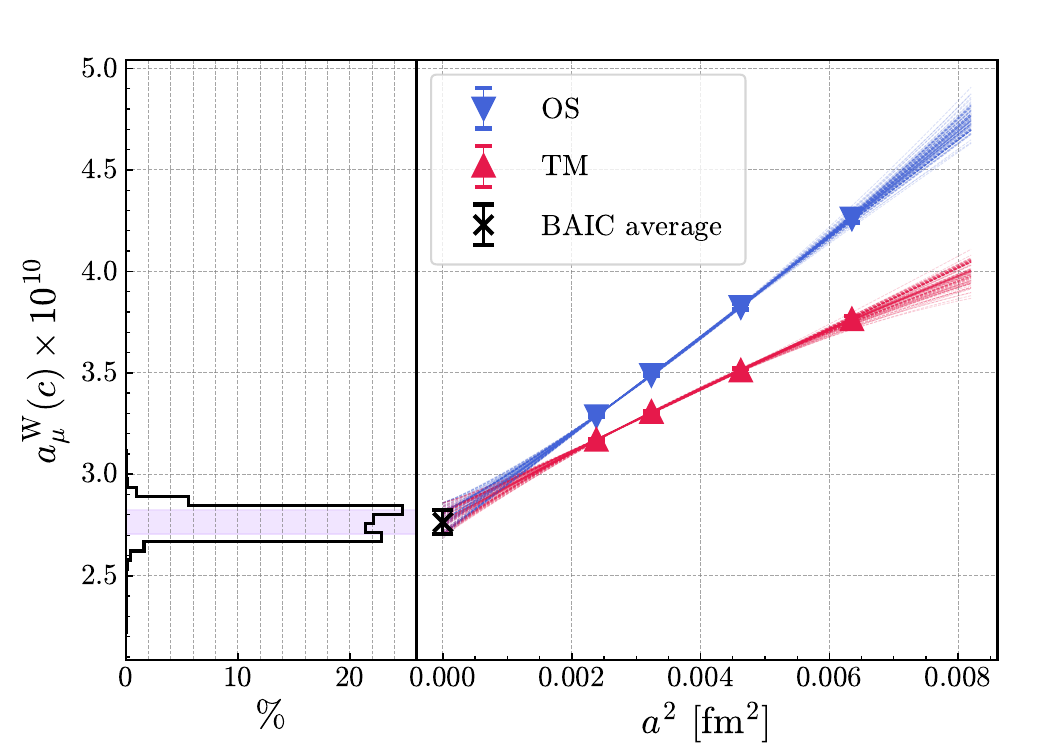}
\caption{
Charm-connected contribution to the Euclidean-time windows in the FLAG scheme.
{\it Left}: Continuum-limit extrapolation of the SD contribution.
{\it Right}: Continuum-limit extrapolation of the W contribution.
In both panels, the TM and OS data are fitted simultaneously with a common
continuum limit and regularization-dependent cutoff effects. The final
continuum values are obtained using the BAIC procedure.
}
\label{fig:charm_windows_SD_W_FLAG}
\end{figure}

We obtain
\begin{align}
 a_{\mu}^{{\rm SD},{\rm FLAG}}(s)
 &=\amuSDsFLAG\times 10^{-10}\,,
\label{eq:strange_SD_FLAG}\\[8pt]
 a_{\mu}^{{\rm SD},{\rm WP25}}(s)
 &=\amuSDsWP\times 10^{-10}\,.
\label{eq:strange_SD_WP25}
\end{align}
The FLAG-scheme result is in good agreement with our previous determination in
Ref.~\cite{ExtendedTwistedMass:2024nyi}, namely $a_{\mu, {\rm ETM-24}}^{{\rm SD}}(s)= 9.063(27)$. We obtain for the $f_{\pi}$-slope of $a_{\mu}^{\rm SD}(s)$ the value
\begin{align}
s_{s}^{\rm SD} = \frac{ d a_{\mu}^{\rm SD}(s)}{d f_{\pi}} = -0.002~(23) \times 10^{-10}\,{\rm MeV}^{-1}~.
\end{align}
Combining Eq.~\eqref{eq:oct_SD_decomposition} with the isotriplet result and
including in the systematic uncertainty the conservative
$3\times10^{-13}$ uncertainty assigned to the disconnected $uds$ term gives
\begin{align}
 a_{\mu}^{{\rm SD},{\rm FLAG}}({\rm oct}) =
\amuSDoctFLAG 
 \times10^{-10}\,,
\label{eq:oct_SD_FLAG}\\[8pt]
 a_{\mu}^{{\rm SD},{\rm WP25}}({\rm oct}) =
\amuSDoctWP
 \times10^{-10}\,.
\label{eq:oct_SD_WP25}
\end{align}
We finally consider the charm-connected contributions. As in the case of the full contribution $a_{\mu}^{\rm HVP}(c)$ discussed in Subsec.~\ref{sec:full_HVP_I0}, finite-volume effects are entirely negligible at the present
accuracy for $L\gtrsim5~{\rm fm}$ in all the three windows
(see also Refs.~\cite{ExtendedTwistedMass:2022jpw,ExtendedTwistedMass:2024nyi}). The continuum extrapolations are based on
the exact fit pool adopted for $a_{\mu}^{\rm HVP}(c)$ in the previous section.\footnote{The fit pool adopted for $a_{\mu}^{\rm HVP}(c)$ is the same as that used for $a_{\mu}^{\rm SD}(I=1)$ and $a_{\mu}^{\rm SD}(s)$.} The
SD and W windows together account for essentially the entire charm-connected
contribution; the LD term is only about $0.1\%$ of the total. Fig.~\ref{fig:charm_windows_SD_W_FLAG} summarizes the continuum-extrapolation in the FLAG scheme for the dominant SD and W contributions. As for the other contributions, the crossed points in black correspond to our final determination after application of the BAIC averaging while the dashed curves corresponds to illustrative continuum-limit fits. For both $a_{\mu}^{\rm SD}(c)$ and $a_{\mu}^{\rm W}(c)$ the cut-off effects are sizable and cut-off effects beyond linear terms in $a^{2}$ are needed to obtain reasonable $\chi^{2}/{\rm dof}$ values. 

The final results for the three windows in the FLAG scheme are
\begin{align}
 a_{\mu}^{{\rm SD},{\rm FLAG}}(c)
 &=\amuSDcFLAG\times 10^{-10}\,,
\label{eq:charm_SD_FLAG_window}\\[8pt]
 a_{\mu}^{{\rm W},{\rm FLAG}}(c)
 &=\amuWcFLAG\times 10^{-10}\,,
\label{eq:charm_W_FLAG_window}\\[8pt]
 a_{\mu}^{{\rm LD},{\rm FLAG}}(c)
 &=\amuLDcFLAG\times 10^{-10}\,,
\label{eq:charm_LD_FLAG_window}
\end{align}
while in the WP25 scheme we obtain
\begin{align}
 a_{\mu}^{{\rm SD},{\rm WP25}}(c)
 &=\amuSDcWP\times 10^{-10}\,,
\label{eq:charm_SD_WP25_window}\\[8pt]
 a_{\mu}^{{\rm W},{\rm WP25}}(c)
 &=\amuWcWP\times 10^{-10}\,,
\label{eq:charm_W_WP25_window}\\[8pt]
 a_{\mu}^{{\rm LD},{\rm WP25}}(c)
 &=\amuLDcWP\times 10^{-10}\,.
\label{eq:charm_LD_WP25_window}
\end{align}
The $f_{\pi}$-slopes of the three charm windows, obtained through the usual linear interpolation, are given by
\begin{align}
 s_c^{\rm SD}
 &=\frac{d a_{\mu}^{\rm SD}(c)}{df_\pi}
 ={0.08}\,(4)\times10^{-10}~{\rm MeV}^{-1}\,,
\label{eq:charm_SD_slope}\\[8pt]
 s_c^{\rm W}
 &=\frac{d a_{\mu}^{\rm W}(c)}{df_\pi}
 ={0.03}\,(2)\times10^{-10}~{\rm MeV}^{-1}\,,
\label{eq:charm_W_slope}\\[8pt]
 s_c^{\rm LD}
 &=\frac{d a_{\mu}^{\rm LD}(c)}{df_\pi}
 ={0.0003}\,(3)\times10^{-10}~{\rm MeV}^{-1}\,.
\label{eq:charm_LD_slope}
\end{align}
We do not repeat the complementary $t_{\rm min}$ analysis for the charm
contribution. In this case, finite-mass effects in the perturbative
contribution over the omitted short-distance interval are sizable, and the
finite-order perturbative estimate shows a significant dependence on the
charm-mass scheme, making this exercise less useful. Since the
integration-generated logarithmically enhanced short-distance cutoff effects
are mass independent, the agreement found in the isotriplet and strange
$t_{\rm min}$ analyses also validates the treatment of these effects in the
charm analysis, where the same fit pool is used.

We finally combine the octet and charm-connected terms according to
Eq.~\eqref{eq:I0_window_decomp}. The residual disconnected contribution
containing at least one charm loop is denoted by
$\delta a_{\mu}^{c,{\rm disc},{\rm w}}$, consistently with the notation used for the
full HVP. As discussed in Subsec.~\ref{sec:full_HVP_I0} for the short-distance and intermediate-window we
assign conservative additional uncertainties of $1.2\times10^{-12}$ and
$6\times10^{-12}$, respectively, to this contribution. For the LD window  we conservatively retain the
$0.2\times10^{-10}$ uncertainty assigned to the full
$\delta a_{\mu}^{c,{\rm disc}}$ term. 

The final FLAG-scheme results for the isosinglet window contributions are
\begin{align}
 a_{\mu}^{{\rm SD},{\rm FLAG}}(I=0)
 &=
\amuSDIzeroFLAG 
 \times10^{-10}\,,
\label{eq:I0_SD_FLAG_final}\\[8pt]
 a_{\mu}^{{\rm W},{\rm FLAG}}(I=0)
 &=
\amuIWIzeroFLAG 
 \times10^{-10}\,,
\label{eq:I0_W_FLAG_final}\\[8pt]
 a_{\mu}^{{\rm LD},{\rm FLAG}}(I=0)
 &=
\amuLDIzeroFLAG 
 \times10^{-10}\,.
\label{eq:I0_LD_FLAG_final}
\end{align}
The corresponding WP25-scheme results are instead
\begin{align}
 a_{\mu}^{{\rm SD},{\rm WP25}}(I=0)
 &=
\amuSDIzeroWP
 \times10^{-10}\,,
\label{eq:I0_SD_WP25_final}\\[8pt]
 a_{\mu}^{{\rm W},{\rm WP25}}(I=0)
 &=
\amuIWIzeroWP
 \times10^{-10}\,,
\label{eq:I0_W_WP25_final}\\[8pt]
 a_{\mu}^{{\rm LD},{\rm WP25}}(I=0)
 &=
\amuLDIzeroWP
 \times10^{-10}\,.
\label{eq:I0_LD_WP25_final}
\end{align}
The SD isoscalar contribution receives comparable contributions from the
octet and charm-connected terms, whereas the charm contribution becomes
progressively less important in the W and, especially, LD windows.

\subsection{Summary and comparison with existing lattice results}
\label{sec:window_summary}

We now collect our results for the individual Euclidean-time windows and compare a subset of the
most relevant contributions with available determinations by other groups. For
the isotriplet contributions, the infinite-volume results are obtained by adding
the finite-volume corrections of
Eqs.~\eqref{eq:I1_SD_FV_correction}--\eqref{eq:I1_LD_FV_correction} to the
corresponding continuum results at $L=L_{\rm ref}$. The octet,
charm-connected, and isoscalar contributions can instead be regarded as
infinite-volume determinations at the present level of precision. Our final results
for all the different contributions are collected in
Table~\ref{tab:window_contributions}. 

\begin{table*}[t]
\centering
\begingroup
\renewcommand{\arraystretch}{1.22}
\begin{tabular*}{0.9\textwidth}{@{\extracolsep{\fill}}lccc@{}}
\toprule
\multicolumn{4}{c}{FLAG scheme} \\
\cmidrule(lr){1-4}
Contribution
& ${\rm w}={\rm SD}$
& ${\rm w}={\rm W}$
& ${\rm w}={\rm LD}$ \\
\midrule
$a_\mu^{\rm w}(I=1)$
 & $\LinfamuSDIoneFLAGT$
 & $\LinfamuIWIoneFLAGT$
 & $\LinfamuLDIoneFLAGT$ \\
\midrule
$a_\mu^{\rm w}({\rm oct})$
 & $\LinfamuSDoctFLAGT$
 & $\LinfamuIWoctFLAGT$
 & $\LinfamuLDoctFLAGT$ \\
$a_\mu^{\rm w}(c)$
 & $\amuSDcFLAGT$
 & $\amuWcFLAGT$
 & $\amuLDcFLAGT$ \\
$\delta a_\mu^{c,{\rm disc},{\rm w}}$
 & $0.000(12)$
 & $0.00(6)$
 & $0.0(2)$ \\
\midrule
$a_\mu^{\rm w}(I=0)$
 & $\LinfamuSDIzeroFLAGT$
 & $\LinfamuIWIzeroFLAGT$
 & $\LinfamuLDIzeroFLAGT$ \\
\midrule
$a_\mu^{\rm w}(s)$
 & $\amuSDsFLAGT$
 & $\amuWsFLAGT$
 & $\amuLDsFLAGT$ \\
$a_\mu^{\rm w}({\rm disc})$
 & $\amuSDdiscT$
 & $\amuIWdiscFLAGT$
 & $\amuLDdiscFLAGT$ \\
\midrule
\multicolumn{4}{c}{WP25 scheme} \\
\cmidrule(lr){1-4}
Contribution
& ${\rm w}={\rm SD}$
& ${\rm w}={\rm W}$
& ${\rm w}={\rm LD}$ \\
\midrule
$a_\mu^{\rm w}(I=1)$
 & $\LinfamuSDIoneWPT$
 & $\LinfamuIWIoneWPT$
 & $\LinfamuLDIoneWPT$ \\
\midrule
$a_\mu^{\rm w}({\rm oct})$
 & $\LinfamuSDoctWPT$
 & $\LinfamuIWoctWPT$
 & $\LinfamuLDoctWPT$ \\
$a_\mu^{\rm w}(c)$
 & $\amuSDcWPT$
 & $\amuWcWPT$
 & $\amuLDcWPT$ \\
$\delta a_\mu^{c,{\rm disc},{\rm w}}$
 & $0.000(12)$
 & $0.00(6)$
 & $0.0(2)$ \\
\midrule
$a_\mu^{\rm w}(I=0)$
 & $\LinfamuSDIzeroWPT$
 & $\LinfamuIWIzeroWPT$
 & $\LinfamuLDIzeroWPT$ \\
\midrule
$a_\mu^{\rm w}(s)$
 & $\amuSDsWPT$
 & $\amuWsWPT$
 & $\amuLDsWPT$ \\ 
$a_\mu^{\rm w}({\rm disc})$
 & $\amuSDdiscT$
 & $\amuIWdiscWPT$
 & $\amuLDdiscWPT$ \\
\bottomrule
\end{tabular*}
\endgroup
\caption{Summary of our results for the individual SD, W, and LD
contributions in the FLAG and WP25 schemes. All entries are in units of
$10^{-10}$. The first uncertainty is statistical, while the second collects
the systematic uncertainties associated with the continuum-limit
extrapolation and finite-volume effects. For $a_\mu^{\rm w}(I=0)$, the second
uncertainty also includes the conservative uncertainty assigned to
$\delta a_\mu^{c,{\rm disc},{\rm w}}$. The uncertainty in square brackets is the
sum in quadrature of the statistical and systematic uncertainties. For
$\delta a_\mu^{c,{\rm disc},{\rm w}}$ and $a_{\mu}^{\rm SD}({\rm disc})$, only a
single uncertainty is quoted, corresponding to the conservative estimate
assigned as described in the text, with no statistical/systematic
separation. The row labelled $a_\mu^{\rm w}({\rm disc})$ denotes the total
disconnected contribution. The estimates of
$\delta a_\mu^{c,{\rm disc},{\rm w}}$ and $a_{\mu}^{\rm SD}({\rm disc})$ are common
to the two schemes.}
\label{tab:window_contributions}
\end{table*}
To complete the flavor decomposition in
Table~\ref{tab:window_contributions}, we have also updated our results for the
strange-connected contribution in the W and LD windows with respect to Ref.~\cite{ExtendedTwistedMass:2024nyi}. The analysis is
repeated in both schemes with the same improvements employed for the full
$a_\mu^{\rm HVP}(s)$ in
Eqs.~\eqref{eq:FLAG_strange_update} and
\eqref{eq:WP25_strange_update}, and the same choice of fit functions for the continuum-limit extrapolation. 
The updated results, reported in
Table~\ref{tab:window_contributions}, are in very good agreement with our
previous FLAG-scheme determinations
$a_{\mu,{\rm ETM-24}}^{\rm W}(s)=27.16(25)\times10^{-10}$ and
$a_{\mu,{\rm ETM-24}}^{\rm LD}(s)=17.32(38)\times10^{-10}$ from
Ref.~\cite{ExtendedTwistedMass:2024nyi}, while their uncertainties are
reduced by approximately $15\%$ and $25\%$, respectively. This improvement
results from the improved determination of the sea-quark-mass mistuning
corrections.

The availability of the light-connected, strange-connected, and octet
contributions allows us to reconstruct the disconnected contribution in the W
and LD windows using the same strategy adopted for the full HVP. For
${\rm w}=\{{\rm W},{\rm LD}\}$, we first determine the charmless disconnected contribution from
\begin{align}
 a_\mu^{\rm w}({\rm disc},uds)
 =a_\mu^{\rm w}({\rm oct})
 -\frac{1}{10}a_\mu^{\rm w}(\ell)
 -a_\mu^{\rm w}(s)\,,
 \qquad {\rm w}=\{{\rm W},{\rm LD}\}\,,
\label{eq:disc_windows_reconstruction}
\end{align}
and then add the conservative estimate of the residual contribution
$\delta a_\mu^{c,{\rm disc},{\rm w}}$. As in the corresponding analysis of the full
HVP, we combine separately the statistical and systematic uncertainties of
the three terms in Eq.~\eqref{eq:disc_windows_reconstruction} in quadrature.
Neglecting their positive correlations with the octet contribution leads to a
conservative uncertainty. In the SD window, instead, both the $uds$
disconnected term and the residual charm-disconnected term have already been
estimated directly in Subsec.~\ref{sec:window_I0}. The results for the disconnected contributions in both FLAG and WP25 scheme 
are included in Table~\ref{tab:window_contributions}.

We now compare our findings with existing lattice determinations~\cite{Wang:2022lkq,
Spiegel:2024dec,
Giusti:2021dvd,
FermilabLatticeHPQCD:2023jof,
Boccaletti:2024guq,
RBC:2018dos,
RBC:2024fic,
FermilabLatticeHPQCD:2024ppc,
ExtendedTwistedMass:2024nyi,
Ce:2022kxy,
ExtendedTwistedMass:2022jpw,
MILC:2024ryz,
RBC:2023pvn,
Borsanyi:2020mff,
Djukanovic:2024cmq,
Aubin:2022hgm,
Kuberski:2024bcj,
Lehner:2020crt}. As in
Subsec.~\ref{sec:full_HVP_summary}, we display our results in both the FLAG and
WP25 schemes. Whenever possible, using information provided in the original
publications or in the WP25 report~\cite{Aliberti:2025beg}, we quote the
results of other groups both in their preferred isospin-symmetric scheme and
in the WP25 and/or FLAG schemes. We use the same graphical convention as in the comparison of the full HVP
in Subsec.~\ref{sec:full_HVP_summary}.

\begin{figure*}[t]
\centering
\includegraphics[width=0.8\linewidth]
{comparison_plots/legend_new.pdf}\\[-15pt]
\includegraphics[width=0.339\linewidth]{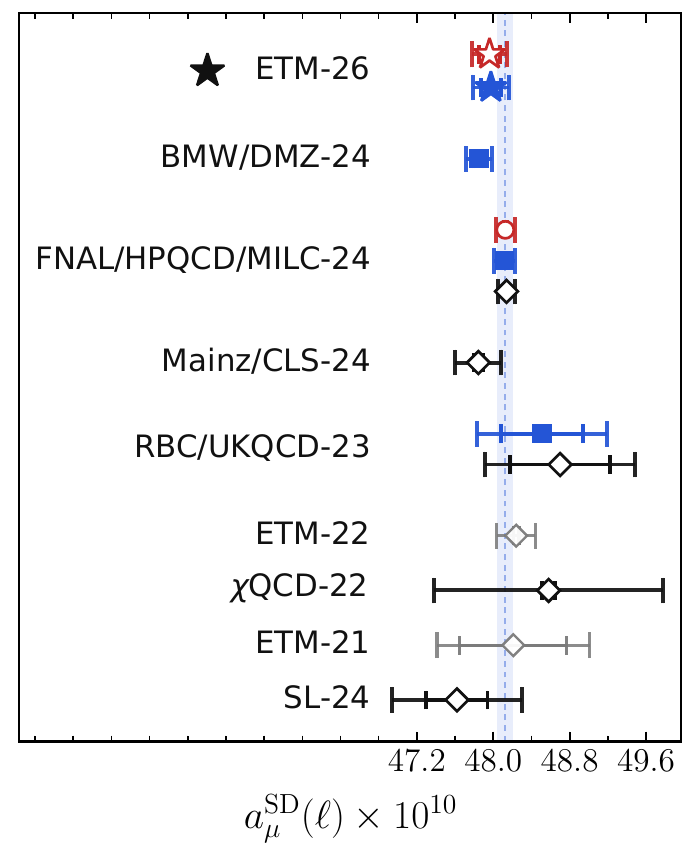}\hspace{-0.25cm}
\includegraphics[width=0.339\linewidth]{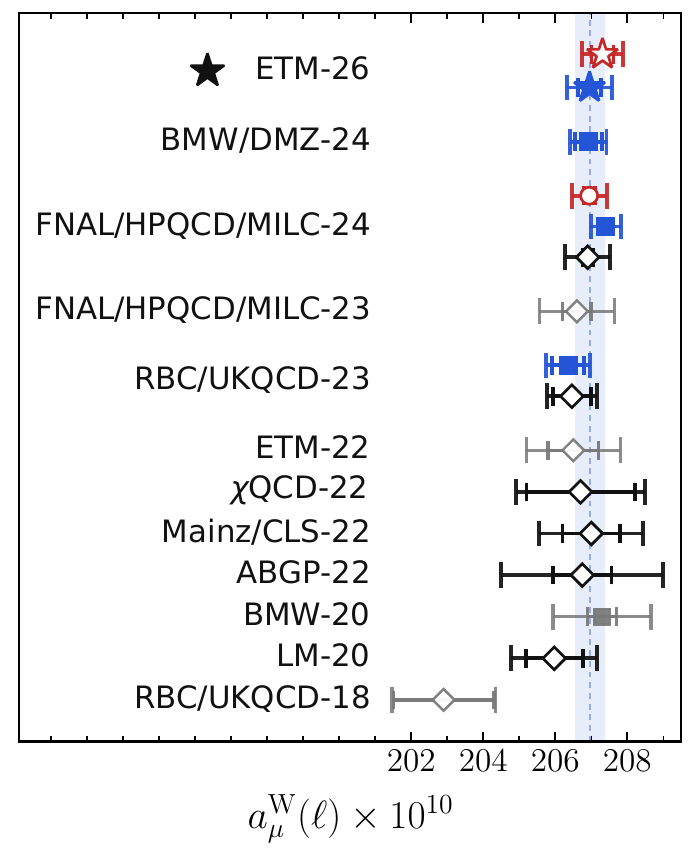}\hspace{-0.25cm}
\includegraphics[width=0.339\linewidth]{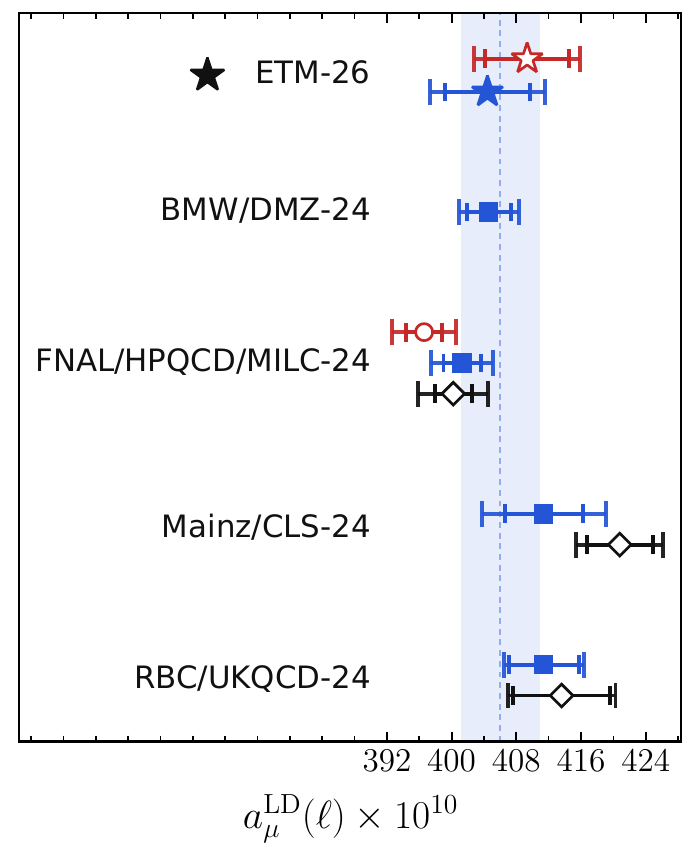}
\caption{Comparison of lattice-QCD determinations of the light-connected SD
(\textit{left}), W (\textit{center}), and LD (\textit{right}) windows. Our results are shown by
star markers, open for the FLAG scheme and filled for the WP25 scheme.
Results from other groups are shown by open circles in the FLAG scheme,
filled squares in the WP25 scheme, and open diamonds in other
isospin-symmetric schemes. Colors provide an additional indication of the
scheme, with red used for FLAG, blue for WP25, and black for the other
schemes. Gray symbols denote superseded determinations. The vertical bands correspond to the WP25 averages~\cite{Aliberti:2025beg}.
References to the quoted papers are as follows:
ABGP-22~\cite{Aubin:2022hgm},
BMW-20~\cite{Borsanyi:2020mff},
BMW/DMZ-24~\cite{Boccaletti:2024guq},
$\chi$QCD-22~\cite{Wang:2022lkq},
ETM-21~\cite{Giusti:2021dvd},
ETM-22~\cite{ExtendedTwistedMass:2022jpw},
FNAL/HPQCD/MILC-23~\cite{FermilabLatticeHPQCD:2023jof},
FNAL/HPQCD/MILC-24~\cite{MILC:2024ryz,FermilabLatticeHPQCD:2024ppc},
LM-20~\cite{Lehner:2020crt},
Mainz/CLS-22~\cite{Ce:2022kxy},
Mainz/CLS-24 (in the left panel)~\cite{Kuberski:2024bcj},
Mainz/CLS-24 (in the right panel)~\cite{Djukanovic:2024cmq},
RBC/UKQCD-18~\cite{RBC:2018dos},
RBC/UKQCD-23~\cite{RBC:2023pvn},
RBC/UKQCD-24~\cite{RBC:2024fic},
SL-24~\cite{Spiegel:2024dec}.
}
\label{fig:comparison_light_windows}
\end{figure*}

We begin with the light-connected contribution,
$a_\mu^{\rm w}(\ell)=\frac{10}{9}a_\mu^{\rm w}(I=1)$. The comparison for all three
windows is shown in Fig.~\ref{fig:comparison_light_windows}.
The central panel of Fig.~\ref{fig:comparison_light_windows} shows a
remarkable agreement among the lattice determinations of the intermediate
window. Our result is fully consistent with the previous calculations and,
relative to our determination in
Ref.~\cite{ExtendedTwistedMass:2022jpw}, improves the precision by more than a
factor of two. The consistency among calculations based on different lattice
actions and analysis strategies is particularly significant for this
benchmark observable, whose uncertainty is now at the few permille level. For the LD contribution, shown in the right panel, we compare with the recent determinations by
Mainz/CLS-24~\cite{Djukanovic:2024cmq},
FNAL/HPQCD/MILC-24~\cite{FermilabLatticeHPQCD:2024ppc},
RBC/UKQCD-24~\cite{RBC:2024fic}, and BMW/DMZ-24~\cite{Boccaletti:2024guq}. In the WP25 scheme, our result
agrees with all of them within one standard deviation. In the FLAG scheme, the
comparison exhibits the same pattern already discussed for the full
light-connected contribution. In particular, the preferred $f_\pi$-based
scheme of Mainz/CLS-24 (which, as already argued, is numerically extremely close to the FLAG scheme, with differences that can hardly be relevant at the present $1\%$ level of precision), is slightly larger than our result in the FLAG scheme by about $1.4\sigma$, while the FNAL/HPQCD/MILC-24 result in the FLAG scheme is $1.8\sigma$ smaller. The FNAL/HPQCD/MILC-24 and Mainz/CLS-24 determinations differ instead by about $3.6\sigma$. 

Our result for the short-distance window, $a_{\mu}^{\rm SD}(\ell)$, shown in the left panel, is also in good overall agreement with existing lattice determinations. Our new result is, however, 
$2.7\times10^{-11}$ smaller than our previous determination in
Ref.~\cite{ExtendedTwistedMass:2022jpw}. Treating the two determinations as
uncorrelated gives a significance of only $1\sigma$, although this is only
a rough estimate because the two analyses partially share gauge configurations. We have traced the shift to two changes that both have a
visible impact on the continuum determination. The first is the inclusion of
the E112 ensemble, which is our finest lattice spacing. The second is the correction
for the residual mistuning of the critical mass, whose systematic treatment
was developed in Ref.~\cite{ExtendedTwistedMass:2024nyi} and was not included
in our earlier window analysis. The C80 ensemble has the largest critical-mass
mistuning, and applying the correction shifts the results in both
regularizations by approximately $3-4\times 10^{-12}$ upwards.
Although the C80 data move upward, the effect on the continuum extrapolation
is in the opposite direction: the corrected point favors a more pronounced curvature in the OS regularization and a slightly reduced curvature in
the TM one, thereby lowering the common continuum-limit value. Together with the
inclusion of the E112 ensemble, this leads to the net downward shift of
$a_\mu^{\rm SD}(\ell)$ quoted above.

\begin{figure*}[t]
\centering
\includegraphics[width=0.8\linewidth]
{comparison_plots/legend_new.pdf}\\[-15pt]
\includegraphics[width=0.339\linewidth]{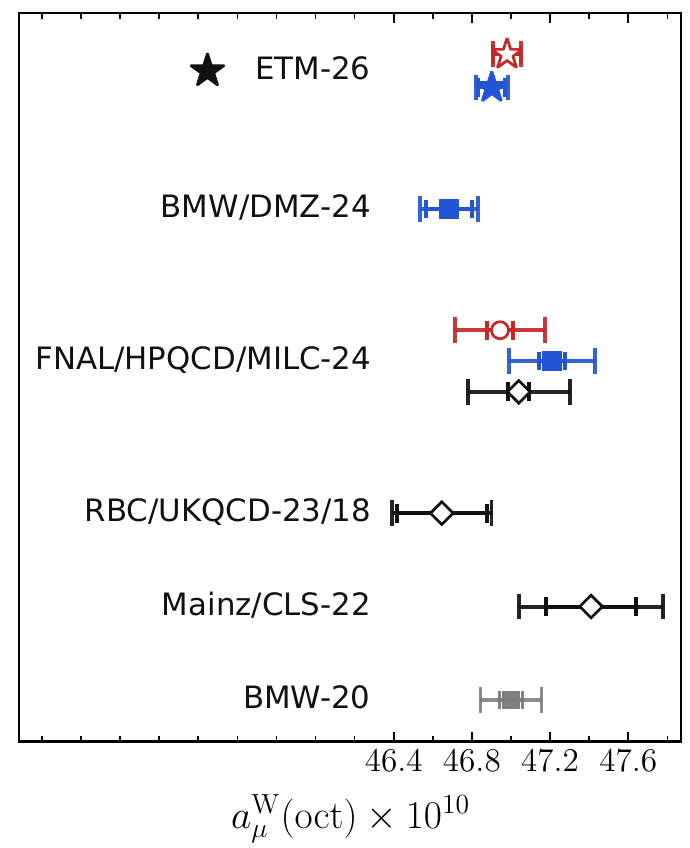}\hspace{-0.25cm}
\includegraphics[width=0.339\linewidth]{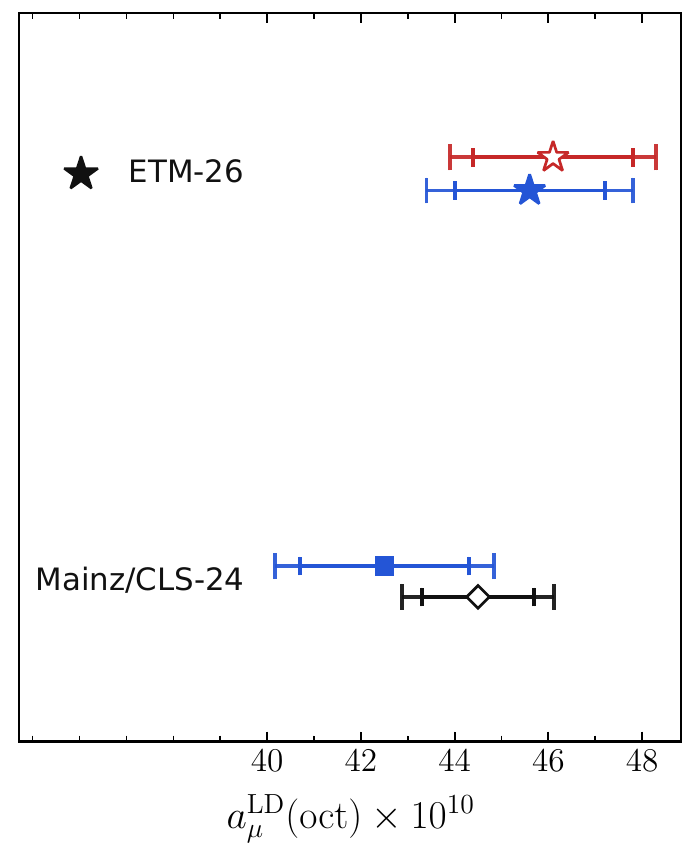}\hspace{-0.25cm}
\includegraphics[width=0.339\linewidth]{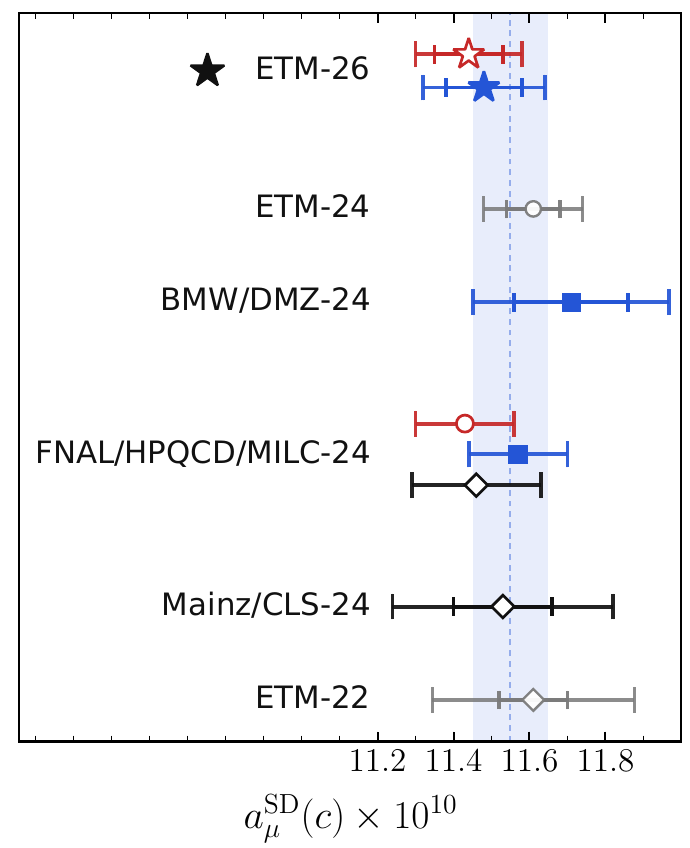}
\caption{Comparison of lattice-QCD determinations of the octet contribution
in the W (\textit{left}) and LD (\textit{center}) windows and of the charm-connected
contribution in the SD window (\textit{right}). Our results are shown by star markers,
open for the FLAG scheme and filled for the WP25 scheme. Results from other
groups are shown by open circles in the FLAG scheme, filled squares in the
WP25 scheme, and open diamonds in other isospin-symmetric schemes. Colors
provide an additional indication of the scheme, with red used for FLAG, blue
for WP25, and black for the other schemes. Gray symbols denote superseded
determinations. The vertical band in the right panel is the WP25 average for $a_{\mu}^{\rm SD}(c)$.
References to the quoted papers are as follows:
BMW-20~\cite{Borsanyi:2020mff},
BMW/DMZ-24~\cite{Boccaletti:2024guq},
ETM-22~\cite{ExtendedTwistedMass:2022jpw},
ETM-24~\cite{ExtendedTwistedMass:2024nyi},
FNAL/HPQCD/MILC-24~\cite{MILC:2024ryz,FermilabLatticeHPQCD:2024ppc},
Mainz/CLS-22~\cite{Ce:2022kxy},
Mainz/CLS-24 (in the central panel)~\cite{Djukanovic:2024cmq},
Mainz/CLS-24 (in the right panel)~\cite{Kuberski:2024bcj},
RBC/UKQCD-18~\cite{RBC:2018dos},
RBC/UKQCD-23~\cite{RBC:2023pvn}.
}
\label{fig:comparison_selected_I0_windows}
\end{figure*}

We now compare the remaining contributions, limiting the discussion to three
representative and particularly relevant quantities in the isoscalar channel:
the octet contribution in the W and LD windows and the charm-connected
contribution in the SD window. The comparison is shown in
Fig.~\ref{fig:comparison_selected_I0_windows}.

For the intermediate-window octet contribution,
$a_{\mu}^{\rm W}({\rm oct})$, shown in the left panel of
Fig.~\ref{fig:comparison_selected_I0_windows}, the recent lattice
determinations show an overall good consistency. In the WP25 scheme, the
BMW/DMZ-24 result~\cite{Boccaletti:2024guq} is smaller than ours by about
$1.3\sigma$, while BMW-20 is fully compatible. The
FNAL/HPQCD/MILC-24 result agrees very well with ours in the FLAG scheme; in the WP25 scheme it is larger
than ours by about $1.3\sigma$.\footnote{For the FNAL/HPQCD/MILC-24
reconstruction in the FLAG and WP25 schemes, we use the correlation
matrix among the light-connected, strange-connected, and disconnected
contributions provided for their preferred $M_\Omega$ scheme. Neglecting
these correlations would reduce the total uncertainties by less than $5\%$.} The reconstructed RBC/UKQCD-23/18 result, obtained by combining the updated
light-connected contribution of Ref.~\cite{RBC:2023pvn} with the
strange-connected and disconnected contributions of
Ref.~\cite{RBC:2018dos}, is slightly smaller than ours, with a difference of
about one standard deviation. The Mainz/CLS-22 result~\cite{Ce:2022kxy}, obtained in their
preferred $f_\pi$-based scheme, which is numerically very close to FLAG, is
instead somewhat larger, by about $1.1\sigma$. Our FLAG- and WP25-scheme
results differ by only $0.078\times10^{-10}$. The comparison of our determination of the LD octet contribution with the
Mainz/CLS-24 result is shown in the central panel. The two calculations are
in good agreement in both the FLAG and WP25 schemes. The Mainz/CLS-24 central
values are slightly smaller than ours, but the differences are within one
standard deviation in both schemes.

Finally, the right panel compares the charm-connected SD contribution, for
which an overall good agreement is observed among the recent lattice
determinations. Our updated central value in the FLAG scheme is lower than
our previous ETM-24 determination
\cite{ExtendedTwistedMass:2024nyi} by about $0.17\times10^{-10}$. This shift
is comparable in size to the total uncertainty of our new result, about
$0.14\times10^{-10}$, and originates from the same improvements in the
treatment of the mistuning corrections discussed for the full
charm-connected contribution in Subsec.~\ref{sec:full_HVP_I0}. The
difference between our FLAG- and WP25-scheme determinations is very small
compared with the current uncertainties.

We finally combine the isotriplet and isoscalar contributions to obtain the
isospin-symmetric results~\footnote{For the LD and W windows, the statistical uncertainties of the $I=1$ and $I=0$ contributions are combined following the same procedure adopted for the full HVP. The ensemble-averaged correlation between the $I=0$ contribution and the OS determination of the $I=1$ contribution is $0.22$ for LD and about $0.8$ for W; for the latter we conservatively set the correlation to unity, which has only a minor impact on the final uncertainty. For the SD window, where the $I=0$ contribution is constructed from $\frac{1}{9}a_\mu^{\rm SD}(I=1)$ together with the strange-, charm-connected and disconnected contributions, the corresponding correlation is accounted for directly by forming the total with $\frac{10}{9}a_\mu^{\rm SD}(I=1)$ and the remaining contributions.}
\begin{align}
 a_\mu^{\rm w}({\rm iso})
 =a_\mu^{\rm w}(I=1)+a_\mu^{\rm w}(I=0)\,,
 \qquad {\rm w}=\{{\rm SD},{\rm W},{\rm LD}\}\,.
\label{eq:window_iso_sum}
\end{align}
For the SD window, the bottom-quark contribution is not completely negligible
at our precision. We therefore include
in the isospin-symmetric SD total, given
in Table~\ref{tab:window_iso_results}, the contribution
$a_\mu^{\rm SD}(b)=0.32\times10^{-10}$, as obtained in
Ref.~\cite{ExtendedTwistedMass:2022jpw} from a perturbative
evaluation using the package RHAD~\cite{Harlander:2002ur}. The
bottom contribution to the W and LD windows is negligible and is not included.

\begin{table*}[t]
\centering
\begin{tabular*}{0.9\textwidth}{@{\extracolsep{\fill}}lccc@{}}
\toprule
\multicolumn{4}{c}{FLAG scheme} \\
\cmidrule(lr){1-4}
Window & $I=1$ & $I=0$ & iso \\
\midrule
SD & $\LinfamuSDIoneFLAGT$ & $\LinfamuSDIzeroFLAGT$ & $68.76(14)(18)[23]$ \\
W  & $\LinfamuIWIoneFLAGT$ & $\LinfamuIWIzeroFLAGT$ & $236.40(36)(45)[57]$ \\
LD & $\LinfamuLDIoneFLAGT$ & $\LinfamuLDIzeroFLAGT$ & $414.5(5.3)(3.9)[6.6]$ \\
\midrule
\multicolumn{4}{c}{WP25 scheme} \\
\cmidrule(lr){1-4}
Window & $I=1$ & $I=0$ & iso \\
\midrule
SD & $\LinfamuSDIoneWPT$ & $\LinfamuSDIzeroWPT$ & $68.81(15)(20)[25]$ \\
W  & $\LinfamuIWIoneWPT$ & $\LinfamuIWIzeroWPT$ & $236.01(36)(49)[60]$ \\
LD & $\LinfamuLDIoneWPT$ & $\LinfamuLDIzeroWPT$ & $409.6(5.3)(4.6)[7.0]$ \\
\bottomrule
\end{tabular*}
\caption{Final isotriplet, isoscalar, and isospin-symmetric results for the
three Euclidean-time windows in the FLAG and WP25 schemes. All entries are in units
of $10^{-10}$. The first uncertainty is statistical, the second is the systematic error, and the uncertainty in square brackets is their
sum in quadrature. In the SD row, the isospin-symmetric total includes the
bottom-quark contribution $a_\mu^{\rm SD}(b)=0.32\times10^{-10}$.}
\label{tab:window_iso_results}
\end{table*}

Our WP25-scheme results in Table~\ref{tab:window_iso_results} are well
compatible with the isospin-symmetric averages quoted in the WP25
report~\cite{Aliberti:2025beg}. In units of $10^{-10}$, the WP25 report quotes
$a_{\mu, {\rm WP25-avg}}^{\rm SD}({\rm iso})=69.06(22)$,
$a_{\mu,{\rm WP25-avg}}^{\rm W}({\rm iso})=236.16(42)$, and
$a_{\mu,{\rm WP25-avg}}^{\rm LD}({\rm iso})=407.9(5.0)$. As a closure test, we also verify that, well within the 
uncertainties, the sum of the independently determined SD, W, and LD windows
reproduces the full HVP result for the isotriplet, isoscalar, and total
isospin-symmetric contributions in both schemes.

For the SD and W windows, the isospin-breaking corrections are separately
provided in the WP25 report~\cite{Aliberti:2025beg} and correspond to the corrections that must be applied to the isospin-symmetric result in the WP25 scheme. Adding them to our results gives
\begin{align}
 a_\mu^{{\rm SD}}
 &=a_\mu^{{\rm SD},{\rm WP25}}({\rm iso})
  +a_\mu^{{\rm SD},{\rm WP25}}({\rm IB})
 =68.85(15)(20)[25]\times10^{-10}\,,
\label{eq:SD_window_full_WP25}\\[8pt]
 a_\mu^{{\rm W}}
 &=a_\mu^{{\rm W},{\rm WP25}}({\rm iso})
  \,+\,a_\mu^{{\rm W},{\rm WP25}}({\rm IB})
 =236.43(36)(49)[61]\times10^{-10}\,,
\label{eq:W_window_full_WP25}
\end{align}
where the first uncertainty is statistical, the second is systematic and
includes the uncertainty on the isospin-breaking correction from
Ref.~\cite{Aliberti:2025beg}, and the value in brackets is the final
uncertainty. We thus determine the SD and W windows with precisions of about
$0.4\%$ and $0.25\%$, respectively. Our results for the short-distance and
intermediate windows are currently among the most precise lattice determination
available. At this level of precision, these quantities should be regarded as
precision observables in their own right, providing stringent,
$g-2$-independent tests of Standard Model predictions against the
experimentally measured $e^+e^-\to{\rm hadrons}$ cross section.

\section{Conclusions}
\label{sec:conclusions}

In this work we have presented a first-principles determination of the
leading-order hadronic-vacuum-polarization contribution to the anomalous
magnetic moment of the muon in isospin-symmetric QCD. The calculation is based
on gauge ensembles generated by ETMC with $N_f=2+1+1$ flavors of
Wilson--clover twisted-mass fermions, pion masses close to the physical one,
four lattice spacings, and spatial extents reaching approximately
$7.7~{\rm fm}$. We organized the calculation of the full HVP in terms of its isospin
components. This decomposition separates the isotriplet channel, which
contains the low-lying two-pion states and therefore exhibits the largest
finite-volume effects and the most pronounced light-quark-mass dependence,
from the isoscalar channel, where these effects are strongly suppressed. The
isotriplet contribution was first extrapolated to the continuum at a common
reference volume and subsequently corrected to infinite volume. The continuum-limit extrapolation has been performed using
two lattice regularizations, referred to throughout the text as the TM and OS
regularizations, which were constrained to approach a common continuum limit.
A blinding procedure was applied, and three
independent analyses were performed to determine this contribution. The analysis
of finite-volume effects has been performed combining direct information from two spatial volumes
with an MLLGS description of the finite-volume two-pion contribution,
validated against the lattice correlators after explicitly accounting in the
model for the charged--neutral pion mass splitting present at nonzero lattice
spacing in twisted-mass QCD. The isoscalar contribution was determined
through its octet, charm-connected, and residual charm-disconnected
components, with the latter found to be numerically irrelevant at the present
precision.

A central feature of the analysis has been the determination of the
isospin-symmetric HVP in two distinct schemes, the FLAG and WP25 schemes. This
required retuning the simulated bare parameters and propagating the
corresponding effects through all observables entering the analysis. Once
compared within the same scheme, our findings are consistent with the
most precise recent lattice determinations; see
Subsecs.~\ref{sec:full_HVP_summary} and \ref{sec:window_summary} for an
extended discussion. The dominant isotriplet contribution is obtained with a
precision of about $1\%$, while the smaller isoscalar contribution has been
determined with a precision of about $2\%$, making it a subleading source of
uncertainty in the final result. We have also provided independent
determinations of the short-, intermediate-, and long-distance Euclidean-time
windows. For
the intermediate-window contribution we have reduced the uncertainty by more than a factor of
two relative to our previous
determination~\cite{ExtendedTwistedMass:2022jpw}.

Our result strengthens the picture that has already emerged from recent lattice calculations. After combining our determination in isospin-symmetric QCD with
the corresponding isospin-breaking correction taken from the WP25
report~\cite{Aliberti:2025beg}, the resulting HVP contribution leads to a
Standard Model prediction for the anomalous magnetic moment of the muon that
is compatible with the experimental world average. At the same time, it
reinforces the discrepancy between lattice-QCD calculations and dispersive
evaluations based on the pre-CMD-3 experimental data.

An important fraction of the computational effort devoted to this work was
required to control the small mistunings of the sea-quark masses and of the
critical mass needed to match the specific hadronic definitions of
isospin-symmetric QCD considered in this study. The dominant effects
associated with the mistuning of the light-quark mass and of the critical
mass were determined on each physical-point gauge ensemble by computing the
corresponding reweighting factors within the linear approximation,
using the full set of gauge
configurations on which the HVP has been evaluated. These corrections were therefore implemented directly from
first principles, without relying on phenomenological models, and the
precision achieved is sufficient not to increase our final uncertainties
significantly. The same strategy proved much less effective for the sea-strange and,
especially, sea-charm mistuning effects. On the large physical volumes used
for the HVP calculation, the corresponding reweighting observables turned out
to be too noisy to be of practical use. Exploiting the fact that their
statistical fluctuations grow approximately as $\sqrt{L^3}$, we therefore
generated dedicated small-volume ensembles comprising several thousand gauge
configurations, used specifically to determine these subleading mistuning
effects with sufficient statistical accuracy. The sea-strange mistuning corrections were found to be negligible for nearly
all observables. The main exception is $f_\pi$, for which the correction
reaches the permille level on the ensembles with the largest strange-mass
mistuning. The sea-charm derivatives were generally determined with
uncertainties comparable to their central values, and no direct effect of the
small charm-mass mistuning was resolved for most observables. Achieving this
level of accuracy was nevertheless essential to prevent the uncertainties
associated with these corrections from significantly inflating the final
errors. A notable exception is provided by the gradient-flow observable $w_0$ (see Appendix~\ref{sec:w0_FLAG}), for which we observe a clear sea-charm-mass dependence
at finite lattice spacing. This dependence is consistent with being a cutoff
effect and vanishes in the continuum limit, as expected from charm-quark
decoupling. It cannot, however, simply be neglected at finite lattice
spacing, because both the magnitude and the sign of the charm-mass mistuning
vary among the ensembles. Controlling these effects has significantly
strengthened the robustness of the calculation.

There remains room to improve the precision already within
isospin-symmetric QCD. Several recent calculations have reduced the
statistical uncertainties originating from the long-distance tail of the
isotriplet correlator through spectral-reconstruction techniques that allow
the contributions of the lowest-lying finite-volume two-pion states to be
determined explicitly
\cite{RBC:2024fic,Djukanovic:2024cmq,FermilabLatticeHPQCD:2024ppc}. Such a
strategy has not yet been deployed in the present analysis. 
We will consider its implementation in a follow-up work since in
twisted-mass QCD, owing to the distortion of the two-pion spectrum at nonzero lattice spacing, the application of such techniques requires handling additional complications, including that associated with the presence of the neutral pion, which requires the calculation of $O(a^2)$ quark-disconnected diagrams.

In parallel, we are pursuing a complete first-principles determination of the
isospin-breaking correction. The leading quark-connected contributions in the
electroquenched approximation have already been determined~\cite{Evangelista:2025dzi}. Current efforts
are focused on the calculation of the corresponding quark-disconnected
isospin-breaking diagrams, as well as on the estimate of electrounquenching effects, generated by the QED correction to the fermionic determinant.
The latter effects have so far been computed from first principles only by
the BMW Collaboration
\cite{Borsanyi:2020mff,Boccaletti:2024guq}, and independent determinations by
other lattice groups would therefore be extremely valuable.

\section*{Acknowledgments}
We thank all members of ETMC for the most enjoyable collaboration.  C.A. and C.S.~acknowledge partial support from the European Joint Doctorate AQTIVATE that received funding from the European Union’s research and innovation programme under the Marie Sklodowska-Curie Doctoral Networks action under the Grant Agreement No 101072344. 
N.K., S.R., and U.W.~acknowledge support by the Swiss National Science Foundation (SNSF) project No.~200020\_208222 and  No.~200021-232288. 
V.L., F.S., G.G., R.F., N.T., L.M., A.E.~and A.D.S.~have been supported by the Italian Ministry of University and Research (MUR) under the grant PNRR-M4C2-I1.1-PRIN 2022-PE2 Non-perturbative aspects of fundamental interactions, in the Standard Model and beyond F53D23001480006 funded by E.U.-NextGenerationEU.
C.A, S.B., A.E. and F.P. acknowledge support from the projects EXCELLENCE/0524/0459 (IMAGE-N) and EXCELLENCE/0524/0017 (MuonHVP), co-financed by the European Regional Development Fund and the Republic of Cyprus through the Research and Innovation Foundation within the framework of the Cohesion Policy Programme “THALIA 2021-2027”.
This project received funding from the European Research Council (ERC) via the project ”LEEX” grant agreement 101170304, funded by the European Union. Views and opinions expressed are however those of the author(s) only and do not necessarily reflect those of the European Union or the European Research Council Executive Agency (ERCEA). Neither the European Union nor the ERCEA can be held responsible for them.
M.G., B.K. and C.U. are supported by the Deutsche Forschungsgemeinschaft (DFG,
German Research Foundation) as part of the CRC 1639 NuMeriQS – project no.~511713970 and no.~528533525.
F.S. is supported by ICSC – Centro Nazionale di Ricerca in High Performance Computing, Big Data and Quantum Computing, funded by European Union -NextGenerationEU and by Italian  Ministry of University and Research (MUR) projects FIS\_00001556 and PRIN\_2022N4W8WR.

The open-source packages tmLQCD~\cite{Jansen:2009xp,Abdel-Rehim:2013wba,Deuzeman:2013xaa,Kostrzewa:2022hsv}, LEMON~\cite{Deuzeman:2011wz}, DD-$\alpha$AMG~\cite{Frommer:2013fsa,Alexandrou:2016izb,Bacchio:2017pcp,Alexandrou:2018wiv}, QPhiX~\cite{joo2016optimizing,Schrock:2015gik} and QUDA~\cite{Clark:2009wm,Babich:2011np,Clark:2016rdz} have been used in the ensemble generation.

We acknowledge the Swiss National Supercomputing Centre (CSCS) access to Alps through the Chronos programme under project ID CH15.
The authors gratefully acknowledge the Gauss Centre for Supercomputing e.V.~(www.gauss-centre.eu) for funding this project by providing computing time on the GCS Supercomputers SuperMUC-NG at Leibniz Supercomputing Centre and JUWELS~\cite{JUWELS, JUWELS-BOOSTER} at Juelich Supercomputing Centre.
The authors acknowledge the Texas Advanced Computing Center (TACC) at The University of Texas at Austin for providing HPC resources (Project ID PHY21001).
The authors gratefully acknowledge PRACE for awarding access to HAWK at HLRS within the project with Id Acid 4886.
We acknowledge the Swiss National Supercomputing Centre (CSCS) and the EuroHPC Joint Undertaking for awarding this project access to the LUMI supercomputer, owned by the EuroHPC Joint Undertaking, hosted by CSC (Finland) and the LUMI consortium through the Chronos programme under project IDs CH17-CSCS-CYP and CH21-CSCS-UNIBE as well as the EuroHPC Regular Access Mode under project ID EHPC-REG-2021R0095.
We are grateful to CINECA and EuroHPC JU
for awarding this project access to Leonardo 
supercomputing facilities hosted at CINECA. 
We gratefully acknowledge EuroHPC JU for the computer
time on Leonardo-Booster provided to us through the 
Extreme Scale Access Call grants EHPC-EXT-2023E01-007 
and EHPC-EXT-2024E01-027.
We gratefully acknowledge
CINECA for the provision of GPU time under the specific initiative INFN-LQCD123 and IscrB S-EPIC.
 We gratefully acknowledge access to the Marvin cluster of the University of Bonn.

\appendix

\clearpage 
\section{Blinding}
\label{sec:blinding}

In this Appendix we describe the blinding procedure applied to the
isotriplet correlator $V^{I=1}(t)$ and report on the  blinded analyses
performed independently by three groups (referred to in the following as group A, B, and C) to determine
$a_{\mu}^{\rm HVP}(I=1)$. We then discuss the results obtained after the
relative unblinding, comment on the main differences among the three
analyses, and summarize the steps that led to the full unblinding of
$a_{\mu}^{\rm HVP}(I=1)$ at the reference spatial extent
$L_{\rm ref}=5.46~{\rm fm}$.

\subsection{Blinding procedure}
\label{sec:A1}

Before distributing the isotriplet correlators to the three analysis groups,
we applied to the bare correlator at given lattice spacing $a$ a group-dependent additive shift of the
form
\begin{align}
V^{I=1}(t)\longrightarrow V^{I=1}_{g,{\rm blind}}(t)
={}&V^{I=1}(t)
+\frac{1}{Z^{2}}
\sum_{n=1}^{k_g}|A_{n,g}|^{2}
\left[
e^{-E_{n,g}t}+e^{-E_{n,g}(T-t)}
\right]\,,
\qquad g=A,B,C\,,
\label{eq:I1_blinding_shift}
\end{align}
where $T$ is the temporal extent of the lattice. The renormalization
constant is $Z=Z_V$ for the OS regularization and $Z=Z_A$ for the TM
regularization. The factor $1/Z^2$ ensures that the same shift is introduced
in the renormalized TM and OS correlators.

The number $k_g$ of fictitious states added to the correlator was chosen by
the person responsible for the blinding and was not communicated to the
analysis groups. For each group, the amplitudes $|A_{n,g}|^2$ and energies
$E_{n,g}$ were generated pseudorandomly within predefined ranges using a
group-specific seed. The energies were restricted to
\begin{align}
0.75~{\rm GeV}\leq E_{n,g}\leq2~{\rm GeV}\,,
\label{eq:I1_blinding_energy_range}
\end{align}
so as to remain above the lowest energy contributing to the physical
isotriplet correlator. The added terms have positive spectral weights, and
therefore preserve the assumptions underlying the bounding analysis. The numbers of fictitious
states, revealed only at the time of the full unblinding, were
\begin{align}
k_A=14\,,
\qquad
k_B=18\,,
\qquad
k_C=16\,.
\label{eq:I1_blinding_number_states}
\end{align}
The presence of such a large number of randomly distributed states makes it
impossible, in practice, to isolate the blinding contribution through a
spectral analysis of the correlator.

The additive and time-dependent blinding adopted in this work has several
useful properties. First, the blinding terms cannot be identified by
inspecting the precise short-distance behavior of the correlator and
comparing it with the correlators used in our previous determination of the
short-distance and intermediate windows
\cite{ExtendedTwistedMass:2022jpw}. This would not be true for a
time-independent multiplicative blinding factor.

Second, finite-volume effects are not modified by the blinding procedure. By construction, the same additive contribution is applied
to correlators at different volumes.\footnote{Up to exponentially suppressed and
numerically completely negligible differences in the backward-propagating term
arising from their different temporal extents $T$.} Therefore, their difference,
and hence the information used in the finite-volume analysis, is left unchanged.

Third, the quark-mass and critical-mass mistuning corrections are
unaffected by the blinding procedure. The additive blinding contribution
is gauge independent and therefore cancels in the gauge-connected
correlation functions entering Eq.~\eqref{eq:sea_derivatives}, as well as
in the differences between correlation functions evaluated at different
values of the valence quark masses and critical mass.

Finally, we note that the previous generation of isotriplet correlators used in
Ref.~\cite{ExtendedTwistedMass:2022jpw} was produced without the
noise-reduction techniques employed in the present calculation. The
statistical noise in the long-distance tail of those correlators prevents a precise knowledge of the shift applied to $a_{\mu}^{\rm HVP}(I=1)$. Indeed this quantity, if determined at fixed $a$ using the correlator data of Ref.~\cite{ExtendedTwistedMass:2022jpw}, which however have not been made available to the three analysis groups, has statistical errors of the order of $15\times 10^{-10}$. In addition, note that the finest lattice spacing ensemble, the E112 ensemble, was not present in the analysis of Ref.~\cite{ExtendedTwistedMass:2022jpw}.

After applying the time-dependent shifts, three different blinded
data sets were distributed to groups A, B, and C. 

\subsection{Relative unblinding of the $I=1$ contribution, comparison and full unblinding}

\begin{figure}
    \centering
    \includegraphics[width=0.6\linewidth]{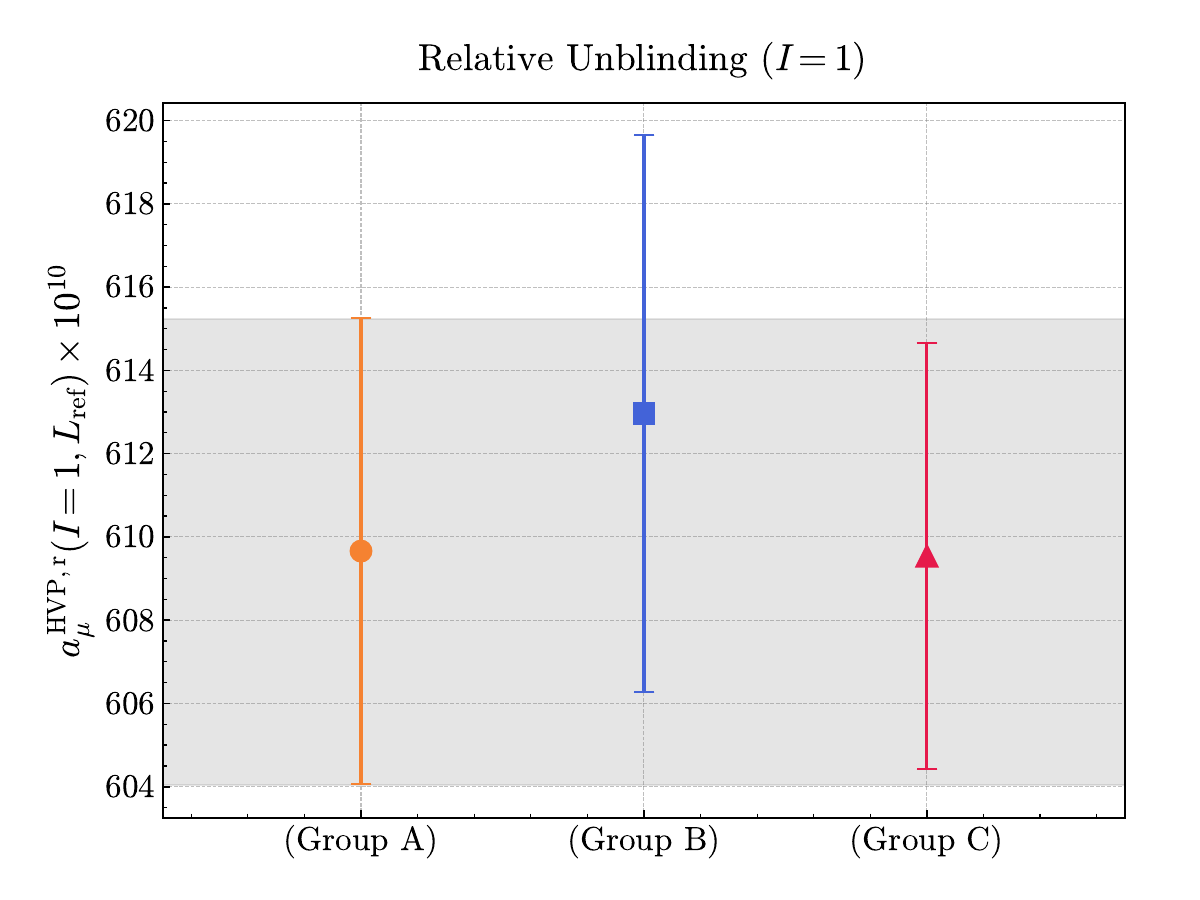}
    \caption{Relative unblinding of the three independent determinations of
    $a_{\mu}^{\rm HVP,FLAG}(I=1,L_{\rm ref})$. Groups A, B, and C analyzed
    the same correlator carrying the group-A blinding shift, so that their
    results could be compared without revealing the physical value. The final
    result quoted in the main text is also shown as a colored band after being translated back
    to the group-A blinding shift.}
    \label{fig:I1_relative_unblinding}
\end{figure}

The three analyses were based on the same main ingredients: tree-level
improvement, application of the quark-mass and critical-mass mistuning
corrections, treatment of the long-distance tail through the bounding
method, interpolation of the B64 and B96 determinations to
$L_{\rm ref}=5.46~{\rm fm}$, and simultaneous continuum extrapolation of
the TM and OS results. The analyses were nevertheless developed
independently and differed in several relevant methodological choices. The
analysis presented in the main text corresponds, apart from negligibly
small refinements, to that developed by group A.

On April 28th 2026, we proceeded to the relative
unblinding of the three determinations. The correlator originally blinded for
group A was also passed to groups B and C, which repeated their complete
analyses on this common data set. In this way, the same unknown additive
blinding shift was present in the data of the three groups, allowing a direct
comparison without revealing the physical value of
$a_{\mu}^{\rm HVP}(I=1)$. The results obtained in the FLAG scheme at
$L=L_{\rm ref}$ are collected in
Fig.~\ref{fig:I1_relative_unblinding}. As shown in the figure, the three
analyses give central values and uncertainties which are in good agreement.

We now discuss the main differences among the three analyses that were most
relevant for selecting the final strategy. The first concerns the
interpolation of the B64 and B96 results to
$L_{\rm ref}=5.46~{\rm fm}$. Groups B and C used an effective finite-volume
dependence proportional to $e^{-M_\pi L}/L$. We found that this form
provides a less accurate effective description of the volume dependence
predicted by the MLLGS model, both with degenerate and nondegenerate pion
masses, in the interval between approximately $5.1$ and $7.7~{\rm fm}$,
which is the range relevant for interpolating between the B64 and B96
ensembles. The $L e^{-M_\pi L}$ ansatz of
Eq.~\eqref{eq:I1_L_interpolation}, used by group A and adopted in the main
analysis, instead provides a very good effective description of the MLLGS
volume dependence over this interval. The groups also adopted somewhat different implementations of the bounding
procedure. In particular, groups B and C explored alternative choices for
the time $t^{*}$ at which the effective mass used in the lower bound was
evaluated, for the criterion used to identify the region where the upper and
lower bounds had converged, and for the extent of the subsequent constant
fit. The variations associated with these choices were found to be
negligible compared with the dominant uncertainties. Another difference
concerns the energy entering the upper bound in
Eq.~\eqref{eq:bounding-V-upper}. Groups B and C used the noninteracting
value $E_{0}^{I=1}=E_{0,{\rm free}}^{I=1}$, whereas group A adopted the more
conservative choice
$E_{0}^{I=1}=0.98\,E_{0,{\rm free}}^{I=1}$, as discussed around
Eq.~\eqref{eq:E0f1_OS}. The latter choice conservatively allows for the
downward shift of the lowest two-pion energy induced by the attractive
two-pion interaction in finite volume. A third difference concerns scale setting and the treatment of mistuning
effects. Group A carried out the complete scale-setting analysis and the
determination of the quark-mass and critical-mass mistuning corrections. The
resulting lattice spacings, quark masses, and direct corrections to the HVP
were then provided to groups B and C, which did not repeat these analyses
independently. The correlations among the tuned quark masses, the critical
mass, and the lattice spacing were supplied explicitly. In the group-A
analysis, these correlations were retained directly through common
bootstrap samples of the underlying raw data, rather than reconstructed
from an estimated correlation matrix.

A further relevant difference concerns the continuum-limit extrapolation.
All three groups fitted the TM and OS determinations simultaneously,
constraining them to approach a common continuum limit while allowing for
regularization-dependent cutoff effects. The construction of the fit pools
and the criteria used to select and combine the models were, however,
different. Group B considered combinations of constant, linear, and quadratic
dependences on $a^2$ for the two regularizations, restricting to fits containing no more than four free parameters. Its fit pool also included
cuts in the data obtained by removing either the coarsest or the
second-coarsest lattice spacing, or one of the two regularizations at
selected lattice spacings. Logarithmically modified $a^2$ terms were not
included. Group C instead retained the complete eight-point data set and
considered constant, linear, quadratic, and logarithmically modified cutoff
effects. In the group-C analysis, the different choices entering both the
bounding procedure and the continuum extrapolation were incorporated into
the BAIC model average. A further feature of the group-C analysis was that only fits for which
all cutoff coefficients differed from zero by at least one standard
deviation were retained in the model average. After discussion, we decided not to impose such a
significance-based selection in the final analysis. The fact that a fit
coefficient is not statistically resolved does not, by itself, invalidate
the corresponding ansatz, nor imply that a fit coefficient vanishes exactly.
A selection based on the fitted significance of individual coefficients
makes the composition of the model pool depend a posteriori on the
statistical fluctuations of the data and may inadvertently exclude
otherwise legitimate fit functions. In the present case, for example, this
criterion would exclude a linear-in-$a^{2}$ description of the TM data. We found it cleaner to define the model pool without conditioning its
composition on the fitted statistical significance of individual cutoff
coefficients, using a restricted set of theoretically motivated ansatzes.

The group-A analysis retained the complete eight-point TM--OS data set and
used the general ansatz of
Eq.~\eqref{eq:I1_continuum_ansatz} to describe the cutoff dependence. The
fit pool consisted of a restricted set of theoretically motivated
ansatzes, including regular $a^2$, logarithmically modified $a^2$, and
$a^4$ cutoff terms, with at most four free parameters, including the common
continuum value. No selection based on the statistical significance of the
fitted cutoff coefficients was applied, and no data cuts 
were included in the final model average.

Given the mild cutoff effects observed in the $I=1$ data, and in particular
the nearly flat behavior of the TM results, with only four lattice spacings
available, after discussion we concluded that removing one of them would
weaken the statistical constraint on the continuum extrapolation and
increase its sensitivity to fluctuations in the remaining points. The
reduced-data analyses performed by group B, obtained by removing either the
coarsest or the second-coarsest lattice spacing, nevertheless provided very
useful cross-checks.

On the basis of these considerations, the analysis developed by group A was
selected as the basis of our final determination.

Following the relative unblinding, a few very small modifications were
introduced in the group-A analysis, mainly to correct minor details in the analysis code. Since all these changes produced only marginal effects on
both the central value and the uncertainty, we do not consider an itemized
discussion of them particularly illuminating. In
Fig.~\ref{fig:I1_relative_unblinding} we show the final FLAG-scheme
result quoted in the main text at $L=L_{\rm ref}$, translated back to match
the group-A blinding. The original group-A result and the final
determination are practically indistinguishable, both in their central
values and in their uncertainties.

Once the final analysis strategy had been fixed, the absolute blinding shift
was revealed. The unblinded result is obtained from the result determined
using the group-A-blinded correlator by subtracting
\begin{align}
\Delta_{\rm BLIND}a_{\mu}^{\rm HVP}(I=1,L_{\rm ref})
\simeq 38.64\times10^{-10}\,.
\label{eq:I1_absolute_blinding_shift}
\end{align}

\section{Details on renormalization in the FLAG scheme}
\label{sec:scale_setting_details}
This Appendix collects methodological details specific to the present analysis.
In particular, it discusses the determination of the derivatives of the renormalization observables relevant for mistuning corrections, as well as the determination of $w_0$ in the FLAG scheme of isospin symmetric QCD, including the corresponding mistuning corrections and continuum extrapolation.

The methodology adopted to determine the derivatives of the renormalization inputs partly builds on Ref.~\cite{ExtendedTwistedMass:2024nyi}.
In that work, the renormalization inputs were interpolated at each iteration in the valence-quark masses so as to reproduce their physical target values.
Although valence-quark-mass derivatives were not introduced explicitly there, the slopes of those linear interpolations coincide with the valence-quark-mass derivatives entering the present formulation.
Since the underlying procedure is unchanged, we refer the interested reader to Ref.~\cite{ExtendedTwistedMass:2024nyi} for details.
Below, we focus instead on the improvements introduced in the present analysis concerning the treatment of the $m_0$ parameter mistuning and  the derivatives of renormalization observables with respect to the sea quark masses.
\subsection{Computation of the $m_0$ mistuning and correction of the renormalization input}
On this matter, we follow the strategy introduced in Ref.~\cite{ExtendedTwistedMass:2024nyi}. The mistuning of the critical mass parameter $m_0$ is determined by solving Eq.~\eqref{eq:syst_B}, which we reproduce here in expanded form for convenience:
\begin{align}
m_{\rm PCAC}^{\rm sim}
+(m_0-m_0^{\rm sim})
\left[
\left.
\frac{d m_{\rm PCAC}}{d m_0}
\right|_{\rm val,sim}
+
\left.
\frac{d m_{\rm PCAC}}{d m_0}
\right|_{\rm sea,sim}
\right]
=0~.
\label{eq:mpcac_correction}
\end{align}
The valence and sea contributions to the derivative are evaluated from correlation functions containing, respectively, pseudoscalar insertions in the valence propagators and pseudoscalar fermionic loops in the sea sector.

While the complete derivative of $m_{\rm PCAC}$ with respect to $m_0$ was not directly available for all ensembles in Ref.~\cite{ExtendedTwistedMass:2024nyi}, both its sea and valence contributions are determined separately on each ensemble in the present analysis. In particular, the valence-quark contribution has now been computed for all ensembles and $d m_{\rm PCAC} / d m_0^{\rm sea}$ for E112 has been evaluated directly on the relevant gauge configurations. 
This provides a fully ensemble-specific determination of the $(m_0-m_0^{\rm sim})$ mistuning corresponding to a given $m_{\rm PCAC}^{\rm sim} \ne 0$, eliminating the additional uncertainties previously introduced to account for missing valence contributions or for derivatives inferred from other ensembles.

For the correction of the renormalization observables $m_\pi^2$ and $f_\pi$, we continue to use the analytic expressions derived in Ref.~\cite{ExtendedTwistedMass:2021qui}, which at leading order in 
small light-quark-mass expansion remove the $O(a^0)$ dependence of 
these quantities on the residual value of $m_{\rm PCAC}$ (given in our
case by $m_{\rm PCAC}^{\rm sim} \neq 0$):
\begin{align}
f_\pi(m_{\rm PCAC}=0)
&=
f_\pi^{\rm sim}
\sqrt{1+\left(\frac{Z_A m_{\rm PCAC}^{\rm sim}}{m_\ell^{\rm sim}}\right)^2}\,,
\label{eq:fpi-anacorr}
\\
m_\pi^2(m_{\rm PCAC}=0)
&=
\frac{({m_\pi^{\rm sim}})^2}
{\sqrt{1+\left(\frac{Z_A m_{\rm PCAC}^{\rm sim}}{m_\ell^{\rm sim}}\right)^2}}\,.
\label{eq:mpi-anacorr}
\end{align}
The key point behind these formulae is that up to
lattice artifacts if $m_{\rm PCAC} \neq 0$ the effective renormalized light-quark mass reads $\hat{m}_\ell = Z_P^{-1} ( (m_\ell^{\rm sim})^2 + Z_A (m_{\rm PCAC}^{\rm sim})^2 )^{1/2}$. Indeed, $m_K$ and especially $m_{D_s}$ are known
to have a much weaker dependence on $\hat{m}_\ell$ than $m_\pi$. Furthermore, the
strange and charm effective quark masses are given (again to $O(a^0)$) by $\hat{m}_{s,c} = Z_P^{-1} ( (m_{s,c}^{\rm sim})^2 + Z_A (m_{\rm PCAC}^{\rm sim})^2 )^{1/2}$. 
Then, it becomes clear that the analogous $m_0$ mistuning corrections to
these two meson masses are expected to be fully
negligible within statistical errors. Indeed, we have explicitly checked that the corrections to $m_K$ and $m_{D_s}$ induced by $m_{\rm PCAC}^{\rm sim} \neq 0$---evaluated by means of the derivatives of $m_K$ and
$m_{D_s}$ with respect to $m_0$---are small and statistically compatible with zero. This is in agreement with the argument above  regarding the $m_{\rm PCAC}$ dependence of the mass of mesons involving strange and charm valence quarks~\cite{ExtendedTwistedMass:2021qui}.

\begin{figure}
\centering
\includegraphics[width=0.48\linewidth]{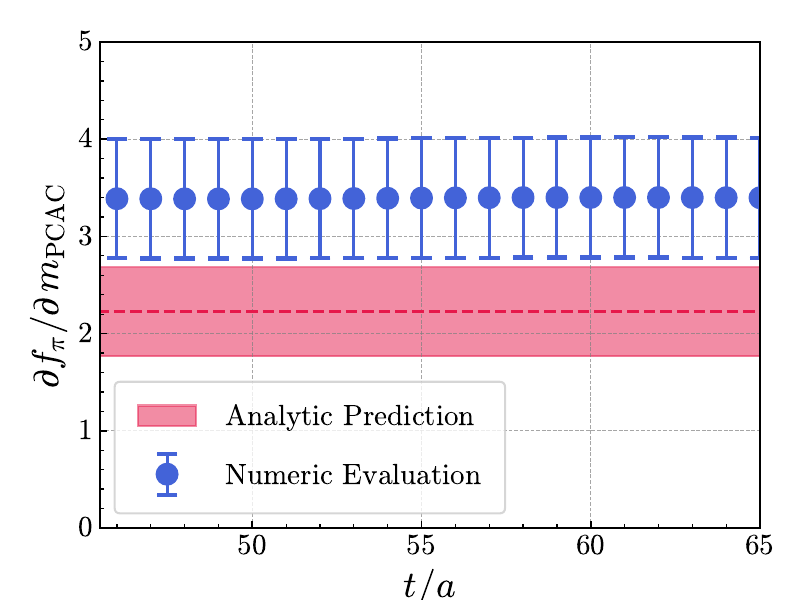}
\includegraphics[width=0.48\linewidth]{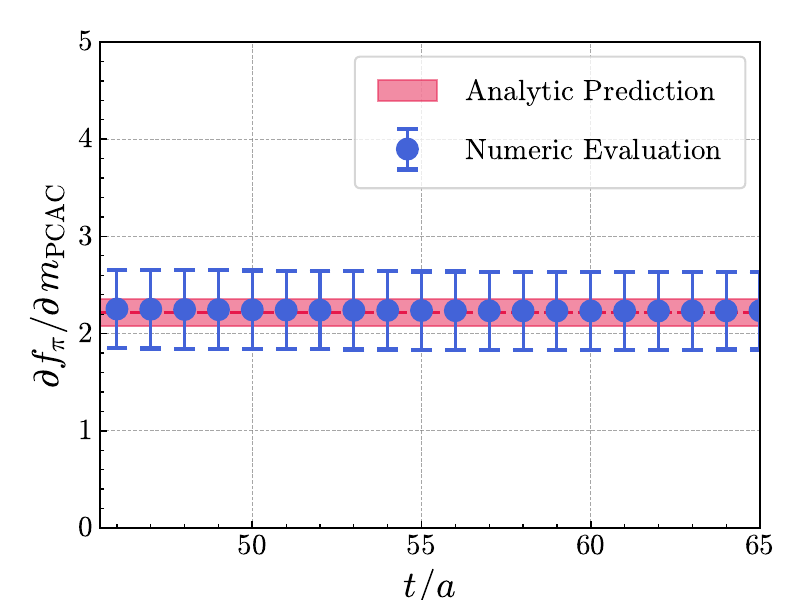}
\caption{Comparison between the numerical and analytic determinations of the $f_\pi$ derivative with respect to the PCAC mass for the two ensembles with the largest deviations from maximal twist, namely C80 (\textit{left}) and E112 (\textit{right}). A mild $1.5\sigma$ tension is observed for C80, while the two determinations are in excellent agreement for E112.}
\label{fig:dfpi_dmpcac}
\end{figure}

The full determination of the derivatives of $f_\pi$ and $m_\pi$ with respect to $m_0$ also allows now the analytic correction formulae in Eqs.~\eqref{eq:fpi-anacorr} and \eqref{eq:mpi-anacorr} to be tested against independent numerical estimates on each ensemble.
Fig.~\ref{fig:dfpi_dmpcac} shows this comparison for the case of $f_\pi$ on the C80 and E112 ensembles, i.e., the ones with the largest deviations from maximal twist. The two determinations of $d f_\pi / d m_{\rm PCAC}$ are in very good agreement for E112, whereas a mild $1.5\sigma$ tension is observed for C80, consistently with the result previously reported in Ref.~\cite{ExtendedTwistedMass:2024nyi}. Nevertheless, the effect of this mild tension on the corrected value of $f_\pi$ is tiny: the results obtained using the analytic formula~(\ref{eq:fpi-anacorr}) and the numerical corrections differ by only $0.5\sigma$, owing to the smallness of $m_{\rm PCAC}^{\rm sim}/m_\ell$ and the fact that $f_\pi^{\rm FLAG} - f_\pi^{\rm sim} \simeq  {\rm O}((m_{\rm PCAC}^{\rm sim})^2) $, with $m_{\rm PCAC}^{\rm sim} \propto (m_0^{\rm sim} - m_0)$.

One can actually prove that in the lattice setup of ETMC~\cite{ExtendedTwistedMass:2022jpw, ExtendedTwistedMass:2024nyi}, to the extent that the value of the Wilson-clover coefficient tuned through a
tadpole-boosted one-loop estimate coincides with its nonperturbative value,
the first derivatives with respect to $m_0$ of both $m_\pi$ and $f_\pi$ vanish up to tiny O($a^3$) cutoff effects at the maximal twist point (where $m_{\rm PCAC}=0$), implying 
$m_\pi^{\rm FLAG} - m_\pi^{\rm sim} \simeq  {\rm O}(m_{\rm PCAC}^2) $
and $f_\pi^{\rm FLAG} - f_\pi^{\rm sim} \simeq  {\rm O}(m_{\rm PCAC}^2) $.
Indeed, at maximal twist a nonzero first derivative of $m_\pi$ with respect to $(m_0^{\rm sim} - m_0) $ would mean a nonvanishing value of the connected insertion of the spacetime integrated $I=1, I_3=0$ pseudoscalar density
in the charged-pion-channel two-point correlator. Based on
a Symanzik expansion analysis of the latter in terms of
continuum QCD correlators, this is possible only to the extent that the Symanzik effective action of the lattice theory contains a parity-odd term, which in turn, owing to the symmetries of maximally-twisted Wilson lattice QCD, can occur only at O($a^3$) if the lattice action is properly O($a$) improved through the appropriate clover term -- or at O($a$) if it is not. At maximal twist the first derivative of $f_\pi$ with respect to $(m_0^{\rm sim} - m_0) $ also vanishes up to $O(a^3)$ terms,
because $f_\pi$ is determined through an exact lattice chiral Ward identity in terms of $2 m_\ell^{\rm sim} \langle 0 | P^+ | \pi^+(\vec{0}) \rangle / m_\pi^2 $ and the pseudoscalar isotriplet density, with no other linear-in-$a$ cutoff effects from the pseudoscalar lattice operator, besides those coming from the lattice action.

The main improvement of the present analysis therefore lies in the ensemble-specific determination of the critical-mass mistuning,
$m_0-m_0^{\rm sim}$,
which is subsequently used to correct $a_\mu^{\rm HVP}$ for the corresponding deviation from maximal twist. 
We give in Table~\ref{tab:m0sim} the values of $am_0^{\rm sim}$, $am_\ell^{\rm sim}$ and $(m_0-m_0^{\rm sim})/m_\ell^{\rm sim}$, together with the ensemble labels as given in Table~\ref{tab:iso_def} (where our determination of the critical-mass parameter $am_0$ is given).

For the renormalization inputs, instead, we retain the analytic corrections of Ref.~\cite{ExtendedTwistedMass:2021qui}, since the numerical estimates based on the newly determined derivatives yield neither significant differences nor an improvement in precision.

\begin{table}[t]

\begin{tabular}{lcccccc} 
ensemble & \(am_{0}^{\mathrm{sim}}\) & $am_\ell^{\rm sim}$& {\((m_0-m_0^{\rm sim})/m_\ell^{\rm sim}\)} & $a\mu_\sigma$ & $a\mu_\delta$ & $c_{sw}$\\[4pt]
\hline\\[0pt]
B64 ~&~ $-0.41388115$ ~& ~0.00072 &{$-0.0087(55)$}           & 0.1315052 ~&~0.1246864~&~1.69	\\[4pt]
B96 ~&~ $-0.41388115$ ~& ~0.00072 &{$-0.0087(55)$}           & 0.1315052 ~&~0.1246864~&~1.69	\\[4pt]
C80 ~&~ $-0.39647041$ ~& ~0.00060 &{\,\,\,\,\,$ 0.0254(54)$} &0.1071460  ~&~0.1065860~&~1.6452\\[4pt]
D96 ~&~ $-0.37608088$ ~& ~0.00054 &{$-0.0158(49)$}           &0.0879110  ~&~0.0862240~&~1.6112	 \\[4pt]
E112~&~ $-0.36133090$ ~& ~0.00044 &{\,\,\,\,\,$ 0.0331(22)$} &0.0777070  ~&~0.0746470~&~1.5792\\[4pt]
\hline
\end{tabular}
\caption{Value of the bare $am_0^{\rm sim}$ and $am_\ell^{\rm sim}$ used in the simulation of the ETMC gauge ensembles, together with the deviation from the critical value $am_0$ defined as the solution of Eq.~\ref{eq:mpcac_correction}.
The definitions of the bare parameters $a\mu_\sigma, a\mu_\delta, c_{sw}$ can be found in \cite{Alexandrou:2018egz} 
or appendix A of \cite{ExtendedTwistedMass:2024nyi}.
\label{tab:m0sim}}
\end{table}

\subsection{Derivatives of the renormalization inputs with respect to sea-quark masses}
In Ref.~\cite{ExtendedTwistedMass:2024nyi}, the sea-quark-mass derivatives were determined using the linearized approximation to determinant reweighting, expressed in terms of scalar fermionic loops in the sea sector. In the present analysis, we instead compute the full determinant reweighting factor associated with a small auxiliary shift of the sea-quark mass, $\delta m$. The corresponding linear response is then extracted numerically by choosing $\delta m=O(10^{-5}/a)$, for which the finite-difference approximation of the derivative is well controlled.

In practice, for each flavor $f=\ell,s,c$, the original correlation function is reweighted from the simulated mass $m_f^{\rm sim}$ to $m_f^{\rm sim}+\delta m$. The observable is extracted both from the original correlator, yielding $\mathcal O$, and from the reweighted correlator, yielding $\mathcal O^{f\textrm{-rew}}$. Its derivative at the simulation point is then estimated as
\begin{equation}
\left.
\frac{d\mathcal O}{d m_f}
\right|_{\rm sea, sim}
\simeq
\frac{\mathcal O^{f\textrm{-rew}}-\mathcal O}
{\delta m},
\qquad f=\ell,s,c.
\label{eq:finite_difference_rew}
\end{equation}
For sufficiently small $\delta m$, this procedure is mathematically equivalent to the linearized reweighting approximation employed in Ref.~\cite{ExtendedTwistedMass:2024nyi}. We have explicitly verified the consistency of the two determinations for all sea-quark-mass derivatives.
We stress that the auxiliary shift $\delta m$ entering Eq.~\eqref{eq:finite_difference_rew} should not be confused with the mistuning of the simulation parameter, $m_f-m_f^{\rm sim}$, which enters the correction to the observable in Eq.~\eqref{eq:mist_correct_obs} and is determined by solving the mistuning system in Eq.~\eqref{eq:syst_A}.

\begin{figure}
    \centering
    \includegraphics[width=0.48\linewidth]{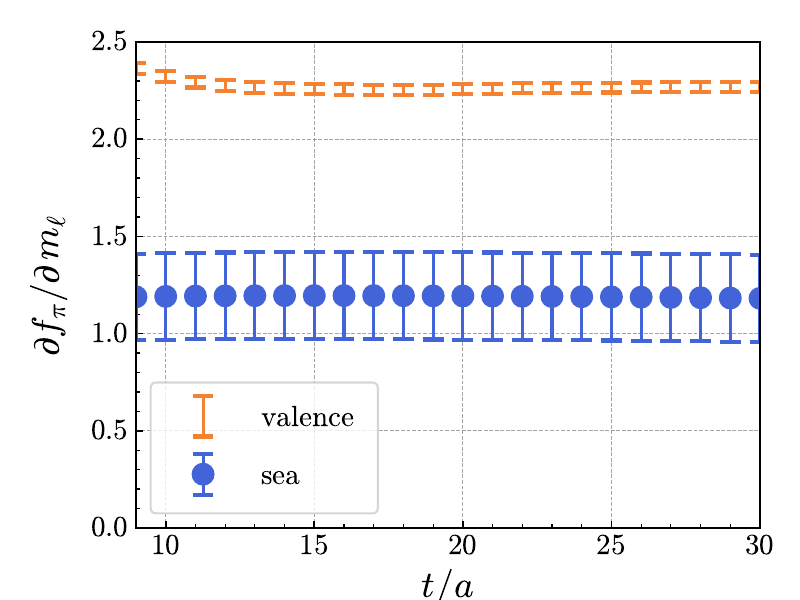}
    \includegraphics[width=0.48\linewidth]{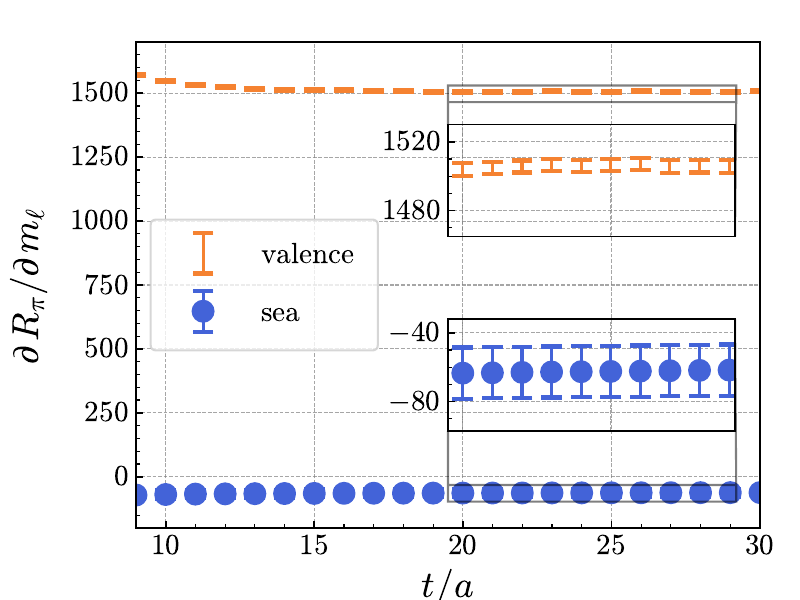}
    \\
    \includegraphics[width=0.48\linewidth]{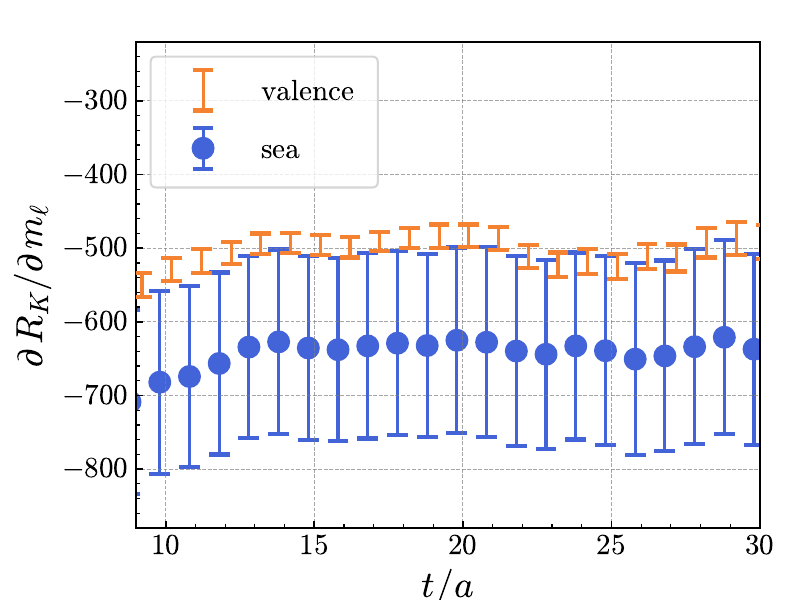}
    \includegraphics[width=0.48\linewidth]{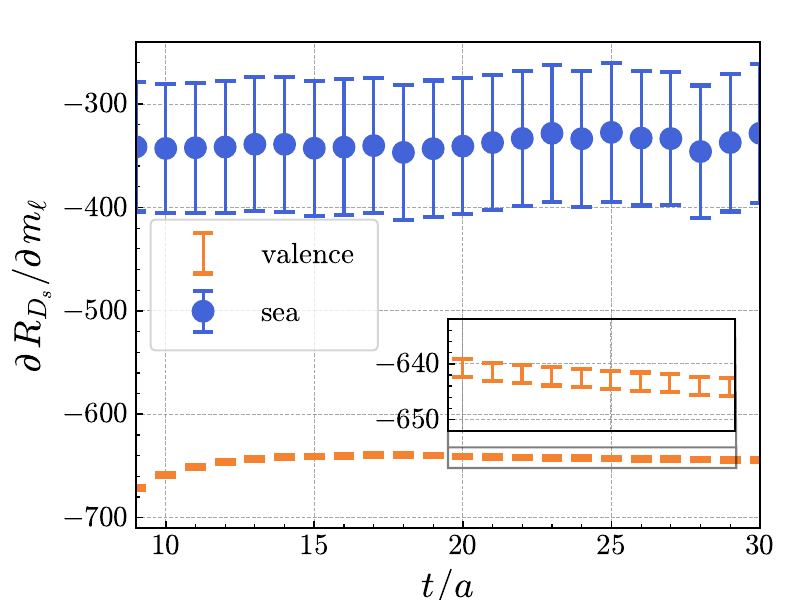}
    \caption{
    Comparison of the sea- and valence-$m_\ell$ derivatives of the renormalization inputs, namely $f_\pi$ ({\it upper left}), $R_\pi$ ({\it upper right}), $R_K$ ({\it lower left}) and $R_{D_s}$ ({\it lower right}), as computed on the B64 ensemble.
    For each observable, the derivative estimator is evaluated at each Euclidean time slice $t/a$.
    When the full $y$-axis range is much larger than the statistical uncertainties, zoomed insets are included to facilitate comparison between the precision of the two determinations. In the $R_K$
    panel, the sea and valence points are slightly offset along the horizontal axis for readability.
    }
    \label{fig:ml_derivatives}
\end{figure}

A significant statistical improvement of the present analysis is that the determinant reweighting factors are evaluated using the full available statistics.
In the light-quark sector, this allows the sea contribution to the derivative to be determined on an ensemble-by-ensemble basis through Eq.~\eqref{eq:finite_difference_rew}.
Although this contribution is considerably less precise than the corresponding valence contribution and therefore dominates the uncertainty of the total light-quark-mass derivative, its effect on the final mistuning correction remains well controlled, since it is multiplied by the small light-quark-mass mistuning, $m_\ell-m_\ell^{\rm sim}$.
The sea-$m_\ell$ contribution can therefore be included directly in the analysis without requiring any additional treatment. Fig.~\ref{fig:ml_derivatives} compares the sea- and valence-$m_\ell$ derivatives for all the renormalization inputs.

The determination is more delicate for derivatives with respect to the strange- and charm-quark sea masses. Although the corresponding mistunings amount to only percent-level corrections relative to the strange- and charm-quark masses, their absolute values exceed the light-quark mistuning by at least one order of magnitude.
Consequently, the statistical uncertainty of the derivative is amplified by a substantially larger factor when propagated to the final mistuning correction.

The statistical fluctuations of reweighted observables exhibit an unfavorable scaling with the lattice volume (see, e.g., Ref.~\cite{RC:2025zoa}), with uncertainties growing approximately as $\sqrt{L^3}$.
This adverse volume dependence is partially offset at finer lattice spacings by the reduced variance of the gauge-field fluctuations.
As a result, while the determinations obtained on the coarser reference-volume ensembles B64 and C80 are overwhelmed by statistical noise, those from D96 and E112 show a physically sensible behavior.
Their uncertainties, however, remain too large for a quantitatively useful determination.

\begin{table}[t]
    \centering
    \begin{tabular}{lcccccc}
        ensemble & $V/a^4$ & $a\,[{\rm fm}]$ & $L\,[{\rm fm}]$ & $M_\pi [{\rm MeV}]$ & ~ sea-$(am_\ell)\times10^3$ & val-$(am_\ell)\times10^3$\\ \midrule
        B24 & ~ $24^3\times48$ ~ & ~ $\sim0.080$ ~ & ~ $\sim1.92$ ~ & ~ $\sim270$ ~ & ~ 2.5 ~ & ~ $[2.5,\, 4.2,\, 5.9]$ \\ 
        B32 & ~ $32^3\times64$ ~ & ~ $\sim0.080$ ~ & ~ $\sim2.56$ ~ & ~ $\sim260$ ~ & ~ 2.5 ~ & ~ $[1.8,\,2.5,\, 4.2,\, 5.9]$ \\[4pt]
        C48 & ~ $48^3\times96$ ~ & ~ $\sim0.068$ ~ & ~ $\sim3.26$ ~ & ~ $\sim250$ ~ & ~ 2.0 ~ & ~ $[2.0,\, 3.4,\, 4.8]$ \\ [4pt]
        \hline
    \end{tabular}
    \caption{Summary of the small-volume ensembles used to determine the sea-quark-mass derivatives with respect to $m_s$ and $m_c$. For each ensemble, measurements are carried out at several values of the valence-light-quark mass, listed in the last column. The lattice spacings at the corresponding values of $\beta$ are determined from the reference-volume ensembles (B64 and C80) in the FLAG scheme. Since these auxiliary ensembles are used only to obtain a reliable estimate of the derivatives, rather than for a high-precision determination of the observables, the limited precision of the scale setting is negligible compared with their statistical uncertainties. The values of the dimensional quantities reported in the table may therefore be regarded as having an uncertainty of approximately $1\%$.}
    \label{tab:small_volume_ens}
\end{table}

To obtain sufficiently precise strange- and charm-quark-mistuning corrections, we therefore determine the corresponding derivatives on the dedicated set of smaller-volume ensembles summarized in Table~\ref{tab:small_volume_ens}, rather than on the reference ensembles listed in Table~\ref{tab:iso_def}.
The derivatives are evaluated at several values of the valence-light-quark mass, reported in the last column of Table~\ref{tab:small_volume_ens}, while keeping the sea parameters fixed. Any dependence on $m_\ell$ is then used to extrapolate the results to the physical light-quark mass.

To compare derivatives obtained at different lattice spacings and volumes, we consider the dimensionless combination
\begin{equation}
\frac{m_\ell}{\mathcal O}
\frac{d \mathcal O}{d m_h^{\rm sea}}
,
\qquad h=s,c,
\end{equation}
where $m_\ell$ is the $\beta$-dependent value of the physical light-quark mass determined on the reference-volume ensembles, listed in Table~\ref{tab:iso_def}. This combination is independent of the quark-mass renormalization factor, since $m_\ell$ and the derivative with respect to $m_h^{\rm sea}$ renormalize with inverse factors, and can therefore be compared directly across ensembles. Any residual differences arise from finite-volume effects, lattice artifacts, or statistical fluctuations.

The analysis strategy described below is applied independently to each renormalization input. For brevity, however,
we illustrate it for the case of the derivatives of $f_\pi$ with respect to sea strange- and charm-quark mass, which is
representative of the general situation and numerically important both in tuning the ratios $R_{\pi,K,D_s}$ to the
target values defining isospin-symmetric QCD and in determining the lattice spacing that enters the analytic kernel for
$a_\mu^{\rm HVP}$. 
The analysis of the sea-$m_s$ and sea-$m_c$ derivatives of $R_{\pi,K,D_s}$ follows closely the one outlined below for the same derivatives of $f_\pi$ and displays very analogous features.\footnote{
In particular,  in the relative derivatives $[R_{\pi,K,D_s}]^{-1}[d R_{\pi,K,D_s} / d m_{s,c}^{\rm sea}]$ the term $- [1/f_\pi] [d f_\pi/ d m_{s,c}^{\rm sea} ] $ turns out numerically dominant with respect to
the term  $ [1/m_{\pi,K,D_s}] [d m_{\pi,K,D_s}/ d m_{s,c}^{\rm sea} ] $.}

\begin{figure}
    \centering
    \includegraphics[width=\linewidth]{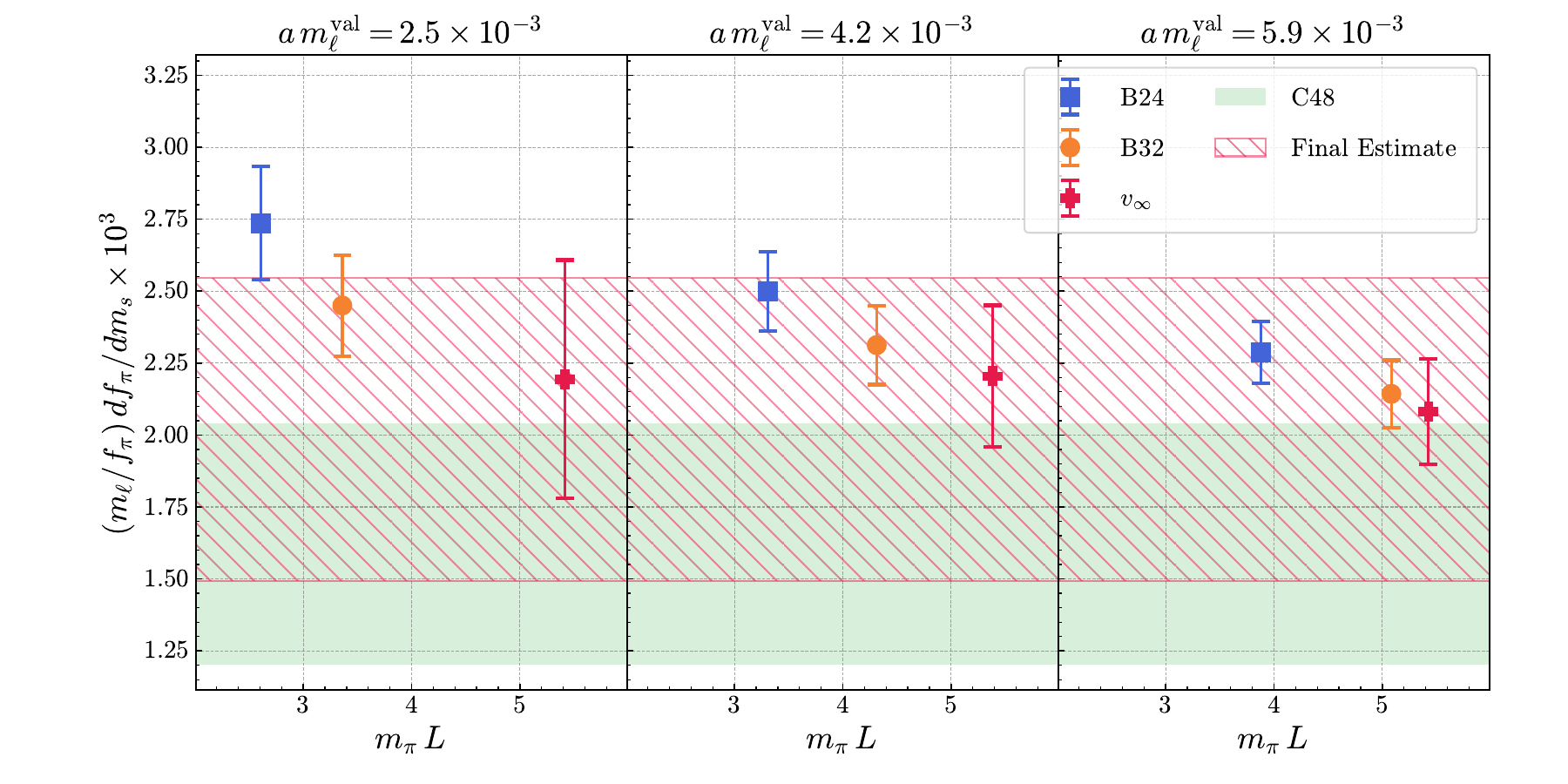}
    \caption{
    Determinations of
    $(m_\ell/f_\pi)\,df_\pi/dm_s$
    at different values of the valence-light-quark mass.
    At fixed valence mass, the B24 and B32 results are fitted using
    $v_\infty+v_1\,e^{-m_\pi L}$, where $v_\infty$ denotes the
    infinite-volume estimate.
    The uncertainty of the final estimate is enlarged so that the corresponding interval also includes the C48 central value.
    }
    \label{fig:fpi_sea-ms_der_small}
\end{figure}

Fig.~\ref{fig:fpi_sea-ms_der_small} shows the normalized derivative of $f_\pi$ with respect to the sea strange-quark mass. The B24 and B32 determinations exhibit a dependence on the spatial volume. At each value of the valence-light-quark mass, this dependence is described using the ansatz
\begin{equation}
v_\infty+v_1\,e^{-m_\pi L},
\end{equation}
where $v_\infty$ denotes the corresponding infinite-volume value. The extrapolated values show no statistically significant residual dependence on the valence-light-quark mass and are therefore combined through a constant fit.
Likewise, no significant valence-mass dependence is resolved on C48, whose determinations are also described by a constant fit.

The infinite-volume result obtained from the B ensembles is broadly compatible with the C48 determination, although a mild residual discrepancy is observed.
We retain the B-ensembles result as the central value and conservatively enlarge its uncertainty so that the resulting interval also includes the C48 central value.
The resulting estimate is then used in the mistuning system.

\begin{figure}
    \centering
    \includegraphics[width=\linewidth]{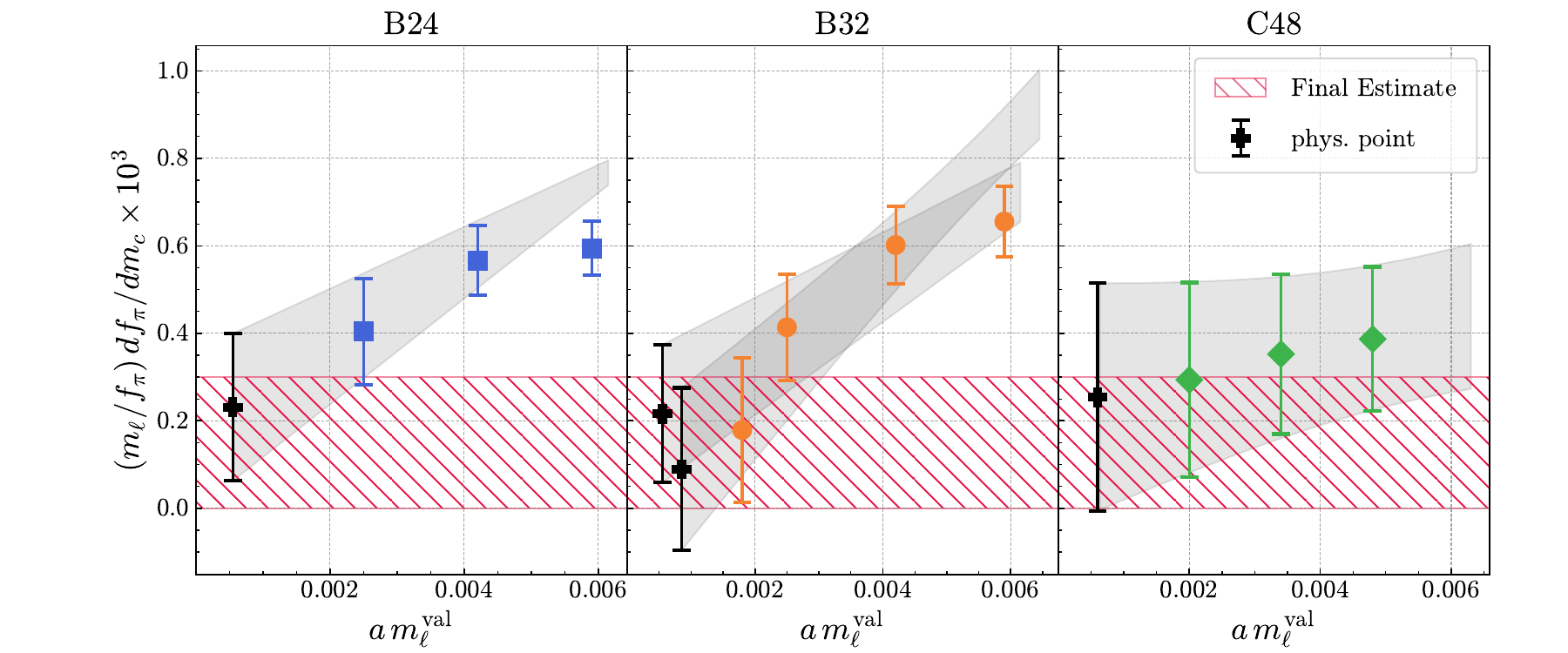}
    \caption{
    Determinations of
    $(m_\ell/f_\pi)\,df_\pi/dm_c$
    on the B24, B32, and C48 auxiliary ensembles as functions of the valence-light-quark mass, together with the corresponding extrapolations to the
    physical point and the final estimate.}
    \label{fig:fpi_sea-mc_der_small}
\end{figure}

Fig.~\ref{fig:fpi_sea-mc_der_small} shows the valence-light-quark-mass dependence of the normalized derivative of $f_\pi$ with respect to the sea charm-quark mass on the three auxiliary ensembles.
The B32 data indicate that a linear description is not adequate over the full range of valence masses. We therefore restrict the physical-point extrapolation to the three lightest valence-mass points. Following the same local-extrapolation strategy, the B24 result is obtained from a linear fit to its last two valence-mass points.
The physical-point estimates obtained from the three auxiliary
ensembles are mutually compatible, and no statistically significant
dependence on either the lattice spacing or the spatial volume can be
resolved at the present level of precision.
We therefore do not interpret the observed
signal as a resolved physical determination. Instead, we
retain its magnitude as a conservative upper estimate of the
correction induced by a mistuning of the sea-charm mass.

\begin{figure}
    \centering
    \includegraphics[width=0.48\linewidth]{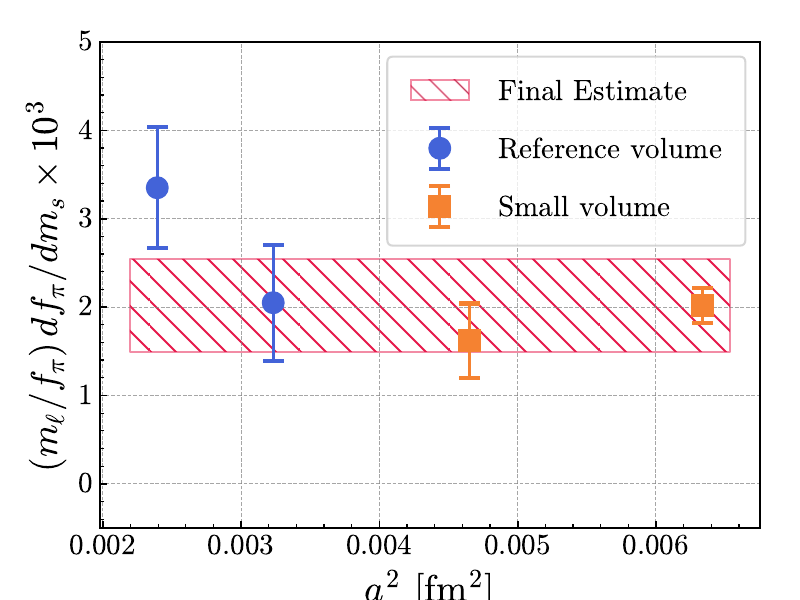}
    \includegraphics[width=0.48\linewidth]{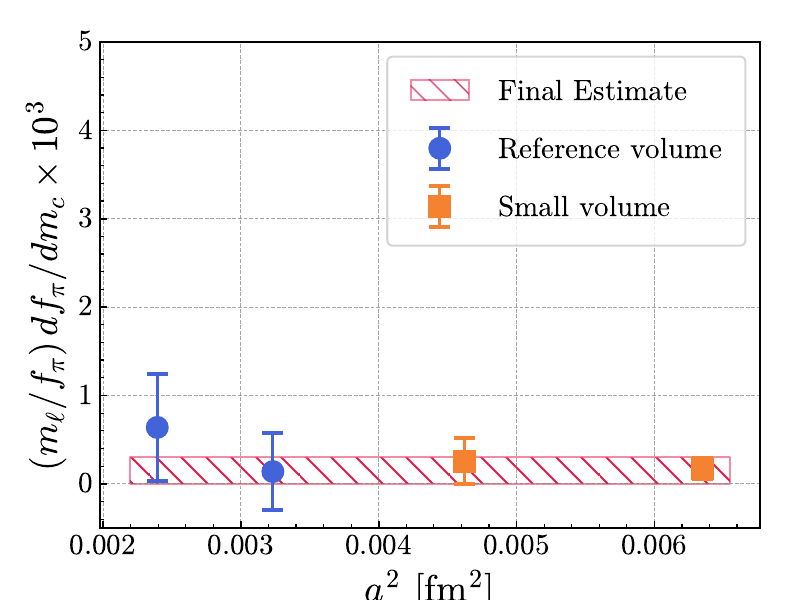}
    \caption{Comparison between the normalized sea-quark-mass derivatives of $f_\pi$ obtained from the auxiliary ensembles and the corresponding direct determinations on the reference-volume ensembles for the sea-$m_s$ ({\it left}) and sea-$m_c$ ({\it right}) derivatives.
    The bands indicate the final estimates used to solve the mistuning system.
    }
    \label{fig:fpi_sea-ders_small-large}
\end{figure}

Fig.~\ref{fig:fpi_sea-ders_small-large} compares, for $f_\pi$, the final estimates obtained from the auxiliary ensembles with the direct determinations on the reference-volume ensembles.

To summarize, the sea-quark-mass derivatives entering the mistuning analysis are treated according to three distinct prescriptions, applied independently to each renormalization input. The sea-$m_\ell$ contribution is determined directly, ensemble by ensemble, from the reference ensembles through Eq.~\eqref{eq:finite_difference_rew} and is included without any additional treatment, since the corresponding mistuning is small. For the sea-$m_s$ derivatives, the central values are obtained from constant fits to the infinite-volume extrapolations determined from B24 and B32. Their uncertainties are conservatively enlarged so that the final intervals also include the corresponding C48 central values, thereby accounting for possible residual cutoff effects.
For the sea-$m_c$ derivatives, heavy-quark decoupling implies a
parametrically suppressed dependence on the charm-quark mass.
We therefore use the auxiliary-ensemble determinations to define
conservative intervals that include zero and all available
small-volume results.

\subsection{Determination of $w_0$ in the FLAG scheme}
\label{sec:w0_FLAG}
In this section we provide the details of the determinations of the gradient-flow scale $w_0$ 
\cite{Luscher:2010iy,BMW:2012hcm}. We calculate $w_0$ and $\sqrt{t_0}$ using
 the standard Wilson action for the gradient-flow
evolution and the symmetrized discretization of the action density as described in \cite{Luscher:2010iy}. 
For each lattice spacing and ensemble the measured value of $w_0$ are quoted in Table~\ref{tab:w0_values}.

\begin{table}[t]
\begin{tabular}{lcc}
ensemble & $w_0/a$ & $\sqrt{t_0}/a$  \\
\hline 
B64    & $2.1261(17)$    & ~ $1.80215(82)$ ~  \\
B96    & $2.1289(08)$    & ~ $1.80355(40)$ ~  \\[4pt]
C80    & $2.5042(15)$    & ~ $2.10922(74)$ ~  \\
C112   & $2.5079(10)$    & ~ $2.11048(46)$ ~  \\[4pt]
D96    & $3.0152(13)$    & ~ $2.51798(52)$ ~  \\
D128   & $3.0162(09)$    & ~ $2.51811(34)$ ~  \\[4pt]
E112   & $3.5064(17)$    & ~ $2.91911(56)$ ~  \\
\hline
\end{tabular}
\caption{Values of $w_0/a$ and $\sqrt{t_0}/a$ measured on the ETMC ensembles considered for this analysis.}
\label{tab:w0_values}
\end{table}

\begin{figure}
    \centering
    \includegraphics[width=0.8\linewidth]{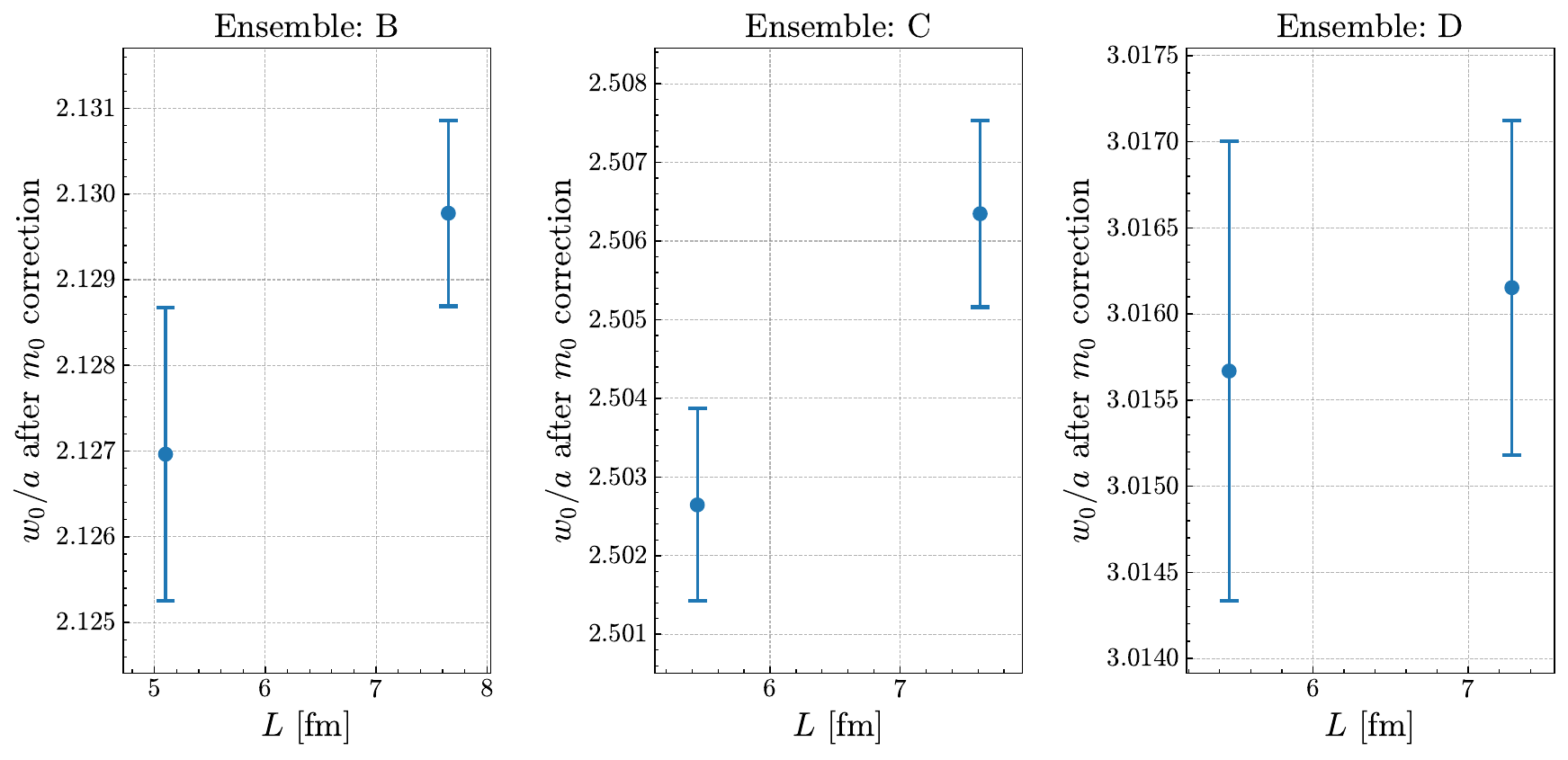}
    \caption{Values of $w_0/a$ computed for the ensembles B64 and B96 (left panel), C80 and C112 (middle panel) and for D96 and D128 (right panel). The data points are corrected for the $m_0$ mistuning.}
    \label{fig:w0_volume}
\end{figure}

The simulated points are first corrected for \(m_{0}\) mistuning adding $(m_0-m_0^{\rm sim})\frac{d w_0/a}{d m_0}\big|_{\rm sim}$. 
In order to quantify the FSE on $w_0$ we consider simulations at different volumes.
In particular in Table~\ref{tab:w0_values} the B64 and the B96 have the same value of the lattice spacing
but different physical volumes, $L\sim5.1~$fm and $L'\sim7.6~$fm. Similarly, the ensembles C80 and C112 have their corresponding linear sizes are $L\sim5.4~$fm and $L'\sim7.6~$fm. Lastly, D96 and D128 have volumes of $L\sim5.4~$fm and $L'\sim7.2~$fm respectively.
When two volumes are available, we compute their weighted average and employ a pull variable to inflate the error according to the formula
\begin{flalign}
\Delta_{\mathrm{FSE}}(w_0)=\max_{a} \left\{ \left|w_0(L)-w_0(L') \right| \mathrm{erf}\left( \frac{\left|w_0(L)-w_0(L') \right|}{\sqrt{2}\sqrt{\Delta_{w_0}(L)^2+\Delta_{w_0}(L')^2}}\right)\right\}\,,
\label{eq:Delta_FVE}
\end{flalign}
where the maximum is computed over the lattice spacings. $\Delta_{w_0}(L)$ is the statistical error of $w_0(L)$ and 
$\mathrm{erf}$ is the error function.
the difference of $w_0/a$ between the different volumes is plotted in Fig.~\ref{fig:w0_volume}.

\begin{figure}
    \centering
    \includegraphics[width=\linewidth]{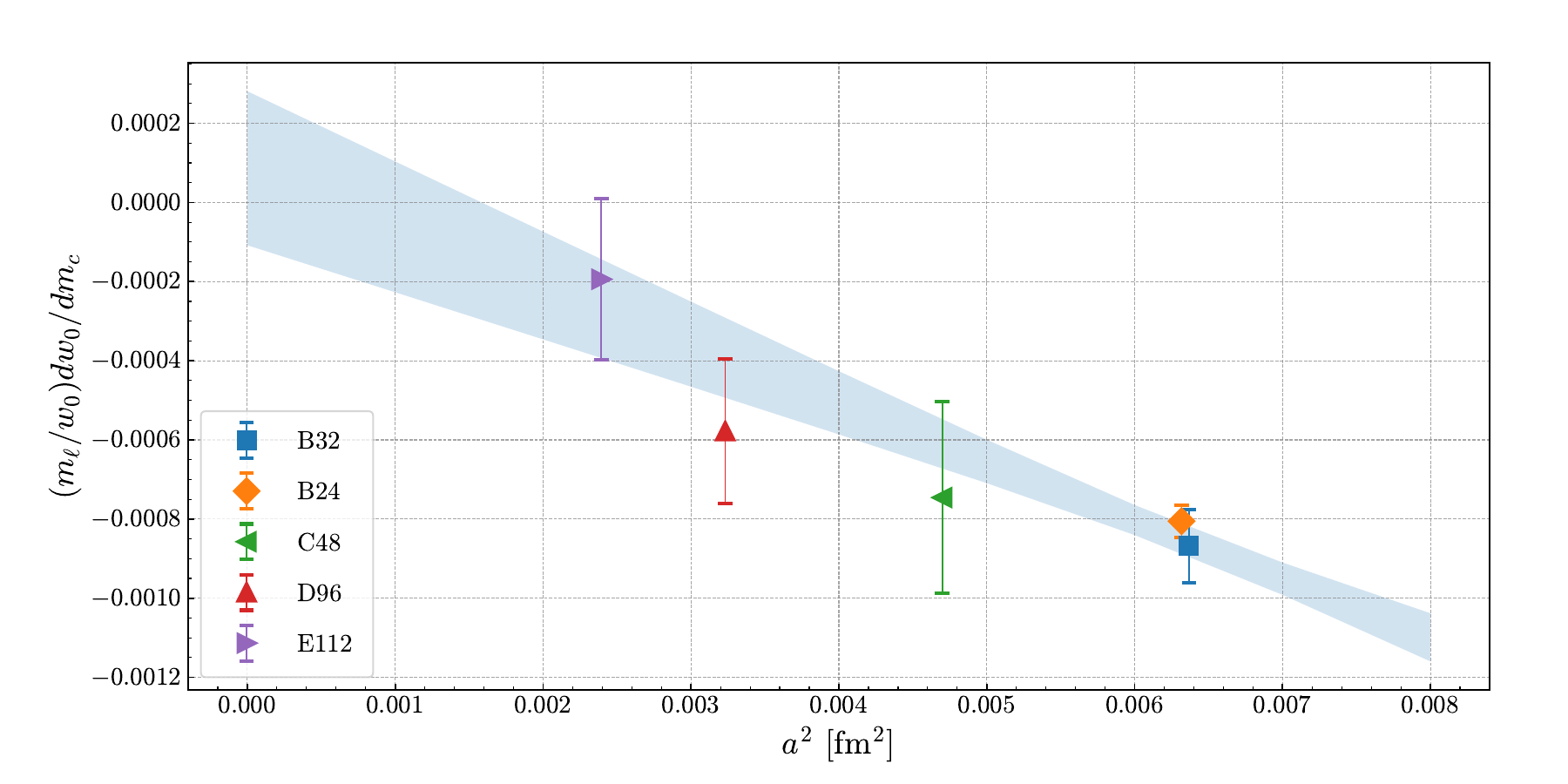}
    \caption{Continuum extrapolation of the normalized derivative \(\frac{\mu_\ell}{w_0}\frac{dw_0}{dm_c}\) as a function of the squared lattice spacing \(a^{2}\). The colored points represent data from various ensembles, and the shaded band indicates the linear fit used to extrapolate to $a=0$.}
    \label{fig:w0_charm_deriv}
\end{figure}

To conclude the process, we correct for the quark-mass mistuning.
The derivative of $w_0$ with respect to the charm-quark mass requires particular attention, as it suffers from substantial lattice artifacts and can produce significant shifts at coarse lattice spacings.
Because the error in the derivative estimate with respect to the charm mass grows both with the lattice spacing (due to increasing gauge field fluctuations) and with the volume (at fixed lattice spacing), on the two coarsest lattice
spacings we also computed the derivative of \(w_{0}\) with respect to the charm-quark mass using our smaller-volume ensembles (B32, B24, and C48). Indeed the 
results for $d w_0 / d m_c$  on the ensembles B64 and C80,
owing to their very large statistical error, add no significant info and are hence
excluded from our analysis.
Fig.~\ref{fig:w0_charm_deriv} shows the derivative $d w_0 / d m_c$ normalized by \(\mu_\ell/w_0\) to obtain a dimensionless quantity with a well-defined continuum limit. We perform a linear fit of \(\frac{\mu_\ell}{w_0}\frac{dw_0}{dm_c}\) to extrapolate our results to the continuum limit.

To estimate \(\frac{dw_{0}}{dm_{c}}\) at each lattice spacing, we take our best fit result evaluated at a given lattice spacing and we divide it by the corresponding prefactor \(\frac{\mu _{\ell }}{w_{0}}\).

\begin{figure}
    \centering
    \includegraphics[width=\linewidth]{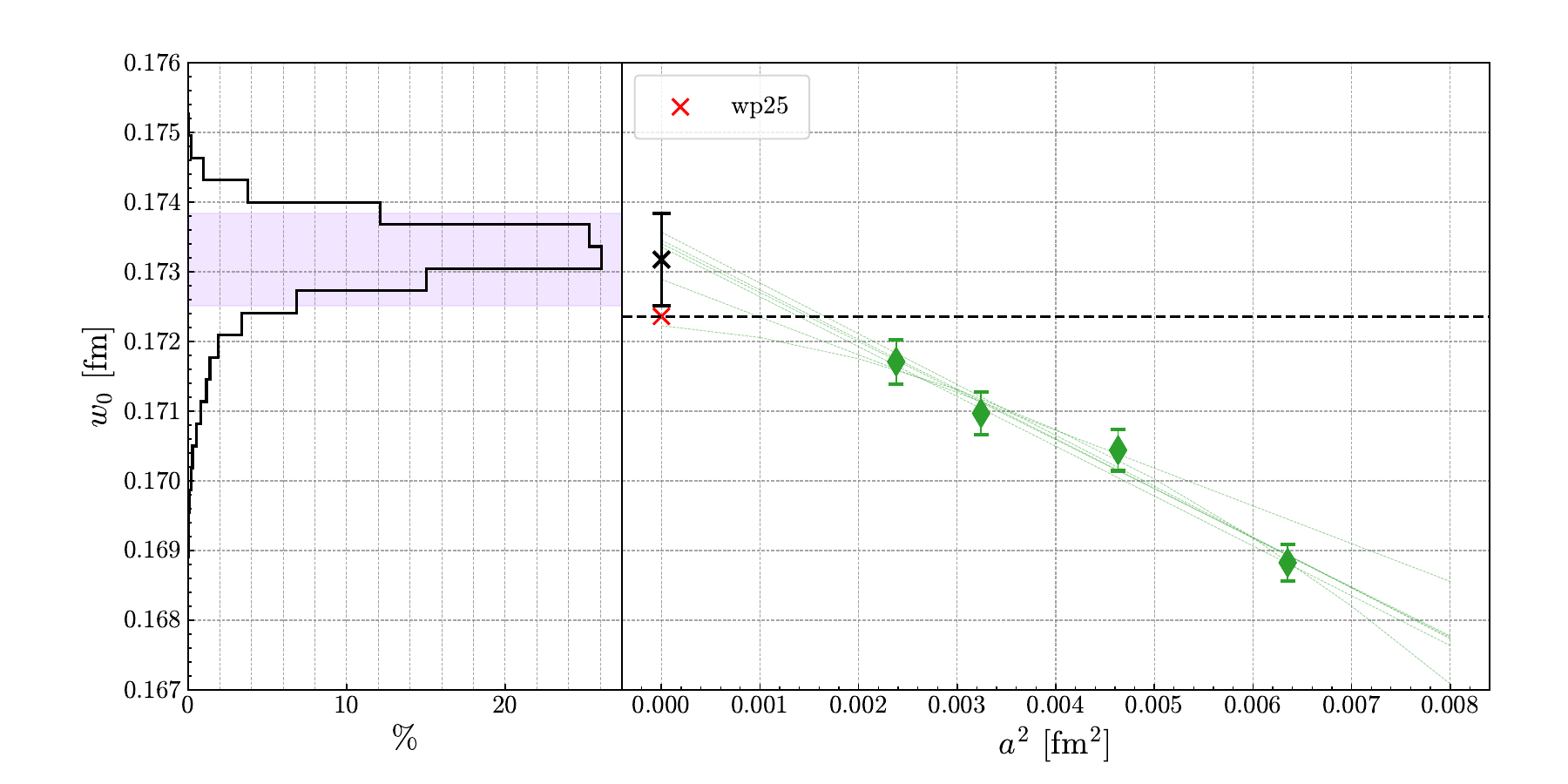}
    \caption{Continuum extrapolation of $w_0$ in the FLAG scheme. The data are corrected for the $m_0$ and the quark-masses mistuning and averaged between different volumes. The dotted lines represent the various fits and the
    BAIC average of the continuum values is plotted as a black cross and on the left we show the histogram of the bootstrap sample for the continuum values weighted according to the BAIC.}
    \label{fig:w0_cont}
\end{figure}

To determine the continuum limit of $w_0$ within the FLAG scheme, we employ the following fit ansatz:
\begin{itemize}
    \item Linear in $a^2$;
    \item Linear in $a^2$ with sequential exclusion of one lattice spacing result at a time;
    \item Linear in $a^2$ with an additional $a^4$ term.
\end{itemize}
The resulting fits are illustrated in Fig.~\ref{fig:w0_cont}. Their BAIC average yields
\begin{equation}
    w_0^{\rm FLAG} =0.17318(50)_{\rm stat}(44)_{\rm syst}[67] ~\rm{fm}\,.
    \label{eq:w0FLAG_val}
\end{equation}
This result for $w_0^{\rm FLAG}$ is in agreement with other lattice determinations, see, e.g., the recent determination by BMW/DMZ~\cite{Boccaletti:2024guq}, and is  slightly larger than the value of $w_0$ defining the WP25 scheme of isospin-symmetric QCD reported in Eq.~\eqref{eq:w0WP25_def}.
Furthermore, the value obtained here agrees with the previous ETMC result in~\cite{ExtendedTwistedMass:2021qui}; while the total quoted error remains unchanged, the analysis presented here provides a much more robust estimate of systematic uncertainties. This improvement comes from the addition of two finer lattice spacings and the treatment of the sea-bare-parameter mistuning.

\section{Details on renormalization in WP25 scheme}
\label{sec:fpi_WP25}
To correct our simulations in order to match the WP25 scheme we replace the 
dimensionless quantities of Eq.~\ref{eq:R-FLAG} with
\begin{align}
P^{\rm WP25}_\pi =
\left[{m_{\pi}^{2}}{w_0^{2}}\right]^{\rm WP25},
\quad
P^{\rm WP25}_K =
\left[{m_{K}^{2}}{w_0^{2}}\right]^{\rm WP25},
\quad
P^{\rm WP25}_{D_s} =
\left[{m_{D_{s}}}{w_0}\right]^{\rm WP25}
\label{eq:R-WP25}
\end{align}
We denote by $P^{\rm sim}_\mathcal{O}$ ($\mathcal{O}=\pi,\,K,\,D_s$) the corresponding products obtained on the simulated ensembles, at the simulated values of the quark masses and gauge coupling $\beta$. 
The value of $w_0$ defining the WP25 scheme of iso-QCD is
\begin{equation}
    w_0^{\rm WP25}=0.17236~\text{fm}\,.
    \label{eq:w0WP25_def}
\end{equation}

Our simulation parameters are actually closer to the FLAG scheme than to the WP25 scheme. In particular, dealing with a larger charm-quark-mass mistuning would induce a larger error on $a_\mu^{\rm HVP}$. Here we explain how we can substantially alleviate this problem by implementing a smooth O($a^2$) lattice deformation of the WP25 scheme, which is labelled as WP25' as long as $a>0$ and goes over to the WP25 scheme as $a \to 0$, and discuss in this scheme our determination of $f_\pi w_0$, yielding in the continuum limit by construction the value of WP25 scheme value of $f_\pi$ in physical units.

\subsection{A convenient implementation of the WP25 scheme}

As anticipated, we employ a smoothly O($a^2$)-modified version of the WP25 scheme (and use the label WP25' for the corresponding quantities at $a>0$) that is obtained by replacing the third input in Eq.~(\ref{eq:R-WP25}) with
\begin{align}
P^{\rm WP25'}_{D_s} =
\left[{m_{D_{s}}}{w_0}\right]^{\rm WP25}+Ca^2
\label{eq:P-WP25'}\; ,
\end{align}
where $C$ is a free parameter to be chosen as discussed below 
and $a$ is the lattice spacing.
In analogy with Eq.~\ref{eq:syst_A} we solve the following system of equations

\begin{align}
\label{eq:syst_AWP25}
\vec{P}^{\rm sim} + (m_{0}-m_{0}^{\rm sim})\frac{d\vec{P}}{dm_{0}}\bigg|_{\rm sim}
+ \sum_{f=\ell,s,c}(m_{f}-m_{f}^{\rm sim})\frac{d\vec{P}}{dm_{f}}\bigg|_{\rm sim}
=\vec{P}^{\rm WP25'}~,
\end{align}
together with Eq.~(\ref{eq:syst_B}) and the equation for the lattice spacing
\begin{align}
\label{eq:syst_CWP25}
a^{\rm WP25'} = {w_{0}^{\rm WP25}}
\bigg[
(w_0/a)^{\rm sim}
+ (m_{0}-m_{0}^{\rm sim})\frac{ d(w_0/a)}{dm_{0}}\bigg|_{\rm sim}
+ \sum_{f}(m_{f}-m_{f}^{\rm sim})
\frac{d(w_0/a)}{dm_{f}}\bigg|_{\rm sim}
\bigg]^{-1}~.
\end{align}
The above system is solved for the four unknowns $\{a,am_{0},am_{\ell},am_{s},am_{c}\}$.
The derivatives with respect to the quark masses and critical mass are computed as discussed in Appendix~\ref{sec:scale_setting_details}. %

The value of the coefficient $C$ in Eq.(\ref{eq:P-WP25'}) is to be chosen such that the mistuning of the simulated charm-quark-mass value with respect to WP25' scheme is much smaller than in the $C=0$ case without introducing any significantly large
lattice artifact in the continuum-limit extrapolation of $f_\pi w_0$. We choose here $C=-5~{\rm fm}^{-2}$, which we found to be a good compromise. 
The value $C=-5~{\rm fm}^{-2}$ shifts the values of $M_{D_s}$ by $\sim-37$~MeV at the coarsest lattice spacing and by $\sim-14$~MeV at the finest lattice spacing.

\begin{figure}
    \centering
    \includegraphics[width=\linewidth]{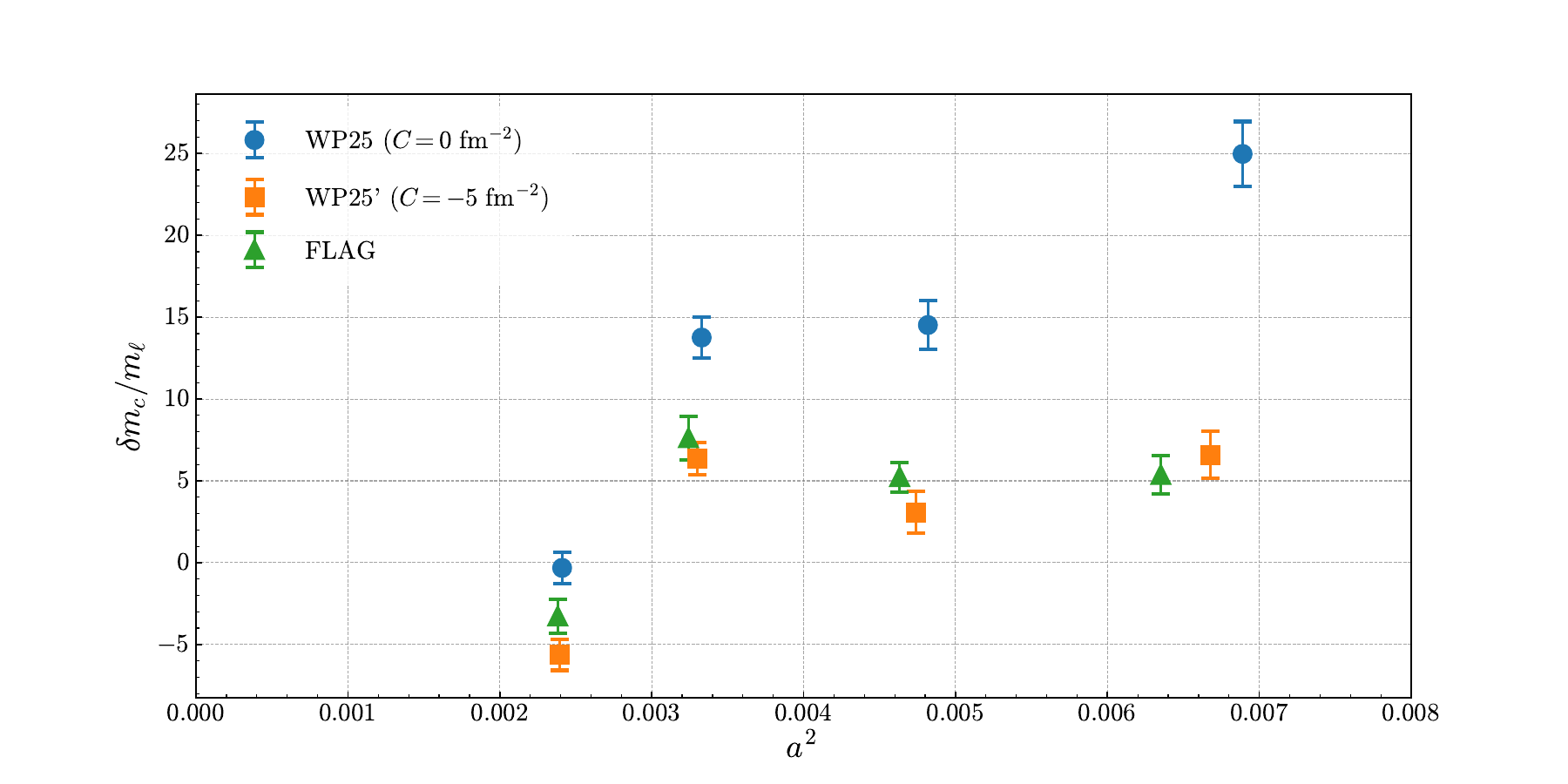}
    \caption{Values of $\delta m_c = m_c-m_c^{sim}$ divided by $m_\ell$ using the FLAG, WP25 and WP25' iso-QCD schemes 
    (see legend) at finite lattice spacing.
    }
    \label{fig:Cvalues}
\end{figure}

The effect of varying $C$ on the charm-quark-mass mistuning is illustrated in Fig.~\ref{fig:Cvalues}: we observe that the value $C=-5~{\rm fm}^{-2}$~ makes the size of the mistuning of the charm mass globally similar to the ones in the FLAG scheme.
The resulting values of the lattice spacings and quark masses, obtained by solving the system of equations above for $C=-5~{\rm fm}^{-2} $, are reported in 
Table~\ref{tab:iso_WP25_values}.

\begin{table}[t]
\begin{tabular}{lccccccc}
ensemble & $V/a^4$ & $a~[\rm fm]$ & $L~[\rm fm]$ &  $am_{\ell}$ & $am_{s}$ & $am_{c}$ & $am_0$  \\
\hline 
\multicolumn{8}{c}{\rule[-0.9em]{0pt}{2.2em} WP25' scheme} \\
\hline \\
B64    & $64^3\times 128$ & ~ $0.08221(18)$ ~  & ~ 5.26 ~ &~ 0.0007110(39)  ~ & ~0.01951(13)~   &  ~0.23558(78) ~ & ~ -0.4138934(46)	~ \\
C80    & $80^3\times 160$    & ~ $0.069030(87)$ ~   & 5.52 &~ 0.0006009(16) ~   &  ~0.016469(66)~  &  ~0.19760(40)	~  & ~ -0.3964534(41)  ~ \\[4pt]
D96    & $96^3\times 192$    & ~ $0.05744(11)$ ~  & 5.51 &~ 	0.0005047(24) ~    &  ~0.013857(62)~  &  ~0.16445(49)	~  & ~ -0.3761252(39) ~  \\[4pt]
E112    & $112^3\times 224$    & ~ $0.048992(61)$ ~  & 5.49 &~ 	0.0004323(13) ~  &  ~0.011812(35)~  &   ~0.14037(29)	~ & ~  -0.3613136(75) ~  \\[8pt]
\hline
\end{tabular}
\caption{ETMC gauge ensembles used to compute $a_{\mu}^{\rm HVP}(I=1)$ and $a_{\mu}^{\rm HVP}(I=0)$ and the values of the lattice spacings and of the bare quark masses fine-tuned in order to match the WP25' definition of isospin-symmetric QCD. The fine tuning is achieved by starting from the simulated gauge ensembles and taking into account the reweighting factors needed to correct the small mistunings of the simulated bare parameters, as discussed in the text.
\label{tab:iso_WP25_values}}
\end{table}

\subsection{Determination of $f_\pi$ in the WP25 scheme}

In the WP25' lattice setup the pion decay constant is an obervable and can be computed and extrapolated to the continuum limit, which corresponds to the WP25 scheme. To do so we consider three definitions of $f_\pi$  that differ only by lattice artifacts. 
The first definition is computed from the  pion correlator given by
 \be
    C_{P_5P_5}^{{\rm TM}}(t) = \frac{1}{L^3} \sum\limits_{x, z} \left\langle 0 \right| P_5^{{\rm TM}} (x) (P_5^{{\rm TM}})^\dag (z) \left| 0 \right\rangle \delta_{t, (t_x  - t_z )} ~ ,
    \label{eq:P5P5}
 \ee
where $P_5^{{\rm TM}} (x) = \overline{q}_\ell(x) \gamma_5 q_\ell(x)$ is a local interpolating pion field with
the Wilson parameters of the two mass-degenerate valence quarks are chosen to have opposite values. 

At large time distances one has
 \be
    C_{P_5P_5}^{{\rm TM}}(t)_{ ~ \overrightarrow{t  \gg a, ~ (T - t) \gg a} ~ } \frac{(\mathcal{Z}_{P_5}^{{\rm TM}})^2}{2m_\pi} \left[ e^{ - m_\pi  t}  + e^{ - m_\pi (T - t)} \right] ~ ,
    \label{eq:larget}
 \ee
so that the pion mass $m_\pi$ and the matrix element $\mathcal{Z}_{P_5}^{{\rm TM}} =  \langle\pi | \overline{q}_\ell \gamma_5 q_\ell | 0 \rangle$ can be extracted from the exponential fit given in the right-hand side of Eq.~\eqref{eq:larget}. 

For maximally twisted fermions the value of $\mathcal{Z}_{P_5}^{{\rm TM}}$ is directly connected to the pion decay constant $f_\pi$ with a Ward-Takahashi-identity~\cite{Frezzotti:2000nk,Frezzotti:2003ni}, namely
 \be
    af_\pi^{{\rm WTI}} = 2 a \mu_\ell \frac{{a^2 \mathcal{Z}_{P_5}^{{\rm TM}} }}{am_\pi ~ \mbox{sinh}(am_\pi)} ~ .
    \label{eq:decayPS}
 \ee
Other definitions of $f_\pi$ can be constructed from the correlator
\be
    C_{A_0P_5}^{{\rm TM}({\rm {\rm OS}})}(t) = \frac{1}{L^3} \sum\limits_{x, z} \left\langle 0 \right| A_0^{{\rm TM}({\rm {\rm OS}})} (x) (P_5^{{\rm TM}({\rm OS})})^\dag (z) \left| 0 \right\rangle \delta_{t, (t_x  - t_z )} ~ ,
    \label{eq:A0P5}
 \ee
where $A_0^{{\rm TM}({\rm OS})} (x) = \overline{q}_\ell(x) \gamma_0\gamma_5 q_\ell(x)$ with the Wilson parameters of the two
valence quarks that have opposite values for the ${\rm TM}$ case and equal value for the ${\rm OS}$ case.
At large time distances one has
\begin{align}
    C_{A_0P_5}^{{\rm TM}({\rm OS})}(t)_{ ~ \overrightarrow{t  \gg a, ~ (T - t) \gg a} ~ } \frac{\mathcal{Z}_{A_0}^{{\rm TM}({\rm OS})}\mathcal{Z}_{P_5}^{{\rm TM}({\rm OS})}}{2m_\pi} \left[ e^{ - m_\pi  t}  - e^{ - m_\pi (T - t)} \right] ~ .
    \label{eq:larget_a0p5}
\end{align}
The matrix element  $\mathcal{Z}_{A_0}^{{\rm TM}({\rm OS})} =  \langle\pi | \overline{q}_\ell \gamma_0\gamma_5 q_\ell | 0 \rangle$ is related to  $f_\pi$ as
\begin{align}
 f_\pi^{{\rm TM}} =Z_V^{\rm RI}\frac{\mathcal{Z}_{A_0}^{{\rm TM}}}{m_\pi}\,, \quad
 f_\pi^{{\rm OS}} =Z_A^{\rm RI}\frac{\mathcal{Z}_{A_0}^{{\rm OS}}}{m_\pi}\,, 
\end{align}
where $Z_V^{\rm RI}$ and $Z_A^{\rm RI}$ are the renormalization constants computed in the RI/MOM scheme~\cite{RCinprep}. The values that we have employed in the present analysis are quoted in Table~\ref{tab:ZV_ZA_RI}.

\begin{table}[t]
    \centering
    \begin{tabular}{c|c|c}
        ensemble & $Z_V^{\rm RI}$ & $Z_A^{\rm RI}$ \\[3pt]
        \midrule 
        B64      & ~0.72316(33) ~  & ~0.76263(23) \\[3pt]
        C80      & ~0.73744(36) ~  & ~0.77363(23) \\[3pt]
        D96      & ~0.75217(28) ~  & ~0.78488(25) \\[3pt]
        E112     & ~0.76392(26) ~  & ~0.79419(22) \\[3pt]
        \hline
    \end{tabular}
    \caption{Values of the renormalization constants $Z_V^{\rm RI}$ and  $Z_A^{\rm RI}$ computed in the RI/MOM scheme for each ensemble.}
    \label{tab:ZV_ZA_RI}
\end{table}

\begin{figure}[t]
\centering
\includegraphics[width=0.8\textwidth]{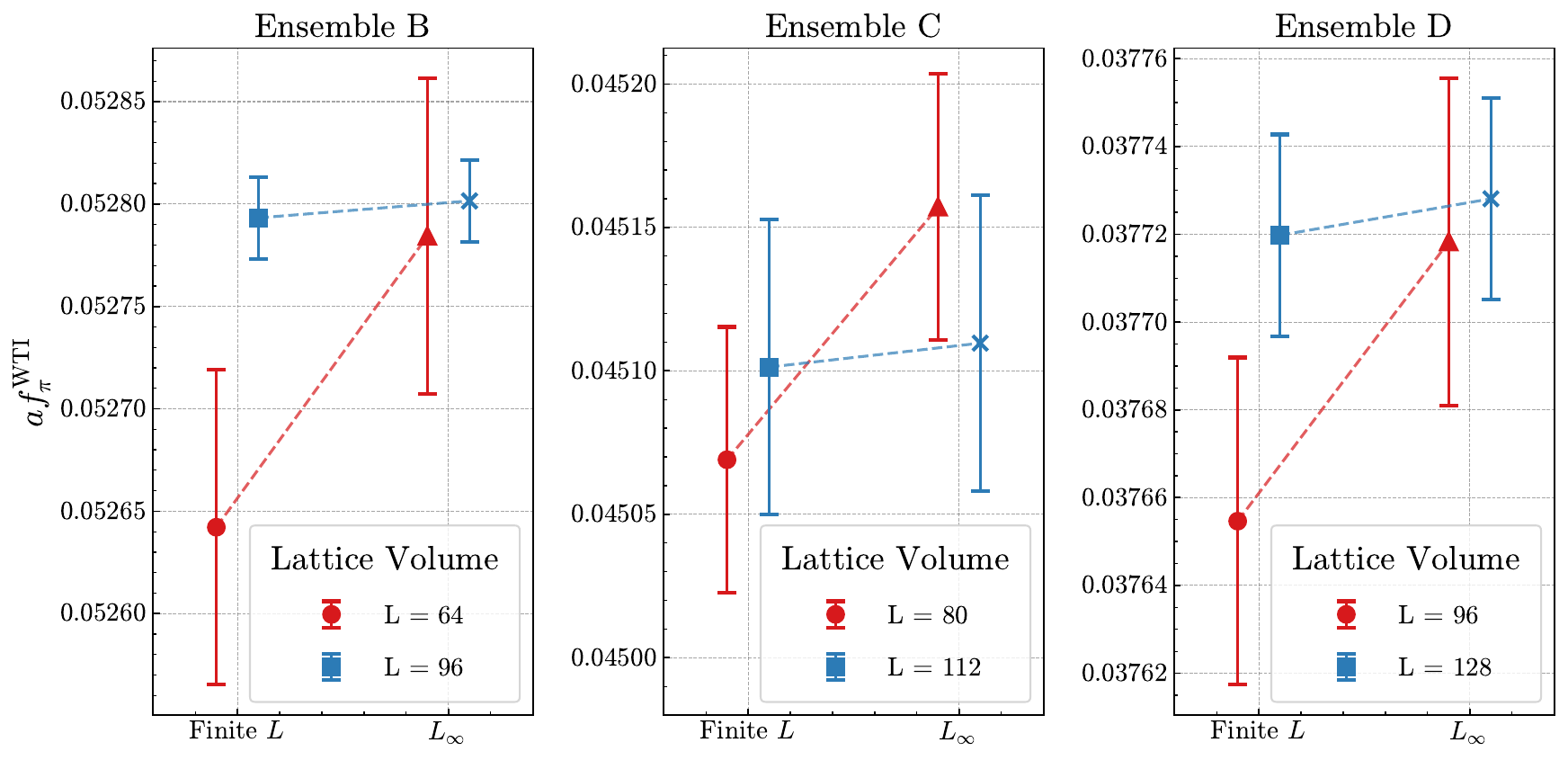}
\caption{Infinite-volume extrapolation of $f_\pi^{{\rm WTI}}$ using the CDH formulae~\cite{Colangelo:2005gd}. 
The extrapolated values show good agreement across different volumes, with the corrections becoming negligible for the largest lattices.
}
\label{fig:fpi_FVE}
\end{figure}

For each ensemble in Table~\ref{tab:iso_WP25_values} we computed $f_\pi$ using the three definitions denoted above as $f_\pi^{\rm WTI}, f_\pi^{{\rm TM}},$ and $f_\pi^{{\rm OS}}$.
The mistuning in the critical mass $m_0$ is corrected using the analytic expression derived in \cite{ExtendedTwistedMass:2021qui}  for the leading, $O(a^0)$, dependence of $f_\pi$ on the $m_{{\rm PCAC}}$ mass, namely
\begin{align}
 f_\pi^{X}(m_{\rm PCAC}=0) =f_\pi^{X}(m_{\rm PCAC})\sqrt{1+(Z_A m_{\rm PCAC}/m_\ell)^2}\,,
\end{align}
where $Z_A$ (see Table~\ref{tab:RCs}) is the renormalization constant of the axial current and $X={\rm WTI}, {\rm TM},  {\rm OS}$.
The data are then extrapolated to the infinite-volume limit using the
 ${\rm SU}(2)$ ChPT relations at NNLO + resummation, i.e., the
Colangelo-Dürr-Haefeli (CDH) formulae~\cite{Colangelo:2005gd}. Those expressions were already validated for our twisted-mass ensembles in \cite{ExtendedTwistedMass:2021qui}.
The effect of such extrapolation can be seen in Fig.~\ref{fig:fpi_FVE} where the extrapolated infinite-volume values are compatible between ensembles with different volumes. Moreover, the effect of the extrapolation is negligible for our larger
volume ensembles.

\begin{figure}[t]
\centering
\includegraphics[width=0.8\textwidth]{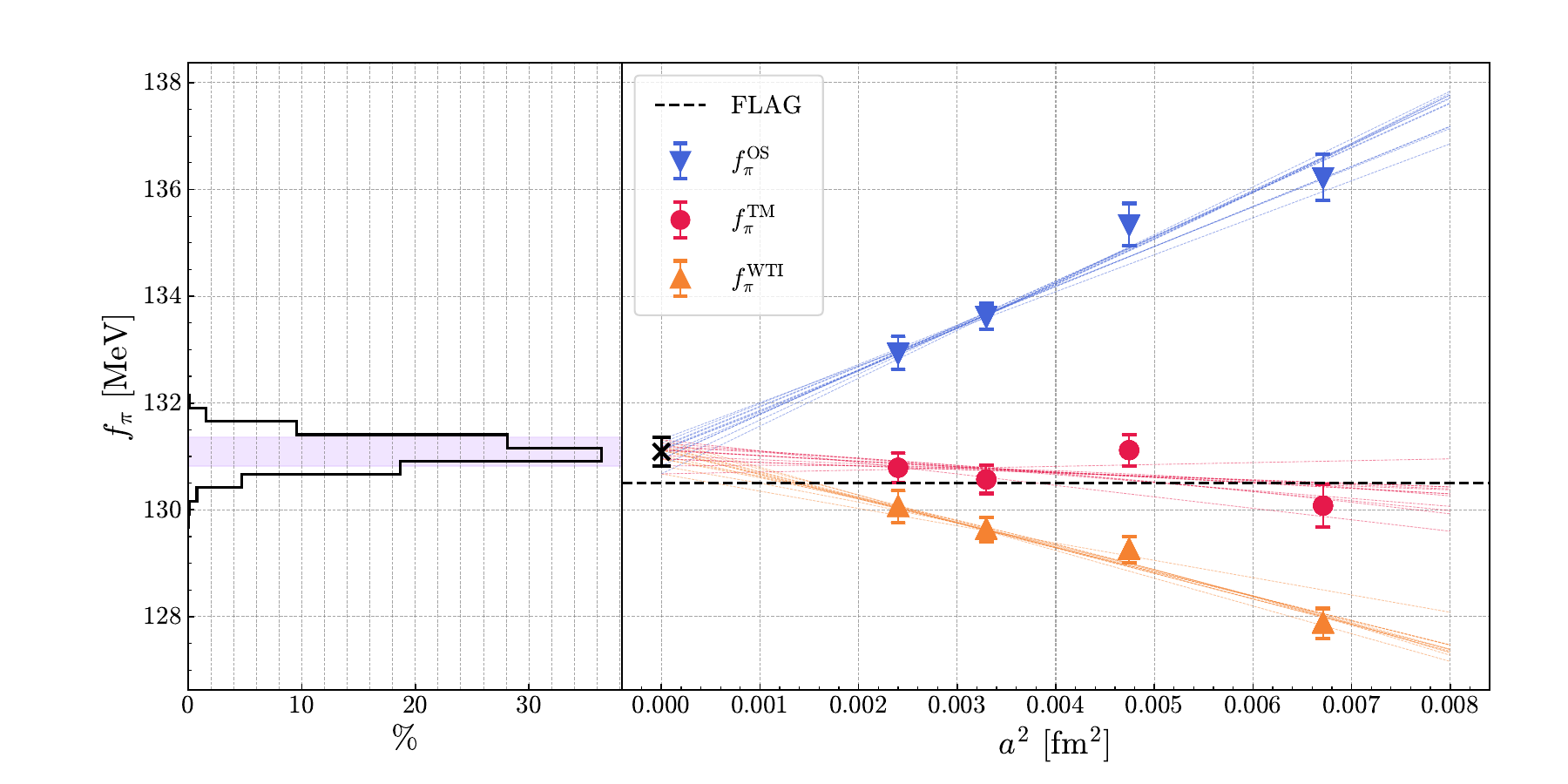}
\caption{Continuum extrapolation of $f_\pi$ in the WP25 scheme. The three different definitions of $f_\pi$ share the same continuum limit. We consider different fit ansatzes for the continuum extrapolation as described in the text. The dotted lines represent the various fits, the
    BAIC average of the continuum values is plotted as a black cross and on the left we show the histogram of the bootstrap sample for the continuum values weighted according to the BAIC. The horizontal dashed line is placed at $f_\pi=f_\pi^{\rm FLAG}$ for comparison.}
\label{fig:cont_fpi_3_reg}
\end{figure}

After the infinite-volume extrapolation we correct the mistuning in the quark masses by computing the derivative with respect to the light-quark mass in the valence and in the sea and the derivative with respect to the strange- and charm-quark masses in the sea. We then corrected the values of $f_\pi$ for the mistuning of the quark masses and the data for each ensemble are extrapolated to the continuum limit.
The three different definitions of $f_\pi$ share the same continuum limit. We consider different fit ansatzes for the continuum extrapolation:
\begin{itemize}
   \item A linear fit in $a^2$ for all three regularizations, evaluated using all lattice spacings, or by omitting either ensemble B or ensemble C;
   \item A linear fit in $a^2$ for all three regularizations,  with an additional \(a^{4}\) term included for one regularization at a time;
   \item A linear fit in $a^2$ for all three regularizations, with additional \(a^{4}\) terms included for two regularizations at a time, testing all combinations of two regularizations with the \(a^{4}\) term and one without;
   \item A linear fit in $a^2$ for all three regularizations, plus an $a^2/|\log(a^2\Lambda_{0}^2)|^{0.42}$ term for one of the regularization at the time. While the value $\gamma=0.42$ is not motivated by the considerations
underlying Eq.~\eqref{eq:I1_continuum_ansatz}, since the observable
considered here does not involve the vector current, we nevertheless retain
it as a reasonable effective choice to assess the impact of possible
logarithmically modified cutoff effects.
\end{itemize} 
The different fits are reported in Fig.~\ref{fig:cont_fpi_3_reg}. The BAIC average of the continuum results gives
\begin{equation}
    f_\pi^{\rm WP25}=131.00(24)_{\rm stat}(12)_{\rm syst}[27]~\rm{MeV} \, .
\end{equation}.

\section{Evaluation of $m_0$ and sea-quark $m_f$ derivatives for $a_\mu^{\rm HVP}$}
\label{sec:app_der_HVP}

In this Appendix, we describe the determination of the 
action parameter mistuning corrections on $a_\mu^{\rm HVP}$. We restrict attention to the quark-connected contributions to the vector correlator and $a_\mu^{\rm HVP}$, because the analogous corrections for quark-disconnected terms are negligible within statistical errors.
For each connected flavor component $V^{f,f; conn}(t)$ of the vector correlator $V(t)$ (in the notation of Eq.~(\ref{eq:Vtot})), with $f=\ell,s,c$, we consider the corrections induced by mistunings of the critical-mass parameter $m_0$ and
of the sea-quark masses $m_\ell$, $m_s$, and $m_c$.
The largest numerical corrections on $a_\mu^{\rm HVP}$ arise from the $m_0$ and $m_\ell$ mistunings in the $V^{\ell,\ell; conn}$ correlator. These corrections,
as well as the valence-$m_s$ and -$m_c$ mistuning effects,
can be evaluated directly on the reference-volume ensembles and are therefore discussed first. Then we discuss the 
sea-$m_{s,c}$ mistuning effects evaluated on smaller-volume gauge ensembles.

We recall that, as already mentioned in the footnote to Eq.~\eqref{eq:HVP_derivatives}, the mistuning
corrections to $a_\mu^{\rm HVP}$ are obtained directly from those for the vector correlator $V(t)$. Indeed the effect of the mistuning of the action parameters on the lattice scale $a$ that
enters the kernel $K(m_\mu t)$ is accounted for separately by evaluating this kernel at the corrected lattice spacing, as obtained from the renormalization procedure of 
Subsec.~\ref{sec:scale setting}

\subsection{$m_0$, $m_\ell$, valence-$m_s$ and valence-$m_c$ mistunings}
\label{secE1}

\begin{figure}
    \centering
    \includegraphics[width=0.48\linewidth]{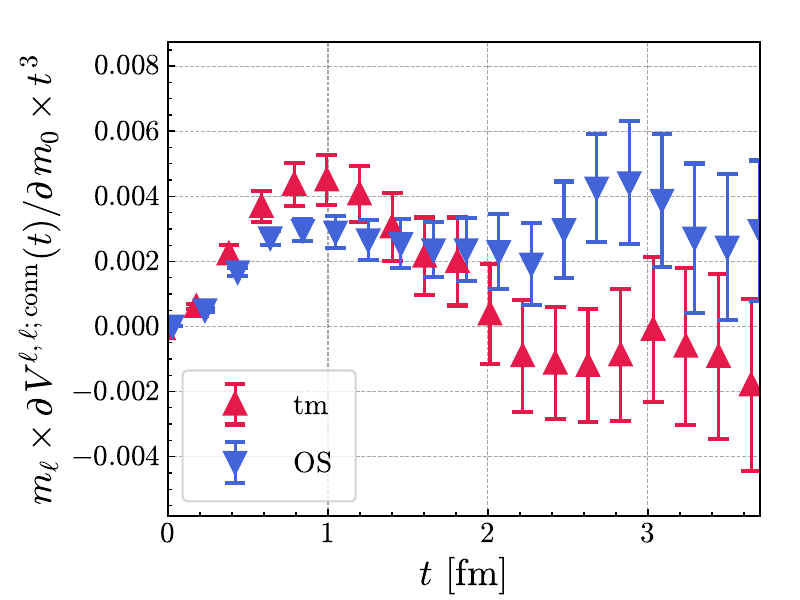}
    \includegraphics[width=0.48\linewidth]{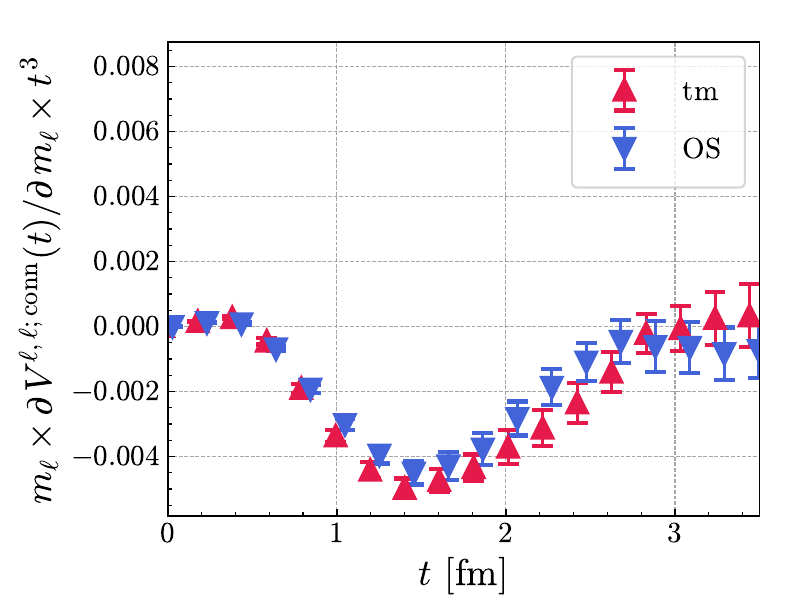}
    \caption{
    Total derivatives of the light-connected vector--vector correlator with
    respect to the critical-mass parameter $m_0$ ({\it left}) and the
    light-quark mass $m_\ell$ ({\it right}), computed on the reference-volume
    C80 ensemble for the TM and OS regularizations.
    The derivatives are shown as functions of the Euclidean time separation
    and are multiplied by $m_\ell\,t^3$ to form dimensionless quantities and
    enhance the visibility of their long-distance behavior.
    For graphical clarity, only one out of every three time slices is
    displayed, and the TM and OS data points are slightly offset along the
    horizontal axis.
    }
    \label{fig:VV_derivatives_m0_ml}
\end{figure}

For a generic flavor component of the vector correlator, the derivative with respect to the valence-$m_0$ is obtained through the pseudoscalar-density insertions schematically
represented by
\begingroup
\newcommand{\pseudoscalarinsa}{%
\raisebox{-0.40\height}{%
\includegraphics[height=2.9em]{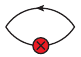}}}
\newcommand{\pseudoscalarinsb}{%
\rotatebox{180}{
\raisebox{-0.60\height}{%
\includegraphics[height=2.9em]{new_plots/fig_val_pseudoscalar.pdf}}}}
\begin{equation}
\frac{d\, V^{f,f;{\rm conn}}}{dm_0}\bigg|_{{\rm val},{\rm sim}}
=
\left\langle
\pseudoscalarinsa + \pseudoscalarinsb
\right\rangle_{\rm sim}\,,
\label{eq:m0_val_derivatives}
\end{equation}
\endgroup
where, for $f=\ell,\,s,\,c$, the solid fermion lines represent quark propagators of flavor $f$, and
the red blobs denote pseudoscalar-density insertions. The derivative of $V^{f,f;{\rm conn}} $ with respect to the
sea-$m_0$ is obtained from the connected (via gluons) correlation function of the two vector currents and the relevant sea-quark pseudoscalar loops, as illustrated in (the second line of) Eq.~\eqref{eq:sea_derivatives}.
As an example, the left panel of Fig.~\ref{fig:VV_derivatives_m0_ml} shows the sum of the valence- and sea-$m_0$ derivatives of the light-connected correlator $V^{\ell,\ell;{\rm conn}}$ on the C80 ensemble, for the TM and OS regularizations.

The derivative of $V^{f,f;{\rm conn}}$ with respect to the valence-quark mass $m_f$ is estimated by evaluating the
vector two-point function at a valence mass slightly displaced from its simulation value and building the corresponding finite difference at fixed gauge background, for $f = \ell, s, c$. A common procedure is thus employed for the light-, strange-, and charm-connected two-point correlators.

The derivatives of $V^{f,f;{\rm conn}}$ with respect to the sea-quark masses are obtained instead by using the reweighting factors as described in Eq.~\eqref{eq:finite_difference_rew}. Different gauge ensembles are used for derivatives with respect to sea-$m_\ell$ and sea-$m_{s,c}$.

The derivative of each connected flavor component with respect to the sea-$m_\ell$ are evaluated on the reference-volume ensembles, with negligible finite-volume
artifacts.\footnote{Owing to the tiny absolute values of the $m_\ell$ mistunings the statistical error on these derivatives has an immaterial impact on the $m_\ell$ mistuning correction to the correlators.} In particular, for light-connected correlator $V^{\ell,\ell;{\rm conn}}$, the derivative with
respect to the sea-$m_\ell$ is combined with that with respect to the valence-$m_\ell$ to obtain the total derivative with
respect to $m_\ell$ that is shown in the right panel of Fig.~\ref{fig:VV_derivatives_m0_ml} for the case of the C80 ensemble.

At each Euclidean time separation, all the derivatives that
are evaluated on the reference-volume ensembles are multiplied
by the corresponding mistunings, and the resulting shifts are directly added to the vector correlator before applying the
bounding procedure to determine the $a_\mu^{\rm HVP}$ contributions, as described in Subsec.~\ref{seq:analysis-and-bounding}. To illustrate the numerical relevance of these corrections, we quote the largest effects found
among the ensembles considered. The C80 ensemble exhibits the largest
$m_0$ mistuning and the largest associated correction. As shown in the left panel of
Fig.~\ref{fig:VV_derivatives_m0_ml}, in the OS regularization the derivative
remains slightly above zero at large Euclidean times, resulting in a $1.1\%$ shift of
the corresponding contribution to $a_\mu^{\rm HVP}$ relative to its
uncorrected value. The largest correction to $a_\mu^{\rm HVP}$ due to the light-quark-mass mistuning occurs instead on the D96 ensemble, where it also amounts to a $\simeq 1.1\%$ shift of the corresponding uncorrected contribution.

As already pointed out in discussing the renormalization inputs, see Appendix~\ref{sec:scale_setting_details}, computing the derivatives of $V^{f,f;{\rm conn}}$ with respect to the sea-$m_s$ and the sea-$m_c$ parameters with an accuracy
sufficient for our goals requires instead a separate procedure, which exploits dedicated small-volume ensembles and is described
below. %

\subsection{sea-$m_s$ and sea-$m_c$ mistunings}
\label{secE2}

The determination of the mistuning corrections to $a_\mu^{\rm HVP}$ associated with the strange- and charm-quark sea masses requires additional care. As noted in Appendix~\ref{sec:scale_setting_details}, the evaluation
through determinant reweighting of stochastic estimators
of the derivatives of a given correlator with respect to a sea-quark mass in general comes with a statistical noise that
increases with the lattice volume. This issue is particularly relevant for the case of the derivatives of $V^{f,f;{\rm conn}}$ with respect to sea-$m_s$ and sea-$m_c$, because
the absolute mistunings of the simulated strange and charm
mass values, although lying at the few percent level, are much larger than those of the light-quark mass. A direct use of determinant reweighting on the reference-volume
ensembles would therefore yield derivatives of the vector
correlators with large statistical uncertainties and hence  a too large error in the corresponding mistuning corrections to 
$a_\mu^{\rm HVP}$.
To get the sea-$m_s$ and sea-$m_c$ mistunings corrections under better statistical control, we make use of the auxiliary small-volume ensembles listed in Table~\ref{tab:small_volume_ens}, where, owing to the reduced volume, the statistical accuracy of the reweighting procedure is sufficient to constrain the dependence of $a_\mu^{\rm HVP}$ on $m_s^{\rm sea}$ and $m_c^{\rm sea}$. 

In evaluating the relevant derivatives $V^{f,f;{\rm conn}}$
we restrict our analysis to the TM regularization. Indeed, on the auxiliary small-volume ensembles, the OS regularization exhibits somehow larger statistical errors and a bit more pronounced dependence on the valence-light-quark mass, which makes the extrapolation of derivative estimators to the target $am_\ell$ more difficult than in the TM regularization case.
For each ensemble and several (heavier-than-physical) values of the valence-light-quark mass, we consider the dimensionless quantity
\begin{equation}
    \frac{m_\ell}{a_\mu^{\rm HVP}(f; t_{\rm cut})}
    \left.\frac{\partial a_\mu^{\rm HVP}(f; t_{\rm cut})}
         {\partial m_h}\right|_{\rm sea,sim},
    \qquad f = \ell, s, c\; , \qquad h=s,c \;,
    \label{eq:sea_mh_norm_deriv}
\end{equation}
where the quark connected $f$-flavor contributions $a_\mu^{\rm HVP}(f; t_{\rm cut})$ is obtained by introducing an upper
Euclidean-time cutoff $t_{\rm cut}$ through the same smooth 
time-modulation prescription as in Eq.~\eqref{eq:Mt_SD}, with $t_0$ replaced by $t_{\rm cut}$. Studying this quantity as a function of $t_{\rm cut}$ allows us to identify an acceptable plateau region in $t_{\rm cut}$ where the statistical error of the integrated sea-quark response is still under control. For the intermediate- and short-distance windows, an analogous analysis, using the appropriate time-modulation functions, is carried out too.

\begin{figure}
    \centering
    \includegraphics[width=\linewidth]{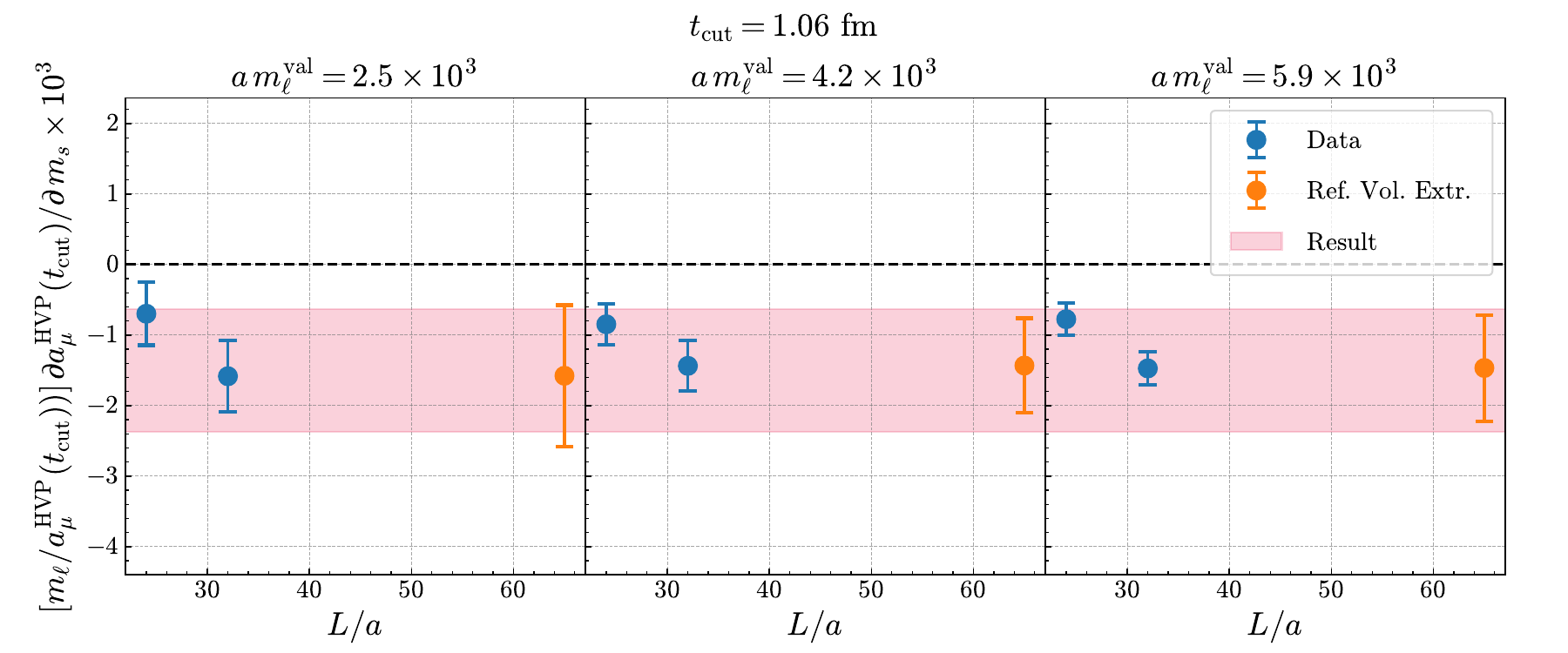}
    \caption{
    Normalized derivative of the light-connected contribution to
    $a_\mu^{\rm HVP}$ with respect to the sea strange-quark mass at $t_{\rm cut}=1.06~{\rm fm}$, 
    $  [ m_\ell / a_\mu^{\rm HVP}(\ell; 1.06~{\rm fm}) ] 
    [\partial a_\mu^{\rm HVP}(\ell; 1.06~{\rm fm}) /
         \partial m_s ] $,
    shown as a function of the spatial extent $L/a$ for
    the three valence-$am_\ell$ values indicated above the panels. Blue points denote the determinations on the B24 and B32 auxiliary ensembles, orange points the corresponding reference-volume estimates, and the pink band the resulting B-small determination.
    }
    \label{fig:Bsmall_extr}
\end{figure}

The strategy of our analysis of the sea-$m_{s,c}$ derivative estimators~(\ref{eq:sea_mh_norm_deriv}) is dictated by the 
fact that even on ensembles of not large volume (the relative
linear size $L$ being about $1.9$~fm, $2.6$~fm and $3.3$~fm
for the ensembles B24, B32 and C48, respectively) the
relative statistical error on the~derivative estimators are large, in part also due to their small mean value, and typically of order 100\%. 
Under these circumstances, the lattice artifacts on $a_\mu^{\rm HVP}(f)$, which in our setup  are below $10\%$ for $f=\ell, s$ and not larger than $\sim 35\%$ for $f=c$, can safely be neglected for the derivative estimators~(\ref{eq:sea_mh_norm_deriv}). Finite-volume effects on the
same estimators, as we show in Fig.~\ref{fig:Bsmall_extr}, have at most a marginal statistical significance at the largest valence-$m_\ell$ (where the statistical noise is smallest) if comparing data from ensembles B24 and B32 and hence, due to their exponential reduction upon increasing $L$, are negliglible within errors for the ensembles B32 and C48. Within statistical errors we also neglect the sea-$m_\ell$ dependence of the estimators~(\ref{eq:sea_mh_norm_deriv}) and extrapolate
them to the target $m_\ell$-value by considering only their valence-$m_\ell$ dependence.

We thus proceed as follows. At the coarser lattice spacing data
for the derivative estimators~(\ref{eq:sea_mh_norm_deriv})
from the ensembles B24 and B32 are available, which at all fixed $t_{\rm cut}$ may be seen either as marginally consistent with each other within statistical errors or as hinting a mild finite-volume dependence. Hence we prefer to avoid introducing a specific finite-volume ansatz (and the associated model dependence) and adopt a conservative prescription that directly incorporates the possible volume dependence into the uncertainty. For each $t_{\rm cut}$ and each valence-$m_\ell$, we take the B32 datum as the central value and enlarge its uncertainty by the pull factor determined from the difference between the B24 and B32 data. 
For the light-connected component of $a_\mu^{\rm HVP}$,
$a_\mu^{\rm HVP}(\ell; t_{\rm cut})$, the prescription above is applied independently for the three available valence-$m_\ell$ values. In the resulting derivative estimators no significant dependence on the valence $m_\ell$ is then observed and we therefore obtain the  ``B-small'' result at each $t_{\rm cut}$ through a 
correlated fit to a constant in $m_\ell$ of the data for the derivative estimators at the three valence-$m_\ell$ values. Fig.~\ref{fig:Bsmall_extr} illustrates the procedure for the
representative example of the derivative of $a_\mu^{\rm HVP}(\ell; 1.06~{\rm fm})$ with respect to $m_s^{\rm sea}$.
At the finer lattice spacing data from the C48 auxiliary ensemble are available, for which within errors again no
statistically significant dependence on the valence $m_\ell$ is observed. We therefore obtain the C48 determination 
of the derivative estimators at each $t_{\rm cut}$ from a fit to a constant in $m_\ell$ of the data at different $m_\ell$ values.

A closely analogous procedure is applied to the sea-$m_{s,c}$
derivative estimators of $a_\mu^{\rm HVP}(s; t_{\rm cut})$ and $a_\mu^{\rm HVP}(c; t_{\rm cut})$. In those cases, only one valence mass is available for each ensemble at (or very close to) the corresponding physical (isoQCD) value. Moreover, the larger values of the valence-quark masses in the vector correlator lead to statistically more precise data than in the case of the derivative estimators of $a_\mu^{\rm HVP}(\ell; t_{\rm cut})$ , thereby making the extraction of results and statistical errors rather straightforward. In the following, we thus focus our discussion on the  sea-$m_{s,c}$ derivative estimators of $a_\mu^{\rm HVP}(\ell; t_{\rm cut})$ , which represents the most delicate case.

\begin{figure}
    \centering
    \includegraphics[width=0.48\linewidth]{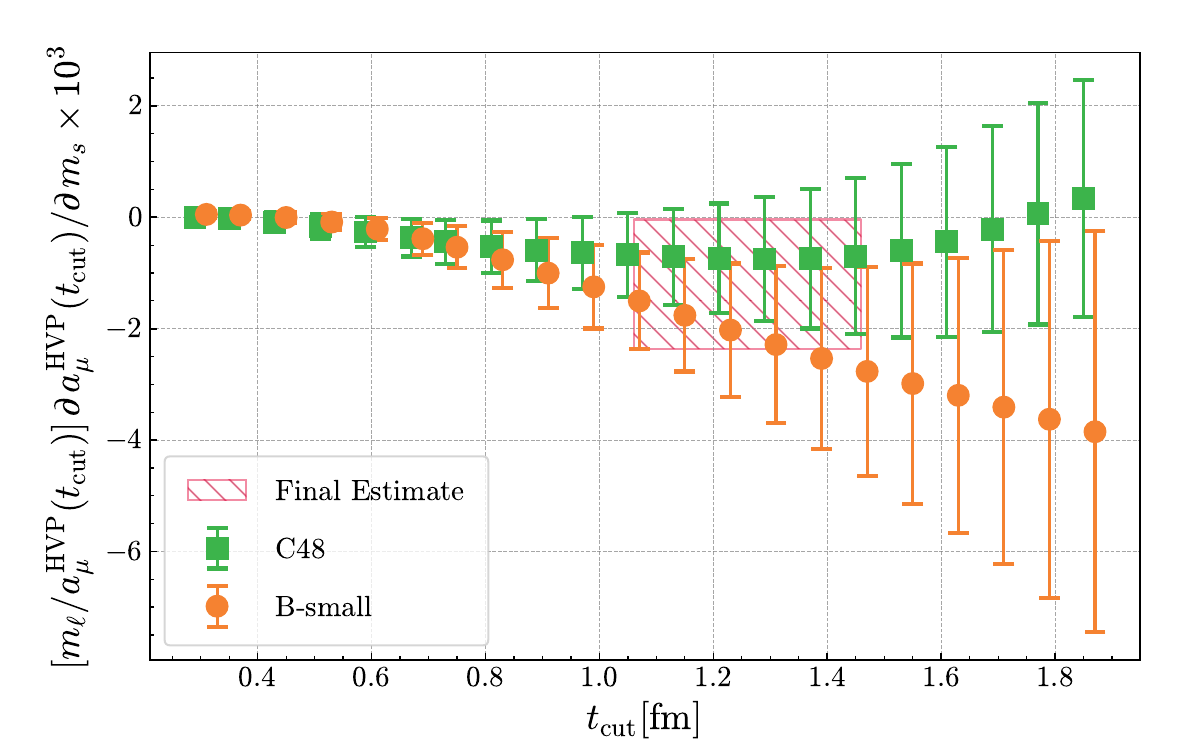}
    \includegraphics[width=0.48\linewidth]{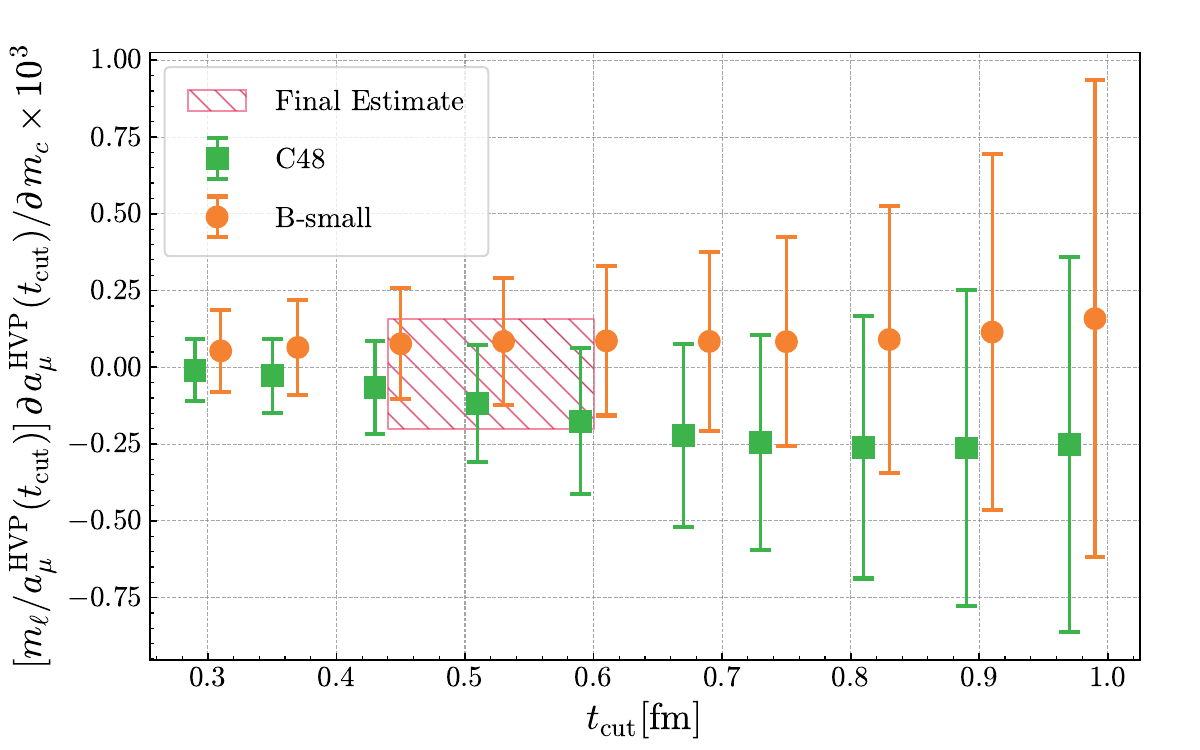}
    \caption{
    Estimators of the normalized derivative of the light-connected $a_\mu^{\rm HVP}$ contribution with respect to the sea-$m_s$, $  [ m_\ell / a_\mu^{\rm HVP}(\ell) ] 
    [\partial a_\mu^{\rm HVP}(\ell) / \partial m_s ] $
    ({\it left}), and to the sea-$m_c$,
    $  [ m_\ell / a_\mu^{\rm HVP}(\ell) ] 
    [\partial a_\mu^{\rm HVP}(\ell) / \partial m_c ] $
    ({\it right}), shown as functions
    of the time cutoff $t_{\rm cut}$. Data points are offset on the horizontal
    axis for readability. The hatched bands indicate the corresponding final estimates adopted in our analysis. The label ``B-small'' in the legend refers to the results obtained from the ensemble B32 and B24 via the procedure described in this appendix (see paragraph about the Fig.~\ref{fig:Bsmall_extr}).
    }
    \label{fig:sea-mh_derivative_small}
\end{figure}

The final estimates of the normalized sea-$m_h$ quark-mass derivative of $a_\mu^{\rm HVP}(f)$, 
\begin{equation}
    \frac{m_\ell}{a_\mu^{\rm HVP}(f)}
    \left.\frac{\partial a_\mu^{\rm HVP}(f)}
         {\partial m_h}\right|_{\rm sea,sim},
    \qquad f = \ell, s, c\;, \qquad h=s,c \; ,
    \label{eq:sea_norm_deriv2}
\end{equation}    
are obtained by studying the $t_{\rm cut}$ dependence of the resulting B-small and C48 determinations. For the reason explained above, in Fig.~\ref{fig:sea-mh_derivative_small} we illustrate the case of $f= \ell$, which also corresponds to the sea-$m_{s,c}$
mistuning correction for the numerically dominating (quark-connected) flavor component of $a_\mu^{\rm HVP}$. In Fig.~\ref{fig:sea-mh_derivative_small} we show the normalized sea-$m_s$ derivative in the left panel and the sea-$m_c$ derivative in the right panel.

For the normalized sea-$m_s$ derivative the C48 determination reaches a stable plateau at approximately $t_{\rm cut}\simeq1~{\rm fm}$, while the ``B-small'' result does not develop an equally clear plateau, rather its statistical uncertainty starts to increase rapidly for $t_{\rm cut}\gtrsim1.4~{\rm fm}$, suggesting a poor statistical resolution of the large time separation tail.
We therefore get our estimate of the sea-$m_s$ derivative by performing separate correlated fits to constant in $t_{\rm cut}$ of the ``B-small'' and C48 data in the range
\begin{equation}
    1.1~{\rm fm} \lesssim t_{\rm cut} \lesssim 1.5~{\rm fm}\,
    \label{eq:tc-ra-s}.
\end{equation}
The final central value is obtained from the average of the two plateau determinations, with the uncertainty enlarged by the pull factor associated with their difference. This prescription conservatively accounts for the residual difference between the
two determinations with no attempt to disentangle possible
small lattice and finite-volume artifacts from sizable statistical errors. Moreover, the lower end of the interval~(\ref{eq:tc-ra-s}) is sufficiently large to account for the on-shell effects of sea-strange quark pairs ($K \bar{K}$ meson states and their excitations) as well as for a large part
of the virtual sea-strange effects up to $t_{\rm cut}$-values where
the statistical error overwhelms any signal.

An analogous, somehow simpler situation is found for the normalized sea-$m_c$ derivative. As shown in the right panel of Fig.~\ref{fig:sea-mh_derivative_small}, no statistically
significant signal is resolved already at short Euclidean times. Extending the integration to larger $t_{\rm cut}$ would therefore predominantly accumulate statistical noise rather than additional signal. We consequently extract the ``B-small'' and C48 plateaus in the interval
\begin{equation}
    0.4~{\rm fm} \lesssim t_{\rm cut} \lesssim 0.6~{\rm fm}\,,
\label{eq:tc-ra-c}
\end{equation}
and combine them following the same conservative prescription adopted for the $m_s^{\rm sea}$ derivative. We note that the lower end of this interval is already more than one order of magnitude larger than the characteristic Euclidean-time scale associated with the on-shell sea-charm contribution, which 
are given by $D \bar{D}$ meson states, and is hence certainly
large enough to account within quoted errors also for off-shell sea-charm effects~\footnote{Here we follow the arguments on which the Mainz/CLS analysis~\cite{Ce:2022kxy} of the effects of sea-charm quenching on $a_\mu^{\rm HVP}$ is based, but we use them only as a cross-check of our data-driven choice of the $t_{\rm cut}$-range~(\ref{eq:tc-ra-c}) in the estimate of a few-percent mistuning of the sea-$m_c$, i.e.\ a  systematic effect different from (and expectedly much smaller than) complete sea-charm quenching.} 
while avoiding the unnecessary accumulation of long-distance statistical noise.

At this stage we find it useful to specify the amount of the resulting sea-$m_{s}$ and sea-$m_c$ mistuning corrections in the cases where the underlying sea-quark-mass mistunings are largest. For the $I=1$ component of $a_\mu^{\rm HVP}$, the largest sea-$m_s$ mistuning with respect to the FLAG scheme
occurs on the C80 ensemble, where the simulated
$m_s^{\rm sim}$ mass differs from its
target value by approximately $3.9\%$. With the procedure described above, the corresponding correction that we apply reads
\begin{equation}
\Delta \big|_{m_s^{\rm sea}}^{\rm mist} a_\mu^{\rm HVP}(I=1)
= 0.5(5)\times 10^{-10}.
\label{eq:spec1}
\end{equation}
The largest charm-quark mistuning is instead found on D96 and is approximately
$2.3\%$. In this case the correction is
\begin{equation}
\Delta \big|_{m_c^{\rm sea}}^{\rm mist} a_\mu^{\rm HVP}(I=1)
= 0.0(1)\times 10^{-10}.
\label{eq:spec2}
\end{equation}
The results in Eqs.~(\ref{eq:spec1})--(\ref{eq:spec2})
illustrate well the conservative nature of our analysis aimed at estimating, or at least constraining, the sea-$m_{s,c}$ mistuning corrections. In line with this feature, we also 
remark that the uncertainty in the shift~(\ref{eq:spec2}),
which corrects for a 2.3\% mistuning in the sea-$m_c$ is only about six times smaller than the O(0.1\%) effect on $a_\mu^{\rm HVP}(I=1)$ that is estimated by the Mainz/CLS group in Ref.~\cite{Ce:2022kxy} for the complete sea-charm quark quenching.

\bibliography{biblio,SM_bib}

\end{document}